\documentclass[A4paper,11pt]{article}

\usepackage{latexsym}
\usepackage{epsfig}
\usepackage{verbatim}
\usepackage{cite}
\usepackage{amssymb}
\usepackage{amsmath}
\usepackage{graphicx}

\newcommand{\ignore}[1]{}
\newtheorem{theorem}{\sc Theorem}[section]
\newtheorem{lemma}[theorem]{\sc Lemma}
\newtheorem{corollary}[theorem]{\sc Corollary}

\newcommand{\eps}{\varepsilon}

\newcommand{\var}{\hbox{\bf var}}

\newcommand{\proofend}{{\medskip\medskip}}
\newcommand{\proof}{{\noindent\em Proof. }}

\newcommand{\defeq}{\mbox{\,$\stackrel{\rm def}{=}$\,}}

\author{
  {\sc Bernard Chazelle}
\thanks{Department of Computer Science,
       Princeton University, 
{\tt chazelle}@{\tt cs.princeton.edu }}
}

\title{The Convergence of Bird Flocking
\thanks{A preliminary version of this work
appeared in {\em Proceedings of ACM-SIAM 
Symposium on Discrete Algorithms
(SODA09)}, January 4--6, 2009, 422--431.
This work was supported in part by NSF grant CCF-0634958
and NSF CCF-0832797.
Categories and Subject Descriptors: F.2.0 {\bf [Analysis of Algorithms and Problem Complexity]}: General.
General Terms: Algorithms, Theory.
Additional Key Words and Phrases: Natural 
Algorithms, Dynamical Systems.  }}

\date{}

\begin{document} \maketitle

\begin{abstract}
We bound the time it takes for a group of birds to reach steady state
in a standard flocking model. 
We prove that (i) within single exponential time
fragmentation ceases and each bird settles on a fixed flying direction;
(ii) the flocking network converges only after a number of steps
that is an iterated exponential of height logarithmic 
in the number of birds.
We also prove the highly surprising result that this bound is optimal. 
The model directs the birds to adjust their
velocities repeatedly by averaging them with their neighbors
within a fixed radius. The model is deterministic,
but we show that it can tolerate a reasonable amount of
stochastic or even adversarial noise. Our methods
are highly general and we speculate that the results
extend to a wider class of models based on undirected flocking
networks, whether defined metrically or topologically.
This work introduces new techniques of broader interest,
including the {\em flight net}, the
{\em iterated spectral shift},
and a certain {\em residue-clearing} argument in circuit complexity.
\end{abstract}

\vspace{1cm}

\section{Introduction}\label{introduction}

What do migrating geese, flocking cranes, 
bait balls of fish, prey-predator systems, and 
synchronously flashing fireflies have in common?
All of them are instances of {\em natural algorithms},
ie, algorithms designed by evolution over
millions of years. By and large, their study has been
the purview of dynamical systems theory within
the fields of zoology, ecology, evolutionary biology, etc.
The main purpose of this work is to show
how combinatorial and algorithmic tools
from computer science
might be of benefit to the study of natural algorithms---in particular,
in the context of collective animal behavior~\cite{ParrishH}.
We consider a classical open question in bird flocking:
bounding the convergence time in a standard neighbor-based model.
We give a tight bound on the number of discrete
steps required for a group 
of $n$ birds to reach steady state.
We prove that, within time exponential in $n$,
fragmentation ceases and each bird settles on a fixed
flying direction. We also show that 
the flocking network
converges after a number of steps
that never exceeds an iterated
exponential of height logarithmic in $n$.
Furthermore, we show that this exotic bound
is in fact optimal. If we view the set of birds
as a distributed computing system, our work establishes
a tight bound on the maximum execution time. Holding
for a large family of flocking mechanisms, it should be thought of
as a {\em busy beaver} type result---or perhaps {\em busy goose}.

The bound is obtained by 
investigating an intriguing ``spectral shift" process, which
could be of independent interest.
In the model, birds forever adjust their
velocities at discrete time steps
by averaging them with their neighbors flying
within a fixed distance. The model is deterministic
but we show that it tolerates a reasonable amount of 
stochastic or even adversarial noise. While, for concreteness,
we settle on a specific geometric model, our methods are quite general and we suspect
the results can be extended to a large class of flocking models,
including topological networks~\cite{balleriniCCCCG}.
The only serious limitation is that the
flocking network must be undirected: this rules out models
where one bird can process information from 
another one while flying in its ``blind spot."

Bird flocking has received considerable attention
in the scientific and engineering literature, including
the now-classical {\em Boids} model of 
Reynolds~\cite{reynolds87,tannerJPI,tannerJPII,tannerJPswitch}.
Close scrutiny has been given to leaderless models
where birds update their velocities
by averaging them out over their nearest neighbors.
Two other rules are often added: one to prevent birds from
colliding; the other to keep them together. Velocity
averaging is the most general and fundamental rule
and, understandably, has received the most attention.
Computer simulations support the 
intuitive belief that, by repeated averaging,
each bird should eventually converge to a
fixed speed and heading.
This has been proven theoretically, but 
how long it takes for the system to converge had
remained an open problem.
The existential question (does the system converge?)
has been settled in many different ways, and it
is useful to review the history briefly.

A ``recurrent connectivity'' assumption stipulates
that, over any time interval of a fixed length,
every pair of birds should be able to communicate
with each other, directly or indirectly via other birds.
Jadbabaie, Lin, and Morse~\cite{jadbabaieLM03}
proved the first of several convergence 
results under that assumption 
(or related ones~\cite{MostaghJD,olfatisaber06,shiWC,tannerJPII}).
Several authors extended these results to variable-length 
intervals~\cite{HendrickxB,liwang2004,Moreau2005}.
They established that the bird group always ends up
as a collection of separate flocks (perhaps only one),
each one converging
toward its own speed and heading.
Some authors have shown how to do away with 
the recurrent connectivity assumption
by changing the model suitably.
Tahbaz-Salehi and Jadbabaie~\cite{tahbaz-jadbabaie}, for example, 
assume that the birds fly on the surface of a torus.
Cucker and Smale~\cite{CuckerSmale1}
use a broadcast model that extends
a bird's influence to the entire
group while scaling it down as a function of distance.
In a similar vein, Ji and Egerstedt~\cite{jiE07} introduce
a hysteresis rule to ensure that connectivity increases over time.
Tang and Guo~\cite{tangG07}
prove convergence in
a high-density probabilistic model.
Recent work~\cite{balleriniCCCCG} suggests a ``topological" rule for
linking birds: a bird is influenced by a fixed number
of its neighbors instead of all neighbors within
a fixed distance. Whether the criteria are metric
or topological, the bulk
of work on leaderless flocking has assumed
neighbor-based consensus rules. We are not aware of 
any bounds on the convergence time.

Our model is a variant of the one
proposed by Cucker and Smale~\cite{CuckerSmale1},
which is itself a holonomic variant
of the classical Vicsek model~\cite{vicsekCBCS95}.
Given $n$ birds
${\mathcal B}_1, \ldots, {\mathcal B}_n$,
represented at time $t$ 
by points $x_1(t),\ldots, x_n(t)$ in $E^3$,
the {\em flocking network} $G_t$
has a vertex for each bird and an edge between
any two of them within distance 1 of each other.
By convention, $G_t$ has no self-loops.
The connected components of $G_t$ are 
the {\em flocks} of the system.
If $d_i(t)$ denotes the number of birds adjacent to 
${\mathcal B}_i$ at time $t$, the total number of birds
within the closed unit disk centered
at ${\mathcal B}_i$ is precisely $d_i(t)+1$.

\vspace{1cm}
\begin{figure}[htb]\label{fig-flock}
\begin{center}
\includegraphics[width=8cm]{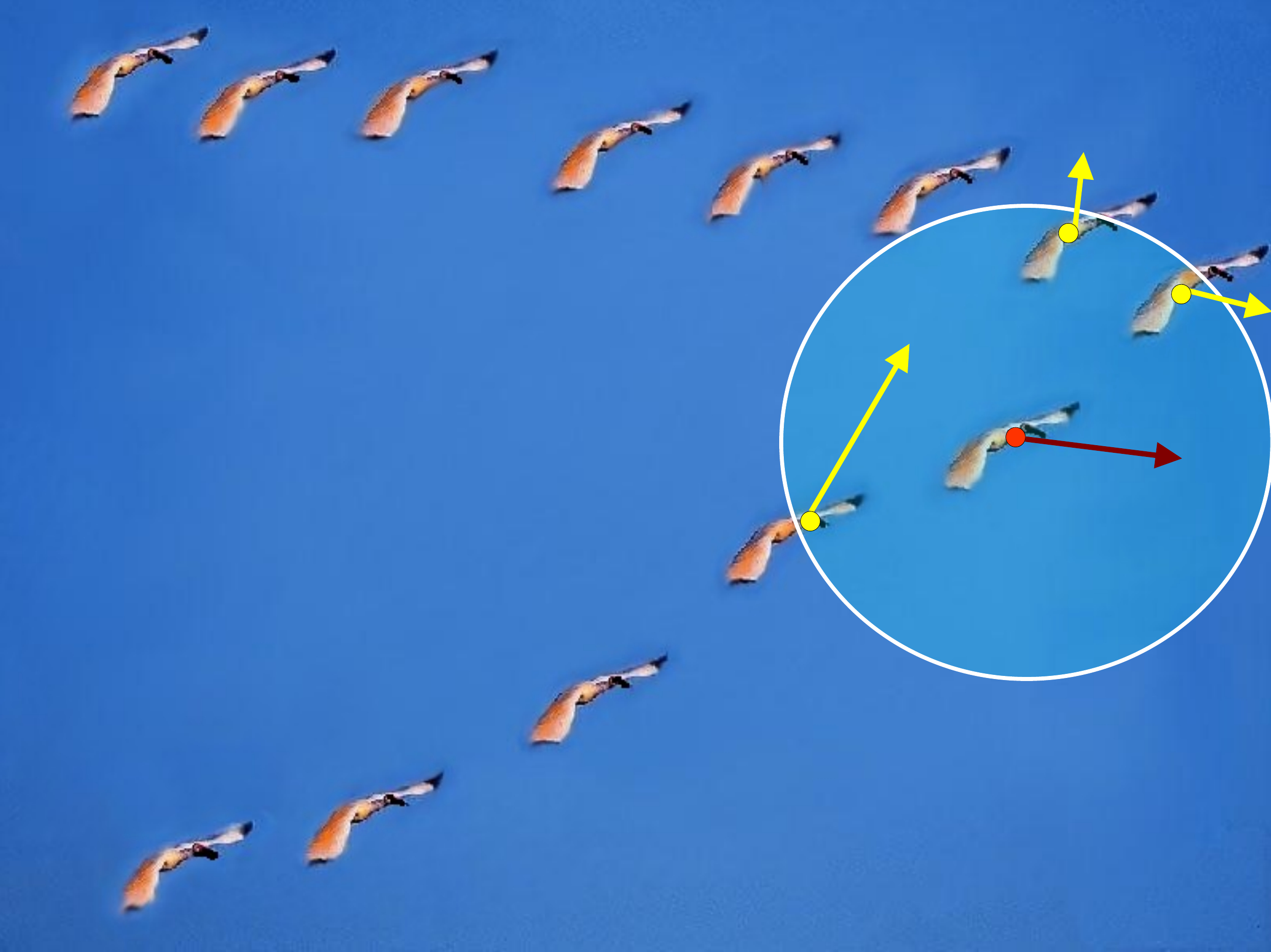}
\end{center}
\caption{\small Each bird updates its velocity 
by averaging it with those of its neighbors
within a unit-radius circle.}
\end{figure}

\paragraph{The Model.}

The input consists of the 
initial position $x(0)$ and velocity $v(1)$.
Both vectors belong to 
${\mathbb R}^{dn}$, for any fixed $d\geq 1$.
For $t\geq 1$ and $1\leq i\leq n$,
$$x_i(t)= x_i(t-1)+ v_i(t),$$
where\footnote{\, We denote 
the coordinates of a vector $x(t)$ 
by $x_i(t)$ and 
the elements of a matrix $X(t)$ (resp. $X_t$)
by $x_{ij}(t)$ (resp. $(X_t)_{ij}$).}
\begin{equation*}
v_i(t+1) - v_i(t) = c_i(t) \sum_{(i,j)\in G_t} (v_j(t) -v_i(t)).
\end{equation*}
The {\em self-confidence} coefficients $c_i(t)$, so named
because they tell us how much a bird is influenced by its
neighbors, are normalized so that $0< c_i(t)d_i(t)< 1$.
(See ``Discussion'' section below for an intriguing interpretation
of these constraints.)
We assume that $c_i(t)$ may vary only
when $G_t$ does; in other words, while
all neighborly relations remain the same,
so do the self-confidence coefficients.
A natural choice of coefficients is the one
used in the classical Vicsek model~\cite{vicsekCBCS95}:
$c_i(t)= (d_i(t)+1)^{-1}$, but we do not make this
restrictive assumption here.

The model captures the simple intuition that, in an effort
to reach consensus by local means, 
each bird should adjust its velocity at each step
so as to be a weighted average of those of its neighbors.
A mechanical interpretation sees in the 
difference
$v_i(t+1) - v_i(t)$ the discrete analogue
of the bird's acceleration, so that,
by Newton's Law, $F=ma$, a bird is
subject to a force that grows in proportion
to the differences with its neighbors.
A more useful take on the model is
to view it as a diffusion process: more precisely, as
the discrete version of the heat equation 
$$\frac{\partial v}{\partial t} = - C_t L_t v,$$
where the Laplacian $L_t$ of the flocking network
$G_t$ is defined by:
\begin{equation*}
(L_t)_{ij} = 
\begin{cases}
d_i(t) & \text{if $i=j$}; \\
\, -1  & \text{if $(i,j)\in G_t$}; \\
\,\,\,\,\, 0  & \text{else}.
\end{cases}
\end{equation*}
and $C_t= \text{diag}\, c(t)$ is 
the self-confidence matrix.
Thus we express the dynamics of the system as
$$
v(t+1)-v(t) = - C_t L_t v(t) \, .
$$
This is correct in one dimension. To deal with 
birds in $d$-space, we use a standard tensor lift.
Here is how we do it. We form the velocity vector $v(t)$ by 
stacking $v_1(t),\ldots, v_n(t)$ together
into one big column vector of dimension $dn$.
Given a matrix $A$, the product\footnote{\, The
Kronecker $A \otimes B$,product of two matrices $A$ and $B$ is the
matrix we get if we replace each $a_{ij}$ by the block $a_{ij}B$.
Formally, if $A$ is $m$-by-$n$ and $B$ is 
$p$-by-$q$, then the product
$A\otimes B$ is the $mp$-by-$nq$ matrix $C$ such that 
$c_{ip+j, kq+l}= a_{i,k}b_{j,l}$.
We will often use, with no further mention, the 
tensor identity $(A \otimes B)(C\otimes D) = AC\otimes BD$.}
$(A\otimes I_d)v(t)$
interlaces into one vector the $d$ vectors
obtained by multiplying $A$ by the vector
formed by the $k$-th coordinate
of each $v_i(t)$, for $k=1,\ldots, d$.
The heat equation would now be written as 
$$v(t+1) = (P(t) \otimes I_d)v(t)\, .$$
where $P(t)= I_n- C_t L_t$.
One can check directly that 
the {\em transition matrix} $P(t)$
is row-stochastic.
In the case of a 3-node path, for example, 
$P(t)$ has the form:
\begin{equation*}
\begin{pmatrix}
1-c_1(t) & c_1(t) & 0 \\
c_2(t) & 1-2 c_2(t) & c_2(t) \\
0 & c_3(t) & 1-c_3(t) 
\end{pmatrix}
.
\end{equation*}

\vspace{2cm}
\begin{figure}[htb]\label{fig-3nodePath}
\hspace{1.5cm}
\includegraphics[width=10cm]{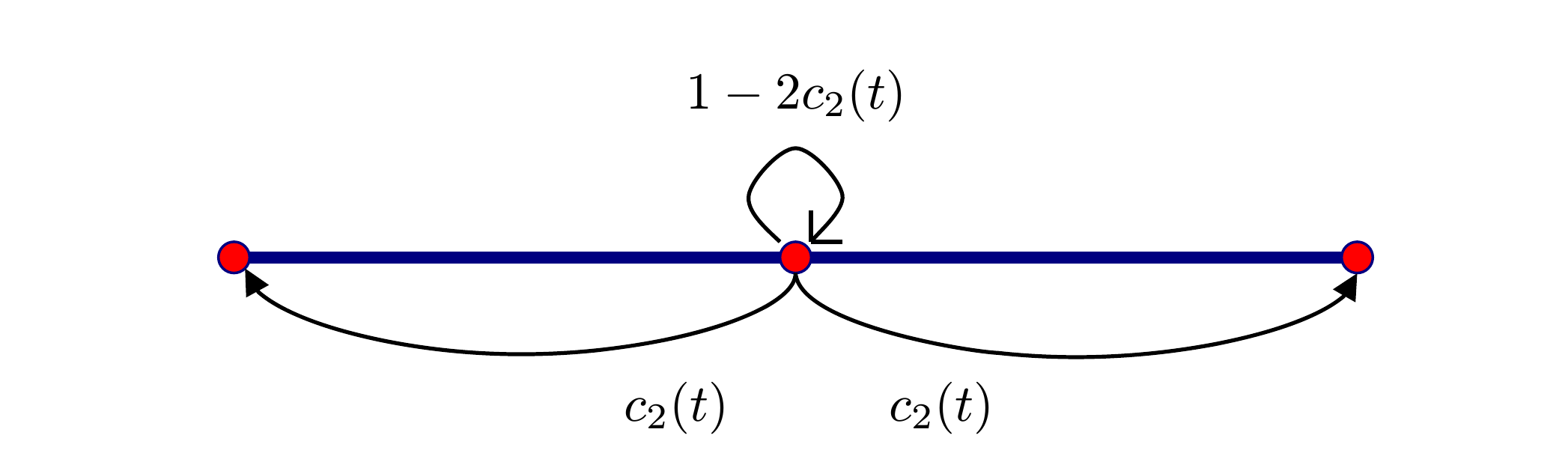}
\begin{quote}
\vspace{0cm}
\caption{\small 
A 3-node flock with the transitions of the middle node
indicated by curved arrows.}
\end{quote}
\end{figure}

The dynamics of flocking is captured by the
two equations of motion:
For any $t\geq 1$,
\begin{equation}\label{modelD}
\begin{cases}
\, x(t)= x(t-1)+ v(t); \\
\, v(t+1)= (P(t)\otimes I_d) v(t).
\end{cases}
\end{equation}

For tie-breaking purposes, we
inject a tiny amount of hysteresis into the system.
As we discuss below, this is necessary for convergence.
Intuitively, the rule prevents edges of
the flocking network from breaking because
of microscopic changes.
Formally, an edge $(i,j)$ of 
$G_t$ remains in $G_{t+1}$ 
if the distance between ${\mathcal B}_i$ 
and ${\mathcal B}_j$ changes by less than
$\varepsilon_{\! h}>0$
between times $t$ and $t+1$. 
We choose $\varepsilon_{\! h}$ to be exponentially small
for illustrative purposes only; in fact,
virtually {\em any} hysteresis rule would work.

\paragraph{The Results.}

To express our main result, we need to define
the fourth level of the Ackermann hierarchy, the
so-called ``tower-of-twos'' function:
$2\uparrow\uparrow 1 =2$ and, for $n>1$,
$2\uparrow\uparrow n= 2^{2\uparrow\uparrow (n-1)}$.
The bird group is said to have reached {\em steady state}
when its flocking network no longer changes.
All the results below hold in any fixed dimension $d\geq 1$.

\begin{itemize}
\item
{\em A group of $n$ birds reaches steady state 
in fewer than $2\uparrow\uparrow (4\log n)$ steps.
The maximum number of switches
in the flocking network of $n$ birds
is at most $n^{O(n^3)}$.
The limit configuration of each bird ${\mathcal B}_i$ 
is of the form $a_i+b_it$, where $a_i,b_i$ are 
$d$-dimensional rational vectors.
After the fragmentation breakpoint $t_f= n^{O(n^3)}$,
network edges can only appear and never vanish.
}
\item
{\em There exists an initial configuration of $n$ birds that
requires more than $2\uparrow\uparrow \log \frac{n}{2}$
steps before reaching steady state. 
The lower bound holds both with and without hysteresis.
}
\end{itemize}

Past the fragmentation breakpoint, 
the direction of each bird is essentially fixed,
so $n^{O(n^3)}$ is effectively the bound for {\em physical} convergence.
(Of course, damped local oscillations typically go on forever.)
{\em Combinatorial} convergence is another matter altogether.
It might take an extraordinarily long time before the network
stops switching. The tower-of-twos' true height
is actually less than $4\log n$, ie, a little better 
than stated above: specifically, the factor $4$ 
can be replaced by $(\log x_0)^{-1}$, where
$x_0$ is the unique real root of $x^5 - x^2 - 1$,
which is about $3.912$.

\begin{figure}[htb]\label{fig-timeline}
\vspace{0.5cm}
\hspace{1.8cm}
\includegraphics[width=10cm]{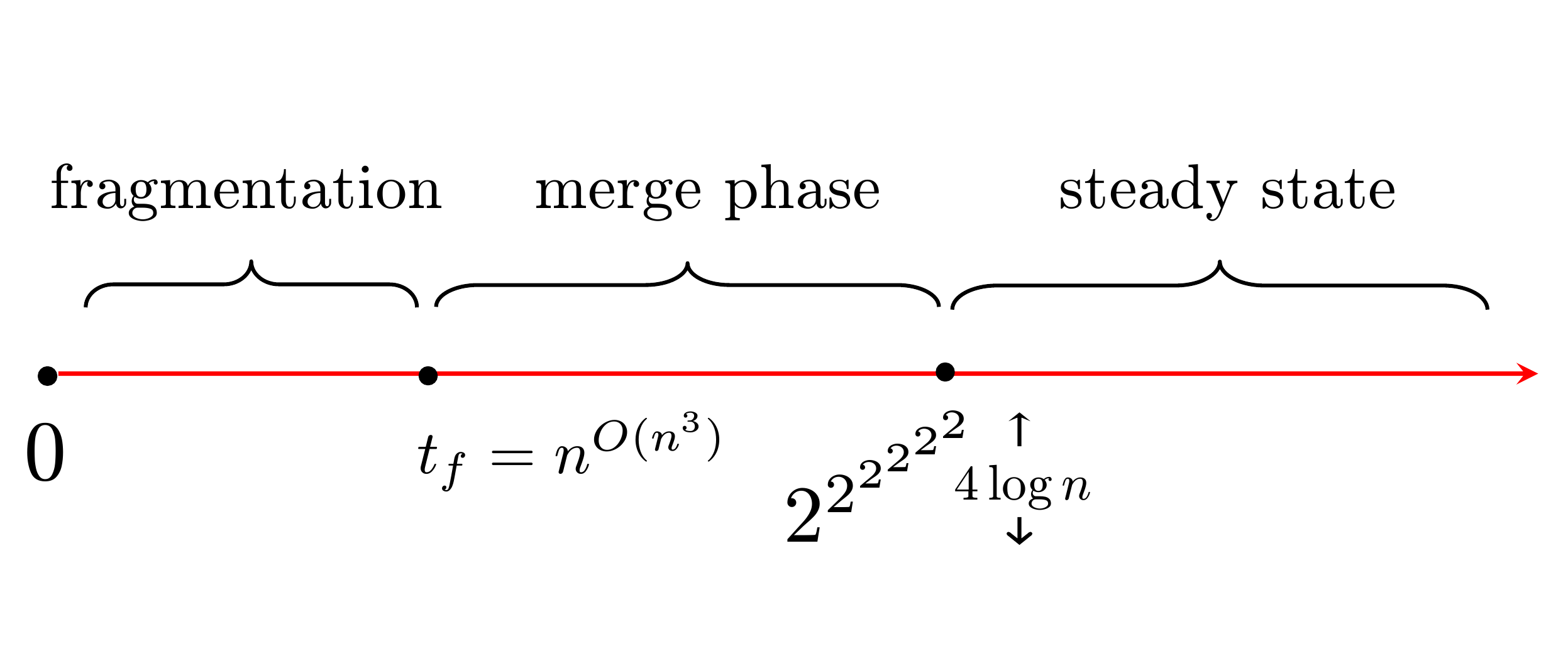}
\begin{quote}
\vspace{0cm}
\caption{\small 
Flocks cease to lose edges
after the fragmentation breakpoint $t_f$
and can only gain new ones. The network
reaches steady state after a tower-of-twos of 
height logarithmic in the number of birds.}
\end{quote}
\end{figure}

\bigskip
\noindent
$\bullet$ {\em How many bits?} \ 
The self-confidence matrices $C_t$ 
are rational with $O(\log n)$ bits per entry.
The bound on the maximum number of network switches 
holds even if the inputs are arbitrary real numbers.
Obviously, there is no hope of bounding 
the convergence time if two birds can be initialized
to fly almost parallel to each other; therefore bounding
the representation size of the input is necessary.
The initial position and velocity of each bird
are encoded as rationals over $\mathfrak{p}$ bits. 
Our results hold for virtually any value of $\mathfrak{p}$.
The dependency on $\mathfrak{p}$ begins to show only
for $\mathfrak{p}\geq n^3$, so this is what we shall assume
when proving the upper bound on the 
convergence time. Keep in mind 
that $\mathfrak{p}$ is only an upper bound
and the actual number of bits does not need to be this long.
In fact, the lower bound requires only
$\log n$ bits per bird.
All computation is exact.
The upper bound\footnote{\, Logarithms to the 
base 2 are written as $\log$ while the natural 
variety is denoted by $\ln$.
For convenience we assume throughout this paper
that $n$, the number of birds,
is large enough. To handle small bird groups, of course,
we can always add fictitious birds that never
interact with anyone.}
of $2\uparrow\uparrow (4\log n)$ 
is extremely robust, and holds for essentially
any conceivable input bit-size
and hysteresis rule.

\bigskip\medskip
\noindent
$\bullet$ {\em Is the lower bound pathological?} \ 
Suprisingly, the answer is no. 
As we mentioned, initial conditions require only
$\mathfrak{p}= O(\log n)$ bits per bird.
Our construction ensures that the hysteresis
rule never kicks in, so the lower bound 
holds whether the model includes hysteresis or not.
The flocks used for the construction
are single paths, and the matrix $P(t)$
corresponds to a lazy random walk with probability
$\frac{1}{3}$ of staying in place.
The lower bound holds in any dimension $d>0$.
Here are the initial positions and velocities for $d=1$:
\begin{equation*}
\begin{cases}
\,
x(0) \, = \Bigl(0,\hbox{$\frac{2}{3}$}, 
2,\hbox{$\frac{8}{3}$}, 
\ldots, 2l, 2l + \hbox{$\frac{2}{3}$}, \ldots,
n-2, n-\hbox{$\frac{4}{3}$}\Bigr)^T;  \\
\\
\, v(1)= 
\Bigl(\, \underset{n}{\underbrace{     
n^{-11}, 0, -n^{-11}, 0, n^{-11}, 0, \ldots, 
n^{-11}, 0, -n^{-11}, 0}}  \, \Bigr)^T.
\end{cases}
\end{equation*}
Flocking obeys two symmetries: one translational;
the other kinetic (or ``relativistic,'' as a 
physicist might say).
The absolute positioning
of the birds is irrelevant and 
adding a fixed vector to each bird's velocity 
has no effect on flocking. In other words, one cannot
infer velocity from observing the evolution of the flocks.
Indeed, only differences between velocities are meaningful.
This invariance under translation in velocity space
implies that slow convergence cannnot be caused by
forcing birds to slow down. In fact, one can trivially
ensure that no bird speed falls below any desired threshold.
The lower bound relies on creating
small angles, not low speeds. (Thus, in particular,
the issue of stalling does not arise.)
To simplify the lower bound proof, 
we allow a small amount of noise into the system.
Within the next $n^{O(1)}$ steps following
any network switch, the velocity 
of an $m$-bird flock may be multiplied
by $I_m\otimes \widehat{\alpha}$, 
where $\widehat{\alpha}$ is the diagonal matrix
with $\alpha=(\alpha_1,\ldots,\alpha_d)$ along
the diagonal and 
rational $|\alpha_i|\leq 1$ encoded over 
$O(\log n)$-bits. The noise-free case corresponds to 
$\alpha_i=1$.
The perturbed velocity at time $t$ 
should not differ from the original one by more
than $\delta_t= \frac{\log t}{t}\, e^{O(n^3)}$
but we allow a number of perturbations
as large as $e^{O(n^3)}$.
This noise model could be enriched considerably
without affecting the convergence bounds, but 
our choice was guided by simplicity.
Note that some restrictions are necessary for convergence;
trivially, noise must be bounded past the last switch
since two flocks flying parallel to each other
could otherwise be forced to merge 
arbitrarily far into the future.
Switching to a noisy model has two benefits: 
one is a more general result, since the same upper bound
on the convergence time 
holds whether the noise is turned on or off;
the other is a simpler lower bound proof. 
It allows us to keep the 
initial conditions extremely simple.
We use only $\log n$ perturbations
and $\delta_t\approx 1/t$, 
so noise is not germane to the
tower-of-twos growth.

\bigskip\medskip
\noindent
$\bullet$ {\em Why hysteresis?} \ 
Network convergence easily implies velocity
convergence, but the converse is not true: velocities
might reach steady state while the network does not. 
Indeed, in~\S\ref{NeedHysteresis}, 
we specify a group of birds that 
alternates forever between one and two flocks
without ever converging.
This is an interesting but somewhat peripheral issue that 
it is best to bypass, as is done in~\cite{jiE07},
by injecting a minute amount of hysteresis into the system.
Whatever one's rule---and, as we mentioned earlier,
almost any rule would work---it must be
sound, meaning that any two birds at distance ever so slightly
away from 1 should have the correct pairing status.
Note that soundness does not follow immediately
from our definition of hysteresis.
This will need to be verified. 
By construction, we know that 
any two birds within unit distance of each other 
at time $t$ are always joined by an edge of
the flocking network $G_t$.
We will show that, if we set 
$\varepsilon_{\! h} = n^{-bn^3}$ for a large 
enough constant $b$, then 
no two birds at distance greater than 
$1+ \sqrt{\varepsilon_{\! h}}$ 
are ever adjacent in $G_t$.

\bigskip\medskip
\noindent
$\bullet$ {\em How robust are the bounds?} \ 
The tower-of-twos bound
continues to hold regardless (almost) of which
hysteresis rule we adopt and how many input
bits we allow.  The assumption 
$\varepsilon_{\! h} = n^{-bn^3}$ 
is introduced for notational convenience;
for example, 
they allow allow us to express
soundness very simply by saying that
no birds at distance greater than 
$1+ \sqrt{\varepsilon_{\! h}}$ should ever be joined
by an edge of the network.
Without the assumptions above, the bounds are more
complicated. For the interested reader, here is what
happens to the number $N(n)$ 
of network switches and the fragmentation breakpoint $t_f$, ie,
the time after which flocks can only merge:
\begin{equation*}
\begin{cases}
\,
N(n) =
n^{O(n^3)} 
(\mathfrak{p} + \log \frac{1}{\varepsilon_{\! h}})^{n-1}
;  \\
\\
\,
t_f =  \hbox{$\frac{1}{\varepsilon_{\! h}}$}\,
n^{O(n^3)} 
2^{O(\mathfrak{p})} 
       (\mathfrak{p}+ \log \hbox{$\frac{1}{\varepsilon_{\! h}}$})^n.
\end{cases}
\end{equation*}

\paragraph{Discussion.}

How relevant are this paper's results?
Why are they technically difficult?
We address these two points briefly.
Our bounds obviously say nothing about 
physical birds in the real world. They merely
highlight the exotic behavior of the 
mathematical models. Although we 
focus on a Cucker-Smale variant, we believe
that the bounds hold for 
a much wider variety of neighbor-based models.
We introduce new techniques that are likely
to be of further interest.
The most promising seems to be the
notion of a ``virtual bird'' flying back in time.
We design a structure, the {\em flight net},
that combines both kinetic and 
positional information in a way that
allows us to use both the geometry
and the algebra of the problem at the same time.
Perhaps the most intriguing part of this work
is the identification of a curious phenomenon,
which we call the (iterated) {\em spectral shift}.

Self-confidence leads to an interesting phenomenon.
Too much of it
prevents consensus but so does too little.
Harmony in a group seems to be helped by
a minimum amount of self-confidence among its members.
Both extreme selfishness and excessive
altruism get in the way of reaching cohesion
in the group. Self-confidence provides a retention mechanism
necessary for reaching agreement. 
The coefficient $c_i(t)d_i(t)$ represents
how much a bird lets itself influenced by its
neighbors. By requiring that it be less than 1,
we enforce a certain amount of self-confidence for
each bird. This idea is not new and
can be found in~\cite{HendrickxB,lorenz05,Moreau2005}.

Besides noise and hysteresis, our model
differs from Cucker-Smale~\cite{CuckerSmale1}
in two other ways.
One is that our flocking networks are not complete graphs:
they undergo noncontinuous
transitions, which create the piecewise linearity of the system.
Another difference is that the transition
matrices of our model are not symmetric. This greatly
limits the usefulness of linear algebra.
The reason why might not be obvious, so 
here is some quick intuition.
Cucker and Smale diagonalize the Laplacian and note
that, since only differences are of interest,
the vectors might as well be assumed to lie
in the space ${\mathbf 1}^\perp$. 
Not only is that space invariant under 
the Laplacian but it contracts at an exponential rate set by
the Fiedler number (the second eigenvalue). 
From this, a quadratic 
Lyapunov function quickly emerges, namely
the energy $v^TL_t v$ of the system.
When the graph is connected, the Fiedler number
is bounded away from 0 by an inverse polynomial, so
differences between velocities decay to 0
at a rate of $2^{tn^{-c}}$ for some constant $c>0$.

In the nonsymmetric case (ours), this approach
is doomed. 
If, by chance, all the transition matrices had the same
left eigenvectors, then the variance of
the time-dependent Markov chain 
sampled at the (common) stationary distribution
would in fact be a valid Lyapunov function, but that
assumption is completely unrealistic.
In fact, it has been 
proven~\cite{jadbabaieLM03,P-07-lyapunov-rev}
that the dynamical systems under consideration
do not admit of any suitable 
quadratic Lyapunov function 
for $n\geq 8$.
Worse, as was shown by Olshevsky
and Tsitsiklis~\cite{P-07-lyapunov-rev},
there is not even any
hope of finding something weaker, such as 
a nonzero positive semidefinite matrix $\Lambda$ 
satisfying, for any allowable transition $v(t)\rightarrow v(t+1)$,
\begin{equation*}
\begin{cases}
\,
\Lambda {\mathbf 1} =0 ;  \\
\, v(t+1)^T \Lambda v(t+1) \leq v(t)^T \Lambda v(t).
\end{cases}
\end{equation*}
Our transition matrices are diagonalizable, 
but the right eigenspace for the subdominant eigenvalues
is not orthogonal to ${\mathbf 1}$
and the maps might not even be globally
nonexpansive: for example, the stochastic matrix
\begin{equation*}
\hbox{$\frac{1}{15}$}\!
\begin{pmatrix}
12 & 3 \\
10 & 5
\end{pmatrix}
\end{equation*}
has the two eigenvalues $1$ and $0.133$; yet it
stretches the unit vector $(1,0)$ to one of
length $1.041$. 
Linear algebra alone
seems unable to prove convergence.
The rationality of limit configurations
is not entirely obvious. In fact,
the iterated spectral shift is reminiscent of
lacunary-series constructions of transcendental numbers,
which is not the most auspicious setting for proving rationality.
This work draws from many 
areas of mathematics and computer science, including Markov chains,
nonnegative matrices, algebraic graph theory,
elimination theory, combinatorics, harmonic analysis, 
circuit complexity, computational geometry,
and of course linear algebra.

\section{A Bird's Eye View of the Proof}

To establish a tight bound on the convergence time,
we break down the proof into four parts, each one using
a distinct set of ideas. We briefly discuss
each one in turn. The first step is to bound the number of
network switches while ignoring all time considerations.
This decoupling allows us to treat the problem as purely
one of information transfer. In one step a bird
influences each one of its neighbors by forcing its
velocity into the computation of these neighbors' new
velocities. This influence propagates to other birds in
subsequent steps in a manner we can easily trace by
following the appropriate edges along the 
time-dependent flocking network.
Because of self-confidence, each bird influences itself
constantly. It follows that once a bird influences another one
(directly or indirectly via other birds) it does so forever, even if
the two birds find themselves forever confined to distinct
connected components. For this reason, influence alone is
a concept of limited usefulness. We need another
analytical tool: {\em refreshed} influence. 
Suppose that, at time $t_0$,
${\mathcal B}_1$ claims influence on ${\mathcal B}_2$. As we
just observed, this claim will hold for all $t>t_0$.
But suppose that we ``reboot" the system at time $t_0+1$
and declare all influences void. We may now ask if
${\mathcal B}_1$ will again claim influence on ${\mathcal B}_2$
at some time $t>t_0$ in the future: in other words, whether a chain
of edges will over time transfer information again from
${\mathcal B}_1$ to ${\mathcal B}_2$ after $t_0$. If yes, we then speak of
refreshed influence. 
Suppose now that ${\mathcal B}_1$ exerts refreshed influence
on ${\mathcal B}_2$ infinitely often: we call such
influence {\em recurrent}.  Although influence
is not a symmetric relation, it is an easy exercise to prove that 
recurrent influence is. 

\vspace{1cm}
\begin{figure}[htb]\label{fig-recurrent}
\begin{center}
\hspace{.2cm}
\includegraphics[width=7cm]{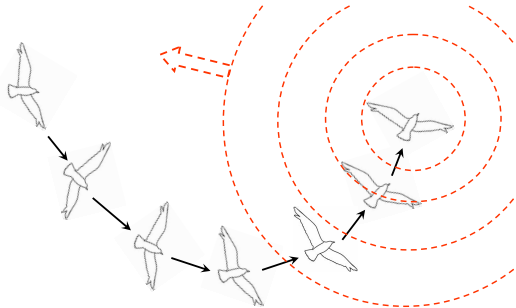}
\end{center}
\begin{quote}
\vspace{0cm}
\caption{\small 
Each bird is influenced by the one pointing to it. 
If this chain
of influence occurs repeatedly (not necessarily with the same
set of intermediate birds), a backward sphere of influence centered
at the end of the chain will begin to propagate backwards
and eventually reach the first bird in the chain.
}
\end{quote}
\end{figure}

This appears to be a principle of general interest.
If political conversations consist of many two-way communications
between pairs of people, with the pairs
changing over time, then the only way $A$ can
influence $B$ repeatedly is if it is itself influenced
by $B$ repeatedly. What makes this fact interesting is that
it holds even if $A$ and $B$ never exchange opinions directly
with each other 
and only a single pairwise communication occurs at any given time.
Self-confidence plays an important role in this phenomenon.
It provides information retention that prevents agents
from being influenced by their own opinions in periodic fashion.
In fixed networks, this avoids the classical oscillation issue
for random walks in bipartite graphs.

In time-dependent networks, the role of self-confidence
is more subtle. To understand it, one must first remember
one fundamental difference between {\em fixed} directed and undirected 
consensus networks (ie, where at each step, the opinion 
at each node $v$ is averaged
over the opinions linked to by the edges from $v$).
In a fixed directed network, the fraction of an agent's opinion
that is measurable at some other node of the network might
be exponentially small in the time elapsed since that
opinion was expressed. This cannot happen in undirected networks:
any fraction of an opinion is either 0 or bounded from
below independently of time. Time-dependent undirected
networks, on the other hand,
are expressive enough to (essentially) simulate fixed directed ones:
time, indeed, can be used to break edge symmetry. 
The benefits of undirectedness are thus lost, and
time-dependent undirected consensus networks can behave much
like fixed directed ones---see~\cite{condonhernek94, CondonL}
for an application of this principle
to interactive proof systems; 
in particular, they can witness
exponential opinion propagation decay.
Adding self-confidence magically prevents such decay.
The idea would appear to warrant special scrutiny
outside of its native habitat of computer science and control theory.

\bigskip\medskip
\noindent
$\bullet$ {\em How many switches?} \ 
Suppose that ${\mathcal B}_1$ exerts recurrent influence on
${\mathcal B}_2$.  We show that, at some point, both birds will join
a connected component of the flocking network and remain
there forever. How many switches can occur before that event?
Let $V_1$ be the set of birds influenced by ${\mathcal B}_1$.
As soon as everyone in $V_1$ has been influenced by
${\mathcal B}_1$, let's 
``reboot" the system and define $V_2$ to be the new set of birds
with refreshed influence from ${\mathcal B}_1$. Obviously
$V_1\supseteq V_2$. Repeating this process leads to 
an infinite nested sequence
$$  V_1\supseteq V_2\supseteq V_3\supseteq \cdots \supseteq V_\infty,$$
where $V_\infty$ contains at least the two birds ${\mathcal B}_1$ 
and ${\mathcal B}_2$. Let $T_k$ be the formation time of $V_k$
and let $\delta_k$ be the difference in velocity between
the two birds at time $T_k$. We wish we could claim a uniform bound, 
$\|\delta_k\|_2< (1-\eps)\|\delta_{k-1}\|_2$, for some 
fixed $\eps>0$ independent of the time difference $T_k-T_{k-1}$.
Indeed, this would show that, for $k$ large enough, the two velocities
are close enough for the hysteresis rule to kick in 
and keep the two birds together in the same flock forever. Of course, since the two
birds need not be adjacent, this argument should be extended to all pairs
of birds in $V_\infty$.
While the inequality $\|\delta_k\|_2< (1-\eps)\|\delta_{k-1}\|_2$
is too much to ask for, we show that $\|\delta_k\|_2\leq \zeta_k$, where
$\zeta_k< (1-\eps)\zeta_{k-1}$. In other words, the velocity
difference between ${\mathcal B}_1$ and ${\mathcal B}_2$ may not
shrink monotonically, but it is bounded by a function that does.
The uniformity of the shrinking, which is crucial, depends critically
on self-confidence and the retention mechanism it implies. 
Technically, this translates into a uniform
lower bound on the nonzero entries of products of stochastic matrices.
This allows us to rescue the previous argument and bound the value
of $k$ such that $V_k=V_\infty$. To bound the number of switches
before time $T_k$, we need to find how many of them can take
place between a reboot at $T_{j-1}$ and the formation of $V_j$.
The key observation is that $V_j$ is formed by a growth process of
smaller flocks (ie, all of them of size less than $n$): we can therefore
set up a recurrence relation and bound the number of switches
inductively.

\bigskip\medskip
\noindent
$\bullet$ {\em How much time between switches?} \ 
Flock behavior between switches is linear, so spectral
analysis provides most of the tools we need to 
bound the inter-switch time. At time $t$, the number of bits needed
to encode the velocity is (roughly) $O(t)$. This means that,
in the worst case, two birds can fly either parallel to each 
or at an angle at least $e^{-O(t)}$. From this we can infer
that, should the birds want to be joined together in the flocking
network after time $t$, this union must happen
within a period of $e^{O(t)}$. Things are more complex if 
the stationary velocities of the two flocks are parallel. 
We need to use root separation bounds for various extension fields 
formed by adjoining to the rationals 
all the relevant eigen-information. Intuitively,
the question we must answer is how long one must wait for
a system of damped oscillators to cross a given real semi-algebraic set
with known parameters.
All of these techniques alone can only yield a
convergence time bound in the form
of a tower-of-twos of height exponential in $n$.
To bring the height down to logarithmic requires two distinct ideas
from computational geometry and circuit complexity.

\bigskip\medskip
\noindent
$\bullet$ {\em How to bring the height down to linear?} \ 
So far, we have only used combinatorics, algebraic
graph theory, linear algebra, and elimination theory.
We use algorithmic ideas from convex geometry to reduce
the height to linear. We lift the birds into 4 dimensions (or $d+1$
in general) by making
time into one of the dimensions. We then prove that, after 
exponential time, birds can only fly almost radially (ie,
along a line passing through the origin). This implies
that, after a certain time threshold, flocks can only merge
and never fragment again. From that point on, reducing the height
of the tower-of-twos to linear is easy.  Our geometric investigation
introduces the key idea of a {\em virtual bird}. The stochastic
transitions have a simple geometric interpretation in terms
of new velocities lying in the convex hulls of previous ones.
This allows us to build an exponential-size {\em flight net}
consisting of convex structures through which all bird
trajectories can be monitored. A useful device is to picture
the 
birds flying back in time with exactly one of them carrying a baton. When a bird
is adjacent to another one in a flock, it may choose to pass
its baton. The trajectory of the baton is identified
as that of a virtual bird. 
Because of the inherent nondeterminism of the process, we may then ask the question:
is there always a virtual bird trajectory that follows a near-straight line?
The answer, obviously negative in the case of actual birds, turns out
to be yes. This is the benefit of virtuality. This fact has 
numerous  geometric consequences bearing on the angular flight motion
of the real birds.

\bigskip\medskip
\noindent
$\bullet$ {\em How to bring the height down to logarithmic?} \ 
It is not so easy to build intuition for the logarithmic height
of the tower-of-twos.\footnote{As a personal aside, let me say that
I acquired that intuition only after
I had established the matching lower bound. For this reason,
I recommend reading the lower bound section before 
the final part of the upper bound proof.} A circuit
complexity framework helps to explain the {\em residue clearing}
phenomenon behind it. To get a tower-of-twos requires
an iterated {\em spectral shift}.
When two flocks meet,
energy must be transferred from the high-frequency range down
to the lowest mode in the power spectrum. 
This process builds a {\em residue}: informally,
think of it, for the purpose of intuition, 
as residual heat generated by the transfer.
This heat needs to be evacuated to make room for further spectral shifts.
The required cooling requires free energy in the form
of previously created spectral shifts. This leads to an inductive
process that limits any causal chain of spectral shifts
to logarithmic length. The details are technical, and the
best way to build one's intuition is to digest the lower bound first.

\bigskip\medskip
\noindent
$\bullet$ {\em How to prove the optimality of the logarithmic height?} \ 
The starting configuration is surprisingly simple. The $n$ birds
stand on a wire and fly off together at various angles. The initial
conditions require only $O(\log n)$ bits per bird. 
The $n$ birds meet in groups of
2, 4, 8, etc, forming a balanced binary tree. Every ``collision"
witnesses a spectral shift that creates flying directions that
are increasingly parallel; hence the longer waits between collisions.
To simplify the calculations, we use the noisy model to
flip flocks occasionally in order to reverse
their flying directions along the $X$-axis. This occurs only
$\log n$ times and can be fully accounted for by the model 
we use for the upper bound. 
Because the flocks are simple paths,
we can use harmonic analysis for cyclic groups to help us resolve all questions
about their power spectra.

\section{The Upper Bound}

We begin with a few opening observations in~\S\ref{Prelim}.
We explore both the algebraic
and geometric aspects of flocking in~\S\ref{AlgGeom}.
We establish a crude convergence bound 
in~\S\ref{Iterat}, 
which gives us a glimpse of the spectral shift.
An in-depth study of its combinatorial
aspects is undertaken in~\S\ref{SpectralS},
from which a tight upper bound follows.
We shall always assume that $\mathfrak{p}\geq n^3$.
To highlight the robustness of the bounds, we 
leave both $\mathfrak{p}$ and $\varepsilon_{\! h}$
as parameters throughout much
of our discussion, thus making it easier
to calculate convergence times for 
arbitrary settings. 
For convenience and clarity,
we adopt the default settings below
in~\S\ref{SpectralS} (but not before).
One should keep in mind that 
virtually any assignment of parameters
would still produce a tower-of-twos.
Let $b$ denote a large enough constant:

\begin{equation}\label{Assumptions}
\text{\sc Default Settings}\ \ \ 
\begin{cases}
\, \mathfrak{p}= n^3 ;\\
\, \varepsilon_{\! h} = n^{-bn^3} .
\end{cases}
\end{equation}
Recall that 
$\mathfrak{p}$
and $\varepsilon_{\! h}$
denote, respectively,
the input bit-size and  
the hysteresis parameter.
With these settings, the fragmentation breakpoint
and the maximum switch count are both
$n^{O(n^3)}$.

\subsection{Preliminaries}\label{Prelim}

We establish a few useful facts about the growth
of the coordinates over time.
It is useful to treat coordinates
as integers, which we can do by expressing
them as fractions sharing the same denominator.
For example, the initial positions and velocities
can be expressed either as $\mathfrak{p}$-bit rationals 
or, more usefully, as $O(\mathfrak{p} n)$-bit {\em CD-rationals},
ie, rationals of the form
$p_i/q$, with the common denominator $q$.
We mention some important properties of
such representations. 
We will also introduce 
some of the combinatorial tools needed
to measure ergodicity. The objective is
to predict how fast backward products 
of stochastic matrices tend to rank-one
matrices.
We treat the general case in this section
and investigate the time-invariant case in the next.

\paragraph{Numerical Complexity.}

The {\em footprint}
of a matrix $A$ is the matrix $\underline{A}$ derived from $A$
by replacing each nonzero entry by $1$.
For $t\geq s$, we use $P(t,s)$ as shorthand for 
$P(t)P(t-1)\cdots P(s)$.
Note that, in the absence of noise,
the fundamental equation~(\ref{modelD})
can be rewritten as
$$
v(t+1)= (P(t,1)\otimes I_d) v(1).
$$
A bird may influence another one over a period of time
without the converse being true; in other words, 
the matrices $P(t,s)$ and 
$\underline{P}(t,s)$ are in general 
not symmetric; the exception is 
$\underline{P}(t)$, which not only is symmetric but
has its diagonal full of ones.
Because of this last property,
$\underline{P}(t,s)$ can never lose
any 1 as $t$ grows, or to put it
differently the corresponding
graph can never lose an edge.
Before we get to the structural properties of $P(t,s)$,
we need to answer two basic questions: how small
can the nonzero entries be and how many bits
do we need to represent them?
As was shown in~\cite{HendrickxB, lorenz05},
nonzero elements of $P(t,s)$
can be bounded uniformly, ie, independently of $t$. 
Note that this relies critically
on the positivity of the diagonals.
Indeed, without the condition 
$c_i(t)d_i(t)< 1$, we could choose 
$P(t)=A$ for even $t$ and $P(t)=B$ for odd $t$, where
\begin{equation*}
A= 
\begin{pmatrix}
0 & 1& 0 \\
1 & 0& 0 \\
0 & 0& 1
\end{pmatrix}
\hspace{3cm}
B= 
\hbox{$\frac{1}{2}$}
\begin{pmatrix}
0 & 2& 0 \\
1 & 0& 1 \\
0 & 2& 0
\end{pmatrix}. 
\end{equation*}
For even $t>0$,
\begin{equation*}
P(t,1)= (AB)^{t/2} =
\begin{pmatrix}
2^{-t/2} & 1- 2^{1- t/2}& 2^{-t/2}  \\
0 & 1& 0 \\
0 & 1& 0
\end{pmatrix}.
\end{equation*}
To understand this process, think of a triangle
with a distinguished vertex called the {\em halver}.
Each vertex holds an amount of money.
At odd steps, the halver splits its amount
in half and passes on each half
to its neighbor; the other vertices, meanwhile,
pass on their full amount to the halver.
The total amount of money in the system remains the same.
At the following 
(even) step, the role of halver is handed to 
another vertex (which one does not matter);
and the process repeats itself.
This alternate sequence of halving and relabeling steps
produces an exponential decay. If each vertex
is prohibited to pass its full amount, however, then 
money travels while leaving a ``trace'' behind.
As we prove below, exponential decay becomes impossible.
This prohibition is the equivalent
of the positive self-confidence built into
bird flocking.

\bigskip

\begin{lemma}\label{x(t)-precision}
$\!\!\! .\,\,$
For any $1\leq s\leq t$, the elements of $P(t,s)$ are 
CD-rationals over $O((t-s+1)n\log n)$ bits.
The nonzero elements are in $n^{-O(n^2)}$.
\end{lemma}
\proof
Each row of $P(t)$ contains rationals with the same
$O(\log n)$-bit denominator, so the matrix $P(t)$ can be
written as $N^{-1}$ times an integer matrix, where
both $N$ and the matrix elements are encoded over $O(n\log n)$ bits.
Each element of $P(t,s)$ is a product of
$t-s+1$ such matrices; hence a matrix with 
$O((t-s+1)n\log n)$-bit integer elements
divided by a common $O((t-s+1)n\log n)$-bit integer.
For the second part of the lemma,
we use arguments from~\cite{HendrickxB, lorenz05}.
Recall that $P(t)= I_n- C_t L_t$, where $C_t$ is 
a diagonal matrix of positive rationals 
encoded over $O(\log n)$ bits, so 
the case $t=s$ is obvious.
Let $\rho(t,s)$ be the smallest positive element 
of $P(t,s)$ and suppose that $t>s$. 

We begin with a few words of intuition.
Because $P(s,t)=P(t)P(t-1,s)$, 
a nonzero entry $p_{ij}(t,s)$ is the expected
value of $p_{kj}(t-1,s)$, for a random $k$ adjacent
to $i$ in $\underline{P}(t)$, or, to be more precise,
in the graph induced by
the nonzero elements of that matrix.
If, for all such $k$, $p_{kj}(t-1,s)>0$, then 
$p_{ij}(t,s)$, being an average of positive numbers, is at least
$\rho(t-1,s)$, and we are done. On the other hand, 
having some $p_{kj}(t-1,s)$ equal to $0$ means
that the edge $(k,j)$ is missing from the ``graph''
$\underline{P}(t-1,s)$. If we now consider the 2-edge path
formed by $(k,i)$ in $\underline{P}(t)$ 
and $(i,j)$ in $\underline{P}(t-1,s)$,
we conclude that at least one of $(i,j)$ or $(k,j)$ is 
a brand-new edge in $\underline{P}(t,s)$. We then use the
fact that such events happen rarely.

\begin{itemize}
\item
Suppose that $p_{kj}(t-1,s)>0$ 
for each $i,j,k$ such that $p_{ij}(t,s)p_{ik}(t)>0$.
Then, for any $p_{ij}(t,s)>0$,
by stochasticity, 
$$p_{ij}(t,s)
=\sum_k p_{ik}(t)p_{kj}(t-1,s)
\geq \Bigl( \sum_k p_{ik}(t) \Bigr) \rho(t-1,s)
= \rho(t-1,s).$$
It follows that 
$\rho(t,s)\geq \rho(t-1,s)$.
\item
Assume now that
$p_{ij}(t,s)p_{ik}(t)>0$ and $p_{kj}(t-1,s)=0$
for some $i,j,k$.
Since $p_{ij}(t,s)$ is positive, so 
is $p_{il}(t)p_{lj}(t-1,s)$ for some $l$; hence
$p_{ij}(t,s)\geq p_{il}(t)p_{lj}(t-1,s) \geq
                    \rho(t-1,s) n^{-O(1)}$.
We show that this drop coincides with
the gain of an 1 in $\underline{P}(t,s)$.
The footprint of $P(t)$ is symmetric, so 
$p_{ki}(t)> 0$ and hence
$$p_{kj}(t,s)= \sum_l p_{kl}(t) p_{lj}(t-1,s)\geq 
p_{ki}(t) p_{ij}(t-1,s)\geq 
n^{-O(1)} p_{ij}(t-1,s).$$
We distinguish between two cases.
If $p_{ij}(t-1,s)$ is positive, then so is $p_{kj}(t,s)$.
Since $p_{kj}(t-1,s)=0$, the matrix $P(t,s)$ has at least
one more positive entry than $P(t-1,s)$;
recall that no entry can become null as we go
from $P(t-1,s)$ to $P(t,s)$.
On the other hand, if $p_{ij}(t-1,s)=0$, our assumption
that $p_{ij}(t,s)>0$ leads us to the same conclusion.
In both cases, $P(t,s)$ differs from
$P(t-1,s)$ in at least one place: this cannot happen
more than $n^2$ times.
\end{itemize}
If we fix $s$ then
$\rho(t,s) \geq \rho(t-1,s)$ for all but at most
$n^2$ values of $t$. For the others, 	
as we saw earlier, 
$p_{ij}(t,s)\geq \rho(t-1,s) n^{-O(1)}$; hence
$\rho(t,s)\geq \rho(t-1,s) n^{-O(1)}$.
\hfill $\Box$
\proofend

The coordinates of $v(1)$ and $x(0)$ can be 
expressed as CD-rationals over $O(\mathfrak{p} n)$ bits.
By the previous lemma, this implies that, 
in the noise-free case,
for $t> 1$, $v(t)= (P(t-1,1)\otimes I_d)v(1)$
is a vector with CD-rational coordinates 
over $O(t n\log n +\mathfrak{p} n)$ bits.
The equation of motion~(\ref{modelD}) yields
\begin{equation*}
x(t)
= x(0)+ \Bigl( 
      (P(t-1,1)+\cdots + P(1,1)+I_n)\otimes I_d \Bigr) v(1).
\end{equation*}
Note that $P(t-1,1)= N^{-1} Q$, where $Q$ is an integer 
matrix with $O(tn\log n)$-bit integer elements
and $N$ is an $O(tn\log n)$-bit integer.
The other matrices are subproducts of
$P(t-1,1)= P(t-1)\cdots P(1)$, 
so we can also express
them in this fashion for the same value of $N$.
It follows that $v(t)$ and $x(t)$ have 
CD-rational coordinates over
$O(t n\log n +\mathfrak{p} n)$ bits.
Adding noise makes no difference asymptotically.
Indeed, bringing all the coordinates
of the scaling vectors $\alpha$ in CD-rational form
adds only $O(n\log n)$ bits to the
velocities at each step.

\begin{lemma}\label{x-v(t)-rationalprecision}
$\!\!\! .\,\,$
For any $t\geq 1$, the vectors
$v(t)$ and $x(t)$ have 
CD-rational coordinates over
$O(t n\log n +\mathfrak{p} n)$ bits.
\end{lemma}

The $\ell_\infty$ norm of the velocity vector
never grows, as transition matrices
only average them out and the noise factors are bounded
by 1: since $\mathfrak{p}\geq n^3$,
it follows that, for any $t\geq 1$,

\begin{equation}\label{|v|Poly}
\|v(t)\|_2= 2^{O(\mathfrak{p})}.
\end{equation}

\paragraph{Ergodicity.}

Ignoring noise, 
the fundamental motion equation~(\ref{modelD})
gives the position of the birds at time $t>1$ as
$x(t)=x(0)+ (P^* (t-1)\otimes I_d)v(1)$, where
$$
P^* (t)=
P(1)+
P(2)P(1)+
P(3)P(2)P(1)+
\cdots +
P(t)\cdots P(2)P(1) . 
$$
Products of the form $P(t)\cdots P(1)$ 
appear in many applications~\cite{seneta06},
including the use of {\em colored random walks}
in space-bounded interactive proof 
systems~\cite{condonhernek94, CondonL}.
One important difference is that random walks
correspond to products that grow by
multiplication from the right while the dynamics
of bird flocking is associated with
backward products: the transition matrices
evolve by multiplication from the left.
This changes the nature of ergodicity.
Intuitively, one would expect (if all goes well)
that these products should look increasingly
like rank-1 matrices. But can the rows continue
to vary widely forever though all in lockstep
(weak ergodicity),
or do they converge to a fixed vector (strong
ergodicity)?
The two notions are equivalent for
backward products but not for 
the forward kind~\cite{seneta06}.
Here is an intuitive explanation.
Backward products keep averaging the rows, so 
their entries themselves tend to converge: 
geometrically, the convex hull of the points
formed by the row keeps shrinking.
Forward products  lack this notion of averaging.
For a simple illustration of the difference, consider
the three stochastic matrices:
\begin{equation*}
A= \frac{1}{2}
\begin{pmatrix}
1 & 1  \\
1 & 1 
\end{pmatrix}
\hspace{2cm} 
B= \frac{1}{2}
\begin{pmatrix}
2 & 0  \\
1 & 1 
\end{pmatrix}
\hspace{2cm} 
C= \frac{1}{4}
\begin{pmatrix}
3 & 1  \\
3 & 1 
\end{pmatrix}.
\end{equation*}
Backward products are given by the simple formula,
\begin{equation*}
\underset{n}{\underbrace{\cdots ABABABAB}}
\,= \, C \, , 
\end{equation*}
for all $n>1$. On the other hand, 
the forward product
tends to a rank-one matrix but never converges:
\begin{equation*}
\,\,\, \underset{n}{\underbrace{ABABABAB\cdots}}
\, = \,
\begin{cases}
\, C \ \ \  \text{even $n>1$} ;  \\
\, A \ \ \  \text{odd $n>0$} ,
\end{cases}
\end{equation*}

\vspace{.5cm}
\begin{figure}[htb]\label{fig-Pshrink}
\hspace{3.5cm}
\includegraphics[width=7cm]{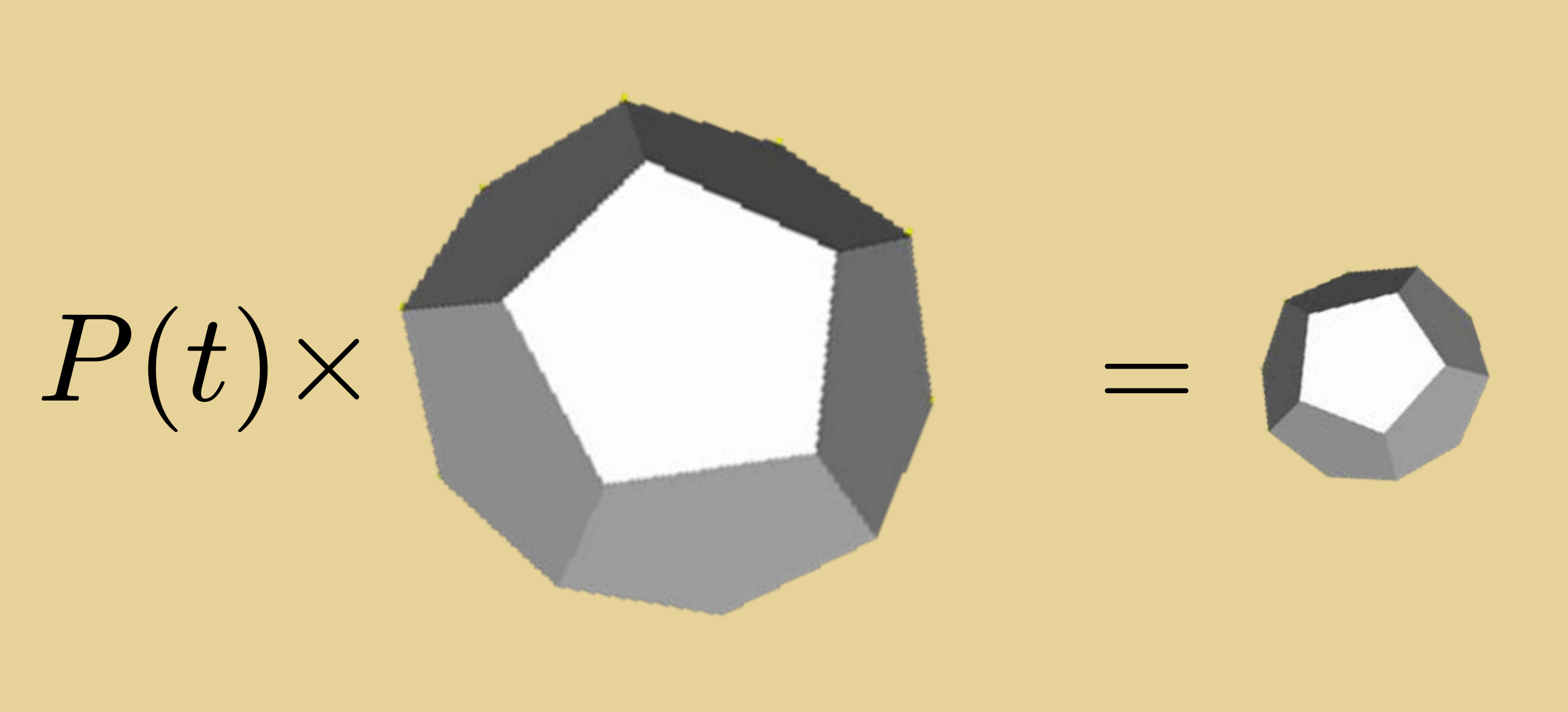}
\vspace{0.5cm}
\begin{quote}
\caption{\small 
Premultiplying a matrix, whose rows are shown as points,
by a stochastic matrix $P(t)$
shrinks its convex hull.
}
\end{quote}
\end{figure}

As we just mentioned, the key to ergodicity for
backward products resides in the convex hull of the rows.
We introduce a family of metrics to measure its ``shrinkage.''
For any $p>1$, let $\tau_p (A)$, the
{\em ergodicity coefficient} of $A$,  denote the 
$\ell_p$-diameter of the convex hull
formed by the rows of a matrix $A$, ie,
$$\tau_p(A)= \max_{i,j} \|A_{i*}- A_{j*}\|_p,$$
where $A_{i*}$ denotes the $i$-th row of $A$.
From the fact that $\ell_p$ is a metric space
for $p>1$, it follows by convexity that
the diameter is always achieved
at vertices of the convex hull.
We extend the definition to $p=1$ but,
for reasons soon to be apparent,
it is important to keep 
the coefficients between 0 and 1, so
we divide the diameter by two, ie,
$$\tau_1(A)= \frac{1}{2} 
      \max_{i,j} \sum_k |a_{ik}-a_{jk}|.$$
To understand why $\tau_p (A)$ relates to ergodicity,
assume that $A$ is row-stochastic.
We observe then that
\begin{equation*}
0\leq \tau_1(A)
= 1- \min_{i,j} \, \sum_{k} \, 
\min\,\{a_{ik}\, , \, a_{jk}\} \leq 1.
\end{equation*}
This follows from the fact that
the distance $|a-b|$
between two numbers $a,b$ is 
twice the difference between their average 
and the smaller one.
There are many fascinating relations between
these diameters~\cite{seneta06}. For our purposes,
the following submultiplicativity result 
will suffice~\cite{lorenz05}.\footnote{\, Submultiplicativity
is not true for $\tau_2$ in general. 
First, to make the notion meaningful, we would need to normalize it
and use $\widehat \tau_2= \tau_2/\sqrt{2}$ instead, to 
ensure that $\widehat\tau_2(A)\leq 1$ for any stochastic $A$.
Unfortunately, $\widehat\tau_2$ is not submultiplicative, as
we easily check by considering a 
regular random walk $A$ on $K_{2,2}$
and checking that $\widehat\tau_2(A^2)> \widehat\tau_2(A)^2$.}

\begin{lemma}\label{ergodicity-submult}
$\!\!\! .\,\,$
Given two row-stochastic matrices $A,B$ that 
can be multiplied, 
$$\tau_2(AB)\leq \tau_1(A)\tau_2(B).$$
\end{lemma}
\proof
Fix the two rows $i,j$ that define $\tau_2(AB)$, and 
let $\alpha = 1- \sum_k \min\{a_{ik},a_{jk}\}$.
Note that $0\leq \alpha\leq \tau_1(A)$.
If $\alpha=0$, then $A_{i*}=A_{j*}$ and 
$\tau_2(AB)=0$, so the lemma holds trivially.
Assuming, therefore, that $\alpha>0$, we derive
\begin{equation*}
\begin{split}
\tau_2(AB)
&= \Bigl\|   \sum_{k} a_{ik}B_{k*} 
               - \sum_{k} a_{jk}B_{k*} \Bigr\|_2 \\
&= \Bigl\|  \sum_{k} (a_{ik} - \min\{a_{ik},a_{jk}\})B_{k*}
   -   \sum_{k} (a_{jk} - \min\{a_{ik},a_{jk}\})B_{k*} \Bigr\|_2
\\
&\leq  \tau_1(A)
   \Bigl\| \hbox{$\frac{1}{\alpha}$} \sum_{k} 
               (a_{ik} - \min\{a_{ik},a_{jk}\})B_{k*}
       -   \hbox{$\frac{1}{\alpha}$} \sum_{k} 
               (a_{jk} - \min\{a_{ik},a_{jk}\})B_{k*}  \Bigr\|_2 .
\end{split}
\end{equation*}
Observe now that the coefficients 
$\alpha^{-1} (a_{ik} - \min\{a_{ik},a_{jk}\})$
are nonnegative and sum up to 1, so the corresponding
sum is a convex combination of the rows of $B$.
The same is true of the other sum; so, 
by convexity, the distance between any two of them cannot
exceed $\tau_2(B)$.
\hfill $\Box$
\proofend

\vspace{0.1cm}
\begin{figure}[htb]\label{fig-diameter}
\hspace{4cm}
\includegraphics[width=5cm]{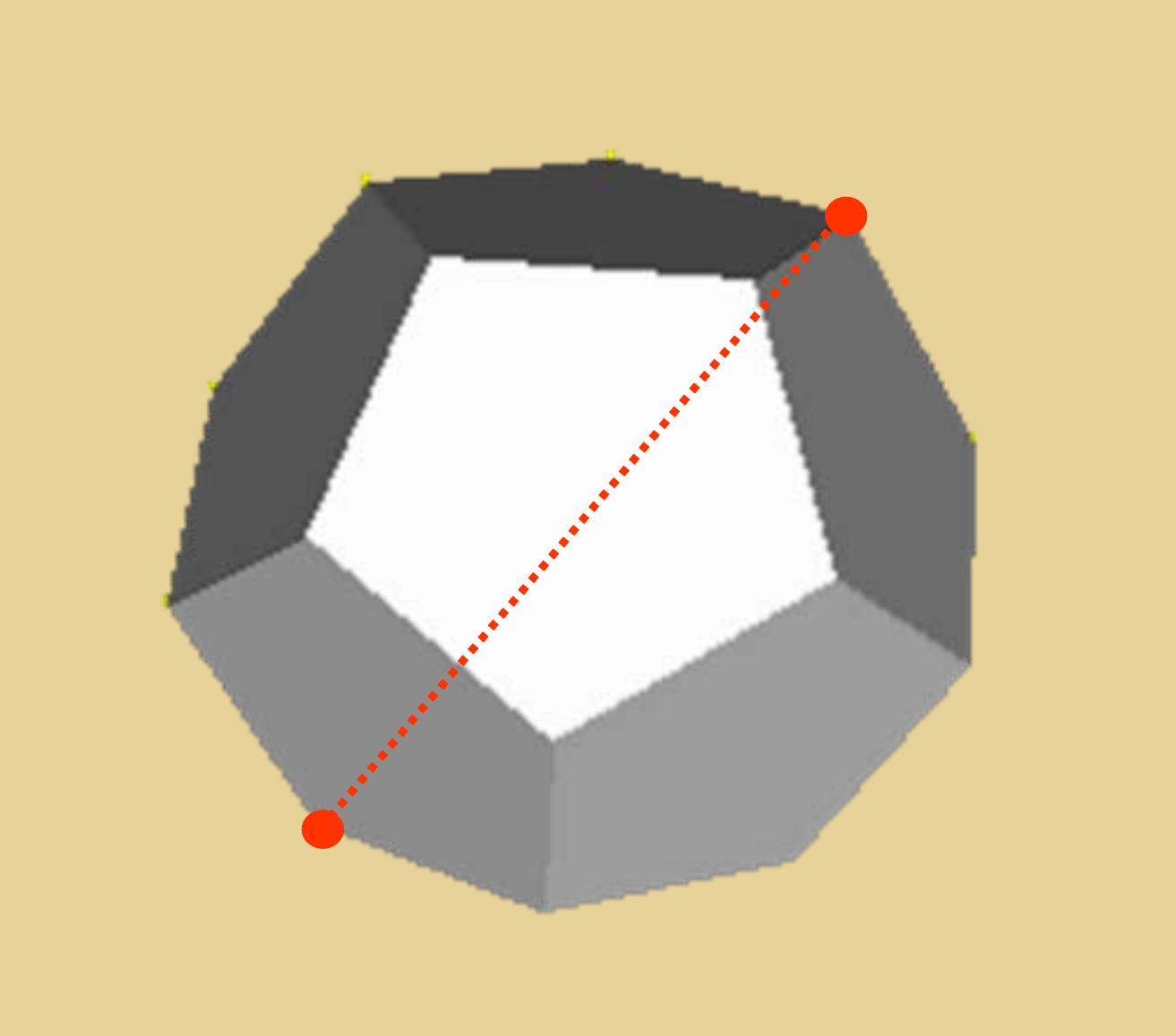}
\vspace{0.4cm}
\begin{quote}
\caption{\small 
$\tau_p(A)$ is the $\ell_p$-diameter of the 
convex hull of the rows of $A$.
}
\end{quote}
\end{figure}

\paragraph{Displacement.}

For future use, we mention an elementary relation between
bird distance and velocity.
The {\em relative displacement} between two birds
${\mathcal B}_i$ and ${\mathcal B}_j$ is defined as
$\Delta_{ij}(t)= 
|\text{\sc dist}_{t}({\mathcal B}_i, {\mathcal B}_j)-
\text{\sc dist}_{t-1}({\mathcal B}_i, {\mathcal B}_j)\,|$,
where the 
distance between two birds
is denoted by $\text{\sc dist}_{t}({\mathcal B}_i, {\mathcal B}_j)
=\|x_i(t)-x_j(t)\|_2$.

\begin{lemma}\label{relativeDisp}
$\!\!\! .\,\,$
For $t\geq 1$, 
$\Delta_{ij}(t) \leq \|v_i(t)- v_j(t) \|_2$.
\end{lemma}
\proof
By the triangle inequality,
$$
\|x_i(t)-x_j(t)\|_2
\leq \|x_i(t-1)-x_j(t-1)\|_2
+ \| x_i(t)-x_i(t-1) - ( x_j(t)-x_j(t-1)) \|_2\, .
$$
Reversing the roles of $t$ and $t-1$ gives us
a similar inequality, from which we find that
$$
|\text{\sc dist}_{t}({\mathcal B}_i, {\mathcal B}_j)-
\text{\sc dist}_{t-1}({\mathcal B}_i, {\mathcal B}_j)\, |
\leq 
\| x_i(t)-x_i(t-1) - ( x_j(t)-x_j(t-1)) \|_2\, .
$$
\hfill $\Box$
\proofend

\subsection{The Algebra and Geometry of Flocking}\label{AlgGeom}

To separate the investigation of network switches
from the time analysis is one of the key ideas of our method.
Our first task, therefore, is to bound the number of times
the flocking network can change, while ignoring how 
long it takes.
Next, we investigate the special case of time-invariant 
networks. In the worst case, the pre-convergence
flying time vastly exceeds the number
of network switches, so it is quite
intuitive that a time-invariant analysis should be critical.
Our next task is then to prove the 
rationality of the limit configuration.
We also show why the hysteresis rule
is sound. We follow this with an in-depth 
study of the convex geometry of flocking.
We define the {\em flight net}, and with it
derive what is arguably 
our most versatile analytical tool:
a mathematical statement that captures the intuition
that flocks that hope to meet in the future must
match their velocities more and more closely over time.
To do this we introduce the key concept of a {\em virtual bird},
which is a bird that can switch identities with its neighbors
nondeterministically.

\paragraph{Counting Network Switches.}

Let $N(n)$ be the maximum number
of switches in the flocking network, ie, the
number of times $t$ such that $P(t)\neq P(t+1)$.
Obviously, $N(1)=0$; 
note that, by our requirement that $C_t$ may vary
only when $G_t$ does, we could use footprints 
equivalently in the definition.
For the sake of our inductive argument,
we need a uniform bound on $N(n)$ over all
initial conditions.
Specifically, we define 
$N(n)$ as the largest number of
switches of an $n$-bird flocking network,
given arbitrary initial conditions: for the purpose
of bounding $N(n)$, 
$x(0)$ and $v(1)$ are {\em any real} vectors,
with $\|v(1)\|_2= 2^{O(\mathfrak{p})}$.
This involves
building a quantitative framework around
the existential analyses
of~\cite{HendrickxB,liwang2004,lorenz05,Moreau2005}.
We now prove the network switching bound claimed
in the ``Results'' section of~\S\ref{introduction}.

\begin{lemma}\label{NetworkChangesUB}
$\!\!\! .\,\,$
The maximum number $N(n)$ of switches in the 
flocking network is bounded by 
$n^{O(n^3)} 
(\mathfrak{p} + \log \frac{1}{\varepsilon_{\! h}})^{n-1}$.
\end{lemma}

\begin{corollary}\label{corol-NetworkChangesUB}
$\!\!\! .\,\,$
Under the default settings~(\ref{Assumptions}),
$N(n)= n^{O(n^3)}$.
\end{corollary}

{\noindent\em Proof of Lemma~\ref{NetworkChangesUB}. }
We begin with the noise-free model.
Fix $s>0$ once and for all.
For $t>s$, let $N(t,s)$ be the number 
of network changes between
times $s$ and $t$, ie, the number of integers
$u$ ($s<u\leq t$) such that 
$\underline{P}(u)\neq \underline{P}(u-1)$.
Since the diagonal of each $P(t)$ is positive, 
$\underline{P}(t,s)$ can never lose a 1 as $t$ grows,
so there exists a smallest $T_1$ such that
$\underline{P}(t,s)= \underline{P}(T_1,s)$ for all $t>T_1$.
Consider the first column and 
let $n_0<\cdots < n_{l_1}\leq n$
be its successive Hamming weights (ie, number of ones);
because $p_{11}(s)\neq 0$, 
$n_0\geq 1$. We define $t_k$ as the smallest $t\geq s$
such that the first column of $P(t,s)$ 
acquires weight $n_k$.
Note that $t_0= s$ and $t_{l_1}\leq T_1$.
How large can $N(t_{k+1}, t_k)$ be,
for $0\leq k<l_1$?
Let $F$ denote the subgraph of $G_{t_k+1}$
consisting of the connected components (ie, flocks)
that include the $n_k$ birds indexed by the first
column of $\underline{P}(t_k,s)$.
Intuitively, at time $t_k+1$,
bird ${\mathcal B}_1$ can claim it has
had influence over the $n_k$ birds since time $t_0$.
At time $t_k+2$,
this influence will spread further to
the neighbors of these $n_k$ birds in $F$.
Note that having been influenced by 
${\mathcal B}_1$ in the past does not imply
connectivity among the $n_k$ birds.

\begin{itemize}
\item
If $F$ contains more than $n_k$ birds then,
at time $t_k+1$, at least one of these extra
birds, ${\mathcal B}_i$, is adjacent in $G_{t_{k+1}}$
to one of the $n_k$ birds, say, ${\mathcal B}_j$.
Then, $p_{ij}(t_k+1)>0$ and $p_{j1}(t_k,s)>0$; hence
$p_{i1}(t_k+1,s)\geq p_{ij}(t_k+1) p_{j1}(t_k,s)>0$.
Since ${\mathcal B}_i$ is not one of the $n_k$ birds,
$p_{i1}(t_k,s)=0$ and the first column
of $\underline{P}(t,s)$ acquires a new 1
between $t_k$ and $t_k+1$.
This implies that
$t_{k+1}= t_{k}+1$ and $N(t_{k+1}, t_k)\leq 1$.
\item
Assume now that $F$ has exactly $n_k$ vertices.
The flocking network $G_{t_k+1}$ consists
of a set of flocks totalling $n_k$ birds
and a separate set of flocks including
the $n-n_k$ others. 
The next $N(n_k) + N(n-n_k)+1$ 
network switches must include
one between the two sets, since
by then we must run out of 
allowable ``intra-switches.''
It follows by monotonicity of $N(n)$ that 
$$N(t_{k+1},t_k)\leq 1+ N(n_k) + N(n-n_k) 
\leq 2N(n-1) +1.$$
\end{itemize}

\vspace{2cm}
\begin{figure}[htb]\label{fig-switch-background}
\hspace{2.7cm}
\includegraphics[width=8cm]{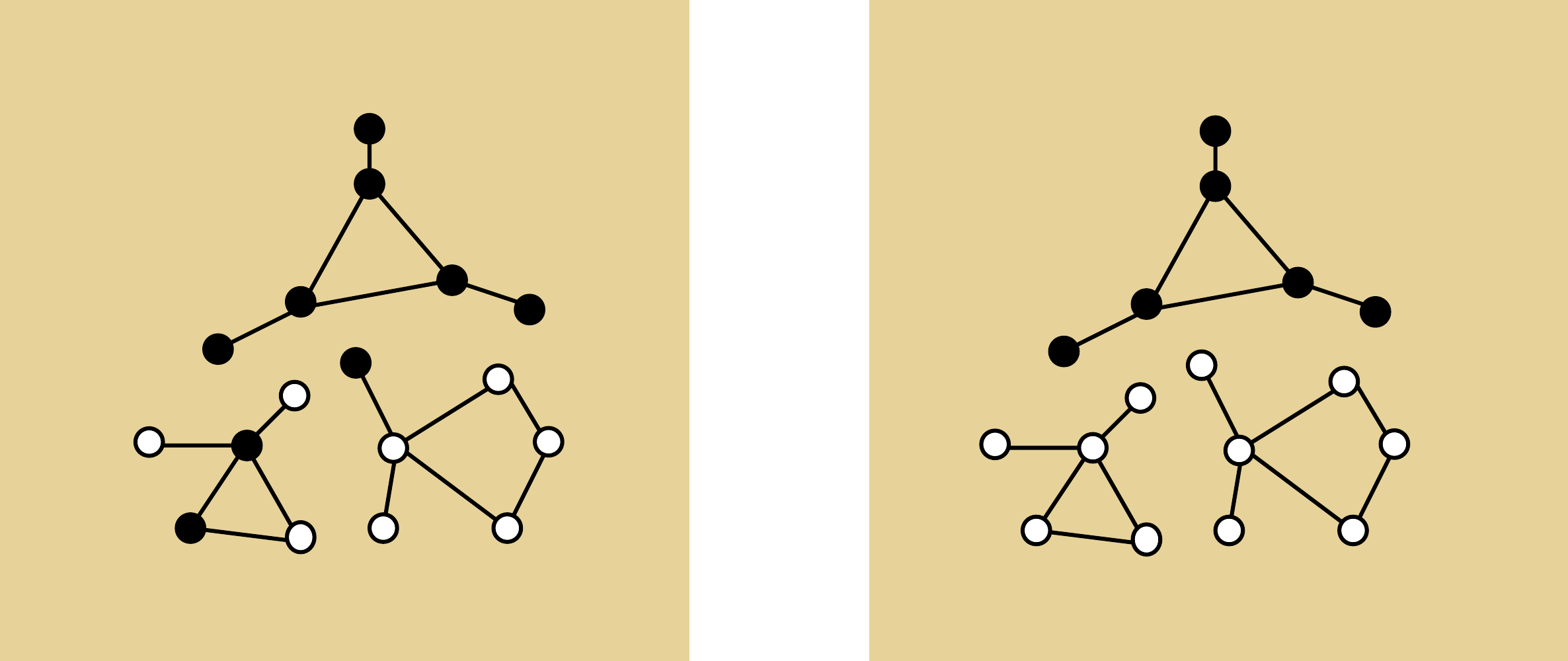}
\begin{quote}
\vspace{0.5cm}
\caption{\small 
The white birds have all been influenced
by ${\mathcal B}_1$: on the left,
they propagate that influence at the next step;
on the right, they have to wait for
flocks to join together before
the influence of ${\mathcal B}_1$ can expand further.
}
\end{quote}
\end{figure}

In both cases, 
$N(t_{k+1},t_k) \leq 2N(n-1) +1$, so summing over
all $0\leq k<l_1$,
$$N(t_{l_1},s)= 
\sum_{k=0}^{l_1-1} N(t_{k+1}, t_k)
\leq 2n N(n-1) + n.$$
Of course, there is nothing special about
bird ${\mathcal B}_{1}$. 
We can apply the same argument for each column
and conclude that the time $T_1$ when the matrix
$\underline{P}(t,s)$ has finally stabilized satisfies
\begin{equation}\label{N(T1)}
N(T_1,s) \leq 2n N(n-1) + n.
\end{equation}
The index set $V_1$ corresponding to the ones in
the first column of 
$\underline{P}(T_1,s)$ is called the first {\em stabilizer}.
For $t>T_1$, no edge of $G_t$
can join $V_1$ to its complement, since this
would immediately add more ones to 
the first column of $\underline{P}(t,s)$.
This means that ${\mathcal B}_{1}$ 
can no longer hope to influence
any bird outside of $V_1$ past time $T_1$.

Relabel the rows and columns so that all
the ones in $\underline{P}(T_1,s)$'s first column appear on top.
Then, for any $t>T_1$, $P(t)$ is a 2-block diagonal matrix
with the top left block, indexed by $V_1\times V_1$,
providing the transitions among the vertices of $V_1$ 
at time $t$. This is a restatement
of our observation regarding $G_t$ and $V_1$. 
Here is why. Since the footprint
of $P(t)$ is symmetric, it suffices to consider
the consequence of a nonzero, nondiagonal entry
in $P(t)$, ie,
$p_{ij}(t)>0$, with $i\not\in V_1$ and $j\in V_1$.
This would imply that 
$$p_{i1}(t,s)\geq p_{ij}(t)p_{j1}(t-1,s)>0,$$
and hence that $i\in V_1$, a contradiction.
Being 2-block diagonal is invariant under composition, so 
$P(t,T_1+1)$ is also a matrix of that type.
Let $A_{|V\times W}$ denote the submatrix of 
$A$ with rows indexed by $V$ and columns by $W$.
Writing $V_0=\{1,\ldots, n\}$, for $t>T_1$,
$$ P_{|V_1\times V_0}(t,s)= 
P_{|V_1\times V_1}(t,T_1+1)P_{|V_1\times V_0}(T_1,s).$$
By setting $s$ to $T_1+1$ we can repeat the same
argument, the only difference being that the 
transition matrices are now 
$|V_1|$-by-$|V_1|$. This leads to 
the second stabilizer $V_2\subseteq V_1$, which,
by relabeling, can be assumed to index the 
top of the subsequent matrices.
We define $T_2$ as the smallest integer 
such that
$\underline{P}_{\,|V_1\times V_1}(t,T_1+1)= 
 \underline{P}_{\,|V_1\times V_1}(T_2,T_1+1)$
for all $t>T_2$.
The set $V_2$ indexes the ones in the first
column of $\underline{P}_{\,|V_{1}\times V_1}(T_2,T_{1}+1)$.
Iterating in this fashion
leads to an infinite sequence of times $T_1<T_2<\cdots $
and stabilizers $V_1\supseteq V_2\supseteq \cdots $
such that, for any $t> T_{k}$,
\begin{multline*}
P_{|V_k\times V_0}(t,s) = 
P_{|V_k\times V_k}(t,T_k+1)
P_{|V_{k}\times V_{k-1}}(T_k,T_{k-1}+1)
\\
\cdots
P_{|V_2\times V_1}(T_2,T_1+1)
P_{|V_1\times V_0}(T_1,T_0+1),
\end{multline*}
where $P_{|V_{i}\times V_{i-1}}(T_i,T_{i-1}+1)$ is 
a $|V_i|$-by-$|V_{i-1}|$ matrix
and $T_0=s-1$. The stabilizers are the sets under refreshed
influence from ${\mathcal B}_1$.
We illustrate this decomposition below:

\begin{equation*}
A= 
\hbox{$\frac{1}{2}$}\!
\begin{pmatrix}
2 & 0& 0 \\
0 & 1& 1 \\
0 & 1& 1
\end{pmatrix} 
\hspace{2cm}
B= 
\hbox{$\frac{1}{2}$}\!
\begin{pmatrix}
1 & 1 & 0 \\
1 & 1 & 0 \\
0 & 0& 2
\end{pmatrix} 
\hspace{2cm}
C= 
\begin{pmatrix}
1 & 0& 0 \\
0 & 1 & 0 \\
0 & 0& 1
\end{pmatrix} .
\end{equation*}
Consider the word $M= CB^3CABABA$. The matrix
$M_{|V_6\times V_0}$ is factored as 
$$
C_{|V_6\times V_5} B_{|V_5\times V_4} B_{|V_4\times V_3}  
(BC)_{|V_3\times V_2} 
(AB)_{|V_2\times V_1} (ABA)_{|V_1\times V_0},
$$
where $V_0=V_1=V_2= \{1,2,3\}$,
$V_3=V_4=V_5=\{1,2\}$ and $V_6=\{1\}$.
The factorization looks like this:
\begin{equation*}
M_{|V_6\times V_0} = 
\begin{pmatrix}
1 & 0
\end{pmatrix} 
\cdot
\hbox{$\frac{1}{2}$} 
\begin{pmatrix}
1 & 1 \\
1 & 1
\end{pmatrix} 
\cdot
\hbox{$\frac{1}{2}$} 
\begin{pmatrix}
1 & 1 \\
1 & 1
\end{pmatrix} 
\cdot
\hbox{$\frac{1}{2}$} 
\begin{pmatrix}
1 & 1 & 0\\
1 & 1 & 0
\end{pmatrix} 
\cdot
\hbox{$\frac{1}{4}$} 
\begin{pmatrix}
2 & 2 & 0 \\
1 & 1 & 2 \\
1 & 1 & 2
\end{pmatrix} 
\cdot
\hbox{$\frac{1}{8}$} 
\begin{pmatrix}
4 & 2 & 2\\
2 & 3 & 3\\
2 & 3 & 3
\end{pmatrix} ,
\end{equation*}
with the infinite nested sequence
$$
V_1=\{1,2,3\} \supseteq\{1,2,3\}
\supseteq
\{1,2\} \supseteq \{1,2\} \supseteq\{1,2\}
\supseteq
\{1\} \supseteq\{1\} \supseteq\{1\}
\cdots
$$

\medskip

What is the benefit of rewriting
the top rows of $P(t,s)$ in such a complicated manner?
The first column of each 
$P_{|V_{i}\times V_{i-1}}(T_i,T_{i-1}+1)$ 
consists entirely of positive entries, so 
the submultiplicativity of the ergodicity
coefficients implies rapid convergence
of the products toward a rank-one matrix.
This has bearing on the relative displacement 
of birds and groupings into flocks.
By Lemma~\ref{x(t)-precision}, 
each entry in the first column of each 
$P_{|V_{i}\times V_{i-1}}(T_i,T_{i-1}+1)$ 
is at least $n^{-O(n^2)}$, so 
half the $\ell_1$-distance between any two rows
is at most $1- n^{-O(n^2)}\leq e^{-n^{-O(n^2)}}$; therefore
$$\tau_1( P_{|V_{i}\times V_{i-1}}(T_i,T_{i-1}+1))
\leq  e^{- n^{-O(n^2)}} .$$
Lemma~\ref{ergodicity-submult} implies that 
$\tau_2(A)\leq 
\tau_1(A)\tau_2(I)\leq \sqrt{2}\, \tau_1(A)$, and 
\begin{equation}\label{tau2(P-vk)}
\begin{split}
\tau_2( P_{|V_k\times V_0}(t,s)  ) 
&\leq 
\sqrt{2}\,
\tau_1( P_{|V_k\times V_k}(t,T_k+1) )
\prod_{i=1}^{k} 
\tau_1( P_{|V_{i}\times V_{i-1}}(T_i,T_{i-1}+1) ) \\
&\leq 
\sqrt{2}\, e^{- k n^{-O(n^2)}}.
\end{split}
\end{equation}
Let $\chi(i,j)$ denote the $n$-dimensional vector with 
all coordinates equal to $0$, except for 
$\chi(i,j)_i=1$ and $\chi(i,j)_j=-1$. 
Note that $$v_i(t)- v_j(t)= 
((\chi(i,j)P(t-1,1))\otimes I_d)v(1);
$$
therefore, by Cauchy-Schwarz and~(\ref{|v|Poly}),
\begin{equation}\label{tau2(vi-vj)CS}
\|v_i(t)- v_j(t) \|_2\leq \sqrt{d}\, \tau_2 ( P(t-1,1) )\|v(1)\|_2
\leq \tau_2 ( P(t-1,1) )2^{O(\mathfrak{p})}.
\end{equation}
If we restrict $i,j$ to $V_k$, we can 
replace $P(t-1,1)$ by $P_{|V_k\times V_0}(t-1,1)$
and write 
\begin{equation*}
\|v_i(t)- v_j(t) \|_2\leq
\tau_2 ( P_{|V_k\times V_0}(t-1,1) )2^{O(\mathfrak{p})}.
\end{equation*}
Setting $k=  n^{b_0 n^2} 
\lceil \mathfrak{p} + \log \frac{1}{\varepsilon_{\! h}}\rceil$ for 
a large enough integer constant $b_0>0$,
we derive from~(\ref{tau2(P-vk)}) that,
for any $t > T_k+1$,

\begin{equation}\label{maxvivjn^n^2}
\max_{i,j\in V_k} 
\|v_i(t)- v_j(t) \|_2
\leq  e^{- kn^{-O(n^2)} + O(\mathfrak{p}) }
< \varepsilon_{\! h}\, .
\end{equation}
By Lemma~\ref{relativeDisp}, it then follows that 
$\Delta_{ij}(t) 
< \varepsilon_{\! h}$.
By the hysteresis rule, this means that 
if birds ${\mathcal B}_i$ and ${\mathcal B}_j$ 
are joined after time $T_k+1$, they will always remain so.
This leaves at most $\binom{|V_k|}{2}$ extra network changes
(final pairings), so the total number is conservatively 
bounded by 
$$ N(T_k, T_{k-1})+\cdots+ N(T_1,1) + \binom{|V_k|}{2} .$$
But~(\ref{N(T1)}) holds for any pair $(T_i, T_{i-1}+1)$, so
$$N(n)< k( 2n N(n-1) + n) + n^2 .$$
Since $N(1)=0$, for all $n>1$,
$$N(n) = n^{O(n^3)}(\mathfrak{p} + 
      \log \hbox{$\frac{1}{\varepsilon_{\! h}}$})^{n-1}.
$$
There is a technical subtlety we need
to address. In the inductive step defining
$N(n-1)$, and more generally $N(n')$ for $n'<n$, 
the initial conditions and element sizes
of the transition matrices should be treated as 
global parameters: they depend on $n$, not $n'$.
In fact, it is safe to treat $n$ as a fixed parameter
everywhere, except in the recurrence~(\ref{N(T1)}).
The key observation is that, as $n'$ decreases,
the bounds provided by~(\ref{tau2(P-vk)}) 
and in the setting of 
$k= n^{b_0 n^2} 
(\mathfrak{p} + \log \frac{1}{\varepsilon_{\! h}})$ 
still provide valid---in fact, increasingly
conservative---estimates as $n'$ decreases.
The noise is handled by reapplying the bound after
each of the $e^{O(n^3)}$ perturbations.
\hfill $\Box$
\proofend

\begin{figure}[htb]\label{fig-infinity-noframe}
\begin{center}
\hspace{0cm}
\includegraphics[width=9cm]{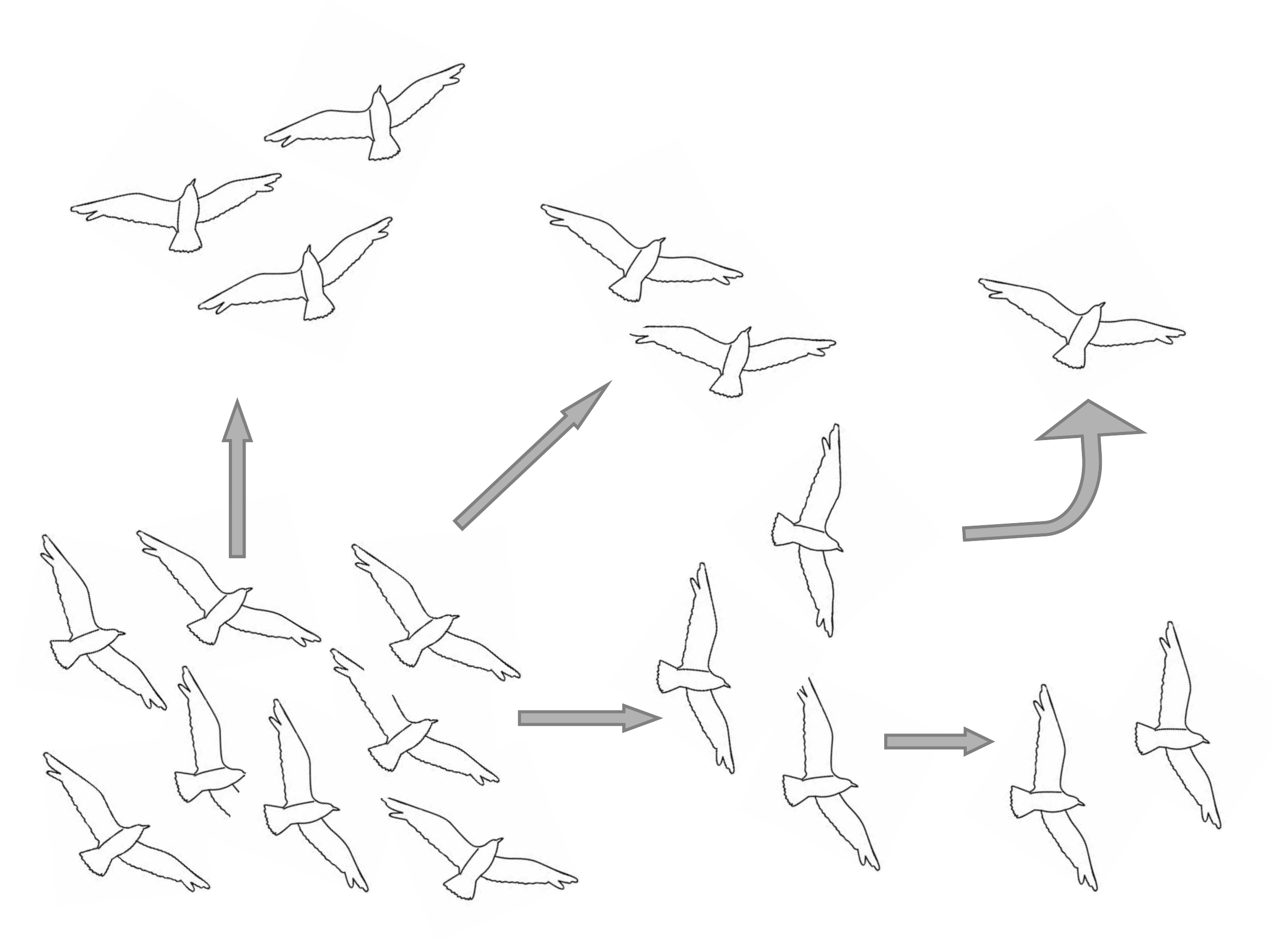}
\end{center} 
\begin{quote}
\vspace{0cm}
\caption{\small 
The arborescence of birds separating
into groups.
}
\end{quote}
\end{figure}

\medskip
\noindent
{\sl Remark 2.1}.
The rationality of positions and velocities 
was never used in the proof.
The only requirement is that the initial velocities
of the birds should have Euclidean norm in $2^{O(\mathfrak{p})}$.

\bigskip\medskip
\noindent
{\sl Remark 2.2}.
The nested sequence $V_1\supseteq V_2\supseteq \cdots $
is infinite but the number of different
subsets obviously is not. The smallest 
stabilizer $V_i$, denoted $V_{k_1}$ to indicate its
relation to ${\mathcal B}_1$, 
cannot be empty since a bird influences itself for ever;
hence $\{1\}\in V_{k_1}$.
If $|V_{k_1}|>1$, then ${\mathcal B}_1$ 
influences all the birds in $V_{k_1}$ recurrently, ie, infinitely often.
In fact, this is true not just of 
${\mathcal B}_1$ but of all $V_{k_1}$, all of whose
birds influence all others in that set recurrently.
The sets $V_{k_1}, \ldots, V_{k_n}$ are 
therefore pairwise disjoint or equal.
This implies a partition of the bird set 
into recurrently self-influencing classes.
One can model the process leading to it
as an arborescence whose root corresponds to the 
first time the set of $n$ birds is split into
two subsets that will no longer influence each other.
Iterating in this fashion produces a tree whose
leaves are associated with the disjoint $V_{k_j}$'s.
Note that the stabilizers $V_1, V_2$, etc, are
specific to ${\mathcal B}_1$ and their counterparts
for ${\mathcal B}_2$ might partly overlap with them
(except for the last one); therefore, the path
in the tree toward the leaf labeled $V_{k_1}$
cannot be inferred directly from the stabilizers.

\paragraph{Time-Invariant Flocking.}

Birds are expected to spend most
of their time flying in fixed flocks.
We investigate this case separately.
The benefit is to derive
a convergence time that is exponentially
faster than in the general case.
In this section, $G_t=G$ is time-invariant;
for notational convenience, we assume 
there is a single flock, ie, $G_t$ is connected.
The flocking is noise-free.
We can express the stochastic matrix
$P$ as $I_n-CL$. 
The corresponding Markov chain is reversible and,
because of connectivity, irreducible. 
The diagonal being nonzero, it is aperiodic, hence ergodic.
The transition matrix $P$ has
the simple dominant eigenvalue $1$ with 
right and left eigenvectors ${\mathbf 1}$ and 
$$\pi = \frac{1}{\hbox{tr}\,C^{-1}}\,C^{-1} \, {\mathbf 1},$$
respectively. Lack of symmetry does not
keep $P$ from being diagonalizable, though it 
denies us eigenvector orthogonality.
Define 
\begin{equation}\label{MP-diagonalized}
M = C^{-1/2} P C^{1/2}= 
C^{-1/2} (I_n-C L) C^{1/2}= I_n- C^{1/2}  L C^{1/2}.
\end{equation}
Being symmetric, $M$ can be diagonalized as 
$\sum_{k=1}^n \lambda_k u_k u_k^T$,
where the $u_k$'s are orthonormal eigenvectors
and the eigenvalues are real.
It follows that $P$ can be diagonalized as well, with 
the same eigenvalues.
By Perron-Frobenius and standard properties 
of ergodic walks~\cite{ChungSpectralGraphTheory,seneta06},
$1=\lambda_1>\lambda_2\geq\cdots\geq \lambda_n> -1$
and $u_1= (\sqrt{\pi_1},\ldots, \sqrt{\pi_n}\,)^T$.
Since $\sum_k u_k u_k^T= I_n$, the following identity
holds for all nonnegative $s$, including $s=0$ (for which
we must assume that $0^0=1$):
\begin{equation}\label{P_N}
P^s= C^{1/2} M^s C^{-1/2}
= {\mathbf 1} \pi^T + \sum_{k=2}^n \lambda_k^s
        C^{1/2} u_k u_k^T  C^{-1/2}.
\end{equation}
The left and right eigenvectors of $P$ for $\lambda_k$ 
are given (in column form) 
by $C^{-1/2} u_k $ and $C^{1/2} u_k$ and,
together, form inverse matrices; in general,
neither group forms an orthogonal basis.
We can bound the second largest eigenvalue by 
using standard algebraic graph theory.
We include a proof for completeness.

\begin{lemma}\label{eigenBounds}
$\!\!\! .\,\,$
If $\mu\,\defeq\, \max_{\,k>1}|\lambda_k|$, then
$\mu\leq 1 -n^{-O(1)}$.
\end{lemma}
\proof
By the $O(\log n)$-bit encoding of $C$,
each diagonal of $P$ is 
at least $n^{-b}$, for some constant.\footnote{\, To simplify
the notation, constants such as $b$ and $c$ 
are reused frequently in the text,
with their values depending on the context.} 
The matrix 
$(1-\frac{1}{2} n^{-b})^{-1} (P-\frac{1}{2} n^{-b} \, I_n)$ 
is stochastic and all of its eigenvalues all lie in $[-1,1]$.
It follows that $\lambda_{n-1}\geq  n^{-O(1)}-1$, 
for any $k>1$.
Observe now that 
$1-\lambda_2$ is the smallest positive
eigenvalue of the normalized Laplacian $C^{1/2}  L C^{1/2}$.
The simplicity of the eigenvalue $0$ (by connectivity) implies
that any eigenvector of 
the normalized Laplacian
corresponding to a nonzero eigenvalue
is normal to $C^{-1/2}{\mathbf 1}$; therefore,
by Courant-Fischer,
$$1-\lambda_2=\min \Bigl\{\, x^T C^{1/2}  L C^{1/2} x \,\,:\,\,\,
{\mathbf 1}^T C^{-1/2}x=0 \,\,\,\, \text{and} \,\,\,\, \|x\|_2=1\, \Bigr\}.$$
Write $y= C^{1/2}x$ and express the system in the equivalent form:
$1-\lambda_2=\min y^T L y$, subject to (i) ${\mathbf 1}^T C^{-1}y=0$ and 
(ii) $\|C^{-1/2}y\|_2=1$.
By using ideas 
from~\cite{ChungSpectralGraphTheory,landauO}, we 
argue that, for some $m$ and $M$, 
by (i), $y_m\leq 0$, for some $m$, and from (ii)
$y_M\geq (\hbox{tr}\,C^{-1})^{-1/2}$.
Since $G$ is connected, there exists a path $\mathcal M$ 
of length at most $n$ joining nodes $m$ and $M$. Thus, by Cauchy-Schwarz,
the solution $y$ of the system satisfies:
\begin{equation*}
\begin{split}
1-\lambda_2 &= y^T L y 
= \sum_{(i,j)\in G} (y_i-y_j)^2
\geq \sum_{(i,j)\in {\mathcal M}} (y_i-y_j)^2 
\geq \frac{1}{n} \Bigl( \sum_{(i,j)\in {\mathcal M}} |y_i-y_j| \Bigr)^2 \\
&\geq \frac{1}{n} |y_M-y_m|^2\geq 
\frac{1}{n (\hbox{tr}\,C^{-1})} = n^{-O(1)}  \, .
\end{split}
\end{equation*}
\hfill $\Box$
\proofend

\noindent
By~(\ref{P_N}), for all $i,j,s>0$,
$(P^{s})_{ij}
\geq \pi_j - 
\sum_{k>1} |\lambda_k|^{s}
\sqrt{\smash[b]{c_i/c_j}}\, |(u_k)_i (u_k)_j|
\geq \pi_j - n^{O(1)} \mu^{s}$.
A similar derivation gives us the corresponding
upper bound; so,\footnote{\, The Frobenius
norm $\|M\|_F$ of a matrix is the 
Euclidean norm of the vector formed by its elements.
The property we will use most often is 
a direct consequence of Cauchy-Schwarz,
$\|Mu\|_2\leq \|M\|_F\|u\|_2$, and more generally
the submultiplicativity of the norm.
}
by Lemma~\ref{eigenBounds},
\begin{equation}\label{Pij^sBounds}
\| P^s - {\mathbf 1} \pi^T \|_F \leq  e^{-s n^{-O(1)}+ O(\log n)}.
\end{equation}
Similarly, for $s> n^{c_0}$, for a constant $c_0$ large enough,
\begin{equation}\label{tau-Spectral-log}
\begin{split}
\tau_1(P^{s})
&=
1- \min_{i,j} \, \sum_{k=1}^n \, 
\min\,\{(P^{s})_{ik}\, , \, (P^{s})_{jk}\} \\
&\leq 1- \sum_{k=1}^n 
(\pi_k - n^{O(1)}  e^{-s n^{-O(1)}})
= n^{O(1)}   e^{-s n^{-O(1)}} < \hbox{$\frac{1}{2}$}\, .
\end{split}
\end{equation}

Given a vector $\xi$ in ${\mathbb R}^n$, consider
the random variable $X$ formed by picking the
$i$-coordinate of $x$ with probability $\pi_i$.
As claimed in the introduction,
the variance of $X$ is a quadratic Lyapunov function.
This is both well known 
and intuitively obvious since
we are sampling from the stationary distribution
of an ergodic Markov chain and then taking one
``mixing'' step: the standard deviation decreases
at a rate given by the Fiedler value.
As was observed in~\cite{P-07-lyapunov-rev},
because the random variable involves only 
$\pi$ and not $P$, any flock switching
that keeps the graph connected
with the same stationary distribution
admits a common quadratic Lyapunov function.
If $\xi= {\mathbf 1}$, then 
obviously, $\var X =0$. We now show that 
the variance decays exponentially fast.

\begin{lemma}\label{lyapunovVariance}
$\!\!\! .\,\,$
\ $\var (PX) \leq \mu^2 (\var X)$.
\end{lemma}
\proof
For any $\xi$, the vector 
$y= (I_n- {\mathbf 1}\pi^T)\xi$ is such that
$C^{-1/2}y$ is orthogonal to
$u_1= (\sqrt{\pi_1},\ldots, \sqrt{\pi_n}\,)^T$.
Therefore the latter lies in the contractive eigenspace
of $M$ and 
$$\|M(C^{-1/2}y)\|_2\leq \mu \|C^{-1/2}y\|_2\, ;$$
hence, by~(\ref{MP-diagonalized}),
\begin{equation*}
\begin{split}
(Py)^TC^{-1}(Py)
&= (y^T C^{-1/2})(C^{1/2}P^T C^{-1/2})
(C^{-1/2}PC^{1/2})(C^{-1/2}y) \\
&= \|MC^{-1/2}y\|_2^2\leq \mu^2  \|C^{-1/2}y\|_2^2 \, .
\end{split}
\end{equation*}
As a result,
$$(Py)^TC^{-1}(Py)\leq \mu^2 y^TC^{-1}y.$$
Since $\pi = (\hbox{tr}\,C^{-1})^{-1}C^{-1} \, {\mathbf 1}$,
$$
\var X= \sum_{i=1}^n \pi_i\Bigl(\xi_i- \sum_i \pi \xi_i\Bigr)^2
= 
\xi^T(I_n- \pi{\mathbf 1}^T)
\frac{C^{-1}}{ \hbox{tr}\,C^{-1} }
(I_n- {\mathbf 1}\pi^T\,)\xi
= y^T\frac{C^{-1}}{ \hbox{tr}\,C^{-1} }y \, .
$$
Because $P$
commutes with $I_n- {\mathbf 1}\pi^T$, 
$$\var (PX) = (Py)^T
\frac{C^{-1}}{ \hbox{tr}\,C^{-1} } (Py)
\leq \mu^2 (\var X),$$
and $\var X$ is 
the desired Lyapunov function.
\hfill $\Box$
\proofend

What both~(\ref{tau-Spectral-log})
and Lemma~\ref{lyapunovVariance} indicate
is that convergence for
a time-invariant flock evolves as
$e^{-tn^{-O(1)}}$, whereas in general
the best we can do is invoke~(\ref{tau2(P-vk)})
and hope for a convergence speed of
the form $e^{-tn^{-O(n^2)}}$, which is
exponentially slower.

\paragraph{The Rationality of Limit Configurations.}

The locations of the birds remain
rational at all times. Does this mean that
in the limit their 
configurations remain so? We prove that this is, indeed,
the case. We do not do this simply out of curiosity.
This will be needed for the analysis of convergence.
We cover the case of a time-invariant connected
network here and postpone the general case for later.
For $t>0$, we define

\begin{equation}\label{Gamma_tSUm}
{\Gamma}_t =  - {\mathbf 1} \pi^T t + \sum_{s=0}^{t-1} P^s. 
\end{equation}
It is immediate that
${\Gamma}_t$ converges 
to some matrix $\Gamma$, 
as $t$ goes to infinity.
Indeed, by~(\ref{P_N}), 
$$
\Gamma = 
\sum_{s\geq 0}( P^s- {\mathbf 1} \pi^T )= 
\sum_{k>1} \hbox{$\frac{1}{1-\lambda_k}$}
\,       C^{1/2} u_k u_k^T  C^{-1/2}.
$$
What is perhaps less obvious is why the limit is rational.
We begin with a simple characterization
of $\Gamma$, which we derive by
classical arguments about fundamental matrices
for Markov chains~\cite{KemenySnell}.
We also provide a more ad hoc characterization
(Lemma~\ref{Gamma-LimitTwo})
that will make later bound estimations somewhat easier.

\begin{lemma}\label{GammaLimit}
$\!\!\! .\,\,$
As $t\rightarrow \infty$, 
${\Gamma}_t$ converges to
${\Gamma} = -{\mathbf 1} \pi^T 
+ (I_n- P + {\mathbf 1} \pi^T)^{-1}$.
\end{lemma}
\proof
Because ${\mathbf 1}$ and $\pi$ are
respectively right and left eigenvectors of $P$ for
the eigenvalue $1$, for any integer $s>0$,
\begin{equation}\label{(P-pi)^s=Ps-pi}
( P- {\mathbf 1} \pi^T )^s
= P^s - {\mathbf 1} \pi^T.
\end{equation}
This follows from the identity
\begin{equation*}
\begin{split}
(P- {\mathbf 1} \pi^T )^s
&= 
P^s+ \sum_{k=0}^{s-1} (-1)^{s-k} \binom{s}{k} P^k
( {\mathbf 1} \pi^T )^{s-k}  \\
&=
P^s+ ( {\mathbf 1} \pi^T )
\sum_{k=0}^{s-1} (-1)^{s-k} \binom{s}{k}
= 
P^s - {\mathbf 1} \pi^T \, .
\end{split}
\end{equation*}
And so, for $t>1$, 
\begin{equation*}
{\Gamma}_t + {\mathbf 1} \pi^T 
 =  I_n 
     + \sum_{s=1}^{t-1} (P^s -{\mathbf 1} \pi^T)
 =  \sum_{s=0}^{t-1} (P -{\mathbf 1} \pi^T)^s \, .
\end{equation*}
Pre-multiplying this identity by the ``denominator''
that we expect from the geometric sum, 
ie, $I_n - P +{\mathbf 1} \pi^T$, we simplify
the telescoping sum, using~(\ref{(P-pi)^s=Ps-pi}) again,
\begin{equation*}
\begin{split}
(I_n - P +{\mathbf 1} \pi^T)
( {\Gamma}_t + {\mathbf 1} \pi^T )
&= (I_n - P +{\mathbf 1} \pi^T)
        \sum_{s=0}^{t-1} (P -{\mathbf 1} \pi^T)^s \\
&=  I_n - (P-{\mathbf 1}\pi^T)^t
=  I_n - (P^t - {\mathbf 1}\pi^T)
\end{split}
\end{equation*}
By~(\ref{P_N}), $P^t$ converges to ${\mathbf 1}\pi^T$
as $t$ goes to infinity, so
$(I_n - P +{\mathbf 1} \pi^T)
( {\Gamma}_t + {\mathbf 1} \pi^T )$
converges to the identity.
This implies that, for $t$ large enough, 
the matrix cannot be singular and, hence, neither can
$I_n - P +{\mathbf 1} \pi^T$.
This allows us to write:
$$
{\Gamma} + {\mathbf 1} \pi^T 
=  (I_n - P +{\mathbf 1} \pi^T)^{-1}.
$$ 
\hfill $\Box$
\proofend

There is another characterization of 
$\Gamma$ without $\pi$ in the inverse matrix.
We use the notation $(Y\,|\,y)$ 
to refer to the $n$-by-$n$ matrix 
derived from $Y$ by replacing its last column 
with the vector $y$.

\begin{lemma}\label{Gamma-LimitTwo}
$\!\!\! .\,\,$
${\Gamma} = 
(I_n- {\mathbf 1} \pi^T \,|\, {\mathbf 0})\,
(I_n-P \,|\, {\mathbf 1}\,)^{-1}$.
\end{lemma}
\proof
Since $\pi$ is a left eigenvector of $P$ for $1$,
${\mathbf 1} \pi^T (I_n-P)=0;$
hence, for $t>0$,
$$
I_n-P^{t} = (I_n+P+\cdots+ P^{t-1})(I_n-P) 
=  (\,{\Gamma}_t + {\mathbf 1} \pi^T t )(I_n-P)
=  {\Gamma}_t\,(I_n-P).$$
As $t\rightarrow \infty$,
$P^t \rightarrow {\mathbf 1} \pi^T$; therefore
${\Gamma}\, (I_n-P) = I_n- {\mathbf 1} \pi^T$. 
Since ${\mathbf 1}$ lies in the kernel of
${\Gamma}_t$, and hence of ${\Gamma}$,
the latter matrix satisfies the relation
\begin{equation}\label{gamma-ip}
{\Gamma}\,(I_n-P\,|\,{\mathbf 1}) = 
(I_n- {\mathbf 1} \pi^T\,|\,{\mathbf 0}).
\end{equation}
The simplicity of $P$'s dominant eigenvalue $1$ implies
that $I_n-P$ is of rank $n-1$.
Since ${\mathbf 1}\in \text{ker}\, (I_n-P)$,
the last column of $I_n-P$ is the negative sum of the others; 
so to get the correct rank the first $n-1$ columns
of $I_n-P$ must be independent.
Note that the vector ${\mathbf 1}$ is not 
in the space they span: if, indeed, it were, we would have
${\mathbf 1}= (I_n-P)y$, for some $y\in {\mathbb R}^n$. Since
$\pi^T (I_n-P)=0$, this would imply that
$1= \pi^T{\mathbf 1}= \pi^T (I_n-P)y= 0$, 
a contradiction.
This is evidence that $(I_n-P\,|\,{\mathbf 1})$ is of full rank,
which, by~(\ref{gamma-ip}), completes the proof.
\hfill $\Box$
\proofend

The motion equation~(\ref{modelD}) becomes,
for $t\geq 1$,
\begin{equation}\label{x(t)-SumP^s}
x(t)= x(0)+ \Bigl(\, \sum_{s=0}^{t-1} P^s \otimes I_d
\,\Bigr) v(1)
\end{equation}
or, equivalently, by~(\ref{Gamma_tSUm}),
\begin{equation}\label{x(t)-Gamma_t}
x(t)= x(0) + 
t (({\mathbf 1} \pi^T) \! \otimes I_d)v(1)
+ (\, {\Gamma}_t \otimes I_d ) v(1).
\end{equation}
We call ${\mathbf m}_\pi[x(t)] =
(\pi^T\! \otimes I_d) x(t)$ the {\em mass center}
of the flock
and the vector ${\mathbf m}_\pi[v(1)]$
its {\em stationary velocity}.
The latter is the first spectral 
(vector) coefficient
of the velocity.
In our lower bound, we will make it the first
Fourier coefficient of the dynamical system.
The mass center drifts in space at constant
speed along a fixed line in $d$-space:
Indeed, $\pi^T\Gamma_t=0$, so
by~(\ref{x(t)-Gamma_t}),
\begin{equation*}
{\mathbf m}_\pi[x(t)]
= 
{\mathbf m}_\pi[x(0)]
+ t 
{\mathbf m}_\pi[v(1)]
\end{equation*}
and
\begin{equation}\label{x=mpi-v1}
x(t)= 
\underset{\text{\em start}}{\underbrace{     
x(0)}}+ 
\underset{\text{\em linear drift}}{\underbrace{     
t ({\mathbf 1} \otimes I_d){\mathbf m}_\pi[v(1)]
}}
+ 
\underset{\text{\em damped oscillator}}{\underbrace{     
(\, {\Gamma}_t \otimes I_d ) v(1)}} \, .
\end{equation}
The oscillations
are damped at a rate of $e^{-tn^{-O(1)}}$.
(We use the term not in
the ``harmonic'' sense but by reference to 
the negative eigenvalues that might cause actual
oscillations.)
Moving the origin to the mass center of the birds,
we express $x(t)$, relative to this moving frame, as 
$$ x^r(t) = x(t) - ({\mathbf 1} \otimes I_d) {\mathbf m}_\pi[x(t)];$$
therefore, by simple tensor manipulation,
\begin{equation}\label{x=xr+stuff}
x(t)= x^r(t)+ 
(({\mathbf 1} \pi^T) \! \otimes I_d)x(0)
+t (({\mathbf 1} \pi^T) \! \otimes I_d)v(1);
\end{equation}
and, by~(\ref{x(t)-Gamma_t}),
$$
x^r(t) = x(t)- (({\mathbf 1} \pi^T)\! \otimes I_d)x(t)
= ((I_n- {\mathbf 1} \pi^T) \! \otimes I_d)x(0)
+ (\, {\Gamma}_t \otimes I_d ) v(1)
$$
and, by Lemma~\ref{GammaLimit},

\begin{lemma}\label{flockConverges}
$\!\!\! .\,\,$
If $G$ is connected, the
relative flocking configuration 
$x^r(t)$ converges to the limit
$$
x^r = 
((I_n- {\mathbf 1} \pi^T) \! \otimes I_d)x(0) +
(\Gamma \otimes I_d ) v(1).
$$
The mass center of the configuration moves
in ${\mathbb R}^d$ at constant speed in a fixed direction.
\end{lemma}

\begin{lemma}\label{GammaCoeff}
$\!\!\! .\,\,$
The elements of $\Gamma$ 
and the coordinates of the limit configuration 
$x^r$ are CD-rationals over 
$O(n\log n)$ and $O(n\log n +\mathfrak{p} n)$ bits,
respectively.
\end{lemma}
\proof
Let $C_b$ denote the $O(n\log n)$-bit long 
product of all the denominators
in the diagonal matrix $C$.
The determinant of $(CL\,|\,{\mathbf 1})$
can be expressed as $C_b^{-1}$ times
the determinant $N$ of 
an $n$-by-$n$ matrix with $O(\log n)$-bit integer elements.
By the Hadamard bound~\cite{yap00},
$N$ is an $O(n\log n)$-bit integer.
For the same reason, each element of 
$\text{adj}\, (CL\,|\,{\mathbf 1})$ is also 
the product of $C_b^{-1}$ with an $O(n\log n)$-bit integer;
therefore,
$$(I_n-P\,|\,{\mathbf 1})^{-1}= 
(CL\,|\,{\mathbf 1})^{-1}= 
\frac{\text{adj}\, (CL\,|\,{\mathbf 1})}
     {\text{det}\, (CL\,|\,{\mathbf 1})} 
$$ 
is of the form $N^{-1}$ times an $O(n\log n)$-bit integer matrix
(since the two appearances of $C_b^{-1}$ cancel out).
The same is true of 
$(I_n- {\mathbf 1} \pi^T\,|\, {\mathbf 0})$:
this is because, trivially, 
$\pi^T= (0,\ldots,0,1)(I_n-P\,|\,{\mathbf 1})^{-1}$.
Therefore, both 
$(I_n- {\mathbf 1} \pi^T\,|\, {\mathbf 0})$
and $(I_n-P\,|\,{\mathbf 1})^{-1}$
are matrices with CD-rational coordinates
over $O(n\log n)$ bits.
Lemma~\ref{flockConverges},
with the formulation of Lemma~\ref{Gamma-LimitTwo}
for $\Gamma$, completes the proof.
\hfill $\Box$
\proofend

This implies that
$x(t)$ tends toward $a+bt$, where $a,b$ are rational vectors.
Since the number of switches and perturbations is
finite, this proves the rationality claim made
in~\S\ref{introduction}.
\hfill $\Box$
\proofend

\paragraph{Soundness of the Hysteresis Rule.}\label{NeedHysteresis}

We begin with a proof that hysteresis is required
to ensure convergence.
We build a 4-bird flock in one dimension,
whose network
cannot converge without a hysteresis rule.
The construction can be trivially
lifted to any dimension.
The speed of the birds will decay exponentially.
In real life, of course, the birds would stall.
But, as we mentioned earlier, we can add a large
fixed velocity to all the birds without altering
the flocking process. Stalling, therefore,
is a nonissue, here and throughout this work.
These are the initial conditions:
\begin{equation*}
\begin{cases}
\,
x(0) \, = 
\hbox{$\frac{1}{16}$}
(0, 8, 21, 29 ); \\
\, v(1)= 
\hbox{$\frac{1}{8}$}
(1, -1, 1, -1).
\end{cases}
\end{equation*}
The flocking network alternates between 
a pair of 2-bird edges and a single 4-bird path, 
whose respective transition matrices are:
\begin{equation*}
\frac{1}{3}
\begin{pmatrix}
1 & 2 & 0 & 0 \\
2 & 1 & 0 & 0 \\
0 & 0 & 1 & 2 \\
0 & 0 & 2 & 1 
\end{pmatrix} 
\hspace{1.8cm}
\text{and}
\hspace{1.8cm}
\frac{1}{3}
\begin{pmatrix}
1 & 2 & 0 & 0 \\
1 & 1 & 1 & 0 \\
0 & 1 & 1 & 1 \\
0 & 0 & 2 & 1 
\end{pmatrix} .
\end{equation*}
The beauty of the initial velocity $v(1)$
is that it is a right eigenvector for both
flocking networks for the same eigenvalue
$-\frac{1}{3}$; therefore,
for $t>0$, $v(t)= (-3)^{1-t} v(1)$ and, 
by~(\ref{modelD}),
\begin{equation}\label{xt=x0+sumv}
x(t)= x(0) + 
\sum_{s=1}^{t} v(s)
= x(0) + \hbox{$\frac{3}{4}$}
\Bigl( 1- (-\hbox{$\frac{1}{3}$})^t \Bigr)v(1).
\end{equation}
It follows that
\begin{equation*}
x_{i+1}(t) - x_{i}(t) =
\begin{cases}
\, \hbox{$\frac{1}{16}$}  
( 5- (-\hbox{$\frac{1}{3}$})^{t-1} )
& \text{if $i=1,3$}; \\
\, 1+ \hbox{$\frac{1}{16}$}  
(-\hbox{$\frac{1}{3}$})^{t-1}
& \text{if $i=2$} . 
\end{cases}
\end{equation*}
The distance between the first and second birds
stays comfortably between 
$\hbox{$\frac{1}{4}$}$ and 
$\hbox{$\frac{1}{2}$}$; same with 
birds ${\mathcal B_3}$ and ${\mathcal B_4}$.
The distance between the middle birds
${\mathcal B_2}$ and ${\mathcal B_3}$
oscillates around 1, so the 
network forever alternates between
one and two connected components.
The pairs 
$({\mathcal B_1}, {\mathcal B_3})$
and 
$({\mathcal B_2}, {\mathcal B_4})$
form fixed inter-bird distances
of $\frac{21}{16}$, so the flocks are always
simple paths.
This proves the necessity of hysteresis.
As we said earlier, virtually any hysteresis rule
would work. Ours is chosen out of convenience.

\vspace{1cm}
\begin{figure}[htb]\label{fig-oscillate-background}
\begin{center}
\hspace{0.3cm}
\includegraphics[width=8.5cm]{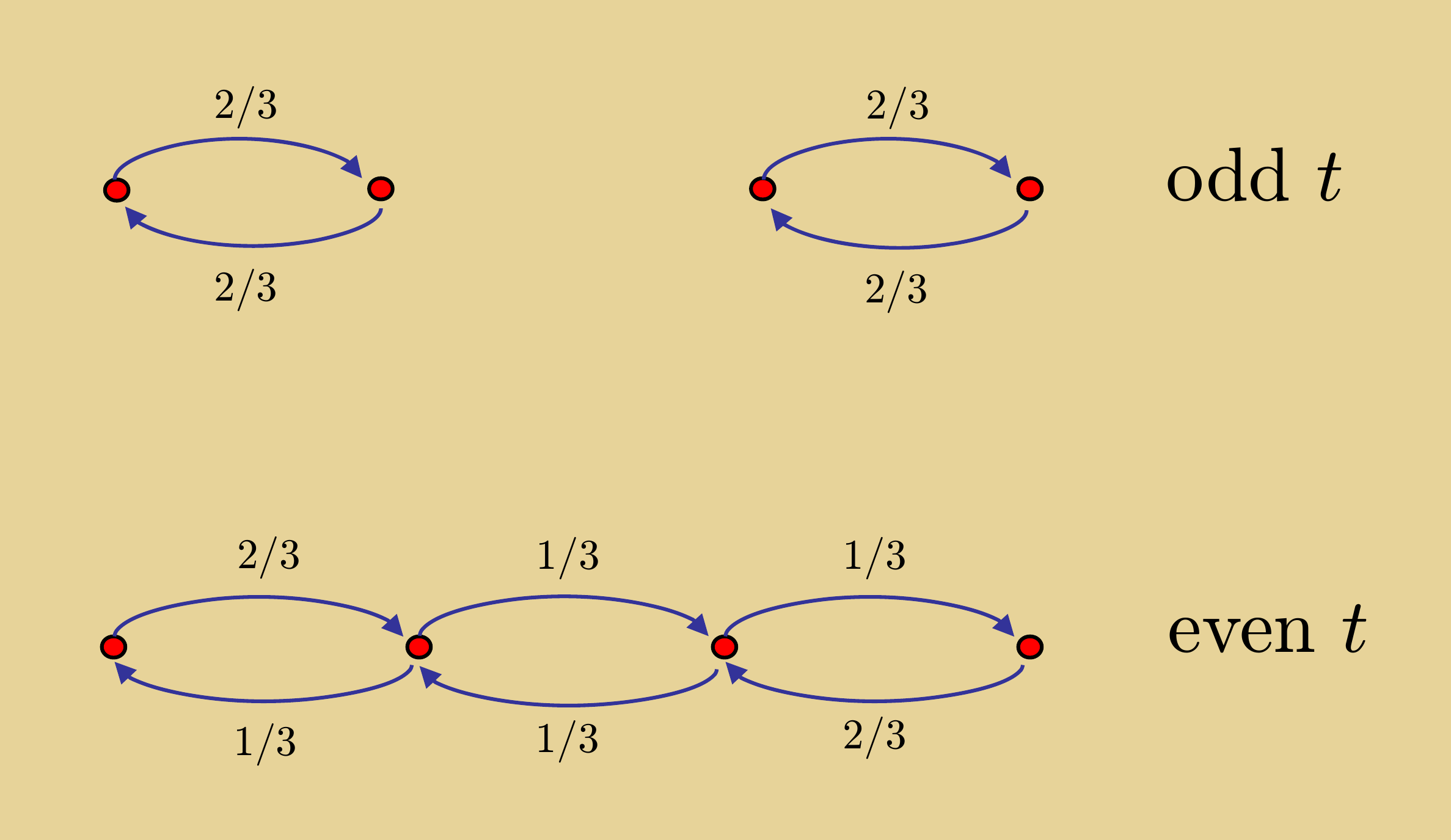}
\end{center}
\begin{quote}
\vspace{0.2cm}
\caption{\small 
The flocking network alternates between
two configurations forever and never converges.
}
\end{quote}
\end{figure}

\begin{lemma}\label{NoSnapTransit}
$\!\!\! .\,\,$
The hysteresis rule is sound:
(i) any two birds within unit distance of each other 
at time $t$ 
share an edge of $G_t$;
(ii) no two birds at distance greater than 
$1+ \gamma \varepsilon_{\! h}$
are ever adjacent in $G_t$, where 
$$\gamma= (\mathfrak{p} + 
\log \hbox{$\frac{1}{\varepsilon_{\! h}}$})^n
\, n^{O(n^3)}.
$$
\end{lemma}

\smallskip

\begin{corollary}\label{corol-NoSnapTransit}
$\!\!\! .\,\,$
Under the default settings~(\ref{Assumptions}),
any two birds within unit distance of each other 
at time $t$ 
share an edge of $G_t$; 
on the other hand, no two birds at distance greater than 
$1+ \sqrt{ \varepsilon_{\! h} }$
are ever adjacent in $G_t$.
\end{corollary}

\smallskip

{\noindent\em Proof of Lemma~\ref{NoSnapTransit}. }
Part (i) is true by definition.
To prove part (ii), assume by contradiction that, at time $t_0$,
two birds ${\mathcal B}_i$ and 
${\mathcal B}_j$ are within unit distance of each
other but further than 1 apart at time $t_0+1$.
Write
\begin{equation}\label{setdeltab_0}
\delta = 
\varepsilon_{\! h} (\mathfrak{p} + 
\log \hbox{$\frac{1}{\varepsilon_{\! h}}$})^n
n^{b_0n^3},
\end{equation}
for some large enough constant $b_0$.
Assume also that the distance is
greater than $1+\delta$ at time $t_1>t_0$
and that, between $t_0$ and $t_1$,
the distance always remains in the interval
$(1, 1+ \delta\,]$ and that the two birds are joined
in $G_t$ for all $t\in [t_0,t_1]$.
Such conditions would violate soundness, so we show
they cannot happen.
Obviously, they imply that the distance between the two birds
never jumps (up or down) by
$\varepsilon_{\! h}$ or more, 
since otherwise the hysteresis rule would 
cease to apply and the edge $(i,j)$ would break.
This means that 
$\Delta_{ij}(t) < \varepsilon_{\! h}$,
for $t_0<t\leq t_1$.

\vspace{1cm}
\begin{figure}[htb]\label{fig-hysteresis-background}
\begin{center}
\hspace{0.2cm}
\includegraphics[width=8cm]{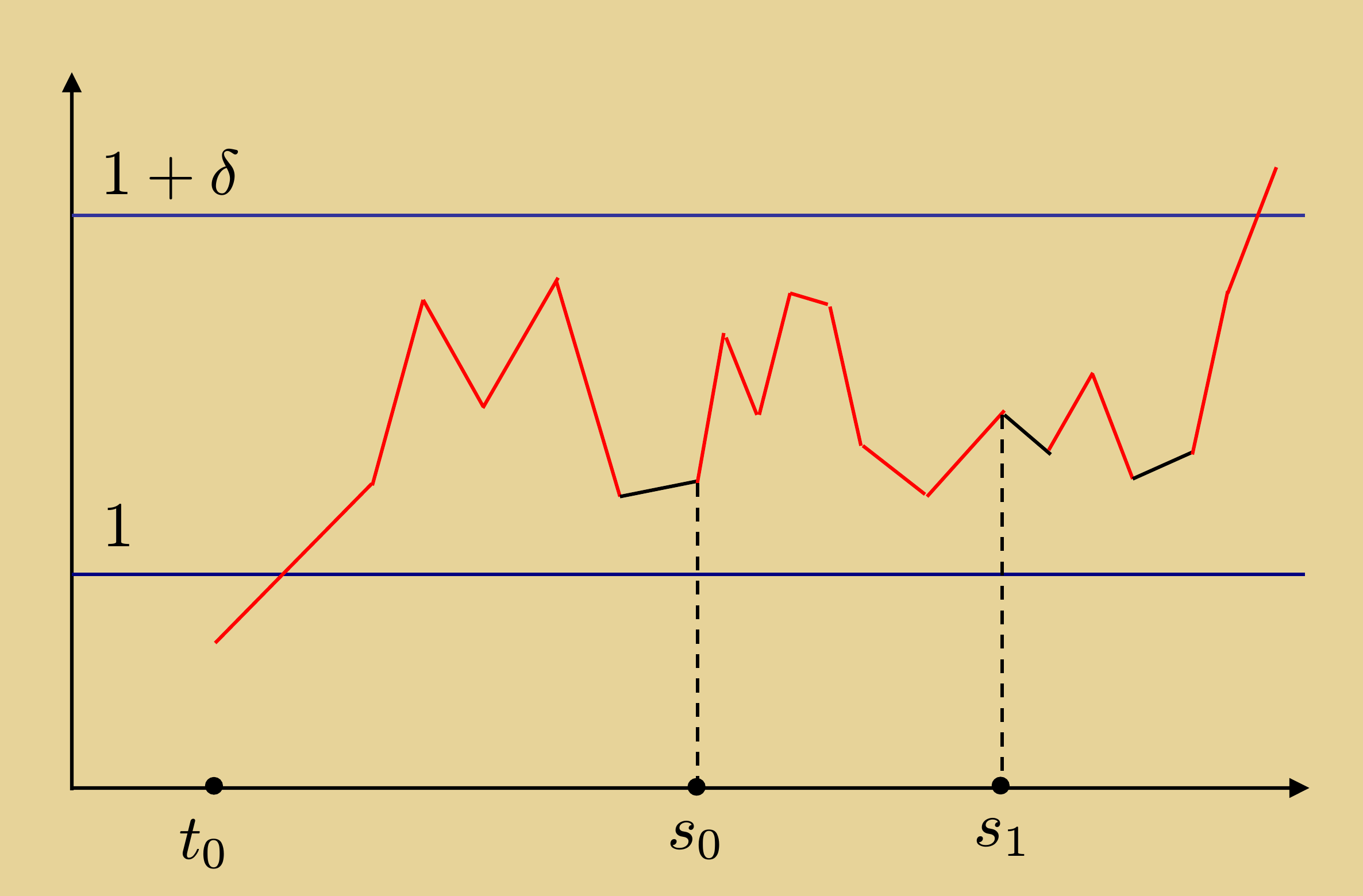}
\end{center}
\begin{quote}
\vspace{0.5cm}
\caption{\small 
The distance between two adjacent birds
cannot exceed 1 by more than $\delta$
before the edge breaks.
}
\end{quote}
\end{figure}

Consider the $t_1-t_0$ relative displacements
in the time interval $[t_0,t_1]$. Together they 
create a displacement in excess of $\delta$.
Let $\kappa= e^{O(n^3)}$ be 
the number of steps witnessing noise.
Mark the unit-time intervals within $[t_0,t_1]$
that are associated
with relative displacements
witnessing a perturbation or a network switch:
there are at most $N(n)+\kappa$ of those, each one
associated with a displacement 
less than $\varepsilon_{\! h}$, so this leaves 
us with a total displacement greater than 
$\delta -  \varepsilon_{\! h} N(n)- \varepsilon_{\! h}\kappa$.
This is contributed by no more than
$N(n)+\kappa+1$ runs of consecutive unmarked unit-time intervals.
By the pigeonhole principle, one of these runs
contributes a total displacement of at least
$(\delta - \varepsilon_{\! h} N(n)- \varepsilon_{\! h}\kappa)
/(N(n)+\kappa+1)$.
If $[s_0,s_1]$ denotes the corresponding time interval
($t_0\leq s_0\leq s_1 \leq t_1$), then 
$G_t$ remains invariant for all $s_0\leq t\leq s_1$
and, by Lemma~\ref{NetworkChangesUB},
\begin{equation}\label{deltaij-geq}
\sum_{t=s_0+1}^{s_1} \Delta_{ij}(t)
\geq 
\frac{\delta -  \varepsilon_{\! h} N(n)- \varepsilon_{\! h}\kappa}
{N(n)+\kappa+1}
\geq 
\delta  n^{-O(n^3)}
(\mathfrak{p} + 
\log \hbox{$\frac{1}{\varepsilon_{\! h}}$})^{1-n}.
\end{equation}
We now show that this displacement is too large
for two birds in the same time-invariant flock for so long.
The edge $(i,j)$ is in the network $G_t$ for 
all $t\in [s_0, s_1]$, so the two birds 
${\mathcal B}_i$ and ${\mathcal B}_j$ 
are in the same flock during that time period.
We already observed that 
$\tau_2(A)\leq \sqrt{2}\, \tau_1(A)$.
By~(\ref{tau2(vi-vj)CS}, \ref{tau-Spectral-log}) and 
Lemmas~\ref{ergodicity-submult},~\ref{relativeDisp},
it follows that, for $s_0< t\leq s_1$,
\begin{equation}\label{Deltaij(t-s0)}
\begin{split}
\Delta_{ij}(t) 
&\leq \|v_i(t)- v_j(t) \|_2
\leq \tau_2( P(t-1,s_0) ) 2^{O(\mathfrak{p})}
\leq  
\tau_1( P^{n^{c_0}}(s_0) )
       ^{\lfloor (t-s_0)n^{-c_0}\rfloor} 2^{O(\mathfrak{p})} \\
&\leq  2^{-\lfloor (t-s_0)n^{-c_0}\rfloor + O(\mathfrak{p})}.
\end{split}
\end{equation}
Technically, the way we phrased it,
our derivation assumes that the flock
that contains the birds ${\mathcal B}_i$ and 
${\mathcal B}_j$ at times $s_0$ through $s_1$
includes all the birds.
This is only done for notational convenience,
however, and the case of smaller flocks can be handled
in exactly the same way.
By~(\ref{deltaij-geq}, \ref{Deltaij(t-s0)})
and the hysteresis rule,
\begin{equation*}
\begin{split}
\delta  n^{-O(n^3)}
(\mathfrak{p} + 
\log \hbox{$\frac{1}{\varepsilon_{\! h}}$})^{1-n}
&\leq 
\sum_{t=s_0+1}^{s_1} \Delta_{ij}(t)
\leq \sum_{t=s_0+1}^{s_1} 
  \min\,\Bigl\{ \,  \varepsilon_{\! h},  
             2^{-\lfloor (t-s_0)n^{-O(1)}\rfloor +O(\mathfrak{p})}\, \Bigr\}
                       \\
&\leq \min_{T>0}\, \{ \,
  T \varepsilon_{\! h}   
+ 2^{-\lfloor T n^{-O(1)} \rfloor+O(\mathfrak{p})}\, \}.
\end{split}
\end{equation*}
Setting
$T= 2^n \lceil
\mathfrak{p} + \log \frac{1}{\varepsilon_{\! h}} \rceil$
leads to
$$
\delta
\leq 
\varepsilon_{\! h} (\mathfrak{p} + 
\log \hbox{$\frac{1}{\varepsilon_{\! h}}$})^n
n^{b_1 n^3},$$
for some positive constant $b_1$ independent of 
the constant $b_0$ used in the definition~(\ref{setdeltab_0})
of $\delta$. Choosing $b_0$ large enough 
thus contradicts our choice of $\delta$.
The two birds therefore cannot be both joined and
apart by more than $1+\delta$.
\hfill $\Box$
\proofend

\paragraph{The Geometry of Flocking: The Virtual Bird.}

Can birds fly in giant loops and come back
to their point of origin? Are there constraints
on their trajectories? 
We show that, after enough time has elapsed,
two birds can be newly joined only if
they fly almost parallel to each other.
We also prove that they cannot can stray 
too far from each other if they want to get
together again in the future. We investigate
the geometric structure of flocking and,
to help us do so, 
we introduce a useful device, the {\em flight net}.

\begin{figure}[htb]\label{fig-net}
\begin{center}
\hspace{.2cm}
\includegraphics[width=7cm]{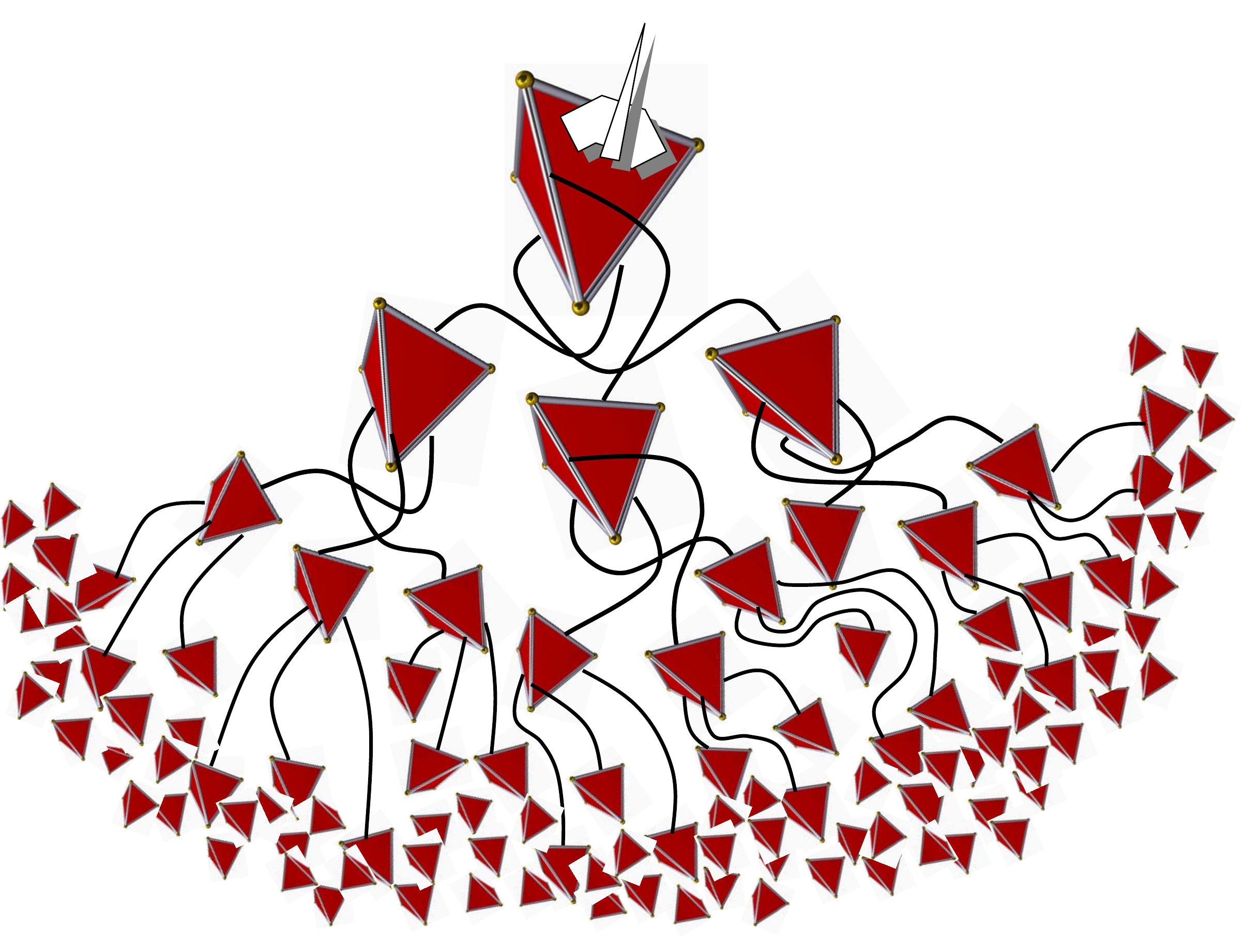}
\end{center}
\begin{quote}
\vspace{0cm}
\caption{\small 
The flight net is formed by joining together
the convex polytopes associated with birds'
new velocities.
}
\end{quote}
\end{figure}

It is convenient to lift the birds into
${\mathbb R}^{d+1}$
by adding time as an extra dimension:\footnote{This is not
a projectivization.}
$x(t)\mapsto (x_1(t),\ldots, x_d(t),t)^T$; 
$v(t)\mapsto (v_1(t),\ldots, v_d(t),1)^T$.
Since $\mathbf 1$ is a right eigenvector, this
lifting still satisfies the equation of motion.
The hysteresis rule kicks in at the 
same time and in the same manner as before;
in fact, the lifting has no bearing whatsoever
on the behavior of the birds.
The {\em angular offset} $\angle (x_i(t),v_i(t))$,
denoted by $\omega_i(t)$, 
plays an important role in the 
analysis.\footnote{\, We use
$x_i(t)$ as both a point and a vector, trusting the context
to make it obvious which is which.}
It represents (roughly) 
how the trajectory of bird ${\mathcal B}_i$ 
deviates at time $t$
from what it would have been had the bird reached
its current position by flying along a straight line.
We will show that the angular offset 
decreases roughly as $(\log t)/t$.
This fact has many important consequences.

Instead of following
a given bird over time and investigating its
trajectory locally, we track an imaginary bird that
has the ability to switch identities with its
neighbors: this {\em virtual bird}
could be ${\mathcal B}_i$ for a while and
then decide, at any time, to become any ${\mathcal B}_j$ 
adjacent to it in the flock. Or, for a rather
implausible but helpful image,
think of a bird passing the baton to any of its
neighbors: whoever holds the baton is the virtual bird.
Its trajectory is highly nondeterministic, as it is
allowed to follow any path in the flight net.
Although in the end
we seek answers that relate to physical birds, 
virtuality will prove to be a very powerful
analytical device. It allows us to
answer questions such as:
Can a virtual bird fly (almost) along
a straight line? How far apart can two birds
get if they are to meet again later? 
Another key idea is to trace the flight path
of virtual birds
backwards in time. This is how we are able to 
translate stochasticity into convexity and 
thus bring in the full power of geometry into the picture.
The translation emanates from this simple 
consequence of the velocity equation,
$v(t)= (P(t-1)\otimes I_d) v(t-1)$:
$$v_i(t)\in \text{Conv}\, \{\,v_j(t-1)\,|\, 
                  (i,j)\in G_{t-1}\,\}.
$$
By iterating in this fashion, we create
the {\em flight net} ${\mathcal N}_i(t)$ 
of bird ${\mathcal B}_i$ at time $t>0$.
It is a connected collection 
of line segments (ie, a 1-skeleton):
${\mathcal N}_i(t)= {\mathcal N}_i(t,K_t)$, where $K_t$
is a large integer parameter. Specifically, we set 
\begin{equation}\label{K_t-setting}
K_t= \lceil n^{b_0}(\mathfrak{p}+ \log t)\rceil
\end{equation}
for a big enough constant $b_0$.
The power of the flight net comes from
its ability to deliver both kinetic
and positional information about the
``genealogy'' of a bird's current state.
Let $K$ be an arbitrary positive integer;
we define ${\mathcal N}_i(t,K)$ inductively
as follows.
The case $t=1$ is straightforward: ${\mathcal N}_i(t,K)$
consists of the single line segment
$x_i(0)x_i(1)$.
Suppose that $t>1$.
We say that time $s$ is {\em critical} if $s\leq K$
or if, during the time interval $[s-K,s]$, there is
a perturbation 
or a network switch, ie, 
the velocity of at least one flock is multiplied by 
by $I_m\otimes \widehat{\alpha}$ or
$G_u\neq G_{u+1}$ for some $u$ ($s-K\leq u\leq s$).

\begin{itemize}
\item
If $t$ is critical, 
then ${\mathcal N}_i(t,K)$ consists of
the segment $x_i(t-1)x_i(t)$, together with 
the translates 
${\mathcal N}_j(t-1,K)+ x_i(t-1)- x_j(t-1)$,
for all $(i,j)\in G_{t-1}$ and $j=i$.
\item
If $t$ is noncritical, then 
${\mathcal N}_i(t,K)$ consists of 
the segment $x_i(t-1)x_i(t)$, together with 
${\mathcal N}_i(t-1,K)$.
\end{itemize}

\begin{figure}[htb]\label{fig-netlocal}
\begin{center}
\hspace{0cm}
\includegraphics[width=8cm]{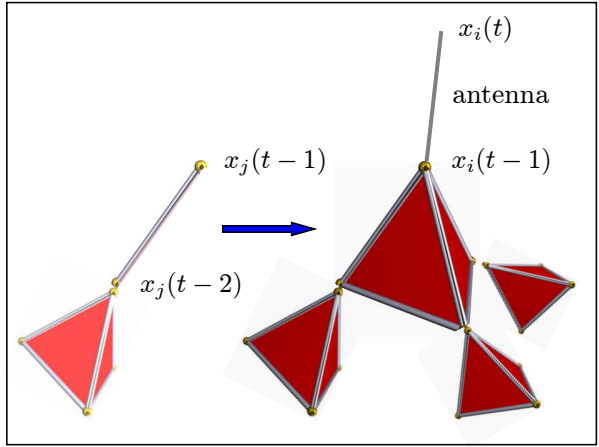}
\end{center}
\begin{quote}
\vspace{0cm}
\caption{\small 
In the critical case, the virtual net
is translated from bird ${\mathcal B}_j$ 
to bird ${\mathcal B}_i$ by the 
baton-passing drift.}
\end{quote}
\end{figure}

Every flight net has an {\em antenna} sitting on top,
which is a line segment extending from
$X_{d+1}=t-1$ to $X_{d+1}=t$ in the case of ${\mathcal N}_i(t,K)$.
In the noncritical case, the antenna is connected on top
of the previous one, ie,
the one for ${\mathcal N}_i(t-1,K)$.
Otherwise, we slide the 
time-($t-1$) flight nets of the adjacent birds so that 
their antennas join with the bottom vertex of the new antenna:
this shift is called the {\em baton-passing drift}.

Here is the intuition. Flying down the top antenna of
the net, the virtual bird hits upon another antenna:
either there is only one to choose from, in which case
it is almost collinear (because
of noncriticality, the corresponding random walk is thoroughly
mixed) or else the virtual bird discovers
a whole bouquet of antennas and picks one of them.
Because the old antenna is a convex combination of the 
new ones, the virtual bird can continue its
backward flight by choosing from a convex cone
of directions: this freedom
is the true benefit of convexity and, hence, stochasticity.
This is when the baton is passed: the virtual
bird changes its correspondence with an actual bird
as it chooses one of these directions.
Because of the translation by $x_i(t-1)- x_j(t-1)$,
this change of correspondence
is accompanied by a
shift of length at most one, what we dub the baton-passing drift.

Viewed from a suitable perspective,
the flight net provides a quasi-convex structure
from which all sorts of metric information can be inferred.
Most important, it yields the crucial {\em Escape Lemma},
which implies that, as time goes by, 
it becomes increasingly easy to predict
the velocity of a bird from its location,
and vice versa. The lemma asserts that
the bird flies in a direction that
points increasingly away from its original position.
We begin with a simple observation.
For any time $t>0$, the $(d+1)$-dimensional vector
\begin{equation}\label{w_i(t)-defn}
w_i(t)= \hbox{$\frac{1}{t}$}x_i(t)
\end{equation}
represents the constant velocity that bird
${\mathcal B}_i$ would need to have
if it were to leave
the origin at time $0$ and be at position $x_i(t)$
at time $t$ while flying in a fixed direction.
Recall that 
that the angular offset $\omega_i(t)$
is $\angle (x_i(t),v_i(t))$; we show that 
it cannot deviate too much from the 
{\em velocity offset} $\|v_i(t)-w_i(t)\|_2$.

\vspace{1cm}
\begin{figure}[htb]\label{fig-angles-background}
\begin{center}
\hspace{.3cm}
\includegraphics[width=7cm]{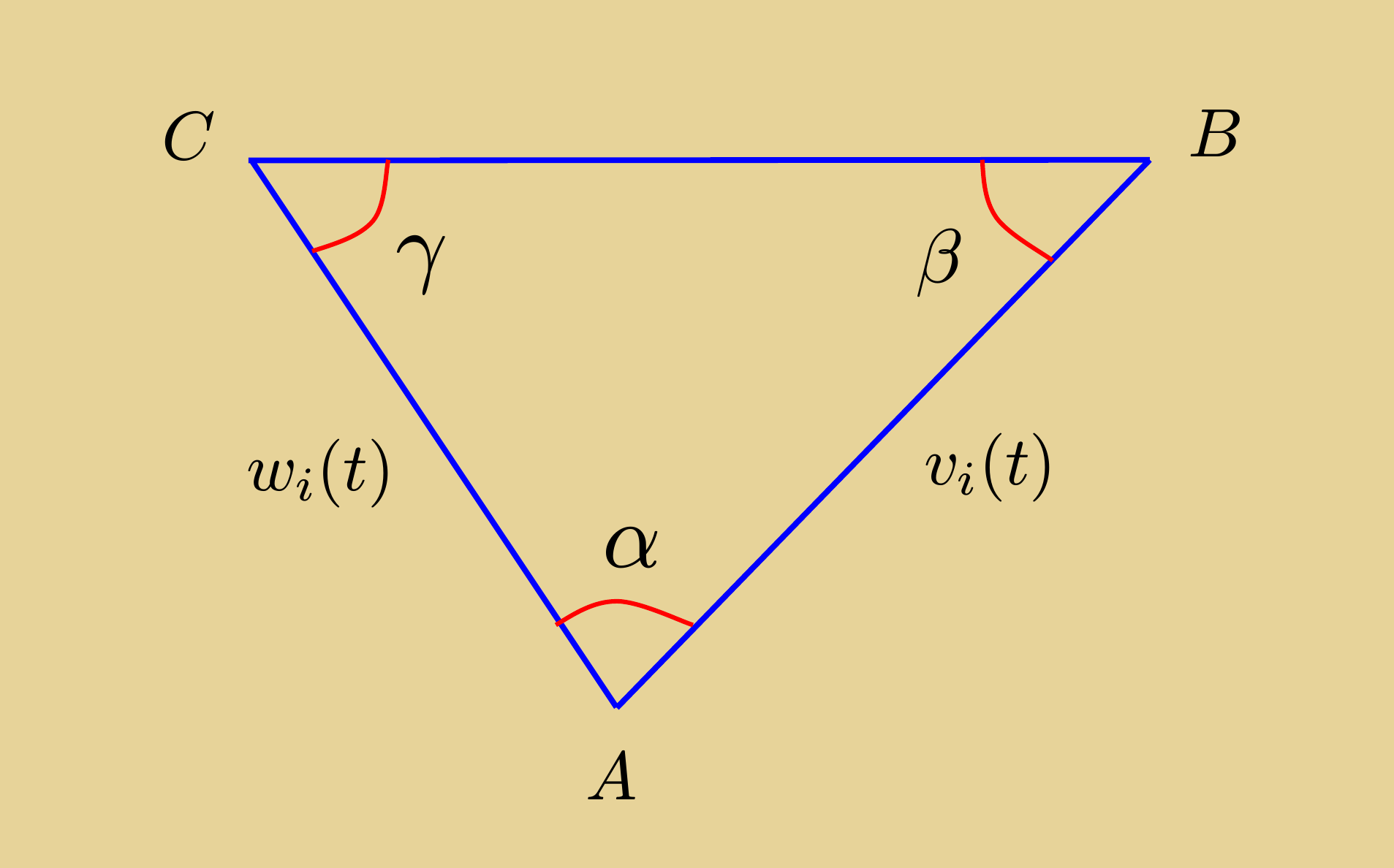}
\end{center}
\begin{quote}
\vspace{0cm}
\caption{\small 
Proving that angular and velocity offsets
are closely aligned.
}
\end{quote}
\end{figure}

\begin{lemma}\label{equivDiffAngle}
$\!\!\! .\,\,$
For any $t>0$,
\begin{equation*}
2^{-O(\mathfrak{p})} \|v_i(t)-w_i(t)\|_2
\leq \omega_i(t)
\leq O(\|v_i(t)-w_i(t)\|_2).
\end{equation*}
\end{lemma}
\proof
Consider the triangle $ABC$ formed
by identifying $\overrightarrow{AB}$ with $v_i(t)$ and 
$\overrightarrow{AC}$ with $w_i(t)$,
and let $\alpha,\beta, \gamma$ be the angles opposite
$BC,CA, AB$, respectively.
Note that $\alpha= \omega_i(t)$
and $\|v_i(t)-w_i(t)\|_2= |BC|$.
Assume that $\beta\leq \gamma$; we omit the
other case, which is virtually identical.
By~(\ref{|v|Poly}), $AB$ and $AC$ 
have length between $1$ and $2^{O(\mathfrak{p})}$;
therefore, if $\alpha\neq 0$ then 
$2^{-O(\mathfrak{p})}\leq  \beta < \pi/2$.
The proof follows from the
law of sines, $|BC|^{-1} \sin\alpha= |AC|^{-1} \sin\beta$.
\hfill $\Box$
\proofend

\begin{lemma}\label{NormalVelInfty}
$\!\!\! .$
{\bf (Escape Lemma)}\ 
For any bird ${\mathcal B}_i$, at any time $t>0$, 
\begin{equation*}
 \omega_i(t)
\leq 
\frac{\log t}{t} 
 \, n^{O(n^3)}
(\mathfrak{p} +\log \hbox{$\frac{1}{\varepsilon_{\! h}}$})^{n-1}
+ 
\frac{1}{t}
\Bigl( 
2^{O(\mathfrak{p})} + 
\mathfrak{p} n^{O(n^3)}  
(\mathfrak{p} +\log \hbox{$\frac{1}{\varepsilon_{\! h}}$})^{n-1}
\Bigr). 
\end{equation*}
\end{lemma}

\medskip

\begin{corollary}\label{corol-NormalVelInfty}
$\!\!\! .\,\,$
Under the default settings~(\ref{Assumptions}),
at any time $t>1$,
$$
\omega_i(t)\leq
\frac{\log t}{t} 
\, 
n^{O(n^3)} .
$$
\end{corollary}

\vspace{1cm}
\begin{figure}[htb]\label{fig-coupling-background}
\begin{center}
\hspace{-0.2cm}
\includegraphics[width=7.5cm]{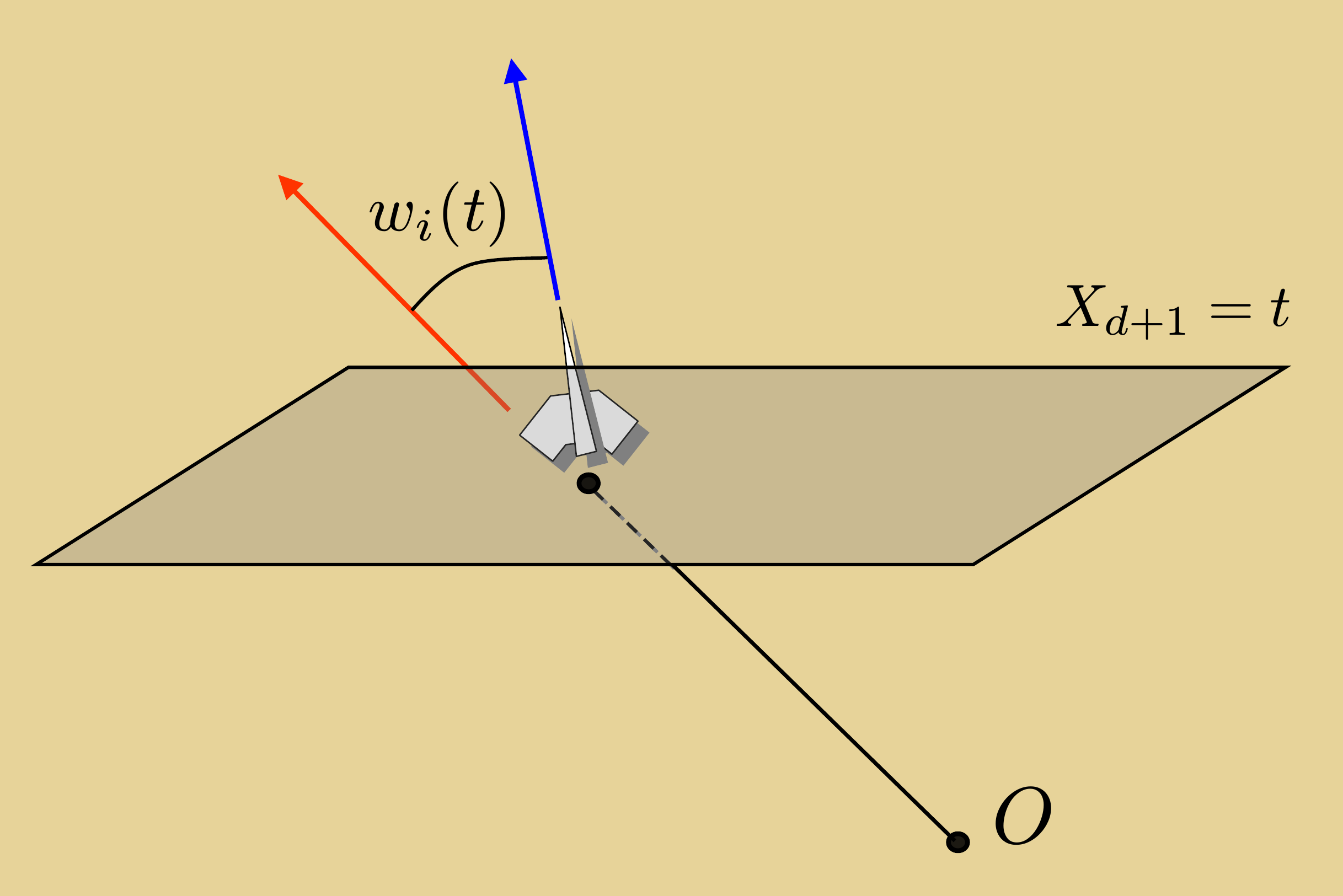}
\end{center}
\begin{quote}
\vspace{-0.0cm}
\caption{\small 
Birds fly increasingly in ``escape'' mode.
}
\end{quote}
\end{figure}

Unlike the other factors in the upper bound, 
the presence 
of $\log t$ is an artifact of the proof and
might not be necessary. 
Our approach is to exploit
the ``convexity'' of single-bird transitions. One should
be careful not to treat flocks as {\em macro-birds} and
expect convexity from stationary velocities. In premixing
states, all sorts of ``nonconvex'' behavior can happen.
For example, consider two flocks in dimension 1, both
with positive stationary velocities. Say the one on the
left has higher speed and catches up with the one on
the right to merge into one happy flock.
It could be the case that the stationary velocity of the
combined flock is negative, ie, the joint flock moves
left even though each one of the two
flocks was collectively moving right prior to
merging. Of course, this a premixing aberration that
we would not expect in the long run.

\bigskip

{\noindent\em Proof of Lemma~\ref{NormalVelInfty}. }
From the initial conditions,
we derive a trivial upper bound of 
$2^{O(\mathfrak{p})}$ for constant $t$, so
we may assume that $t$ is large enough
and $\omega_i(t)>0$.
The line passing through $x_i(t)$ in the direction
of $v_i(t)$ intersects the hyperplane $X_{d+1}=0$
in a point $p$ at distance from the origin,
$\|p\|_2= \Omega( t \omega_i(t) )$.
Recall that the bird ${\mathcal B}_i$
started its journey at distance 
$2^{O(\mathfrak{p})}$ from the origin.
If it had flown in a straight line, then 
we would have $p= x_i(0)$, hence
$\omega_i(t)= \frac{1}{t} 2^{O(\mathfrak{p})}$, and
we would be done.
Chances are the bird did not fly straight, however.
If not, then we exhibit a virtual bird that 
(almost) does, at least in the sense that it does
not get much closer to the origin at time $0$ that
a straightline flight would.
The idea is to use the flight net
to follow the trajectory of a virtual bird that 
closely mimics a straight flight 
from $p$ to $x_i(t)$.

Some words of intuition.
If all times were critical and no perturbation ever took place,
then it would be
easy to prove by backward induction that, 
for all $0\leq s <t$, the segment
$p x_i(t)$ intersects each hyperplane $X_{d+1}=s$
in a point that lies within the convex hull
of ${\mathcal N}_i(t)\cap \{ X_{d+1}=s \}$.
This would imply that $p$ lies in the convex hull
of the birds at time $0$, which again would
give us the same lower bound on $\omega_i(t)$
as above (modulo the baton-passing drift).
In fact, it would be possible
to trace a {\em shadow path} from $x_i(t)$ down the 
flight net that leads to a virtual bird at time $0$ that is even
further away from the origin than $p$. 
(We use here a fundamental property of convexity, that
no point can be further to a point in
a convex polytope than to all of its vertices.)
Unfortunately, this convexity
argument breaks down because of the net's
jagged paths over noncritical time periods.
The jaggedness is so small, 
however, that it provides us enough
``quasi-convexity'' to rescue the argument.

\vspace{.5cm}
\begin{figure}[htb]\label{fig-shadowpath}
\begin{center}
\hspace{1cm}
\includegraphics[width=7cm]{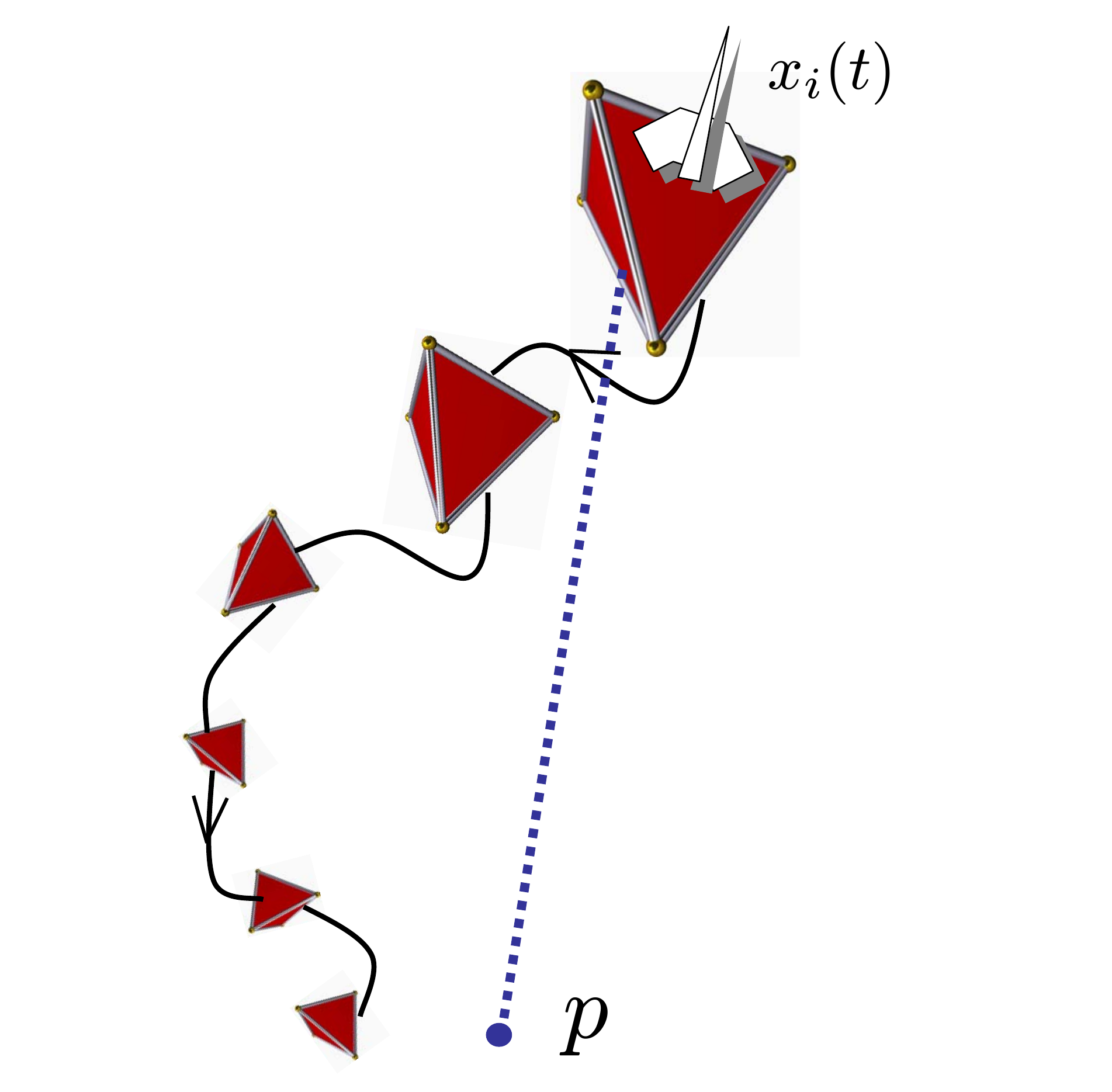}
\end{center}
\begin{quote}
\vspace{-0.4cm}
\caption{\small 
The shadow path attempts to follow the segment $p x_i(t)$ closely.
}
\end{quote}
\end{figure}

First we describe the shadow path; then we show why it works.
Instead of handling convexity 
in ${\mathbb R}^{d+1}$, we will 
find it easier to do this in projection.
By Lemma~\ref{equivDiffAngle}, there exists
a coordinate axis, say $X_1$,
such that 
\begin{equation}\label{angleVelDiff}
0<\omega_i(t)= O( v_i(t)_1 - w_i(t)_1 ).
\end{equation}
Note that we may have to reverse
the sign of $v_i(t)_1 - w_i(t)_1$, but this is immaterial.
The shadow path 
$x_i^{\tt v}(t), x_i^{\tt v}(t-1),\ldots, x_i^{\tt v}(0)$
describes the flight of the virtual bird
${\mathcal B}^{\,\tt v}_i$ backwards in time. 
The first two vertices are 
$x_i^{\tt v}(t)= x_i(t)$ and $x_i^{\tt v}(t-1)= x_i(t-1)$.
This means the virtual bird flies
down the topmost edge of 
${\mathcal N}_i(t)$, ie,
in the negative $X_{d+1}$ direction.
Next, the following rule applies for $s=t,t-1,\ldots, 2$:

\begin{itemize}
\item
If $s$ is noncritical, 
${\mathcal N}_i(t)$ has a single edge $y_{s-2}y_{s-1}$, with
$(y_{s-2})_{d+1}=s-2$.
The virtual bird flies down $y_{s-2}y_{s-1}$
and we set $x_i^{\tt v}(s-2)= y_{s-2}$ accordingly.
\item
If $s$ is critical, 
${\mathcal N}_i(t)$ has one or 
several edges $y^k_{s-2}y_{s-1}$, with
$(y^k_{s-2})_{d+1}=s-2$.
The virtual bird follows
the edge with maximum $X_1$-extant, ie,
the one that maximizes $(y_{s-1})_1 - (y^k_{s-2})_1$.
(Recall that, although neither 
$y_{s-1}$ nor $y^k_{s-2}$ might be the position
of any actual bird, their difference  
$y_{s-1} - y^k_{s-2}$ is the 
velocity vector $v_j(s-1)$ of some ${\mathcal B}_j$.)
We set $x_i^{\tt v}(s-2)= y^k_{s-2}$.
\end{itemize}

\vspace{.5cm}
\begin{figure}[htb]\label{fig-virtual-background}
\begin{center}
\hspace{-0.3cm}
\includegraphics[width=7cm]{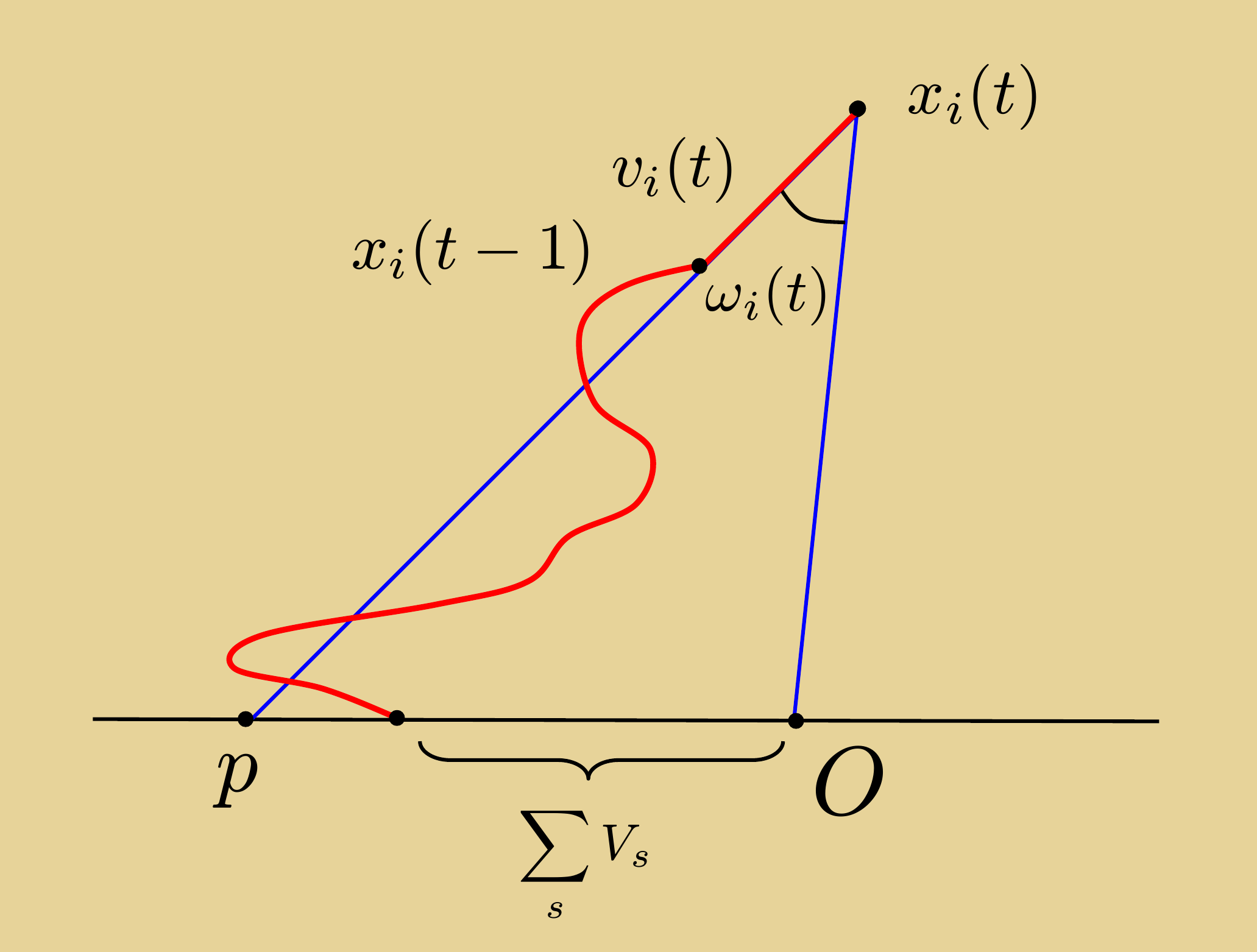}
\end{center}
\begin{quote}
\vspace{0cm}
\caption{\small 
Following the red shadow path.
}
\end{quote}
\end{figure}

The virtual bird thus moves down the flight net
back in time until it lands at $X_{d+1}=0$.
The resulting collection of $t+1$ vertices
forms the shadow path of 
the virtual bird ${\mathcal B}^{\,\tt v}_i$ at time $t$.
Naturally, we define the velocity of ${\mathcal B}^{\,\tt v}_i$
at time $s>0$ as 
$v_i^{\tt v}(s)= x_i^{\tt v}(s)- x_i^{\tt v}(s-1)$.
Note that $v_i^{\tt v}(t)= v_i(t)$.
To prove that the shadow path does not stray far from
the straightline flight from $x_i(t)$ to $p$, we 
focus on the difference 
\begin{equation}\label{VsDefn}
V_s= v_i^{\tt v}(s)_1- w_i(t)_1 \, ,
\end{equation}
for $s\geq 1$. If we could show that $V_s$ is always nonnegative
then, measured in projection along the $X_1$ axis,
the virtual bird would fly back in time
even further away from the origin that it would
if it flew straight from $x_i(t)$ to the hyperplane
$X_{d+1}=0$ in the direction of $-v_i(t)$.
Except for the fact that a virtual bird at time $0$ may not
share the location of any actual bird (an issue we will
address later), this would entirely rescue our initial argument. 
We cannot quite ensure 
the nonnegativity of $V_s$, but we come close enough
to serve our purposes.

Consider an interval $[r,s]$ consisting
entirely of noncritical times (hence $r>K_t$).
The flock that contains
the virtual bird ${\mathcal B}^{\,\tt v}_i$
is invariant between times $r-K_t$ and $s$
and undergoes no perturbation
during that period; furthermore,
${\mathcal B}^{\,\tt v}_i$ has the same
incarnation as some fixed ${\mathcal B}_j$ 
during the time period $[r-1,s]$.
If $\chi(j)$ denotes the $n$-dimensional vector with 
all coordinates equal to $0$, except for $\chi(j)_j=1$, then, 
for $r-1\leq u\leq s$,
$$v_j(u)=  ((\chi(j)^T P^{u-r+K_t}) \otimes I_d)v(r-K_t).$$
We abuse notation and restrict $P$
and $v(r-K_t)$ to the flock of 
${\mathcal B}_j$ and not to all of
$G_{r-K_t}=\cdots = G_s$.
By~(\ref{|v|Poly}, \ref{Pij^sBounds}), we find that
\begin{equation*}
\begin{split}
| v_j(u)_1 - ({\mathbf m}_\pi[v(r-K_t)])_1 |
&\leq 
\|
  ( (  (\chi(j)P^{u-r+K_t})\otimes I_d ) 
   - (\pi^T\otimes I_d)) v(r-K_t)
\|_2   \\
&\leq 
\sqrt{d} \,
\| P^{u-r+K_t} -  {\mathbf 1} \pi^T \|_F 
\| v(r-K_t) \|_2 \\
&\leq 
 e^{-(u-r+K_t) n^{-O(1)}+ O(\mathfrak{p}+\log n)}.
\end{split}
\end{equation*}
We conclude that
\begin{equation*}
\begin{split}
|V_{r-1}- V_s|
&= | v_i^{\tt v}(r-1)_1  - v_i^{\tt v}(s)_1 |
= |v_{j}(r-1)_1- v_{j}(s)_1| \\
&\leq 
|v_{j}(r-1)_1- ({\mathbf m}_\pi[v(r-K_t)])_1|
+ 
|v_{j}(s)_1- ({\mathbf m}_\pi[v(r-K_t)])_1| ;
\end{split}
\end{equation*}
hence, using $\mathfrak{p}\geq n^3$,
\begin{equation}\label{VrsBound}
|V_{r-1}- V_s|
\leq e^{-K_t n^{-O(1)}+O(\mathfrak{p})}.
\end{equation}

As usual, $\kappa= e^{O(n^3)}$ denotes 
the number of steps witnessing noise.
Suppose now that $s>1$ is critical.
If no perturbation occurs at time $s-1$,
then $v_i^{\tt v}(s)$ is a convex combination of
the vectors of the flight net 
joining $X_{d+1}=s-2$ to $X_{d+1}=s-1$.
By construction, it follows that 
$$v_i^{\tt v}(s-1)_1 \geq v_i^{\tt v}(s)_1.$$
If the vector is perturbed by $\zeta$, then 
$$v_i^{\tt v}(s-1)_1 \geq v_i^{\tt v}(s)_1 - \zeta_1
\geq v_i^{\tt v}(s)_1 - \delta_{s-1},$$
where 
$\delta_t =  \frac{\log t}{t}\, e^{O(n^3)}$
(the perturbation bound).
In both cases, therefore,
$V_{s-1}\geq V_s - \delta_{s-1}$.
Let $\mathfrak{C}$ be the number of critical times.
By~(\ref{VrsBound}), for all $1\leq s\leq t$,
$$
V_s\geq V_t- \mathfrak{C} e^{-K_tn^{-O(1)} + O(\mathfrak{p})} - 
\sum_{u=s}^{t-1} \delta_u \, .
$$
Summing over all $s$,
$$
\sum_{s=1}^t V_s
\geq t V_t- 
(t-1) \mathfrak{C} e^{-K_tn^{-O(1)} + O(\mathfrak{p})} - 
\sum_{s=1}^{t-1} s \delta_s \, .
$$
Since, by assumption,
$\delta_s = 0$ at all but $\kappa$ places,
$$\sum_{s=1}^{t-1} s \delta_s 
=  \kappa e^{O(n^3)}\log t.$$
By~(\ref{angleVelDiff}), 
$V_t= v_i^{\tt v}(t)_1- w_i(t)_1
= \Omega( \omega_i(t) )$; therefore,
\begin{equation}\label{SumVs}
 \omega_i(t)
= O(V_t)
= \frac{O(1)}{t} \Bigl| \sum_{s=1}^t V_s \Bigr|
+ \mathfrak{C}  e^{-K_tn^{-O(1)} + O(\mathfrak{p})} 
+  \frac{\log t}{t} \, \kappa e^{O(n^3)} .
\end{equation}
By~(\ref{w_i(t)-defn}, \ref{VsDefn}),
\begin{equation*}
\begin{split}
\sum_{s=1}^t V_s
&= 
\sum_{s=1}^t \Bigl\{ x_i^{\tt v}(s)_1- x_i^{\tt v}(s-1)_1
       - w_i(t)_1 \Bigr\} 
= x_i^{\tt v}(t)_1 - x_i^{\tt v}(0)_1 - t w_i(t)_1  \\
&= x_i^{\tt v}(t)_1 - x_i(t)_1 - x_i^{\tt v}(0)_1  
= - x_i^{\tt v}(0)_1  \, .
\end{split}
\end{equation*}
Since $x_i^{\tt v}(0)_1$ is the position of a virtual
bird at time 0, it is tempting to infer that it is
also the position of some actual bird at that time; hence
$|x_i^{\tt v}(0)_1|= 2^{O(\mathfrak{p})}$.
This is not quite true because adding together
the velocity vectors ignores 
the baton-passing drift, ie,
the displacements caused
by switching birds. At critical times, the virtual
bird gets assigned a new physical bird that
is adjacent to its currently assigned feathered creature.
Recall how the net
${\mathcal N}_j(t-1,K)$ is translated by
$x_i(t-1)- x_j(t-1)$. Since $(i,j)\in G_{t-1}$,
this causes a displacement of at most 1.
Note that unlike the 
velocity perturbations, whose effects are 
multiplied by time, the drift is additive.
This highlights the role of the flight net 
as both a kinetic and a positional object.
Summing them up, we find that 
$|x_i^{\tt v}(0)_1|\leq  \mathfrak{C} + 2^{O(\mathfrak{p})}$; hence
\begin{equation}\label{Vs-xi0_1}
\begin{split}
\Bigl| \sum_{s=1}^t V_s \Bigr|
\leq  \mathfrak{C} + 2^{O(\mathfrak{p})}.
\end{split}
\end{equation}
Recall that a time is critical if
there exists either a perturbation
or a network switch in the past $K_t$ steps.
Recall~(\ref{K_t-setting}) that
$K_t= \lceil n^{b_0}(\mathfrak{p}+ \log t)\rceil$ for a large
enough constant $b_0$.
By Lemma~\ref{NetworkChangesUB},
this bounds the number of critical times by 
$$ \mathfrak{C} \leq K_t (N(n)+\kappa) 
\leq (\mathfrak{p}+ \log t)
n^{O(n^3)} (\mathfrak{p} +
\log \hbox{$\frac{1}{\varepsilon_{\! h}}$})^{n-1},
$$
and the lemma follows from~(\ref{SumVs}, \ref{Vs-xi0_1}).
\hfill $\Box$
\proofend

We mention a few other corollaries of Lemma~\ref{NormalVelInfty}
that rely on the model's assumptions. Again, recall
that the sole purpose of these assumptions is
to alleviate the notation and help one's intuition.

\begin{corollary}\label{corol-anglechange}
$\!\!\! .\,\,$
Under the default settings~(\ref{Assumptions}),
at any time $t>1$,
a bird turns by an angle 
$\angle (v_i(t),v_i(t+1))$ that is at most
$$\frac{\log t}{t}\, n^{O(n^3)}.
$$
\end{corollary}
\proof
By~(\ref{|v|Poly})
and $\delta_t= \frac{\log t}{t}\, e^{O(n^3)}$,
no bird can take a step
longer than $2^{O(\mathfrak{p})}$, therefore
the angle between the vectors $x_i(t)$ and $x_i(t+1)$
is at most $\frac{1}{t} 2^{O(\mathfrak{p})}$.
As a result,
\begin{equation*}
\begin{split}
\angle (v_i(t),v_i(t+1))
&\leq 
\angle (v_i(t), x_i(t))+ 
\angle (x_i(t), x_i(t+1)) +
\angle (x_i(t+1), v_i(t+1)) \\
&= \omega_i(t)
+\angle (x_i(t),x_i(t+1))
+\omega_i(t+1),
\end{split}
\end{equation*}
and the proof follows from
Corollary~\ref{corol-NormalVelInfty}.
The property we are using here is 
the triangle inequality for angles: equivalently,
the fact that, among the 3
angles around a vertex of a tetrahedron in ${\mathbb R}^3$,
none can exceed the sum of the others.
Even though the birds live in higher
dimension, our implicit argument involves only 3
points at a time and therefore belongs in ${\mathbb R}^3$.
\hfill $\Box$
\proofend

\begin{corollary}\label{DriftBound}
$\!\!\! .\,\,$
Under the default settings~(\ref{Assumptions}),
if two birds are adjacent in the flocking
network at time $t>1$, their
distance prior to $t$ always remains within
$n^{O(n^3)}\log t$.
\end{corollary}
\proof
For reasons discussed above, 
any two birds are within
distance $2^{O(\mathfrak{p})}$ after 
a constant number of steps,
so we may assume that $t$ is large enough.
Consider the time 
$s$ that maximizes the distance $R_s$, 
for all $s\in [0,t-1]$,
between the points $x_i(s)$ and $p=(s/t) x_i(t)$ in
the hyperplane $X_{d+1}=s$.
For the same reason, we may assume that $s>1$.
By Corollaries~\ref{corol-NormalVelInfty},~\ref{corol-anglechange},
\begin{equation}\label{xi-vi(s+1)}
\angle (x_i(s),v_i(s+1) )
\leq \omega_i(s)+ \angle (v_i(s),v_i(s+1) )
\leq \frac{\log s}{s}\, n^{O(n^3)}.
\end{equation}
Set up an orthogonal coordinate system in the plane
spanned by $O,p,x_i(s)$: $O$ is the origin;
the $X$-axis lies in the hyperplane $X_{d+1}=0$
and runs in the direction from
$p$ to $x_i(s)$; 
the $Y$-axis is normal to $OX$ in the $O,p,x_i(s)$ plane.
By~(\ref{|v|Poly}),
the $Y$-coordinate $p_Y$ of $p$ satisfies
$$s \leq p_Y\leq s 2^{O(\mathfrak{p})}.$$
Let $Y=X \tan \alpha$ and $Y=X\tan \beta$ be the two lines
through the origin passing through $x_i(t)$
and $x_i(s)$, respectively.
Setting $Y=p_Y$ we find that  
$p_X =p_Y/\tan \alpha$ 
and $x_i(s)_X= p_X/\tan \beta$; therefore
$$R_s\leq 
\Bigl| \frac{1}{\tan\beta} - \frac{1}{\tan\alpha} \Bigr| s
2^{O(\mathfrak{p})}
\leq 
\frac{\sin(\alpha-\beta)}{(\sin\alpha)(\sin\beta)}\, s
\, 2^{O(\mathfrak{p})} \, .$$
By construction, the velocity $v_i(s+1)$
cannot take the bird ${\mathcal B}_i$ 
outside the elliptical cylinder
that is centered at the line $(O,x_i(t))$ with 
the point $x_i(s)$ on its boundary
and that intersects $X_{d+1}=0$ in a disk
of radius $R_s= |px_i(s)|$.
It follows that the normal projection $w$ of 
$v_i(s+1)$ on the $(X,Y)$-plane forms
an angle $\gamma$ with $x_i(s)$ at least 
equal to the angle between the two lines
$Y=X\tan \alpha$ and $Y=X\tan \beta$,
which is $\alpha-\beta$.
By~(\ref{xi-vi(s+1)}), therefore,
$$
\alpha-\beta
\leq \gamma
\leq \angle (x_i(s),v_i(s+1)
\leq \frac{\log s}{s}\, n^{O(n^3)}.
$$
Birds are at most 
$2^{O(\mathfrak{p})}$ away from the origin at time $0$
and, by~(\ref{|v|Poly}), take no step larger than
that bound. It follows that 
both $\alpha$ and $\beta$ are at least
$2^{-O(\mathfrak{p})}$, therefore
$$R_s \leq 2^{O(\mathfrak{p})} \, n^{O(n^3)}\log t
\, .
$$
If two birds ${\mathcal B}_i$ 
and ${\mathcal B}_j$ share an edge in a flock
at time $t$, then 
$\|x_i(t)- x_j(t)\|_2\leq 1$; so,
by the triangle inequality,
at any time $1< s\leq t$, 
$$\|x_i(s)- x_j(s)\|_2\leq 
2^{O(\mathfrak{p})} \, n^{O(n^3)}\log t
+ \frac{s}{t} \, \|x_i(t)- x_j(t)\|_2,$$
which, by the default settings~(\ref{Assumptions}),
proves the lemma.
\hfill $\Box$
\proofend

\vspace{1cm}
\begin{figure}[htb]\label{fig-far-background}
\begin{center}
\hspace{0.2cm}
\includegraphics[width=8cm]{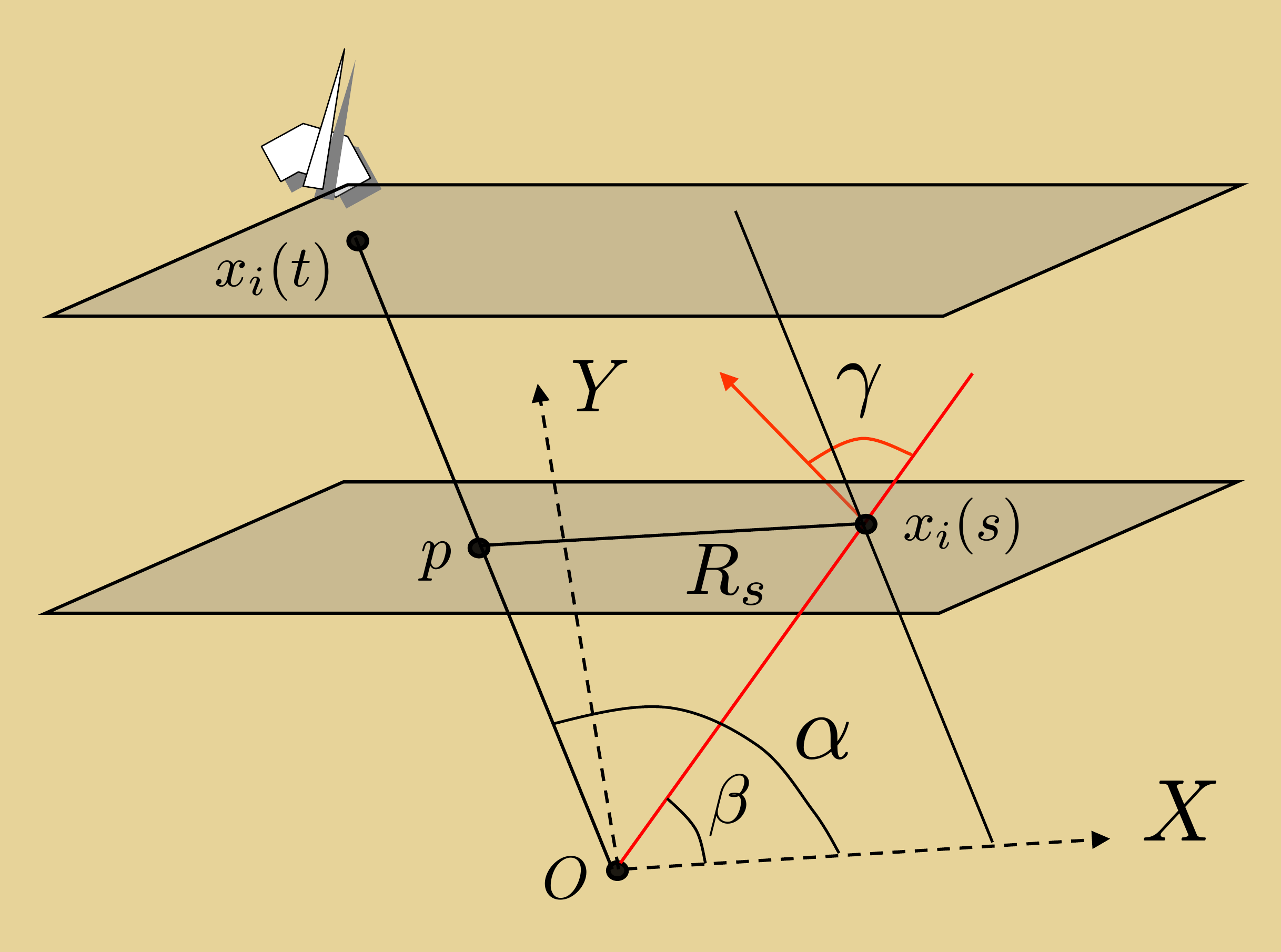}
\end{center}
\begin{quote}
\vspace{-0.2cm}
\caption{\small 
Two birds can't stray too far from each other
if they're ever to meet again.
}
\end{quote}
\end{figure}

Suppose that birds ${\mathcal B}_i$
and ${\mathcal B}_j$ are distance at most $D$
at time $t>0$. (No assumption is made whether
they belong to the same flock or 
whether~(\ref{Assumptions}) holds.)
By~(\ref{w_i(t)-defn}) and
Lemmas~\ref{relativeDisp},~\ref{equivDiffAngle},~\ref{NormalVelInfty},
\begin{equation}\label{Deltaij-logt/t-pD}
\begin{split}
\Delta_{ij}(t) 
&\leq \|v_i(t)- v_j(t) \|_2 \\
&\leq \|v_i(t)- \hbox{$\frac{1}{t}$} x_i(t) \|_2
+ \|v_j(t)- \hbox{$\frac{1}{t}$}x_j(t) \|_2 
+ \hbox{$\frac{1}{t}$}\|x_i(t)- x_j(t) \|_2
\\
&\leq 
(\omega_i(t) + \omega_j(t) ) 2^{O(\mathfrak{p})}
+ \hbox{$\frac{D}{t}$} \, .
\end{split}
\end{equation}

\begin{corollary}\label{VelocityDiff}
$\!\!\! .\,\,$
Under the default settings~(\ref{Assumptions}),
at any time $t>1$, the difference in stationary
velocities between two distinct flocks joining
into a common one at time $t+1$
has Euclidean norm at most
$\hbox{$\frac{\log t}{t}$} \,  n^{O(n^3)}$.
\end{corollary}
\proof
The stationary velocity of a flock
is a convex combination of its
constituents' individual velocities,
so the difference in stationary
velocities cannot exceed, length-wise, the maximum
difference between individual ones.
By~(\ref{|v|Poly}) and the connectivity of flocks,
the distance
at time $t$ between any two birds 
in the common flock at time $t+1$ 
cannot exceed
$D= n+ 2^{O(\mathfrak{p})}$.
The lemma follows from~(\ref{Deltaij-logt/t-pD})
and Corollary~\ref{corol-NormalVelInfty}.
\hfill $\Box$
\proofend

We define the {\em fragmentation breakpoint} $t_f$ as
\begin{equation}\label{tfeqn}
t_f =  \hbox{$\frac{1}{\varepsilon_{\! h}}$}
c_f^\mathfrak{p} n^{c_f n^3} 
   (\mathfrak{p}+ \log \hbox{$\frac{1}{\varepsilon_{\! h}}$})^n,
\end{equation}
where $c_f$ is a large enough constant.
Setting $D=1$ in~(\ref{Deltaij-logt/t-pD}),
we find that, by hysteresis and 
the Escape Lemma, 
the edges of $G_t$ can break only if $t< t_f$.
Past the fragmentation breakpoint, flocks can only merge.

\begin{lemma}\label{entermergephase}
$\!\!\! .\,\,$
At any time $t\geq t_f$, the flocking network
$G_t$ may gain new edges but never lose any.
\end{lemma}

The Escape Lemma tells us that,
after the fragmentation breakpoint,
birds fly almost in a straightline
and both their positions and velocities
can be predicted with low relative error.
From a physical standpoint, {\em they have already converged.}
The flocking network may still change, however.
It may keep doing so for an unbelievably long time.
This is what we show in the next section.
Note that, under the default settings of~(\ref{Assumptions}),
the fragmentation breakpoint $t_f$ is $n^{O(n^3)}$.

\subsection{Iterated Exponential Growth}\label{Iterat}

To pinpoint the exact convergence time requires
some effort, so it is helpful to break down the task
into two parts. We begin with a proof that
the flocking network reaches steady state
after a number of steps equal to a tower-of-twos
of linear height. This allows us to present
some of the main ideas and prepare the grounds for
the more difficult proof of the logarithmic height
in~\S\ref{SpectralS}.
The main tools we use in this section are 
the rationality of limit configurations and
root separation bounds from elimination theory.
Our investigation focuses on the post-fragmentation phase,
ie, $t>t_f$. We do not yet
adopt the assumptions of~(\ref{Assumptions});
in particular, we use
the definition of $t_f$ given in~(\ref{tfeqn}).

\begin{lemma}\label{duringmergephase}
$\!\!\! .\,\,$
Consider two birds adjacent at time $t$ but not $t-1$.
Assume that the flocks that contain them
remain invariant and noise-free
during the period $[t_1,t-1]$, where
$t_f< t_1< t-1$.
If, at time $t-1$, the birds are in different
flocks with distinct stationary velocities,
then $t\leq n^{O(t_1 n)}$; otherwise,
$t\leq t_1 2^{n^{O(1)}}$.
\end{lemma}
\proof
Assume that the flocking network $G_t$
stays invariant during the period $[t_1,t-1]$.
Consider two birds ${\mathcal B}_i$ and ${\mathcal B}_j$ 
that are adjacent in $G_t$ but not 
during $[t_1,t-1]$. The two birds
may or may not be in the same flock at time $t-1$.
Let the flock for ${\mathcal B}_i$ 
(resp. ${\mathcal B}_j$) consist of 
$m$ (resp. $m'$) birds: $m=m'$ if the birds
are in the same flock, else $m+m'\leq n$. 
By abuse of notation,
we use the terminology 
of~(\ref{P_N}), ie, $P$, $\pi$, $C$, $u_k$, $\lambda_k$,
as well as $v(t)$, 
to refer to the flock of $m$ birds, and we add primes
to distinguish it from the flock of ${\mathcal B}_j$.
We wish to place an upper bound on $t-t_1$.
Let $\chi(i)$ denote the $m$-dimensional vector with 
all coordinates equal to $0$, except for $\chi(i)_i=1$.
By~(\ref{P_N}, \ref{x(t)-SumP^s}), for $t> t_1$,
\begin{equation*}
\begin{split}
x_i(t)
&= x_i(t_1)+ \Bigl(\,  \sum_{s=0}^{t-t_1-1} 
             (\chi(i)^T P^s) \otimes I_d \,\Bigr) v(t_1+1) \\
&= 
x_i(t_1)+ (t-t_1)y+ \sum_{k=2}^m 
    \frac{\,\,\,\, 1- \lambda_k^{t-t_1}}{1-\lambda_k} \, \Phi_k,
\end{split}
\end{equation*}
where
\begin{equation}\label{yPhikident}
\begin{cases}
\, y= (\pi^T \otimes I_d)v(t_1+1) 
= {\mathbf m}_\pi[v(t_1+1)] \, ; \\
\, \Phi_k =   ((\chi(i)^T C^{1/2} u_k u_k^T  C^{-1/2}) 
\otimes I_d) v(t_1+1).
\end{cases}
\end{equation}

\medskip
\noindent
Note that, by~(\ref{P_N}, \ref{Gamma_tSUm}), 
\begin{equation*}
\begin{split}
\sum_{k=2}^m  \frac{1}{1-\lambda_k} \, \Phi_k
&=\lim_{t\rightarrow\infty}
\, \sum_{s=0}^{t-1}  
\, \sum_{k=2}^m  \lambda_k^s
((\chi(i)^T C^{1/2} u_k u_k^T  C^{-1/2}) 
\otimes I_d) v(t_1+1) \\
&=\lim_{t\rightarrow\infty}
\, \sum_{s=0}^{t-1}  
\, ((\chi(i)^T(P^s - {\mathbf 1} \pi^T))
\otimes I_d) v(t_1+1) \\
&=  ((\chi(i)^T \Gamma ) \otimes I_d) v(t_1+1) \, ;
\end{split}
\end{equation*}
therefore,
\begin{equation*}
x_i(t)= x_i(t_1) + ((\chi(i)^T \Gamma ) \otimes I_d) v(t_1+1)
+ (t-t_1)y 
- \sum_{k=2}^m  \lambda_k^{t-t_1} \frac{\Phi_k}{1-\lambda_k} \, .
\end{equation*}
Adding primes to distinguish between
the flocks of ${\mathcal B}_i$ and ${\mathcal B}_j$
(if need be), we find that 
\begin{equation}\label{aij-ABPsi}
x_i(t)- x_j(t) = A + B(t-t_1) 
- \sum_{k=1}^{m_0}   \Psi_k \,\mu_k^{t-t_1}, 
\end{equation}
where
\begin{itemize}
\item[(i)]
$A=  x_i(t_1) - x_j(t_1) + 
(( \chi(i)^T \Gamma ) \otimes I_d) v(t_1+1)
-
(( \chi'(j)^T \Gamma') \otimes I_d) v'(t_1+1)$:
By Lemma~\ref{x-v(t)-rationalprecision},
the vectors $v(t_1+1)$, $v'(t_1+1)$,
$x_i(t_1)$, and $x_j(t_1)$ 
have CD-rational coordinates over 
$O(t_1 n\log n +\mathfrak{p} n)$ bits, which is 
also $O(t_1 n\log n)$, since, 
by~(\ref{tfeqn}), $t_1>t_f>\mathfrak{p}$.
In view of Lemma~\ref{GammaCoeff}, this
implies that the same is true
of the vector $A$.
\item[(ii)]
$B= y-y'$: 
The stationary distribution
$\pi = (\hbox{tr}\,C^{-1})^{-1}C^{-1} \, {\mathbf 1}$
is a CD-rational vector over $O(n\log n)$ bits.
Together with Lemma~\ref{x-v(t)-rationalprecision},
this implies that $B$ has CD-rational
coordinates over 
$O(t_1 n\log n)$ bits; hence
either $B=0$ or 
$\|B\|_2 \geq n^{-O(t_1 n)}$.
\item[(iii)]
$\mu_1\geq \cdots \geq \mu_{m_0}$:
Each $\mu_k$ is an eigenvalue 
$\lambda_l$ or $\lambda_l'$ ($l,l'>1$)
and $|\mu_k|<1$. Their number $m_0$ is either $m-1$ 
(if the two birds ${\mathcal B}_i$ and ${\mathcal B}_j$
belong to the same flock) or $m+m'-2$, otherwise.
\item[(iv)]
Each $\Psi_k$ is a $d$-dimensional vector
of the form $\Phi_l/(1-\lambda_l)$ 
or $-\Phi_l'/(1-\lambda_l')$. 
Since the eigenvalues are bounded away from 1
by $n^{-O(1)}$ (Lemma~\ref{eigenBounds}), 
it follows from~(\ref{|v|Poly}), the 
submultiplicativity of the Frobenius norm, and 
$\mathfrak{p}\geq n^3$ that 
$\|\Psi_k\|_2= 2^{O(\mathfrak{p})}$.
In the same vein, we note for future reference that 
\begin{equation}\label{Psi-k-mu-UB}
\|\sum_{k=1}^{m_0} \Psi_k \,\mu_k^{t-t_1} \|_2 
\leq e^{-(t-t_1)n^{-O(1)}+ O(\mathfrak{p})}
= 2^{O(\mathfrak{p})}.
\end{equation}
\end{itemize}
We distinguish among three cases:

\bigskip\medskip
\noindent
{\bf Case I.}\ \ 
$B\neq 0$:
The two flocks must be distinct,
for having the two birds in the same flock
would imply that $\pi=\pi'$ and 
$v(t_1+1)= v'(t_1+1)$; hence $y=y'$.
By (i, ii), $\|A\|_2 \leq n^{O(t_1n)}$ 
and $\|B\|_2\geq n^{-O(t_1 n)}$. 
If the two birds are to be joined in $G_t$,
then $\text{\sc dist}_{t}({\mathcal B}_i, {\mathcal B}_j)
= \|x_i(t)-x_j(t)\|_2\leq 1$.
By~(\ref{tfeqn}),
$t_1>t_f > \mathfrak{p}$; hence
$2^{O(\mathfrak{p})}= n^{O(t_1n)}$. It follows 
from~(\ref{yPhikident}, \ref{Psi-k-mu-UB})
that $t-t_1 \leq  n^{O(t_1n)}$.
Note that, for the lower bound of $n^{-O(t_1 n)}$ 
on $\|B\|_2$ to be tight, the flock would
have to be able to generate numbers
almost as small as 
Lemma~\ref{x-v(t)-rationalprecision} allows.
For this to happen, energy must shift
toward the dominant eigenvalue. This spectral shift
occurs only in a specific context, which we examine
in detail in the next section.

\bigskip\medskip
\noindent
{\bf Case II.}\ \ 
$B= 0$ and $\|A\|_2 \neq 1$:
By (i), $\|A\|_2$ is 
bounded away from 1 by $n^{-O(t_1 n)}$.
It follows from~(\ref{aij-ABPsi}, \ref{Psi-k-mu-UB}) 
and the triangle inequality that
\begin{equation*}
\begin{split}
| \, \|x_i(t)- x_j(t)\|_2 -1 \, |
&\geq |\|A\|_2 -1| - 
| \, \|x_i(t)- x_j(t)\|_2 - \|A\|_2 \, |  \\
&\geq 
n^{-O(t_1 n)}
- \|\hbox{$\sum_k \Psi_k \,\mu_k^{t-t_1} $} \|_2 \\
&\geq 
n^{-O(t_1 n)} -
e^{-(t-t_1)n^{-O(1)}+ O(\mathfrak{p})}.
\end{split}
\end{equation*}
Since $t_1> \mathfrak{p}$, this 
implies that, for a large
enough constant $b_0$,
the distance between
the two birds remains bounded away from 1 by 
$n^{-O(t_1 n)}$ at any time $s\geq t_1 n^{b_0}$.
Not only that, but the sign of 
$\text{\sc dist}_{s}({\mathcal B}_i, {\mathcal B}_j) -1$
can no longer change after time $t_1 n^{b_0}$.
Indeed, for any $s\geq t_1 n^{b_0}$,
the distance between times $s-1$ and $s$ 
varies by an increment of 
$\Delta_{ij}(s)$, where, 
by~(\ref{Psi-k-mu-UB}),
\begin{equation*}
\begin{split}
\Delta_{ij}(s)
&=
| \, \|x_i(s)- x_j(s)\|_2 -
  \, \|x_i(s-1)- x_j(s-1)\|_2 | \\
&\leq 
\|\hbox{ $\sum_k \Psi_k \,\mu_k^{s-1-t_1} $}\|_2 
+ \|\hbox{ $\sum_k \Psi_k \,\mu_k^{s-t_1} $}\|_2  \\
&\leq e^{-(s-t_1)n^{-O(1)}+ O(\mathfrak{p})}
 \leq e^{-t_1 n^2}.
\end{split}
\end{equation*}
With $n$ assumed large enough, this ensures that,
past time $t_1 n^{b_0}$,
the distance can never cross the value 1.
Thus, if the two birds have not gotten within distance
1 of each other by time $t_1 n^{b_0}$, they never 
will---at least while their respective flocks remain invariant.
We conclude that $t\leq t_1 n^{O(1)}$.

\bigskip\medskip
\noindent
{\bf Case III.}\ \ 
$B= 0$ and $\|A\|_2 = 1$:
The distance between the two birds tends toward 1. 
The concern is that the two birds 
might stay safely away from
each other for a long period of time
and then suddenly decide to get close
enough to share an edge.
The rationality of the limit configuration
is insufficient to prevent this.
Only a local analysis of the convergence
can show that a long-delayed pairing is impossible.
We wish to prove that, if 
$\text{\sc dist}_{s}({\mathcal B}_i, {\mathcal B}_j)$
is to fall below 1 for $s>t_1$, this
must happen relatively soon.
Recall that, by~(\ref{aij-ABPsi}), 
$$
x_i(s)- x_j(s) = A - \sum_{k=1}^{m_0}   \Psi_k \,\mu_k^{s-t_1}, 
$$
where $A$ is a unit vector. We investigate the 
behavior of the birds' distance locally around 1.
$$
\|x_i(s)- x_j(s)\|_2^2 = 1 -
2 \sum_{k}   A^T \Psi_k \,\mu_k^{s-t_1}
+ \sum_{k,k'}  \Psi_k^T \Psi_{k'} \, (\mu_k \mu_{k'})^{s-t_1}.
$$
Let $1>\rho_1>\cdots > \rho_N>0$ be the distinct
nonzero values among $\{|\mu_k|, |\mu_k \mu_{k'}|\}$ ($N<n^2$).
These absolute values may appear with a plus or minus sign
(or both) in the expression above, so we rewrite it as 
\begin{equation}\label{xij-A-UpsilonSimple}
\|x_i(s)- x_j(s)\|_2^2-1 =
\sum_{k=1}^{N} \Upsilon_{k} \,\rho_k^{s-t_1},
\end{equation}
where each 
$$\Upsilon_{k}= \Upsilon_{k}^+ + (-1)^s\, \Upsilon_{k}^-$$
corresponds to a distinct $\rho_k$.
We distinguish between odd and even values of $s$ so as to
keep each $\Upsilon_{k}$ time-invariant. 
We assume that $s$ is even and skip the odd case 
because it is similar.
Of course, we may also assume that each 
$\Upsilon_{k}= \Upsilon_{k}^+ + \Upsilon_{k}^-$
is nonzero.
We know that $\sum_k \Upsilon_{k} \,\rho_k^{s-t_1}$ tends to 0
as $s$ goes to infinity,
but the issue is how so.
To answer this question, we need bounds on eigenvalue gaps
and on $|\Upsilon_{k}|$.
Tighter results could be obtained from
current spectral technology, but they would not make
any difference for our purposes, so we settle for
simple, conservative estimates. 

\begin{lemma}\label{BoundUpsilonk}
$\!\!\! .\,\,$
For all $k>1$ and $k\geq 1$, respectively,
$$
\rho_{k}\leq (1- 2^{-n^{O(1)}})\rho_1
\hspace{1cm}\text{and}\hspace{1cm}
2^{-t_1 2^{n^{O(1)}}} \leq|\Upsilon_{k}| = 
2^{O(\mathfrak{p})}.
$$
\end{lemma}
\proof
We begin with the eigenvalue gap.\footnote{\, 
For the purpose of this lemma, 
we again abuse notation
by letting $P$ and $n$ pertain to the
flock of either one of the two birds.
This will help the reader keep track of the
notation while, as a bonus, releasing $m$ as a variable.}
For this we use a conservative version of
Canny's root separation bound~\cite{canny88,yap00}:
given a system of $m$ integer-coefficient polynomials in $m$ variables
with a finite set of complex solution points, 
any nonzero coordinate has modulus
at least 
\begin{equation}\label{root-sep-bound}
2^{-\ell D^{O(m)}},
\end{equation}
where $D-1$ is the maximum degree of any polynomial
and $\ell$ is the number of bits needed
to represent any coefficient.
Any difference $\rho_k-\rho_{l}$ can expressed
by a quadratic polynomial, $z= z_1z_2 - z_3z_4$,
where each $z_i$ is either $1$ or the root of 
the characteristic polynomial
$\det\,(P- \lambda I_n)$.
The elements of $P$ are CD-rationals over $O(n\log n)$ bits,
so by the Hadamard bound~\cite{yap00}
the roots of $\det\,(P- \lambda I_n)$ 
are also those a polynomial of degree $n$
with integer coefficients over $O(n^2\log n)$ bits;
therefore,
$m\leq 5$; $D=n+1$; and 
$\ell= n^{O(1)}$. This proves that the
minimum gap between two $\rho_k$'s is
$2^{-n^{O(1)}}$. Since $\rho_1<1$, we find that,
for $k>1$,
\begin{equation*}
\rho_k\leq (1- 2^{-n^{O(1)}})\rho_1,
\end{equation*}
which proves the first part of the lemma.

By~(iv), $\|\Psi_l\|_2 = 2^{O(\mathfrak{p})}$; therefore,
by Cauchy-Schwarz and the inequalities $\rho_k<1$
and $\mathfrak{p}\geq n^3$, 
the same bound of $2^{O(\mathfrak{p})}$
applies to any $|\Upsilon_{k}|$, which
proves the second upper bound of the lemma.
We now prove that $|\Upsilon_{k}|$ cannot be too small.
Recall that it is the sum/difference of inner products 
between vectors in $\{ A, \Psi_h \}$.
We know from~(iv) that $\Psi_h$ is 
of the form $\Phi_l/(1-\lambda_l)$ 
or $-\Phi_l'/(1-\lambda_l')$. We assume
the former without loss of generality.
By~(\ref{P_N}, \ref{Gamma_tSUm}),
$$\Gamma = \sum_{r=2}^n \,
     \sum_{s\geq 0} \, 
       \lambda_r^s
        C^{1/2} u_r u_r^T C^{-1/2}.
$$ 
In view of~(iv) and~(\ref{yPhikident}), it then follows that
\medskip
\begin{equation*}
\begin{split}
\Psi_h&= \frac{\Phi_l}{1-\lambda_l}
= \frac{1}{1-\lambda_l}
  \Bigl\{ (\chi(i)^T C^{1/2} u_l u_l^T C^{-1/2})\otimes I_d \Bigr\} v(t_1+1) \\
&= \frac{1}{1-\lambda_l}
  \Bigl\{ (\chi(i)^T C^{1/2} u_l u_l^T C^{-1/2} C^{1/2} u_l u_l^T C^{-1/2}
      )\otimes I_d \Bigr\} v(t_1+1) \\
&=  \sum_{r=2}^n
  \frac{1}{1-\lambda_r}
 \Bigl\{ (\chi(i)^T C^{1/2} u_r u_r^T C^{-1/2} C^{1/2} u_l u_l^T C^{-1/2}
      )\otimes I_d \Bigr\} v(t_1+1) \\
&= \sum_{r=2}^n \,
     \sum_{s\geq 0} \, 
 \Bigl\{ (\chi(i)^T \lambda_r^s
        C^{1/2} u_r u_r^T C^{-1/2} C^{1/2} u_l u_l^T C^{-1/2}
           )\otimes I_d \Bigr\} v(t_1+1) \\
&=   ((\chi(i)^T \Gamma C^{1/2} u_l u_l^T C^{-1/2}
           )\otimes I_d ) v(t_1+1) 
=   ((\chi(i)^T \Gamma)\otimes I_d )W,
\end{split}
\end{equation*}

\medskip
\noindent
where $W= ((C^{1/2} u_l u_l^T C^{-1/2} )\otimes I_d ) v(t_1+1)$.
By Lemma~\ref{x-v(t)-rationalprecision},
$v(t_1+1)$ is a vector with CD-rational
coordinates over $O(t_1n\log n)$ bits; 
remember that $t_1> \mathfrak{p}$.
By Lemma~\ref{GammaCoeff},
the elements of $\Gamma$ are CD-rationals
encoded over $O(n\log n)$ bits.
Any coordinate of $\Psi_h$ can thus be written
as a sum $\sum_i$ of at most $n^2$ terms of the 
form $R_i\alpha_i y_i z_i$, where:
\begin{itemize}
\item
All the $R_i$'s are
products of the form $\Gamma_{\star \star}v_{\star}(t_1+1)$,
hence CD-rationals over $O(t_1n\log n)$ bits;
\item
$\alpha_i$ is the
square root of a rational $c_\star/c_\star$ over $O(\log n)$ bits;
\item
$y_i, z_i$ are two coordinates of $u_l$.
Recall that, by~(\ref{MP-diagonalized}),
$u_l$ is a unit eigenvector of $C^{-1/2} P C^{1/2}$.
\end{itemize}
By (i), $A$ is a vector with CD-rational
coordinates over $O(t_1n\log n)$ bits.
It follows that $\Upsilon_k$ is a sum $\sum_i$ of 
$n^{O(1)}$ terms of the form 
$S_i \gamma_i y_i z_i y_i' z_i'$:
\begin{itemize}
\item
All the $S_i$'s are 
CD-rationals over $O(t_1n\log n)$ bits;
\item
$\gamma_i$ is the square root of an $O(\log n)$-bit rational, ie,
a number of the form $\sqrt{(c_\star/c_\star)(c_\star/c_\star)}$;
\item
$y_i,z_i,y_i',z_i'$ are coordinates of the eigenvectors
(or 1, to account for $A^T\Psi_h$).
\end{itemize}
It is straightforward (but tedious) to set up 
an integer-coefficient algebraic
system over $m=n^{O(1)}$ variables that includes 
$\Upsilon_k$ as one of the variables. The number of equations
is also $m$ and the maximum degree is $n$.
All the coefficients are integers over
$O(t_1n\log n + n^{O(1)})$ bits.
Rather than setting up the system in full, let us briefly review
what it needs to contain: 
\begin{enumerate}
\item
$\Upsilon_k$ is a sum of $n^{O(1)}$ quintic monomials 
$S_i \gamma_i y_i z_i y_i' z_i$;
where the $S_i$'s are CD-rationals
over $O(t_1n\log n)$ bits.
\item
Each $\gamma_i$ is of the form $\sqrt{a/b}$, where
$a,b$ are $O(\log n)$-bit integers. We express it by
the equation $b \gamma_i^2 = a$.
(This yields two roots, 
but any solution set is fine as long as it
is finite and contains those we want.)
\item
The $y_i, z_i, y_i', z_i'$ are 
coordinates of the eigenvectors $u_l$ of 
$C^{-1/2} P C^{1/2}$.
We specify them by first 
defining the eigenvalues $\lambda_1,\ldots, \lambda_n$
and
\begin{equation*}
\begin{cases}
\, \det\,(P- \lambda_i I_n)=0 ; \\
\, C^{-1/2} P C^{1/2}u_i=\lambda_i u_i ;
\hspace{1.5cm} (1\leq i<j\leq n) \\
\, \|u_i\|_2^2=1$, and $u_i^Tu_j=0.
\end{cases}
\end{equation*}
The issue of multiplicity arises.
If the kernels of the various 
$P- \lambda_i I_n$ are not of dimension $1$, we 
must throw in cutting planes to bring down
their sizes.
We add in coordinate hyperplanes to the mix until
we get the right dimension. We then repeat this process
for each multiple eigenvalue in turn. 
(Of course, we do all this prior
to forming the vectors $\Psi_h$.)
We rewrite each eigensystem as 
$P v_i =\lambda_i v_i$, where $v_i= C^{1/2}u_i$,
and again we square the latter set of 
equations to bring them in polynomial form.
\end{enumerate}
Once we reduce all the rational coefficients
to integers, we can use the separation
bound~(\ref{root-sep-bound}),
for $m= n^{O(1)}$, $D=n+1$, and 
$\ell= O(t_1n\log n + n^{O(1)})$, which is 
$O(t_1n\log n)$.
This gives us a bound on the modulus of any nonzero coordinate
of the solution set; hence on $|\Upsilon_k|$.
\hfill $\Box$
\proofend

By~(\ref{xij-A-UpsilonSimple}), it follows from the lemma that 
$\|x_i(s)- x_j(s)\|_2^2 -1
= \Upsilon_{1} \,\rho_1^{s-t_1} ( 1 + \zeta)$,
where
$$ 
|\zeta|
\leq e^{-(s-t_1)2^{-n^{O(1)}}+ t_1 2^{n^{O(1)}}}
= o(1),
$$
for $s\geq t_1 2^{n^{b_1}}$, with $b_1$ being a large enough constant.
The same argument for odd values of $s$ shows that,
after $t_1 2^{n^{b_1}}$, either 
$ \|x_i(s)- x_j(s)\|_2^2$ stays on one side of 1
forever or it constantly alternates (at odd and even times).
Since the birds are joined in $G_t$
but not in $G_s$ ($t_1\leq s< t$), it must
be the case that $t\leq t_1 2^{n^{O(1)}}$.
This concludes Case III.

Putting all three results together, we find that the bound 
from Case I is the most severe, $t \leq  n^{O(t_1n)}$,
while Case II is the most lenient.
When the two birds are in the same flock
at time $t-1$, however, the
bound from Case III takes precedence.
\hfill $\Box$
\proofend

Lemmas~\ref{entermergephase} and~\ref{duringmergephase}
show that all network switches take place
within the first $t_\infty= 2\uparrow\uparrow O(n)$ steps.
Perturbations occur within $n^{O(1)}$ steps of
a switch and do not affect 
Lemma~\ref{x-v(t)-rationalprecision}.
The previous argument thus still applies 
and shows that the same upper bound 
also holds in the noisy model.
After time $t_\infty$, the flocking network
remains invariant. 
By virtue of~(\ref{x=xr+stuff}),
the limit trajectory of the birds within a given flock
is expressed as
$$
x(t)= x^r + 
(({\mathbf 1} \pi^T) \! \otimes I_d)x(t_\infty)
+(t-t_\infty)
(({\mathbf 1} \pi^T) \! \otimes I_d)v(t_\infty+1),
$$
where the stationary distribution $\pi$ refers to
the bird's flock (and therefore should be annotated
accordingly).

\subsection{Tower of Logarithmic Height}\label{SpectralS}

We prove that the tower-of-twos has height less than $4\log n$.
To simplify the notation (a decision whose wisdom
the reader will soon come to appreciate), we 
now adopt the assumptions of~(\ref{Assumptions}).
As we discussed earlier, this
means setting the fragmentation breakpoint $t_f= n^{f_0 n^3}$
for some large enough constant $f_0$.
The improvement rests on a more
careful analysis of the merges subsequent to 
the fragmentation breakpoint $t_f$.
Note that in the proof of Lemma~\ref{duringmergephase}
the bottleneck lies in Case I: specifically, in the 
lower bound on $\|B\|_2$ and the upper bound
on $\|A\|_2$. The latter can be improved easily
by invoking the Escape Lemma.
To get around $\|B\|_2$ requires more work.
Recall from~(\ref{x=mpi-v1}) that the position
vector of one flock is given by
$$
x(t)= a+ bt + (\, {\Gamma}_t \otimes I_d ) v,
$$
where the matrix ${\Gamma}_t$ describes
a damped oscillator.
The stationary velocity $b$ is formed by the
first spectral coordinates, one for each
dimension, associated with the eigenvalue $1$. The 
oscillator involves only 
the spectral coordinates 
corresponding to the subdominant
eigenvalues ($|\lambda_k|<1$).

\paragraph{The Combinatorics of the Spectral Shift.}

The reason flocks take longer to merge into 
larger flocks is that
they fly in formations increasingly parallel
to one another.
The term $bt$ grows linearly in $t$, so 
an iterated exponential growth can only come
from the oscillator. Of course, the 
angle between the flight directions of 
two flocks is given by the stationary velocities.
Therefore, for the angles to inherit 
an exponentially decaying growth,
it is necessary to {\em transfer} the 
fast-decaying energy of the oscillators to the stationary
velocities themselves. In other words,
the collision between two flocks must witness a spectral
shift from the ``subdominant'' eigenspace to
the stationary velocities.
Small angles are achieved by getting two
stationary velocities to be very close to each other.
Indeed, the spectral shift does not cause a decay of
the velocities themselves but of pairwise
differences. Recall that flocking is invariant
under translation in velocity space; so any interesting
phenomenon can be captured only by differences.

Let $b$ be the stationary
velocity of the new flock formed by
two flocks joining together after flying on
their own during $t$ steps. Let $b'$ be
the stationary velocity resulting from two 
other flocks flying in similar conditions.
The spectral shift will ensure that the
difference $b-b'$ has Euclidean norm
$e^{-tn^{-O(1)}}$, ie, exponentially small
in the flight time.
One should think of it as an energy transfer
from the subdominant eigenspaces to the stationary
velocities.
The challenge is to show that this transfer can occur
only under certain conditions that greatly
restrict its occurence. This requires
a combinatorial investigation of the spectral shift.

\vspace{1cm}
\begin{figure}[htb]\label{fig-spectralshift1}
\begin{center}
\hspace{0.1cm}
\includegraphics[width=6cm]{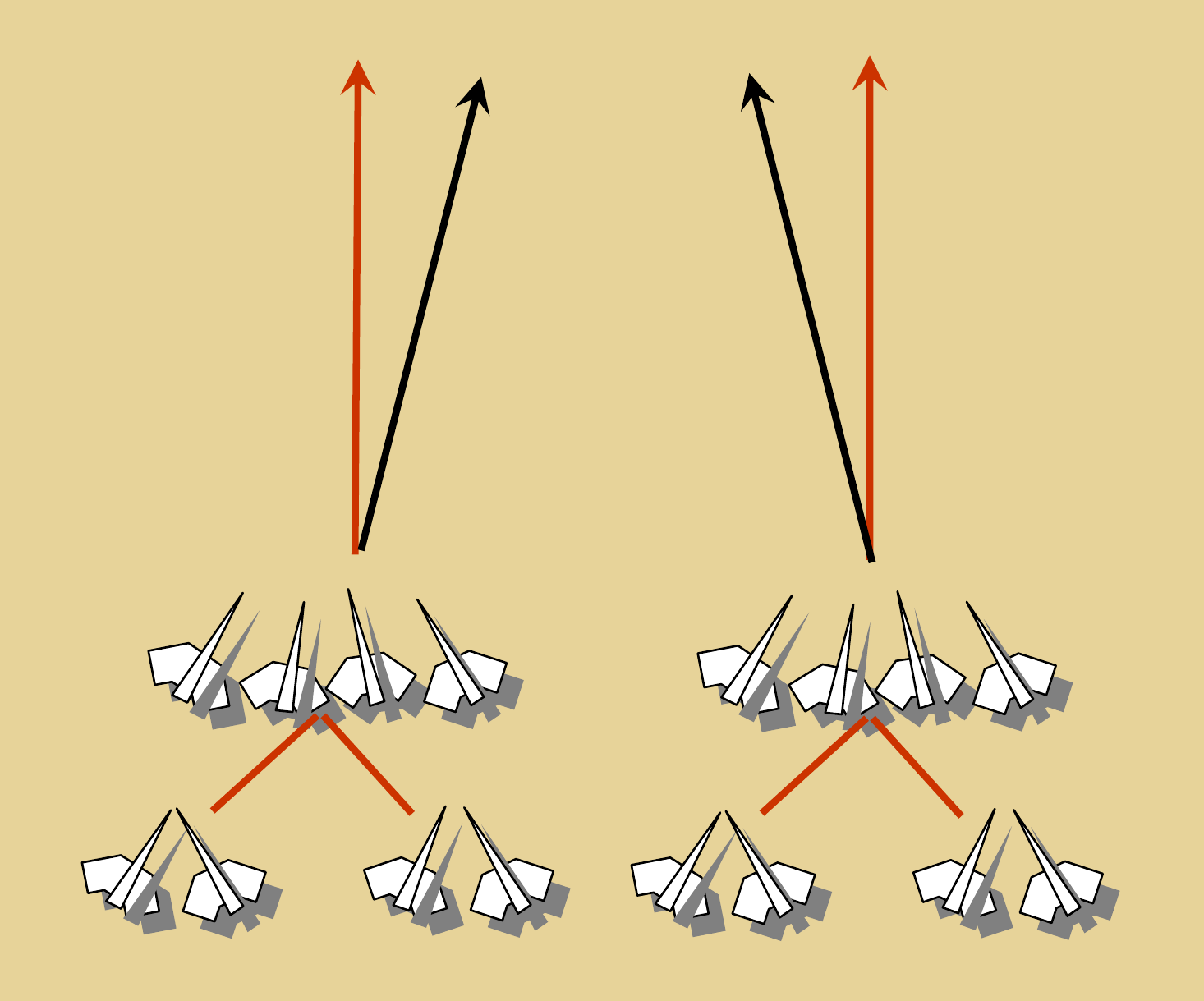}
\end{center}
\begin{quote}
\vspace{0.4cm}
\caption{\small 
Without spectral shift, the 
difference between stationary velocities
becomes null and the two flocks never meet.
The spectral shift resupplies the stationary
velocities with the fast-decaying energy located in
the subdominant part of the spectrum. This causes
a slight inflection of the trajectory (black lines).
}
\end{quote}
\end{figure}

We model the sequence of post-fragmentation breakpoint
merges by a forest $\mathcal F$:
each internal node $a$ corresponds 
to a flock $F_a$ of $n_a$ birds 
formed at time $t_a > t_f$. 
If $a$ is a leaf of $\mathcal F$, then its formation
time $t_a$ is at most $t_f$.
A node with at least two children
is called {\em branching}.
A nonbranching node
represents the addition of edges within
the same flock. Our analysis will focus 
on branching nodes with no more than two children.
In general, of course, this number can be arbitrarily high, 
as several flocks may come together to merge
simultaneously. We will see later how to break down
multiple aggregation of this form
into pairwise merges.

Let $L(t)$ denote the minimum value of $n_a$,
the number of birds in $F_a$, over all branching nodes $a$
and all initial conditions subject to~(\ref{Assumptions}), 
such that $t_a\geq t$.
Our previous upper bound shows that 
$L(t)=\Omega(\log^*t)$. We strengthen this:

\begin{lemma}\label{L(t)-bound}
$\!\!\! .\,\,$
$L(t)\geq (1.1938)^{\log^*t - O(\log\log n)}$,
where $x_0\approx 1.1938$
is the unique real root of $x^5 - x^2 -1$.
\end{lemma}

This implies that the last merge must take
place before time $t$ such that $L(t)\leq n$; hence
$t\leq 2\uparrow\uparrow (3.912\log n)$.
By Lemma~\ref{duringmergephase},
multiplying this quantity by $2^{n^{O(1)}}$
suffices to account for the network switches following
the last merge. As observed earlier, noise 
has no effect on this bound.
This proves the upper bound claimed in~\S\ref{introduction}.
\hfill $\Box$
\proofend

\paragraph{The Intuition.}

Think of the group of birds as a big-number
engine. How many bits can $n$ birds encode
in their velocities at the last network switch?
The previous argument shows that this
number cannot exceed a tower-of-twos
of linear height. We show that in fact the height is only
logarithmic. What keeps this number down is
the presence of {\em residues}.
We begin with a toy problem that has no direct connection
to bird flocking but illustrates the notion of residue.
Consider an $n$-leaf binary tree whose nodes are 
associated with polynomials in ${\mathbb R}[X]$.
Each leaf is assigned its own polynomial of degree 1.
The polynomial $p_v$ at an internal node $v$ 
is defined recursively by combining those
at the children $u,w$: 
$$ p_v= p_u \oplus \, p_w = 
p_u + p_w +
(p_u-p_w)x^{2^{h(p_u-p_w)}},$$
where $h(p)$ is $0$ if $x=0$ not a root of $p$
and $h(p)$ is its multiplicity otherwise;
in other words, it is the lowest degree among
the (nonzero) monomials of $p$.
How big can the degree of $p_{\text{root}}$ be?
It is immediate to achieve a degree that is
a tower-of-twos of logarithmic height.
Take a complete binary tree and assign 
the polynomial $(-1)^{l(v)} x$ to a leaf $v$, where
$l(v)$ is the number of left turns from the root to the leaf.
We verify by induction that the polynomials
at level $k$ are of the form $\pm c_k x^{d_k}$,
where $d_1=1$ and, for $k>1$,
$$
c_k x^{d_k} 
= \pm \,  2c_{k-1} x^{d_{k-1}+ 2^{d_{k-1}}}.
$$
This shows that $d_k = d_{k-1}+ 2^{d_{k-1}}$; hence
the stated tower-of-twos of logarithmic height.
Couldn't we increase the height by choosing a
nonbalanced tree? The answer is no, but not for
the obvious reason.  The ``obvious reason'' would be
that to go from a node $u$ of degree $d$ to a parent
of degree $2^d$ requires not just one but two children $u,v$
of degree $d$. Nice idea. Unfortunately, it is not true:
\begin{equation}\label{exp-growth-deg}
  x^d \oplus \, 0=   x^d + x^{d+2^d}.
\end{equation}
Note, however, that if we try to repeat this trick we get
$$ \Bigl( x^d + x^{d+2^d} \Bigr) \oplus \, 0=   
x^{d}\Bigl(1 + x^{2^d}\Bigr)^2,
$$
which increases the degree by only a constant factor.
The reason for this is that during 
the exponential jump in~(\ref{exp-growth-deg})
the polynomial inherited a {\em residue}, 
ie, the ``low-degree'' monomial $x^d$, which 
will hamper future growth until it is removed.
But to do so requires another ``big-degree'' child.
This residue-clearing task is what keeps the tower's 
height logarithmic. We prove this below.

\begin{theorem}\label{residue-clear}
$\!\!\! .\,\,$
A tree of $n$ nodes can produce only polynomials
of degree at most $2\uparrow\uparrow O(\log n)$.
\end{theorem}
\proof
Let $L(d)$ be the minimum number of leaves
needed to produce at the root a polynomial
of degree at least $d$.
We prove that
\begin{equation}\label{L(d)-toyproblem}
L(d)\geq 2^{\Omega(\log ^*d)},
\end{equation}
from which the theorem follows.
Let $v$ be the root of the smallest $n$-leaf tree
that achieves $d_v\geq d$, where 
$d_v$ denotes the degree of $p_v$.
Let $u, w$ be the children of $v$, with
$d_u\geq d_w$, and let $y,z$ be the children
of $u$ with $d_y\geq d_z$.
We assume that $d$ is large enough, so all these nodes exist.
Note that 
\begin{equation}\label{d-du-dx-d}
d_v\leq d_u+2^{d_u}
\hspace{1cm}
\text{and}
\hspace{1cm}
d_u\leq d_y+2^{d_y}.
\end{equation}
Assume that 
\begin{equation}\label{L(d)-assump}
\begin{cases}
\, d_z, d_w < \log\log\log d_v; \\
\, d_y< \sqrt{d_u}< d_u< \sqrt{d_v}.$$
\end{cases}
\end{equation}
In view of~(\ref{d-du-dx-d}), this shows that 
$d_u>d_w$ and $d_y>d_z$.
This first inequality implies that
$d_u+ 2^{h(p_u-p_w)}= d$; therefore, by~(\ref{L(d)-assump}),
$h(p_u-p_w)> \frac{1}{2}\log d_v$.
In other words,
\begin{equation}\label{pu:sqrt-d}
p_u= p_w+ x^{\lceil (\log d_v)/2\rceil} q_u \, , 
\end{equation}
for some polynomial $q_u\neq 0$.
Repeating the same line of reasoning at node $u$, we derive
the identity $d_y+ 2^{h(p_y-p_z)}= d_u$ from
the strict inequality $d_y>d_z$.
It follows from~(\ref{d-du-dx-d}, \ref{L(d)-assump}) that 
$$d_z< \log\log\log d_v< 
\hbox{$\frac{1}{2}\log d_u$}
< h(p_y-p_z)
\leq \log d_u < \hbox{$\frac{1}{2}\log d_v$}.
$$
This implies two things: first, 
by $d_z< h(p_y-p_z)$, the polynomial
$p_y$ has a monomial $q$
of degree $h(p_y-p_z)$; second,
that degree is strictly 
between $\log\log\log d_v$ and
$\frac{1}{2}\log d_v$.
A quick look at the formula
$$ p_u= p_y + p_z +
(p_y-p_z)x^{2^{h(p_y-p_z)}}$$
shows that $p_u$ also contains $q$:
indeed, by $d_z< h(p_y-p_z)$, it must be the case that
$p_y + p_z$ contains the monomial $q$;
on the other hand, the minimum degree in 
$(p_y-p_z)x^{2^{h(p_y-p_z)}}$ exceeds 
$h(p_y-p_z)$. This proves the presence of $q$ in $p_u$,
which contradicts~(\ref{pu:sqrt-d}).
This, in turn, means that (\ref{L(d)-assump}) cannot hold.
The monomial $q$ of degree 
$h(p_y-p_z)$ is the 
{\em residue} that the big-number engine must clear
before it can continue exponentiating degrees.
Since $d_y> \frac{1}{2}\log\log d$,
at least one of these two
conditions applies for any large enough $d$:
\begin{equation*}
L(d) \geq
\begin{cases}
\,  L(\hbox{$\frac{1}{2}\log\log d$}) + L(\log\log\log d) \, ; \\
\,  L( d^{1/4} ) + 1 \, .
\end{cases}
\end{equation*}
We use the monotonicity of $L$ to reduce
all the cases to the two above.
The lower bound~(\ref{L(d)-toyproblem}) follows by induction.
\hfill $\Box$
\proofend

\paragraph{Clearing Residues.}

Recall that $t_a>t_f$ is the time at which the flock
$F_a$ is formed at node $a$ of 
$\mathcal F$ after the fragmentation breakpoint 
$t_f= n^{f_0 n^3}$.
With the usual notational convention,
it follows from~(\ref{modelD}, \ref{P_N}) that,
in the absence of noise, for $t\geq t_a$,
\begin{equation*}
\begin{split}
v_a(t)
&= (P^{t-t_a} \otimes I_d) v_a(t_a) \\
&= ({\mathbf 1}_{n_a} \otimes I_d){\mathbf m}_a 
+ \sum_{k>1} \lambda_k^{t-t_a}
       (( C^{1/2} u_k u_k^T  C^{-1/2}) \otimes I_d)v_a(t_a), 
\end{split}
\end{equation*}
where ${\mathbf m}_a = (\pi_a^T \otimes I_d)v_a(t_a)$ is
the stationary velocity of the flock $F_a$, ie, the 
$d$-dimensional vector of first spectral coordinates.
As usual, it is understood that $P, C, \lambda_k, u_k$, etc,
are all defined with respect to the specific flock $F_a$
and not the whole group of $n$ birds.
We subscript ${\mathbf 1}$ with the flock size
for convenience. 
By~(\ref{Assumptions}), $\mathfrak{p}= n^3$; hence,
by~(\ref{|v|Poly}, \ref{Pij^sBounds}),
\begin{equation}\label{vat-p-lambda}
\|v_a(t) - ({\mathbf 1}_{n_a} \otimes I_d){\mathbf m}_a \|_2
\leq e^{-(t-t_a) n^{-O(1)}+ O(n^3)} .
\end{equation}
By the general form of the stationary distribution
$\pi_a$ as $(\hbox{tr}\,C^{-1})^{-1}C^{-1} \, {\mathbf 1}_{n_a}$,
its coordinates are CD-rationals over $O(n\log n)$ bits.
So, by Lemma~\ref{x-v(t)-rationalprecision}, 
each coordinate of ${\mathbf m}_a$ 
is an irreducible CD-rational $p_a/q_a$, where 
the number of bits needed for $p_a$ and $q_a$ 
is $O(t_a n\log n + \mathfrak{p} n)= O(t_a n\log n)$.
We denote the maximum bit length over all
$d$ coordinates by $\ell({\mathbf m}_a)$.
The following holds even in the noisy model:

\begin{equation}\label{ell(ma)}
\ell({\mathbf m}_a)= O(t_a n\log n).
\end{equation}

Consider a flock $F_{c}$ associated
with a branching 
node $c$ of $\mathcal F$: let $a$ and $b$ be the two children of 
$c$ in $\mathcal F$ (hence $n_c=n_a+n_b$)
and assume that $t_a\geq t_b$ 
and that no node of the forest $\mathcal F$ 
has more than two children, ie, flocks merge only two 
at a time.\footnote{\, The simultaneous merging
of more than two flocks can be dealt with by breaking ties
arbitrarily. Since there are fewer than $n$ merges, 
this means that in our calculations time might be off by at most
an additive term less than $n$. 
One can verify that 
this discrepany has no real effect
on any of the derivations and conclusions presented below.}
By Corollary~\ref{VelocityDiff},
the difference in stationary
velocities between $F_a$ and $F_b$ satisfies
\begin{equation}\label{diff-vel-timeUB}
\|{\mathbf m}_a - {\mathbf m}_b \|_2
\leq \hbox{$\frac{\log t_c}{t_c}$}\,  n^{O(n^3)}.
\end{equation}
If the difference is null, then by Cases II, III
of the previous analysis ($B=0$),
$t_c=t_a 2^{n^{O(1)}}$.
Otherwise, 
by~(\ref{ell(ma)}) and the equivalent bound for
$\|{\mathbf m}_b\|_2$, 
\begin{equation}\label{diff-ma-mabLB}
\|{\mathbf m}_a - {\mathbf m}_b \|_2 
\geq n^{-O(t_a n)}.
\end{equation}
The two inequalities~(\ref{diff-vel-timeUB}, \ref{diff-ma-mabLB})
yield an upper bound on $t_c$. 
By our treatment of Cases II, III 
in the proof of Lemma~\ref{duringmergephase},
we conclude that, whether ${\mathbf m}_a= {\mathbf m}_b$ or not,

\begin{equation}\label{Exponent-Growth}
t_c \leq n^{O(t_a n)}.
\end{equation}

This leads to our earlier
$\Omega(\log^*t)$ bound on $L(t)$.
It is essentially a new derivation of
our previous result. We now see how to improve it.
Let ${\mathcal F}_o$ be the forest derived
from ${\mathcal F}$ by removing all nonbranching internal nodes
and merging the adjacent edges in the obvious way.
Our assumption implies that each internal node of
${\mathcal F}_o$ has exactly two children.
Let $a_0,\ldots, a_k$ ($k>1$) be an ascending path 
in ${\mathcal F}_o$ and let $b_i$ denote
the unique sibling of $a_i$.
The following lemma assumes the noisy model.
Its proof is postponed.

\begin{lemma}\label{a01--ak}
$\!\!\! .\,\,$
Assume that $2^{2^{t_f}}< \log\log\log t_{a_k}< t_{a_0}^4<t_{a_1}
<\log t_{a_k}$.
Then, 
$t_{b_{i_0}}\geq \sqrt{\log\log t_{a_0}}$, for some $0\leq i_0 <k$.
\end{lemma}

\vspace{1cm}
\begin{figure}[htb]\label{fig-treeresidue}
\begin{center}
\hspace{.3cm}
\includegraphics[width=8cm]{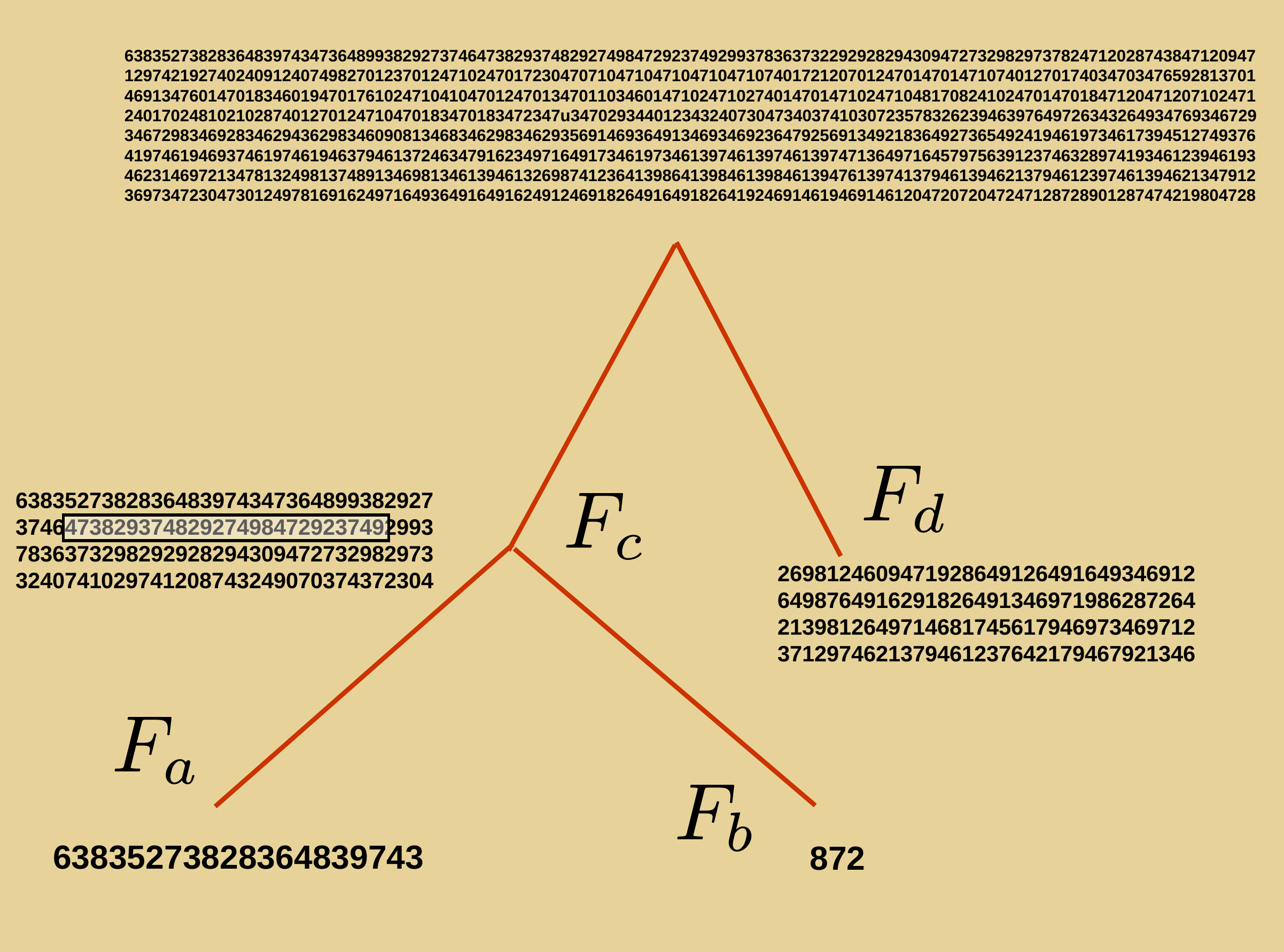}
\end{center}
\begin{quote}
\vspace{0cm}
\caption{\small 
A big flock $F_a$ may join with a small one, $F_b$, to form
a flock $F_c$ that produces 
a much larger number than either one could manufacture
on their own. This, however, cannot be repeated in the next step.
To create a
bigger number at the parent flock of
$F_c$, the residual heat in $F_c$ (numbers in box) must be evacuated, which itself
requires free energy that can only be provided by
a flock $F_d$ that roughly matches $F_c$ in size.
In this way, an abundance of spectral shifts
forces balance into the forest ${\mathcal F}$.
}
\end{quote}
\end{figure}

\paragraph{The Recurrence.}

We set up a recurrence relation on $L(t)$
to prove the lower bound of Lemma~\ref{L(t)-bound}, ie,
$L(t)\geq (1.1938)^{\log^*t - O(\log\log n)}$.
Let $t_0= 2\uparrow\uparrow \lfloor \log\log n\rfloor$.
It is assumed as usual that $n$ is large enough.
For $t\leq t_0$, we have the trivial lower bound
$L(t)\geq 1$ (choose the constant in the big-oh
to be larger than 1), so we may assume that $t> t_0$.
The child $b$ of a node $c$ (both defined with
respect to ${\mathcal F}_o$)
is called {\em near} if $t_b> (\log t_c)^{2/3}$.

\begin{lemma}\label{near-always}
$\!\!\! .\,\,$
Any internal node $c$ of ${\mathcal F}_o$ such that
$t_c\geq 2^{2^{t_f}}$ has at least one near child.
\end{lemma}
\proof
By~(\ref{Exponent-Growth}), we know that
$c$ has a child $b_0$ in the original forest
${\mathcal F}$ such that 
$t_c = n^{O(t_{b_0} n)}$.
We exhibit a near child $b$ for $c$.
If $b_0$ is branching, set $b=b_0$;
otherwise, set $b$ to the nearest branching descendent
of $b_0$. By Lemma~\ref{duringmergephase},
the formation times of 
any node in ${\mathcal F}$
and its nonbranching parent differ by at most a factor
of $2^{n^{O(1)}}$.  Perturbations make
no difference since they occur within
polynomial time of a switch.
Since ${\mathcal F}$ has fewer than
$n^2$ nodes and 
$t_c\geq 2^{2^{t_f}}$, with 
$t_f= n^{f_0 n^3}$,
$$
t_b\geq 2^{-n^{O(1)}} t_{b_0}
   \geq 2^{-n^{O(1)}} \log t_c
    > (\log t_c)^{2/3}.
$$
\hfill $\Box$
\proofend

Let $c_0$ be an arbitrary node of ${\mathcal F}_o$
such that 
\begin{equation}\label{t-c0>t>t0}
t_{c_0}\geq t > 
t_0= 2\uparrow\uparrow \lfloor \log\log n\rfloor.
\end{equation}
By the previous lemma, 
we can follow a descending path 
in ${\mathcal F}_o$
of near children
$c_0,c_1,\ldots, c_l$, where 
$t_{c_l}< 2^{2^{t_f}}\leq t_{c_{l+1}}$.
Because $t_0$ is so much
greater than $t_{c_l}$, the path has more than
a constant number of nodes---in fact, at least on 
the order of $\log\log n$.
For future use, we note that

\begin{equation}\label{Cond1-a01-ak}
2^{2^{t_f}}< \log\log\log t_{c_0}.
\end{equation}

\begin{lemma}\label{ck-exists}
$\!\!\! .\,\,$
There exists $k>1$ such that
$$
\log\log\log t_{c_0} < t_{c_k}^4< t_{c_{k-1}}
< \log t_{c_0}\, .
$$
\end{lemma}
\proof
By~(\ref{Cond1-a01-ak}) and Lemma~\ref{near-always},
there exists some $c_j$ in ${\mathcal F}_o$
such that
$$(\log\log t_{c_0})^{2/3}< t_{c_j}< \log t_{c_0}.$$
Suppose now that all the nodes $c_i$, for $i=j+1,j+2,\ldots,l$,
satisfy $t_{c_i}^4\geq t_{c_{i-1}}$.
Since there are most $n$ nodes along the path from
$c_0$ to $c_l$ in ${\mathcal F}_o$, 
then, by~(\ref{Cond1-a01-ak}) again,
\begin{equation}\label{tf-tch-cont}
2^{2^{t_f}} > t_{c_l}\geq t_{c_j}^{4^{-n}}
> (\log\log t_{c_0})^{4^{-n-1}}
> 2^{2^{2^{t_f/2}}}.
\end{equation}
This contradiction proves the existence
of some node $c_k$ ($j<k\leq l$) such that
$$t_{c_k}^4< t_{c_{k-1}}< \log t_{c_0}.$$
The argument used in~(\ref{tf-tch-cont})
shows that the smallest such $k$ satisfies,
via~(\ref{Cond1-a01-ak}),
$$t_{c_{k-1}}\geq t_{c_j}^{4^{-n}}
> (\log\log t_{c_0})^{4^{-n-1}}
> 2^{2^{t_f}}.
$$
Another application of the inequality above,
$t_{c_{k-1}}> 2^{2^{t_f}}$, allows us to invoke
Lemma~\ref{near-always}. By virtue of $t_{c_0}$ being
so big~(\ref{Cond1-a01-ak})
and $c_k$ being a near child of $c_{k-1}$ 
(by construction), 
$$t_{c_k}^4> (\log t_{c_{k-1}})^{8/3} >
4^{-8n}(\log\log\log t_{c_0})^{8/3} >
\log\log\log t_{c_0}.
$$
\hfill $\Box$
\proofend

We now prove Lemma~\ref{L(t)-bound}.
Setting $a_i=c_{k-i}$ for $i=0,\ldots, k$, 
together with~(\ref{Cond1-a01-ak}), the lemma
sets the conditions of Lemma~\ref{a01--ak}.
This shows that
$t_{a_0}> (\log\log\log t_{a_k})^{1/4}$
and, conservatively,
$$t_{b_{i_0}}
> (\log\log \log\log\log t_{a_k})^{1/3}.
$$
Nodes $a_0$ and $b_{i_0}$ are roots of disjoint 
subtrees, so the number of leaves below $a_k$ is at least
that of those below $a_0$ added to those below $b_{i_0}$.
Since $L$ is a monotone function and,
by~(\ref{t-c0>t>t0}), 
$a_k$ is an arbitrary node such 
that $t_{a_k}\geq t$, 
$$ L(t) \geq  L( (\log\log\log t)^{1/4} )
           +  L( (\log\log\log\log\log t)^{1/3} ),
$$
for $t>t_0= 2\uparrow\uparrow \lfloor \log\log n\rfloor$,
and $L(t)\geq 1$ for $t\leq t_0$.
We solve the recurrence without the exponents,
and then show that ignoring them makes 
no asymptotic difference.
Define $L^*(t)=1$ for $t\leq t_0$ and, for any
$t>t_0$, 
$$ L^*(t) =  L^*( \log\log\log t )
           + L^*( \log\log\log\log\log t ).
$$
Given the bound we are aiming for, we can 
round off $t$ down to the next tower-of-twos.
If $L^*(t)=M(\sigma)$, where $\sigma = \log^*t$,
we can rewrite the recurrence relation as
$$M(\sigma)= M(\sigma-3)+  M(\sigma-5),$$ where
$M(\sigma)=1$ for $\sigma \leq \log^*t_0$.
Quite clearly, $M(\sigma)$ upper-bounds the maximum 
number $n_s$ of leaves
in a binary tree ${\mathcal T}^*$ where: (i) each left edge
is labeled $3$ and each right edge $5$;
and (ii) the sum of the labels along any path
is at most $s= \log^*t - \log^*t_0$.
Note that ${\mathcal T}^*$ is binary: the constraint
that each internal node should have exactly two nodes
does not follow from the definition and is therefore added.
We seek a lower bound of the form $c x^s$. This means 
that $x^s\geq x^{s-3} +  x^{s-5}$, for $s\geq 5$
and $cx^s\leq 1$ else.
The characteristic equation is
$$
x^5 - x^2 -1 =0.
$$
We choose the unique real root 
$x_0\approx 1.19385$; this leads to $c=x_0^{-5}$.
This shows that $n_s\geq x_0^{s-5}$; hence,
$$
L^*(t)
\geq x_0^{\log^*t - \log\log n -5}.
$$
It is obvious that the binary tree 
$\mathcal T$ associated with the recurrence for $L(t)$
embeds in ${\mathcal T}^*$ with the same root.
We claim that it is not much smaller:
specifically, no leaf in 
$\mathcal T$ has more than a constant number of
descendents in ${\mathcal T}^*$. 
This implies immediately that
$$
L(t)
\geq x_0^{\log^*t - O(\log\log n)},
$$
which proves Lemma~\ref{L(t)-bound}.
\hfill $\Box$
\proofend

To prove our claim, we show that no path in
${\mathcal T}^*$ extends past its counterpart
in ${\mathcal T}$ by more than a constant number of nodes.
We model simultaneous, parallel walks down the trees
as a collaborative game between two players,
Bob and Alice, who take turns. Initially, both of them share
the same value 
$$t_A=t_B=t>t_0.$$
In one round,
Bob modifies his current value by 
taking iterated logs.
He is entitled to up to 5 logarithm iterations;
in other words, he can set
$t_B\leftarrow \log t_B$ or 
$$t_B\leftarrow \log\log\log\log\log t_B,$$ or anything
in-between.
Alice mimics Bob's move but then completes
it by taking a fractional power;
for example, if Bob opts for, say,
$\log\log t_B$, then Alice resets her value to 
$(\log\log t_A)^\alpha$, where $\alpha$
is a number between $\frac{1}{4}$ and 1.
To summarize, Bob chooses the number of log
iterations and Alice chooses $\alpha$: they can
change these parameters at each round.
A player's {\em score} is the number of rounds before
his or her value falls below (or at) $t_0$. Alice's score cannot
be higher than Bob's, so the latter is expected
to play the last rounds on his own.

The joint goal of the players is to maximize their
score differential.
Regardless of either player's strategy, we show that
Bob's score never exceeds Alice's by more than a constant.
This follows directly from the 
next two lemmas, whose proofs we postpone.

\begin{lemma}\label{L(t)vsL*(t)-game}
$\!\!\! .\,\,$
The score differential is maximized when Bob
always selects the single-iterated log rule
and Alice follows suit with $\alpha=\frac{1}{4}$;
in other words, 
$t_B\leftarrow \log t_B$ and 
$t_A\leftarrow (\log t_A)^{1/4}$.
\end{lemma}

With the strategy of the lemma,
Bob's score is $\log^*t- \log^* t_0$.
Within an additive constant, Alice's score is
at least the minimum $h$ such that $c_h\geq t$, where
$c_i$ is defined by $c_0=t_0^4$ and, for $i>0$, 
$c_i= 2^{4c_{i-1}}$.
To see why, note that the
inverse of the function $z\mapsto (\log z)^{1/4}$
is $z\mapsto 2^{z^4}$; taking logarithms on both sides
gives the recurrence on $c_i$.

\begin{lemma}\label{L(t)vsL*(t)}
$\!\!\! .\,\,$
For $t>t_0$, $\ \min\{\,h\,|\,  c_h\geq t\, \} \geq 
\log^* t - \log^* t_0 - O(1)$.
\end{lemma}

This validates our claim that no path in
${\mathcal T}^*$ extends past its counterpart
in ${\mathcal T}$ by more than a constant number of nodes.
This fills in the missing part
in the proof of Lemma~\ref{L(t)-bound}
and establishes the upper bound on the convergence
time claimed in~\S\ref{introduction}.

\paragraph{Proof of Lemma~\ref{a01--ak}.}

We begin with a few technical facts.
Recall from the ``Clearing Residues'' section  
that the flock $F_{c}$ is associated
with a branching 
node $c$ of $\mathcal F$ and that
$a$ and $b$ are its two children in
$\mathcal F$; furthermore,
$t_a\geq t_b$ and $t_a>t_f$, where $t_f= n^{f_0 n^3}$.
Assume that the velocity vector of $F_a$ at time $t_a$ is
of the form
\begin{equation}\label{va(ta)}
v_a(t_a)= ({\mathbf 1}_{n_a} \otimes I_d)\widetilde{\mathbf m}_a 
      + (u_a \otimes I_d)\mu_a + \zeta_a \, , 
\end{equation}
where $u_a\in {\mathbb R}^{n_a}$,
$\mu_a\in {\mathbb R}^{d}$, and,
for some real $\tau$,
\begin{equation}\label{ua-wa-ta}
\begin{cases}
\, 2^{t_f}\leq \tau \leq t_a^{1/3}; \\
\, \ell(\widetilde {\mathbf m}_a) = O(\log\log \tau); \\
\, \|u_a\|_\infty=1 
\hspace{.2cm} \&  \hspace{.2cm} 
u_a\geq 0 ;\\
\, e^{-\tau n^{O(1)}}\leq \|\mu_a\|_2
\leq \hbox{$\frac{1}{\tau}$}\, ; \\
\, \|\zeta_a\|_2 \leq e^{-\tau^2 n^{-O(1)} + n^{O(1)}}.
\end{cases}
\end{equation}
Note that the $d$-dimensional rational
vector $\widetilde {\mathbf m}_a$
is not defined as the stationary velocity 
${\mathbf m}_a$ of $F_a$, though it plays essentially
the same role.
The term $(u_a \otimes I_d)\mu_a$ 
creates the {\em residue} $\|\mu_a\|_2$
of $F_a$. Unless $F_b$ can ``destroy'' this residue
when it joins with $F_a$,
one should not expect the flock formation time
to grow exponentially. The crux is then to show
that only a flock $F_b$ 
with many birds can perform such a task.
The following result says that, if the flock $F_b$ 
settles too early, its effect on the residue of $F_a$ 
is negligible. The conditions on $F_c$ stated below differ
slightly from those for $F_a$ to make them closed under
composition. The lemma below also covers the case
$n_b=0$, when the transition from $F_a$ to $F_c$ is
involves the addition of an edge within the same flock.
(Here, too, we assume without loss of generality
that these additions occur only one at a time within
the same flock.) We postpone the proof of this result.

\begin{lemma}\label{abc1}
$\!\!\! .\,\,$
Suppose that $F_a$ undergoes no perturbation.
If node $b$ is well defined, then 
assume that $t_b< \log\log \tau$.
Whether node $b$ exists or not,
$$v_c(t_c)= ({\mathbf 1}_{n_c} \otimes I_d)\widetilde {\mathbf m}_a 
+ (u_c \otimes I_d)\mu_c + \zeta_c\, ,$$
where 
\begin{equation*}
\begin{cases}
\, \|u_c\|_\infty=1
\hspace{.2cm} \&  \hspace{.2cm} 
 u_c\geq 0 \, ;\\
\, \|\mu_a\|_2\, n^{-O(1)}\leq 
\|\mu_c\|_2 \leq 
\|\mu_a\|_2 \, ; \\
\, \|\zeta_c\|_2 \leq n \|\zeta_a\|_2 + e^{-\tau^2}.
\end{cases}
\end{equation*}
Furthermore, if node $b$ is well defined, then 
${\mathbf m}_{b} = \widetilde{\mathbf m}_{a}
\neq {\mathbf m}_a$.
\end{lemma}

\medskip
\noindent
{\sl Remark 2.3}.
It might be helpful to explain, at an intuitive level,
the meaning of the three
terms in the expression for $v_{a}(t_a)$, or 
equivalently $v_{c}(t_c)$:
$\widetilde {\mathbf m}_a$ is a low-precision
approximation of the stationary velocity
${\mathbf m}_a$; the 
vector $(u_a \otimes I_d)\mu_a$
creates the residue;
the remainder $\zeta_a$ is an error term.
The term $\widetilde {\mathbf m}_a$ is a 
low-resolution component of the velocity that
any other flock $F_b$ has to share if it 
is to create small angles with $F_a$ (the key to high
flock formation times) 
Think of it as a shared velocity caused by,
say, wind affecting all flocks in the same way.
This component must be factored out from the analysis
since it cannot play any role in engineering small angles.
This is a manifestation of the relativity principle
that only velocity {\em differences} matter.
To create small angles with $F_a$, 
incoming flocks $F_b$ must attack
the residue vector $(u_a \otimes I_d)\mu_a$. 
Of course, they could potentially take turns doing so.
Informally, one should read the
inequalities of the lemma as 
a repeat of~(\ref{ua-wa-ta}).
The lemma states a closure property:
unless $F_b$ brings many bits to the table
(via a formation time at least $\log\log \tau$),
conditions~(\ref{ua-wa-ta}) will still hold.
These conditions prevent the creation of small angles
between flocks, and hence of huge formation times.
In other words, flocks that settle too early cannot
hope to dislodge the residue $\|\mu_a\|_2$.  
The reason is that this residue is shielded in three ways:
first, it is too big 
for the error term $\zeta_a$ to 
interfere with it---compare 
$e^{-\tau n^{O(1)}}$ with $e^{-\tau^2 n^{-O(1)} + n^{O(1)}}$;
second, it is too small to be
affected by $\widetilde {\mathbf m}_a$---compare 
$\frac{1}{\tau}$ with a rational
over $O(\log\log\tau)$ bits;
third, all of its coordinates have the same sign
($u_a\geq 0$), so taking averages among them
cannot cause any cancellations.
This form of ``enduring'' positivity is
the most remarkable aspect of residues.

\medskip

By~(\ref{ua-wa-ta}), the lemma's bounds imply that 
$$
e^{-\tau n^{O(1)}}\leq \|\mu_c\|_2
\leq \hbox{$\frac{1}{\tau}$}
\hspace{.5cm} \&  \hspace{.5cm} 
\|\zeta_c\|_2 \leq e^{-\tau^2 n^{-O(1)} + n^{O(1)}},
$$
which brings us back to~(\ref{ua-wa-ta}).
If $c$ has a (unperturbed) parent $c'$ and sibling $b'$, then
we can apply the lemma again. 
Note that composition will
always be applied for the same value of $\tau$, ie, one is
that is not updated at each iteration.
In other words, the first two lines of~(\ref{ua-wa-ta}),
unlike the last three, 
are global inequalities that do not change with
each iteration. 
This closure property is not foolproof.
First, of course, we need to ensure that 
$t_{b'}< \log\log \tau$.
More important, 
we lose a polynomial factor at each iteration,
which is conveniently hidden in the big-oh notation.
So we may compose
the lemma only $n^{O(1)}$ times if we are
to avoid any visible loss 
in the bounds of~(\ref{ua-wa-ta}).
Since the forest has fewer than $n^2$ nodes, this means
that, as long as its conditions are met,
we can compose the lemma with ancestors of $c$ to
our heart's content and still get the full
benefits of~(\ref{ua-wa-ta}).

The provision that $b$ might not be well defined allows us
to handle nonbranching switch nodes 
with equal ease.
Recall that $v_c(t_c)$ is the velocity leading to time $t_c$, ie,
before the flock $F_b$ has had a chance to infuence it.
The provision 
in question might thus appear somewhat vacuous.
Its power will come from allowing us to 
apply the lemma repeatedly with no concern
whether a node has one of two children.
A related observation is that 
nowhere shall we use the fact that $t_a$ is 
the actual formation time of $F_a$. It could be
replaced in~(\ref{va(ta)}) by any 
$t_a'$ strictly between $t_a$ and $t_c$.
We thus trivially
derive a ``delayed'' version of Lemma~\ref{abc1}.
We summarize its two features:
(i) Lemma~\ref{abc1} can be composed iteratively
as often as we need to;
(ii) node $a$ need not be an actual
node of $\mathcal F$ but one introduced artificially
along an edge of $\mathcal F$.

What if $F_a$ undergoes a perturbation between $t_a$ and $t_c$?
Then the flock $F_a$ sees its velocity
multiplied by $I_{n_a}\otimes \widehat{\alpha}$,
where $\widehat{\alpha}$
is the diagonal matrix
with $\alpha=(\alpha_1,\ldots,\alpha_d)$ along
the diagonal and 
rational $|\alpha_i|\leq 1$ encoded over 
$O(\log n)$-bits.
Observe that the two matrices
$P_a\otimes I_d$ and 
$I_{n_a}\otimes \widehat{\alpha}$ commute;
therefore the perturbation can be assumed to
occur at time $t_a$. This means that, in lieu of~(\ref{va(ta)}), 
we have, using standard tensor rules,
\begin{equation*}
\begin{split}
v_a(t_a)
&=
(I_{n_a}\otimes \widehat{\alpha})
({\mathbf 1}_{n_a} \otimes I_d)\widetilde{\mathbf m}_a 
 + (I_{n_a}\otimes \widehat{\alpha})
(u_a \otimes I_d)\mu_a + 
(I_{n_a}\otimes \widehat{\alpha})
\zeta_a \\
&=
({\mathbf 1}_{n_a} \otimes \widehat{\alpha})\widetilde{\mathbf m}_a 
+ (u_a \otimes \widehat{\alpha})\mu_a + 
(I_{n_a}\otimes \widehat{\alpha})
\zeta_a \,.
\end{split}
\end{equation*}
Bringing it in the format of~(\ref{va(ta)}), we find that
$$
v_a(t_a)= ({\mathbf 1}_{n_a} \otimes I_d)
\widetilde{\mathbf m}_a
      + (u_a \otimes I_d)\mu_a + \zeta_a \, , 
$$
with the new assignments:
\begin{equation*}
\begin{cases}
\, \widetilde{\mathbf m}_a \leftarrow
 \widehat{\alpha}\, \widetilde{\mathbf m}_a \, ;\\
\, u_a \leftarrow u_a \, ;\\
\, \mu_a \leftarrow
 \widehat{\alpha} \, \mu_a \, ;\\
\, \zeta_a \leftarrow
(I_{n_a}\otimes \widehat{\alpha})
\zeta_a \,.
\end{cases}
\end{equation*}
It is immediate that the conditions of~(\ref{ua-wa-ta}) 
still hold: the only difference is that
$$\ell(\widetilde {\mathbf m}_c) 
\leq \ell(\widetilde {\mathbf m}_a) + O(\log n).$$
By~(\ref{ua-wa-ta}), 
$\log\tau\geq t_f= n^{f_0 n^3}$,
so $\ell(\widetilde {\mathbf m}_c)$
stays in $O(\log\log \tau)$ as long as 
the number of compositions is $O(n^3)$,
which it is. We summarize these observations:

\begin{lemma}\label{abc2}
$\!\!\! .\,\,$
Let $c_0,\ldots, c_l$ be an ascending path in
${\mathcal F}$ and let $d_i$ be the sibling, if any, of $c_i$.
Assume that $c_0$, possibly an artificial node,
satisfies the conditions of node $a$ in~\text{\rm (\ref{ua-wa-ta})}
and that $t_{d_i}< \log\log \tau$ for all $d_i$.
Then, 
$$v_{c_i}(t_{c_i})= ({\mathbf 1}_{n_{c_i}} 
   \otimes I_d)\widetilde {\mathbf m}_{c_i} 
+ (u_{c_i} \otimes I_d)\mu_{c_i} + \zeta_{c_i}\, ,$$
where 
\begin{equation*}
\begin{cases}
\, \ell(\widetilde {\mathbf m}_{c_i}) 
= O(\log\log \tau); \\
\, \|u_{c_i}\|_\infty=1
\hspace{.2cm} \&  \hspace{.2cm} 
 u_{c_i}\geq 0 \, ;\\
\, e^{-\tau n^{O(1)}}\leq \|\mu_{c_i}\|_2
\leq \hbox{$\frac{1}{\tau}$}\, ; \\
\, \|\zeta_{c_i}\|_2 \leq e^{-\tau^2 n^{-O(1)} + n^{O(1)}}.
\end{cases}
\end{equation*}
For all $d_i$,
${\mathbf m}_{d_i} = \widetilde{\mathbf m}_{c_i}
\neq {\mathbf m}_{c_i}$.
\end{lemma}

We are now equipped with the tools we need to 
prove Lemma~\ref{a01--ak}.
Recall that $a_0,\ldots, a_k$ ($k>1$) is an ascending path 
in ${\mathcal F}_o$ and $b_i$ denotes
the unique sibling of $a_i$. (Note that
$a_0\cdots a_k$ is a path in 
${\mathcal F}_o$ whereas, in Lemma~\ref{abc2},
$c_0\cdots c_l$ is a path in ${\mathcal F}$.)
Also, 
$$
2^{2^{t_f}}< \log\log\log t_{a_k}< t_{a_0}^4<t_{a_1}
<\log t_{a_k}. $$
Assume, by contradiction, that $t_{b_i}< \sqrt{\log\log t_{a_0}}$
for $i=0,\ldots, k-1$.
As we observed earlier, 
Lemma~\ref{duringmergephase} ensures that,
regardless of noise, the 
ratio between the formation times of 
any node in ${\mathcal F}$
and that of its nonbranching parent 
is at least $2^{-n^{O(1)}}$.
Since there are fewer than $n^2$ switches, this
implies that $F_{a_0}$ can undergo switches or
perturbations only between $t_{a_0}$ and $t_{a_0} 2^{n^{O(1)}}$.
Because $t_{a_1}> t_{a_0}^4> 2^{2^{t_f}}$, with 
$t_f= n^{f_0 n^3}$,
this shows that the entire 
time interval $[\frac{1}{2}t_{a_1}, t_{a_1})$ is 
free of switches and noise.
Let $a$ be the last node in 
${\mathcal F}$ from $a_0$ to $a_1$
and let $c_0$ be the artificial parent of $a$
corresponding to
the flock $F_{a}$ at time $t_{a_1}-1$: we set
$n_{c_0}=n_{a_0}$ and $t_{c_0}= t_{a_1}-1$.
The bound in~(\ref{vat-p-lambda})
ensures that the oscillations in the
flock $F_{c_0}$ are heavily damped. Indeed,
\begin{equation}\label{va0-1=mzeta}
v_{c_0}(t_{c_0}) = 
({\mathbf 1}_{n_{c_0}} \otimes I_d){\mathbf m}_{c_0}
+ \zeta_{c_0} \, , 
\end{equation}
where, because of the magnitude of $t_{a_1}$,
\begin{equation}\label{zetaUB}
\|\zeta_{c_0}\|_2\leq e^{-(t_{a_1}/2-1) n^{-O(1)}+ 
O(n^3)}
\leq e^{-t_{a_1} n^{-O(1)}}
       \leq e^{-\tau^2}.
\end{equation}
where  $\tau= \hbox{$\frac{1}{2}$}t_{a_1}^{1/3}$.
The rest of the sequence $\{c_i\}$ is now entirely
specified. In particular, note that
$c_1=a_1$ and $d_0=b_0$. By extension,
${\mathbf m}_{c_0}= {\mathbf m}_{a}$; so,
by~(\ref{diff-vel-timeUB}),
$$
\|{\mathbf m}_{c_0} - {\mathbf m}_{b_0} \|_2
\leq \hbox{$\frac{\log t_{a_1}}{t_{a_1}}$}\,  n^{O(n^3)}
< \hbox{$\frac{1}{\tau}$} \, .
$$
therefore, ${\mathbf m}_{c_0} = {\mathbf m}_{b_0} + \mu_{c_0}$,
where
\begin{equation}\label{mu_a0<1/tau}
\|\mu_{c_0}\|_2
< \hbox{$\frac{1}{\tau}$} \, .
\end{equation}
As we shall see, 
the presence of the square $\tau^2$ in the exponent 
of~(\ref{zetaUB}) ensures that the oscillations
of $F_{c_0}$ are too small to 
interfere with the residue $\|\mu_{c_0}\|_2$.
Writing $\widetilde{\mathbf m}_{c_0} = {\mathbf m}_{b_0}$,
it follows from~(\ref{va0-1=mzeta}) that 
$$
v_{c_0}(t_{c_0}) = 
({\mathbf 1}_{n_{c_0}} \otimes I_d)\widetilde{\mathbf m}_{c_0}
+ ({\mathbf 1}_{n_{c_0}} \otimes I_d)\mu_{c_0}
+ \zeta_{c_0}\, ,  $$
which matches~(\ref{va(ta)}), 
with $u_{c_0}= {\mathbf 1}_{n_{c_0}}$.
Since all the nodes $d_i$ are of the form $b_{j_i}$, 
$$t_{d_i}< \sqrt{\log\log t_{a_0}}
< \log\log\tau.$$
Thus, we will be able to apply
Lemma~\ref{abc2} once we verify 
that all conditions in~(\ref{ua-wa-ta}) are met:

\begin{itemize}
\item
$[\,2^{t_f}\leq \tau \leq t_{c_0}^{1/3}\,]$:  This follows from our
setting $\tau= \hbox{$\frac{1}{2}$}(t_{c_0}+1)^{1/3}$
and our assumption that $t_{a_1}> 2^{2^{t_f}}$.
\item
$[\,\ell(\widetilde{\mathbf m}_{c_0})= O(\log\log \tau)\,]$:
Because $\tau > 2^{2^{t_f-2}}$,
$$\sqrt{\log\log t_{a_0}}\, n\log n
< (\log\log t_{a_0})^{2/3}= o(\log\log \tau).
$$
The desired bound follows from~(\ref{ell(ma)}):
$$
\ell(\widetilde{\mathbf m}_{c_0})=
\ell({\mathbf m}_{b_0})= 
O(t_{b_0} n\log n)= 
O(\sqrt{\log\log t_{a_0}}\, n\log n)
< \log\log \tau.
$$ 
\item
$[\,e^{-\tau n^{O(1)}}\leq \|\mu_{c_0}\|_2
\leq \hbox{$\frac{1}{\tau}$}\,]$: 
The upper bound comes from~(\ref{mu_a0<1/tau}).
For the lower bound, 
note that ${\mathbf m}_{c_0}= {\mathbf m}_{a}$,
with $t_a \leq t_{a_0} 2^{n^{O(1)}}$.
Another application of~(\ref{ell(ma)}) 
shows that 
$$\ell({\mathbf m}_{c_0})= 
O(t_a n\log n)< t_{a_0}^{7/6} < \tau. $$
We just saw that $\ell({\mathbf m}_{b_0}) < \log\log \tau$,
so 
$\mu_{c_0}= {\mathbf m}_{c_0} - {\mathbf m}_{b_0}$
is a $d$-dimensional vector with
rational coordinates over fewer than $2\tau$ bits.
The lower bound follows from the fact that
$\mu_{c_0}\neq 0$.
By Lemma~\ref{duringmergephase},
the stationary velocities ${\mathbf m}_{c_0}$
and ${\mathbf m}_{b_0}$ cannot be equal, otherwise
the two flocks $F_a$ and $F_{b_0}$ could not
take so long to meet at time $t_{a_1}$. Indeed,
the time elapsed would be at least $t_{a_1}- t_a$,
(since $t_a> t_{b_0}$), which would 
greatly exceed the limit
of $t_a 2^{n^{O(1)}}$ allowed.
\item
$[\,\|u_{c_0}\|_\infty=1 
\hspace{.2cm} \&  \hspace{.2cm} 
u_{c_0}\geq 0
\hspace{.2cm} \&  \hspace{.2cm} 
\|\zeta_{c_0}\|_2 \leq e^{-\tau^2 n^{-O(1)} + n^{O(1)}}\,]$:
The bounds follow from~(\ref{zetaUB}) 
and $u_{c_0}= {\mathbf 1}_{n_{a_0}}$.
\end{itemize}
Let $c_l$ be the node $a_k$.
By applying Lemma~\ref{abc2} at $c_l$, we find that
${\mathbf m}_{b_{k-1}} = \widetilde{\mathbf m}_{c_{l-1}}$.
Applying the same lemma now at node $c_{l-1}$ shows that
$$v_{c_{l-1}}(t_{c_{l-1}})= ({\mathbf 1}_{n_{c_{l-1}}} 
   \otimes I_d)\widetilde {\mathbf m}_{c_{l-1}} 
+ (u_{c_{l-1}} \otimes I_d)\mu_{c_{l-1}} + \zeta_{c_{l-1}}\, ,$$
where 
\begin{equation*}
\begin{cases}
\,\|\mu_{c_{l-1}}\|_2 \geq e^{-\tau n^{O(1)}}  \; \\
\, \|u_{c_{l-1}}\|_\infty=1
\hspace{.2cm} \&  \hspace{.2cm} 
 u_{c_{l-1}}\geq 0 \, ;\\
\, \|\zeta_{c_{l-1}}\|_2 \leq e^{-\tau^2 n^{-O(1)} + n^{O(1)}}.
\end{cases}
\end{equation*}
The lemma also allows us to express the stationary velocity 
at $c_{l-1}$:
\begin{equation*}
\begin{split}
{\mathbf m}_{c_{l-1}} 
&= (\pi_{c_{l-1}}^T \otimes I_d)v_{c_{l-1}}(t_{c_{l-1}}) \\
&= (\pi_{c_{l-1}}^T \otimes I_d)
(({\mathbf 1}_{n_{c_{l-1}}} \otimes I_d)
   \widetilde{\mathbf m}_{c_{l-1}} 
      + (u_{c_{l-1}} \otimes I_d)\mu_{c_{l-1}} + \zeta_{c_{l-1}}) \\
&= {\mathbf m}_{b_{k-1}} 
 + (\pi_{c_{l-1}}^T u_{c_{l-1}} \otimes I_d)\mu_{c_{l-1}} 
 + (\pi_{c_{l-1}}^T \otimes I_d) \zeta_{c_{l-1}}.
\end{split}
\end{equation*}
By the triangle inequality, it follows that
\begin{equation*}
\begin{split}
\|{\mathbf m}_{c_{l-1}} - {\mathbf m}_{b_{k-1}} \|_2
&\geq
\| (\pi_{c_{l-1}}^T u_{c_{l-1}} \otimes I_d)\mu_{c_{l-1}} \|_2
- \| (\pi_{c_{l-1}}^T \otimes I_d) \zeta_{c_{l-1}} \|_2  \\
&\geq
\pi_{c_{l-1}}^T u_{c_{l-1}} \|\mu_{c_{l-1}}\|_2
- \| (\pi_{c_{l-1}}^T \otimes I_d)\|_F\| \zeta_{c_{l-1}} \|_2 \\
&\geq
\min_i\{(\pi_{c_{l-1}})_i\} e^{-\tau n^{O(1)}} 
- \sqrt{d}\, e^{-\tau^2 n^{-O(1)} + n^{O(1)}}
\geq  e^{-\tau n^{O(1)}}.
\end{split}
\end{equation*}
By~(\ref{diff-vel-timeUB}),
$$
\|{\mathbf m}_{c_{l-1}} - {\mathbf m}_{b_{k-1}} \|_2
\leq \hbox{$\frac{\log t_{a_k}}{t_{a_k}}$}\,  n^{O(n^3)};
$$
therefore, since $t_{a_k}>2^{t_f}> n^{n^4}$,
$$
t_{a_k}\leq
\|{\mathbf m}_{c_{l-1}} - {\mathbf m}_{b_{k-1}} \|_2^{-2}
\leq  e^{\tau n^{O(1)}}\leq  e^{\tau^{1.5}},
$$
which contradicts our assumption that
$\tau= \hbox{$\frac{1}{2}$}t_{a_1}^{1/3}
< (\log t_{a_k})^{1/3}$.
\hfill $\Box$
\proofend

\paragraph{Proof of Lemma~\ref{abc1}.}

Using the shorthand 
$u^a= P_a^{t_c-t_a} u_a$ and 
$\zeta^a = (P_a^{t_c-t_a} \otimes I_d)\zeta_a$, 
we express the velocity of the flock $F_a$ at time $t_c$.
From $$v_a(t_c)= (P_a^{t_c-t_a} \otimes I_d)v_a(t_a),$$
we find that, by~(\ref{va(ta)}),
\begin{equation}\label{va(tc)}
\begin{split}
v_a(t_c)
&= (P_a^{t_c-t_a} \otimes I_d)
({\mathbf 1}_{n_a} \otimes I_d)\widetilde{\mathbf m}_a 
      + (P_a^{t_c-t_a} \otimes I_d)(u_a \otimes I_d)\mu_a
   + \zeta^a \\
&= ({\mathbf 1}_{n_a} \otimes I_d)\widetilde{\mathbf m}_a 
+  (u^a \otimes I_d)\mu_a + \zeta^a.
\end{split}
\end{equation}
Because $P_a$ is an averaging operator,
$\| P_a^{t_c-t_a} u_a \|_\infty\leq \|u_a \|_\infty=1$.
The vector $u_a$ is nonnegative, so 
\begin{equation*}
\begin{split}
\| P_a^{t_c-t_a} u_a \|_\infty
&\geq \hbox{$\frac{1}{n_a}$} \| P_a^{t_c-t_a} u_a \|_1
= \hbox{$\frac{1}{n_a}$} {\mathbf 1}_{n_a}^T 
P_a^{t_c-t_a} u_a 
\geq \hbox{$\frac{1}{n_a}$} \pi_a^T P_a^{t_c-t_a} u_a  \\
&\geq \hbox{$\frac{1}{n_a}$} 
\pi_a^T u_a 
\geq \hbox{$\frac{1}{n_a}$} \min_i\{(\pi_a)_i\}\|u_a \|_\infty
\geq n^{-O(1)}.
\end{split}
\end{equation*}
Similarly, by convexity,
\begin{equation*}
\begin{split}
\| (P_a^{t_c-t_a}\otimes I_d) \zeta_a \|_2 
&\leq  
\sqrt{d n_a} \, \| (P_a^{t_c-t_a}\otimes I_d) \zeta_a \|_\infty \\
&\leq 
\sqrt{d n_a} \, \|\zeta_a\|_\infty \leq \sqrt{d n_a}\,\| \zeta_a\|_2 \, ;
\end{split}
\end{equation*}
therefore,
\begin{equation}\label{u^a-zeta^a-cond}
\begin{cases}
\, n^{-O(1)}\leq \|u^a\|_\infty \leq 1 
\hspace{.2cm} \&  \hspace{.2cm} 
u^a \geq 0 ; \\
\, \|\zeta^a\|_2 \leq n \| \zeta_a\|_2  \, .
\end{cases}
\end{equation}

\bigskip\medskip
\noindent
{\bf Case I.}\ \ 
Node $b$ is well defined and $t_b< \log\log \tau$:
Since, by~(\ref{ua-wa-ta}),
$t_c> t_a \geq \tau^3 \geq 8^{t_f}$, 
with $t_f= n^{f_0 n^3}$,
$$
-(t_c-t_b) n^{-O(1)}+ \Theta(n^3)
\leq - \tau^2 ;$$
so, by applying~(\ref{vat-p-lambda}) 
to the flock $F_b$, we find that 
$$\|v_b(t_c) - ({\mathbf 1}_{n_b} \otimes I_d){\mathbf m}_b \|_2
\leq e^{-(t_c-t_b) n^{-O(1)}+ O(n^3)};
$$
hence
$$v_b(t_c)= ({\mathbf 1}_{n_b} \otimes I_d){\mathbf m}_b 
+ e^{-\tau^2} z_c\, ,$$
where $\|z_c\|_2\leq 1$.
It follows from~(\ref{va(tc)}) that

\begin{equation}\label{vc-3parts}
\begin{split}
v_c(t_c)
=
\begin{pmatrix}
v_a(t_c) \\
v_b(t_c)
\end{pmatrix}
&=
\begin{pmatrix}
({\mathbf 1}_{n_a} \otimes I_d) \widetilde{\mathbf m}_a \\
({\mathbf 1}_{n_b} \otimes I_d){\mathbf m}_b 
\end{pmatrix}
+
\begin{pmatrix}
(u^a \otimes I_d)\mu_a \\
0
\end{pmatrix}
+
\begin{pmatrix}
\zeta^a \\
e^{-\tau^2} z_c
\end{pmatrix}
\\
&=
\begin{pmatrix}
({\mathbf 1}_{n_a} \otimes I_d) \widetilde{\mathbf m}_a \\
({\mathbf 1}_{n_b} \otimes I_d){\mathbf m}_b 
\end{pmatrix}
+
\left\{
\begin{pmatrix}
u^a \\
0
\end{pmatrix}
\otimes I_d
\right\}\mu_a
+
\begin{pmatrix}
\zeta^a \\
e^{-\tau^2} z_c
\end{pmatrix}.
\end{split}
\end{equation}
By~(\ref{va(ta)}),
the stationary velocity of $F_a$ is equal to 
\begin{equation}\label{ma-matilde-diff}
\begin{split}
{\mathbf m}_a 
&= (\pi_a^T \otimes I_d)v_a(t_a)
= (\pi_a^T \otimes I_d)
(({\mathbf 1}_{n_a} \otimes I_d)\widetilde{\mathbf m}_a 
      + (u_a \otimes I_d)\mu_a + \zeta_a) \\
&= \widetilde{\mathbf m}_a 
      + (\pi_a^T u_a)\mu_a 
      +  (\pi_a^T \otimes I_d) \zeta_a.
\end{split}
\end{equation}
By the triangle inequality, it follows that
\begin{equation*}
\begin{split}
\|{\mathbf m}_a - \widetilde{\mathbf m}_a\|_2
&\geq
\pi_a^T u_a \|\mu_a \|_2
- \| (\pi_a^T \otimes I_d)\|_F\| \zeta_a \|_2 \\
&\geq
\min_i\{(\pi_a)_i\} e^{-\tau n^{O(1)}} 
- \sqrt{d}\, e^{-\tau^2 n^{-O(1)} + n^{O(1)}}
\geq  e^{-\tau n^{O(1)}};
\end{split}
\end{equation*}
which shows that
\begin{equation}\label{ma-neq-matildea}
{\mathbf m}_a \neq \widetilde{\mathbf m}_a \, .
\end{equation}
Note also that, by~(\ref{ma-matilde-diff}),
\begin{equation*}
\begin{split}
\|\widetilde{\mathbf m}_a - {\mathbf m}_b \|_2
&\leq 
\|\widetilde{\mathbf m}_a - {\mathbf m}_a \|_2
+ 
\|{\mathbf m}_a - {\mathbf m}_b \|_2
\\
&\leq 
\pi_a^T u_a\|\mu_a\|_2
      +  \|\pi_a^T \otimes I_d\|_F\| \zeta_a\|_2
+ 
\|{\mathbf m}_a - {\mathbf m}_b \|_2 \, .
\end{split}
\end{equation*}
We bound each term on the right-hand side:
by~(\ref{ua-wa-ta}) and Cauchy-Schwarz,
$$\pi_a^T u_a\|\mu_a\|_2
\leq \hbox{$\frac{1}{\tau}$}
\|\pi_a\|_2 \|u_a\|_2 
\leq \hbox{$\frac{1}{\tau}$}
\sqrt{n_a}\, \|u_a\|_\infty 
\leq 
\hbox{$\frac{1}{\tau}$}\sqrt{n}\, .
$$
By~(\ref{diff-vel-timeUB})
and $t_c> \tau^3\geq 8^{t_f}$,
$$
\|{\mathbf m}_a - {\mathbf m}_b \|_2
\leq \hbox{$\frac{1}{\tau}$}.
$$
Also, 
$\| \pi_a^T \otimes I_d \|_F = O(1)$ and, 
by~(\ref{ua-wa-ta}),
$\|\zeta_a\|_2 \leq e^{-\tau^2 n^{-O(1)} + n^{O(1)}}$;
therefore 
$$\|\widetilde{\mathbf m}_a - {\mathbf m}_b \|_2
< \sqrt{\hbox{$\frac{1}{\tau}$}} \, .
$$
By~(\ref{ell(ma)}), our assumption that 
$t_b< \log\log \tau$ implies that 
$$\ell({\mathbf m}_b)= O(n(\log n)\log\log \tau)
< (\log\log \tau)^2.
$$
Since, by~(\ref{ua-wa-ta}),
$\ell(\widetilde {\mathbf m}_a) = O(\log\log \tau)$,
the squared distance $\|\widetilde{\mathbf m}_a - {\mathbf m}_b \|_2^2$
is a rational over $O(\log\log \tau)^2$ bits:
being less than $1/\tau$ implies that it is actually zero; hence
$\widetilde{\mathbf m}_a = {\mathbf m}_b$, 
as claimed in the lemma.
We verify from~(\ref{u^a-zeta^a-cond}) that
\begin{equation*}
\mu_c \, \defeq\, \mu_a \|u^a\|_\infty
\hspace{1cm}
\text{and}
\hspace{1cm}
u_c \, \defeq\, 
\begin{pmatrix}
u^a\\
0
\end{pmatrix}
\|u^a\|_\infty^{-1}
 \, 
\end{equation*}
satisfy the conditions of the lemma.
By~(\ref{vc-3parts}),
$$
v_c(t_c)= 
({\mathbf 1}_{n_c} \otimes I_d) \widetilde{\mathbf m}_a
+ (u_c \otimes I_d)\mu_c + \zeta_c  \, ,
$$
where 
\begin{equation*}
\zeta_c= 
\begin{pmatrix}
\zeta^a \\
e^{-\tau^2} z_c
\end{pmatrix} \, .
\end{equation*}
By~(\ref{u^a-zeta^a-cond}) and 
$\|z_c\|_2\leq 1$,
the lemma's condition on $\zeta_c$ is trivially satisfied.

\bigskip\medskip
\noindent
{\bf Case II.}\ \ 
Node $b$ is not defined:
We set $\zeta_c= \zeta^a$; 
$\mu_c= \mu_a \|u^a\|_\infty$; and
$u_c= u^a \|u^a\|_\infty^{-1}$.
This matches the identity~(\ref{va(tc)})
with the one claimed in the lemma.
\hfill $\Box$
\proofend

\paragraph{Proof of Lemma~\ref{L(t)vsL*(t)-game}.}

Suppose that Bob does not always follow the
single-iterated log rule. We show how to force him
to do so without decreasing the score differential.
If Bob uses the rule $t_B\leftarrow \log\log t_B$,
then Alice follows up with 
$t_A\leftarrow (\log\log t_A)^{\alpha}$.
Let us break this round into two parts:
\begin{enumerate}
\item
$t_B\leftarrow \log t_B$ and $t_A\leftarrow \log t_A$;
\item
$t_B\leftarrow \log t_B$ and $t_A\leftarrow (\log t_A)^{\alpha}$.
\end{enumerate}
We proceed similarly for higher log-iterations
and apply the modification systematically.
This transformation increases the scores
of the players but it does not change their difference.
Finally, we apply one last transformation to the new game, which
is to convert all of Alice's moves into
$t_A\leftarrow (\log t_A)^{1/4}$. This can only
increase the score differential.
\hfill $\Box$
\proofend

\paragraph{Proof of Lemma~\ref{L(t)vsL*(t)}.}

Consider the two recurrence relations: 
$$a_0(x)=b_0(x)=x,$$
and, for $h>0$,
\begin{equation*}
\begin{cases}
\, a_h(x)= 2^{a_{h-1}(x)} \\
\, b_h(x)= 2^{b_{h-1}(x)}+2.
\end{cases}
\end{equation*}
Recall that $c_h$ is defined by 
$c_0=t_0^4$ and, for $h>0$, 
$c_h= 2^{4c_{h-1}}$.
We verify by induction that, for any $h>0$, 
\begin{equation*}
c_h= 2^{2^{b_{h-1}(4\log t_0+ 2)}}.
\end{equation*}
To prove the inequality we seek,
$$\min\{\,h\,|\,  c_h\geq t\, \} \geq \log^* t 
    - \log^* t_0 - O(1),$$
where $t>t_0$, we may assume that
$t>2^{t_0}$, otherwise the result is trivial.
The assumption implies that the minimum $h$ is positive;
therefore it suffices to prove that, 
for all $h\geq 0$,
\begin{equation}\label{ah-bh-compare}
b_h(4\log t_0+ 2)\leq a_h(4\log t_0+ 4).
\end{equation}
We see by induction that, for  
all $h\geq 0$, $x\geq 2$, and $\varepsilon>0$, 
\begin{equation}\label{ah(x)+eps}
a_h(x)+\varepsilon
\leq a_h(x+\varepsilon 2^{-h}).
\end{equation}
The case $h=0$ is obvious, so consider $h>0$.
Note that, for any $y\geq 2$,
$$ 2^y+\varepsilon \leq 2^{y+\varepsilon/2},$$
which follows from
$$
\ln(1+ \varepsilon 2^{-y})\leq \varepsilon 2^{-y}
\leq \hbox{$\frac{\ln 2}{2}$} \varepsilon.$$
Since $a_{h-1}(x)\geq 2$, this shows that
$$
a_h(x)+\varepsilon
= 2^{a_{h-1}(x)} +\varepsilon
\leq 2^{a_{h-1}(x)+\varepsilon/2}
\leq 2^{a_{h-1}(x+\varepsilon 2^{-h})}
= a_h(x+\varepsilon 2^{-h}),
$$
which proves~(\ref{ah(x)+eps}).
Next, we show by induction that, for all $h\geq 0$ 
and $x\geq 2$, 
\begin{equation}\label{bh(x)ah}
b_h(x)\leq a_h(x+2-2^{1-h}).
\end{equation}
The case $h=0$ again being obvious, assume that $h>0$.
By~(\ref{ah(x)+eps}), 
\begin{equation*}
\begin{split}
b_h(x)
&= 2^{b_{h-1}(x)}+2
\leq    2^{a_{h-1}(x+2-2^{2-h})}+2  \\
&\leq    a_h(x+2-2^{2-h})+2 \leq  a_h(x+2-2^{1-h}),
\end{split}
\end{equation*}
which establishes~(\ref{bh(x)ah});
and hence~(\ref{ah-bh-compare}).
\hfill $\Box$
\proofend

\section{The Lower Bound}\label{LB-section}

We specify initial positions and velocities for $n$ birds,
using only $O(\log n)$-bits per bird, and prove that 
their flock network converges only after a number of steps
equal to a tower-of-twos of height $\log n$.
Our proof is entirely constructive.
The hysteresis assumption of the model
is not used and, in fact, the lower bound
holds whether the model includes hysteresis or not.
Our construction is in two dimensions, $d=2$, but it works 
for any $d>0$.
The $n$ birds all start from the $X$-axis (think
of them on a wire), and fly in the $(X,Y)$-plane, 
merging in twos, fourths,
eights, etc, until they form a single connected flock.
This process forms a {\em fusion tree} $\mathcal T$
of height $\log n$.
(We assume throughout this section that
$n$ is a large odd power of two.)
Every flock formed in the process is a single path.
The transition matrix is that of a lazy symmetric 
random walk with, at each node, a probability $\frac{1}{3}$
of staying put.

\vspace{1cm}
\begin{figure}[htb]\label{fig-fusiontree}
\begin{center}
\includegraphics[width=8cm]{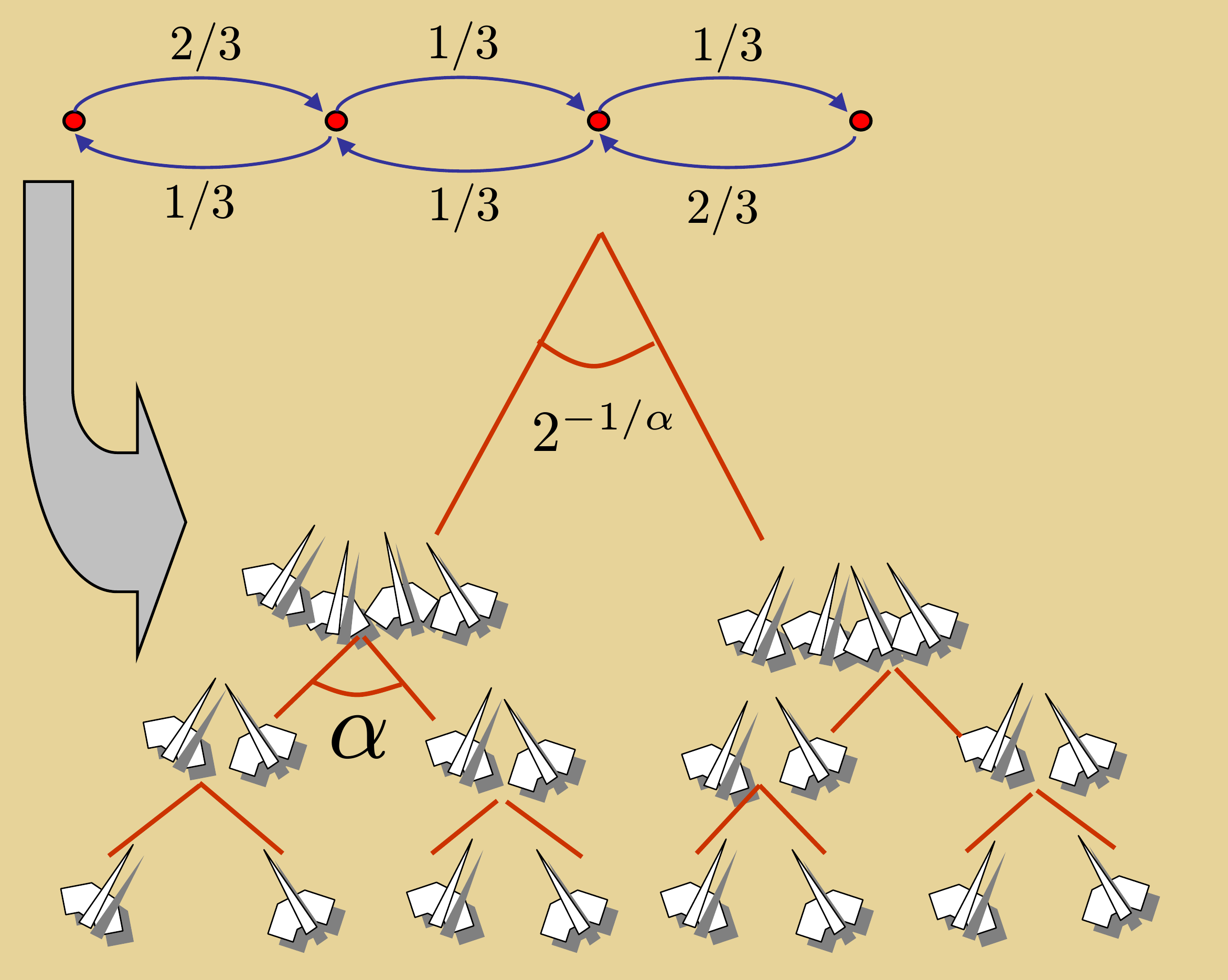}
\end{center}
\caption{\small 
Birds join in flocks of size $2$, $4$, $8$, etc,
up in the fusion tree, 
each time flying in a direction closer to the $Y$-axis.
The angle decreases exponentially at each level,
so the time between merges grows accordingly.
The big arrow indicates the Markov chain
corresponding to a 4-bird flock.
At each state, the probability of staying put 
is $\frac{1}{3}$, with 
the remaining $\frac{2}{3}$ being distributed
uniformly among the outgoing edges.
}
\end{figure}

Initially, 
the velocity of each bird has its $Y$-coordinate equal to 1.
Since averaging these velocities will only produce 1,
the birds move up away from
the $X$-axis forever at constant speed 1.
We can then factor out the $Y$ coordinates
and focus our entire investigation on the birds' projections
on the $X$-axis. In fact, we might as well view the birds
as points moving along the $X$-axis and joining into
edges when their distance is 1 or less.
In other words, 
we let $x(t)$ denote
the vector $(x_1(t), \ldots, x_n(t))$ and let
$v(t)= x(t)- x(t-1)$.
The coordinates of the velocity vector
$v(t)$ will quickly decrease, but we should not
be mistaken into thinking that the birds slow down
accordingly. Because of the
$Y$ motion, all the birds will always fly at speeds very near 1.
Let $c$ be a large enough odd integer: one will
easily check that $c=11$ works, but no effort was made to
find the smallest possible value. We leave 
$c$ as a symbol to make it easier to follow the derivations.

\begin{equation*}
\hbox{\sc Initial Conditions}\ 
\begin{cases}
\,
x(0) \, = \Bigl(0,\hbox{$\frac{2}{3}$}, 
2,\hbox{$\frac{8}{3}$}, 
\ldots, 2l, 2l + \hbox{$\frac{2}{3}$}, \ldots,
n-2, n-\hbox{$\frac{4}{3}$}\Bigr)^T;  \\
\\
\, v(1)= 
\Bigl(\, \underset{n}{\underbrace{     
n^{-c}, 0, -n^{-c}, 0, n^{-c}, 0, \ldots, 
n^{-c}, 0, -n^{-c}, 0}}  \, \Bigr)^T.
\end{cases}
\end{equation*}

Each nonleaf node $a$ of the fusion tree~$\mathcal T$
has associated with it a flock of $2^j$ birds
whose network is a single path: the index $j>0$
is also the height of the node.
The flock $F_a$ at node $a$ is formed at 
a time $t_j$ that depends only on the height in~$\mathcal T$;
by convention, $t_1=0$.
Given a node $a$ at height $j>0$,
we denote by $v^{a}$ the $2^j$-dimensional
velocity vector of the flock $F_a$ at time $t_j$ and
by ${\mathbf m}_a$ its stationary velocity.
For $j>2$, if $l$ and $r$ denote the left and 
right children of $a$, respectively, then 
$v^{l}= v^{r}$. In other words, two
sibling flocks start out with the same initial
velocity. At time $t_j$, because of noise called {\em flipping},
the velocity vectors of these flocks will 
have evolved into
${\mathcal L}\, v^{l}$ and $- {\mathcal L}\, v^{r}$, 
respectively, where ${\mathcal L}$ is
a linear transformation specific to that sibling pair.
This implies that
$$v^{a}=
      \begin{pmatrix}
\,\,\,\,   {\mathcal L}\, v^{l}\\
         - {\mathcal L}\, v^{r}
      \end{pmatrix}.
$$
The stationary velocity
of the flock $F_a$ satisfies
\begin{equation}\label{StatVelSigmaj=}
{\mathbf m}_a
= \hbox{$ \frac{1}{2^j-1} $}
(\, \overset{2^j}{\overbrace{
\hbox{$\frac{1}{2}$},1,\ldots, 1, \hbox{$\frac{1}{2}$}}}\,)
v^{a}\, .
\end{equation}
The initial conditions provide the velocity
vectors of the 2-bird flocks at height 1 one step
after $t=0$. It follows that, if
$a$ is a node at height $j=1$, 
the stationary velocity ${\mathbf m}_a$
is equal to $\frac{1}{2} (-1)^{k+1} n^{-c}$, 
where $k$ is the rank of $v$ among the nodes
at height 1 from left to right.
For consistency, we must set 
$v^{a}= (-1)^{k}n^{-c}(1,-2)^T$.
This choice is dictated by the initial conditions
set above, so that, for any $j\geq 1$, 
the velocity of the flock at $v$
at time $t$ ($t_j\leq t< t_{j+1}$)
is equal to $P_j^{t-t_j} v^{a}$, 
where 
\begin{equation}\label{P_j=1/3Matrix}
P_j = 
\frac{1}{3}
\underset{2^j}{\underbrace{      
\begin{pmatrix}
1 & 2 & 0 & 0 &\dots & 0 \\
1 & 1 & 1 & 0 &\dots & 0 \\
0 & 1 & 1 & 1 &\dots & 0 \\
\vdots & \ddots &   &   &\ddots& \vdots \\
0 & \dots & 0 & 1 & 1    & 1 \\
0 & \dots & 0 & 0 & 2    & 1 
\end{pmatrix}
}} .
\end{equation}
At height 2 and above, some flocks undergo
a {\em velocity flip} at chosen times: this means
that the sign of their current velocity is reversed
and it becomes $-P_j^{t-t_j} v^{a}$
at time $t$. By abuse of notation, we say that
the node flips: it is instantaneous
and does not count as an averaging transition. When does this happen and why?
Fix an integer $f=3$. Again, we leave this constant
as a symbol for clarity.

\medskip

\begin{quotation}
\noindent
{\sc Flipping Rule}: \ 
It applies at time $t= t_j+n^f$ 
to any flock of a left child
of even height $j>1$ 
and to any flock of a right child
of odd height $j>2$.
\end{quotation}

\vspace{1cm}
\begin{figure}[htb]\label{fig-flip}
\hspace{2.5cm}
\includegraphics[width=8cm]{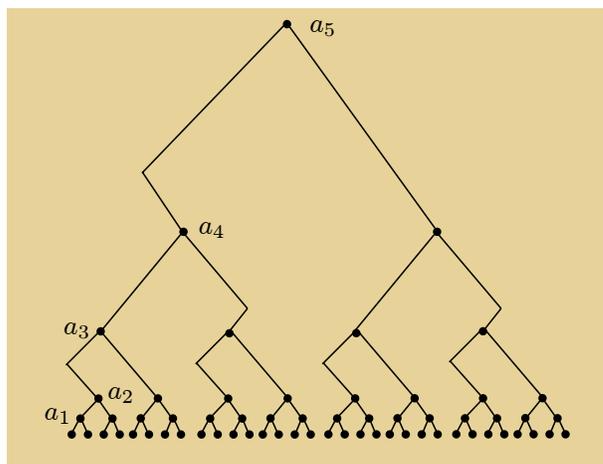}
\caption{\small 
Flipping alternates between left and right.
Leaves were added to indicate that nodes at height 1
correspond to 2-bird flocks.
}
\end{figure}

\medskip
\noindent
Flips are convenient to make flocks collide. We show later
that they conform to the noisy flocking model.
At height 2 and higher, any two sibling nodes
$l$ and $r$ are assigned the same velocity vector
$v^l=v^r$. Their corresponding flocks
evolve in parallel for $n^f$ steps, 
like two identical copies. Then, one of them ``flips''
(which one, left or right, depends on the height 
in the tree), meaning that the two velocity vectors
become opposite of each other.
The flip type alternates between
left and right as we go up the tree.
Although flipping has only a trivial effect on
velocities, which decays over time, we must
be careful that it does not break flocks apart. We
could rely on hysteresis to prevent this from
happening, but as we said earlier we seek a lower bound
that holds whether hysteresis is present or not.
That is why we introduce the lag $n^f$.
The averaging
operations act like glue and the glue needs to dry up
before changing direction.

Up to sign, $v^{a}$ depends only on
the height $j$ of node $a$, so we focus our attention on the
left spine of the tree, denoted
$a_1,\ldots, a_{\log n}$ in ascending order.
The exact behavior of every flock in the system
can be found in replica
either at a node $a_j$ or at a sibling
of such a node. That is why, when checking the 
structural integrity of the flocks, it is not quite
enough to concentrate on the left spine:
we must also check the right children hanging off of it.
For any $1\leq j<\log n$, we define $\theta_j= t_{j+1}-t_j$
as shorthand for the lifetime of the flock $F_{a_j}$.
Our task is two-fold. 
First, we must show that 
$|{\mathbf m}_{a_j}|$ decreases very fast: 
we prove that (roughly)
$$|{\mathbf m}_{a_j}|
< e^{-\Omega(|{\mathbf m}_{a_{j-1}}^{-1}|)},$$
which implies that $\theta_j$ is exponentially
larger than $\theta_{j-1}$; hence the 
logarithmic tower-of-twos lower bound.
Second, we must prove the integrity of the scheme: 
that each flock remains a single path over its
lifetime; that two flocks meet when and where they should; 
that flipping fits within the model; etc.

\subsection{The Early Phases}\label{TakeOff}

The proofs are technical but one can develop
some intuition for the process they mean to explain
by working out the calculations 
for ${\mathbf m}_{a_{j}}$ ($j=1,2,3$) explicitly.
At time $t=t_1= 0$, the network consists of the edges
$(1,2), (3,4),\ldots, (n-1,n)$.
We already saw that the 2-bird flock 
$({\mathcal B}_{1}, {\mathcal B}_{2})$ 
has initial and stationary velocities
\begin{equation}\label{sigma1}
v^{a_1}= n^{-c}
     \begin{pmatrix}
      -1\\ \,\,\,\, 2 
     \end{pmatrix}
\ \ \ \ \ \ \ 
\text{and}
\ \ \ \ \ \ \ 
{\mathbf m}_{a_{1}}=  \hbox{$\frac{1}{2}$} n^{-c}.
\end{equation}

\paragraph{Flying at Height 1.}

Because the velocity at time $t$ captures
the motion {\em ending} at $t$, 
the velocity of the flock
$({\mathcal B}_{1}, {\mathcal B}_{2})$ 
at time $1$ is $P_1 v^{a_1}$.
By~(\ref{x(t)-SumP^s}), for $t>0$,
$$x(t)= x(0)+ \sum_{s=0}^{t-1} P^s v(1),$$
which gives us
\begin{equation*}
\begin{pmatrix}
x_1(t)\\
x_2(t)
\end{pmatrix}
= 
\begin{pmatrix}
x_1(0)\\
x_2(0)
\end{pmatrix}
+
\sum_{s=0}^{t-1}P_1^s 
(P_1 v^{a_1})
= 
\hbox{$\frac{2}{3}$}
\begin{pmatrix}
0\\
1
\end{pmatrix}
+
\sum_{s=0}^{t-1}P_1^s 
\begin{pmatrix}
n^{-c}\\
0
\end{pmatrix}.
\end{equation*}
Diagonalizing $P_1$ shows that, for any integer $s>0$,
\begin{equation*}
P_1^s = 
\hbox{$\frac{1}{2}$}
\begin{pmatrix}
1 \\
1  
\end{pmatrix}
\begin{pmatrix}
1 & 1 
\end{pmatrix}
+
\hbox{$\frac{1}{2}$}
(-3)^{-s}
\begin{pmatrix}
\,\,\,\, 1 \\
-1  
\end{pmatrix}
\begin{pmatrix}
1 &\! -1
\end{pmatrix}.
\end{equation*}
It follows that, for $0=t_1<t\leq t_2$,
\begin{equation}\label{x1(t)equals}
\begin{cases}
x_1(t)
= 
\frac{t}{2}n^{-c}
+ \frac{1}{2}n^{-c}
\sum_{s=0}^{t-1} (-3)^{-s} 
=
\frac{1}{2}n^{-c}\,
(t+\frac{3}{4} + \frac{1}{4} (-3)^{1-t}) ; \\
x_2(t)
=
\hbox{$\frac{2}{3}$} +
\frac{t}{2}n^{-c}
- \frac{1}{2}n^{-c}
\sum_{s=0}^{t-1} (-3)^{-s}
= 
\frac{2}{3}+ 
\frac{1}{2}n^{-c}\,
(t-\frac{3}{4} - \frac{1}{4} (-3)^{1-t}).
\end{cases}
\end{equation}
Note that ${\mathcal B}_{1}$ always
stays to the left of ${\mathcal B}_{2}$ and
their distance is
\begin{equation}\label{x12(t2)}
x_2(t)- x_1(t)
= 
\hbox{$\frac{2}{3}$}-
\hbox{$\frac{3}{4}$} n^{-c}\,
(1- (-\hbox{$\frac{1}{3}$})^{t}).
\end{equation}
Left to their own devices, the two birds
would slide to the right at speed
${\mathbf m}_{a_{1}}$, plus or minus an exponentially vanishing term;
their distance would oscillate around 
$\hbox{$\frac{2}{3}$}- \hbox{$\frac{3}{4}$} n^{-c}$
and converge exponentially fast, with the 
oscillation created by the negative eigenvalue.
This is what happens until the flock
at $a_1$ begins to interact with its ``sibling'' 
flock to the right, $({\mathcal B}_{3}, {\mathcal B}_{4})$.
The latter's velocity vector is $(-n^{-c},0)^T$ 
at time $t=1$ and, for 
$t_1<t\leq t_2$,
\begin{equation}\label{x3x4(t)brace}
\begin{cases}
x_3(t)
= 2
-\frac{1}{2}n^{-c}\,
(t+\frac{3}{4} + \frac{1}{4} (-3)^{1-t}) ; \\
x_4(t)
=
\frac{8}{3}
- \frac{1}{2}n^{-c}\,
(t-\frac{3}{4} - \frac{1}{4} (-3)^{1-t}).
\end{cases}
\end{equation}
The stationary velocity of $({\mathcal B}_{3}, {\mathcal B}_{4})$
is $-{\mathbf m}_{a_{1}}= - \hbox{$\frac{1}{2}$} n^{-c}$,
but the flock is {\em not} the mirror
image of $({\mathcal B}_{1}, {\mathcal B}_{2})$, a situation
that would bring the flocking to an end.
In particular, note that the diameter of the flock~is
\begin{equation}\label{x34(t2)}
x_4(t)- x_3(t)
= 
\hbox{$\frac{2}{3}$}+
\hbox{$\frac{3}{4}$} n^{-c}\,
(1- (-\hbox{$\frac{1}{3}$})^{t}), 
\end{equation}
which always exceeds 
that of $({\mathcal B}_{1}, {\mathcal B}_{2})$
for all $t>0$.
The diameters of both flocks oscillate 
around $\frac{2}{3}$ but in phase opposition: indeed, their sum
remains constant.
Both 2-bird flocks drift toward each other at
distance\footnote{The linearity in $t$
is due to an accidental
cancellation that will not occur for bigger flocks.}
$
x_3(t)- x_2(t)
=
\hbox{$\frac{4}{3}$} -  t n^{-c} .
$
This implies that 
$t_2= t_1+ \theta_1= \lceil 
     \hbox{$\frac{1}{3}$} n^{c} \, \rceil$.
Because $n$ is an odd power of two and $c$ is odd,
$n^c= 2 \pmod{6}$;
hence,  
$\lceil \frac{1}{3}  n^{c}\rceil= \frac{1}{3} (n^{c} +1)$
and 
\begin{equation}\label{theta1}
t_2= \theta_1 = \hbox{$\frac{1}{3}$} (n^{c} +1)
= 1 \pmod{2}
\end{equation}
We conclude that
\begin{equation}\label{x23(t2)}
x_3(t_2)- x_2(t_2)
= 1- \hbox{$\frac{1}{3}$} n^{-c}.
\end{equation}

\begin{figure}[htb]\label{fig-4bird}
\vspace{1cm}
\begin{center}
\includegraphics[width=8cm]{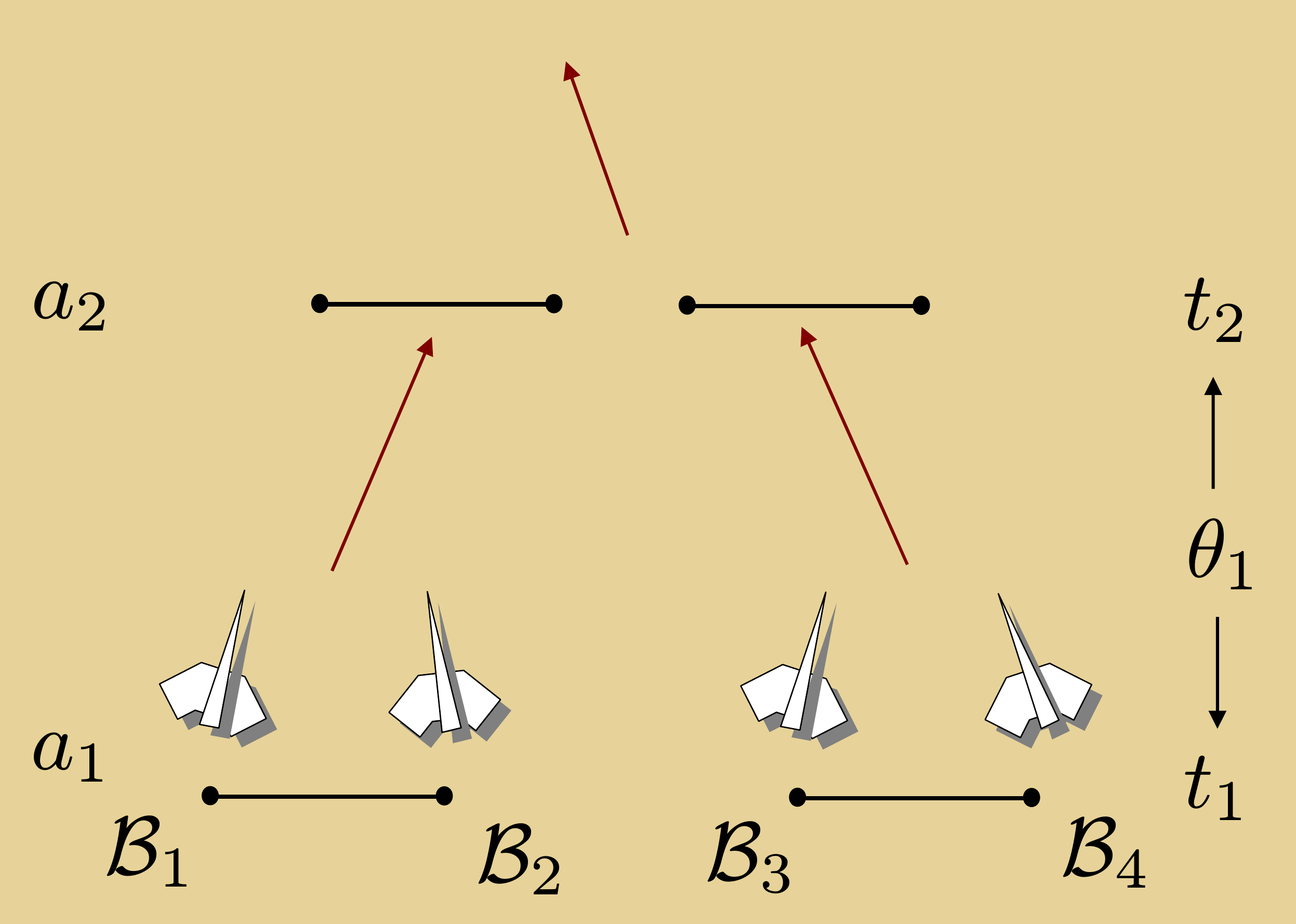}
\end{center}
\begin{quote}
\vspace{0cm}
\caption{\small 
The 4-bird flock is formed at time $t_2$ and
acquires a negative stationary velocity ${\mathbf m}_{a_{2}}$.}
\end{quote}
\end{figure}

\noindent
The definition of flip nodes suggests 
a cyclic process with period 2 that is inherent
to the flocking process.
At time $t_2$, the flock at $a_2$ is formed with the initial velocity
\begin{equation}\label{Deltav2}
v^{a_2} =
\begin{pmatrix}
\,\,\,\, P_1^{t_2-t_1}
v^{a_1}
\\
- P_1^{t_2-t_1}
v^{a_1}
\end{pmatrix} 
=
\begin{pmatrix}
\,\,\,\, P_1^{\theta_1-1}
\begin{pmatrix}
n^{-c}\\
0
\end{pmatrix} \\
\\
- P_1^{\theta_1-1}
\begin{pmatrix}
n^{-c}\\
0
\end{pmatrix}
\end{pmatrix} 
=
\hbox{$\frac{1}{2}$}
n^{-c}
\begin{pmatrix}
\,\,\,\, 1+(-3)^{1-\theta_1} \\
\,\,\,\, 1-(-3)^{1-\theta_1} \\
-1-(-3)^{1-\theta_1} \\
-1+(-3)^{1-\theta_1}
\end{pmatrix} .
\end{equation}
By~(\ref{StatVelSigmaj=}), 
the stationary velocity for the 4-bird flock is
$\frac{1}{3}(\frac{1}{2}, 1, 1, \frac{1}{2}) v^{a_2}$;
hence, by~(\ref{sigma1}, \ref{theta1}),
\begin{multline}\label{sigma2}
0> {\mathbf m}_{a_{2}} = 
\hbox{$\frac{1}{3}$}
( \hbox{$\frac{1}{2}$}, 1 , 1, \hbox{$\frac{1}{2}$} )
v^{a_2} =
\hbox{$\frac{1}{2}$}n^{-c} (-3)^{-\theta_1}
\\
=
- \hbox{$\frac{1}{2}$} n^{-c}
(\hbox{$\frac{1}{3}$})^{(n^c+1)/3}
\geq - e^{-\Omega( {\mathbf m}_{a_{1}}^{-1} )}.
\end{multline}
This inequality gives an inkling of the kind
of exponential decay we envision as we go up the
fusion tree. Note that ${\mathbf m}_{a_{2}} <0$, which means
that the flock is drifting in the wrong direction:
that is why $a_2$ is a flip node.

\paragraph{Flying at Height 2.}

Again, by~(\ref{x(t)-SumP^s}),
for $t_2<t\leq t_2+n^f< t_3$,
$$
\begin{pmatrix}
x_1(t)\\
\vdots\\
x_4(t)
\end{pmatrix}
= 
\begin{pmatrix}
x_1(t_2)\\
\vdots\\
x_4(t_2)
\end{pmatrix}
+
\sum_{s=0}^{t-t_2-1}P_2^{s+1} v^{a_2},
\ \ \ \ \ \ 
\text{with}\ \ \
P_2 = 
\hbox{$\frac{1}{3}$}\begin{pmatrix}
1 & 2 & 0 & 0 \\
1 & 1 & 1 & 0 \\
0 & 1 & 1 & 1 \\
0 & 0 & 2 & 1
\end{pmatrix}.
$$
By straightforward diagonalization, we find that,
for any integer $s>0$,
\begin{multline}\label{P2s-vec}
P_2^s = 
\hbox{$\frac{1}{6}$}
\begin{pmatrix}
1 \\
1 \\
1 \\
1
\end{pmatrix}
(1,2,2,1) 
+
\hbox{$\frac{1}{6}$}
\Bigl(\frac{2}{3}\Bigr)^{s}
\begin{pmatrix}
\,\,\,\, 2 \\
\,\,\,\, 1 \\
-1 \\
-2
\end{pmatrix}
(1,1,-1,-1) \\
+
\hbox{$\frac{1}{6}$}
(-3)^{-s}
\begin{pmatrix}
\,\,\,\, 1 \\
-1 \\
\,\,\,\, 1 \\
-1
\end{pmatrix}
(1,-2,2,-1);
\end{multline}
therefore, 
\begin{multline}\label{x1-x4-noflip}
\begin{pmatrix}
x_1(t)\\
\vdots\\
x_4(t)
\end{pmatrix}
=
\begin{pmatrix}
x_1(t_2)\\
\vdots\\
x_4(t_2)
\end{pmatrix}
+ 
{\mathbf m}_{a_{2}} (t-t_2)
\begin{pmatrix}
1 \\
1 \\
1\\
1
\end{pmatrix}
+ \hbox{$\frac{1}{8}$} n^{-c}
\begin{pmatrix}
\,\,\,\, 11 \\
\,\,\,\, 5 \\
-5\\
-11
\end{pmatrix}
\\
+ n^{-c}
\Bigl(\frac{2}{3}\Bigr)^{t-t_2+1}
\begin{pmatrix}
-2 \\
-1 \\
\,\,\,\, 1\\
\,\,\,\, 2
\end{pmatrix}
+ \hbox{$\frac{1}{24}$} n^{-c}
(-3)^{t_2-t}
\begin{pmatrix}
-1 \\
\,\,\,\, 1 \\
-1\\
\,\,\,\, 1
\end{pmatrix}.
\end{multline}
It follows from~(\ref{x12(t2)}, \ref{x34(t2)}) that,
for $t_2<t\leq t_2+n^f$,
both $x_2(t)- x_1(t)$ and
$x_4(t)- x_3(t)$ are
$\frac{2}{3}\pm O( n^{-c} )$; therefore, the two
end edges of the 4-bird flock are {\em safe},
which we define as being of length less than 1
(so as to belong to the flocking network) but
greater than $\frac{1}{2}$ (so as to avoid edges joining
nonconsecutive birds).
The middle one, $(2,3)$, is more problematic.
Its length is
$$
x_3(t)- x_2(t)
=
x_3(t_2)- x_2(t_2)
- \hbox{$\frac{1}{12}$} n^{-c} \,
( 15- 16 (\hbox{$\frac{2}{3}$})^{t-t_2}
+ (-3)^{t_2-t}).
$$
We can verify that
$$15- 16 (\hbox{$\frac{2}{3}$})^{t-t_2} 
+ (-3)^{t_2-t}\geq 0,$$
for all $t> t_2$, 
which, by~(\ref{x23(t2)}), 
shows that the distance between the two
middle birds ${\mathcal B}_{2}, {\mathcal B}_{3}$
always lies comfortably between 
$1- (\frac{1}{3}+ O(1))n^{-c}$ and $1-\frac{1}{3} n^{-c}$.
The upper bound is both lucky and intuitive: lucky because
the edge starts with length very near 1 and it could
easily be perturbed and break up;
intuitive because the two flocks have
inertia when they bump into each other and one
expects the edge $(2,3)$ to act like a spring 
being compressed, 
thereby shrinking during the initial steps.

\begin{figure}[htb]
\vspace{1cm}
\begin{center}
\includegraphics[width=7cm]{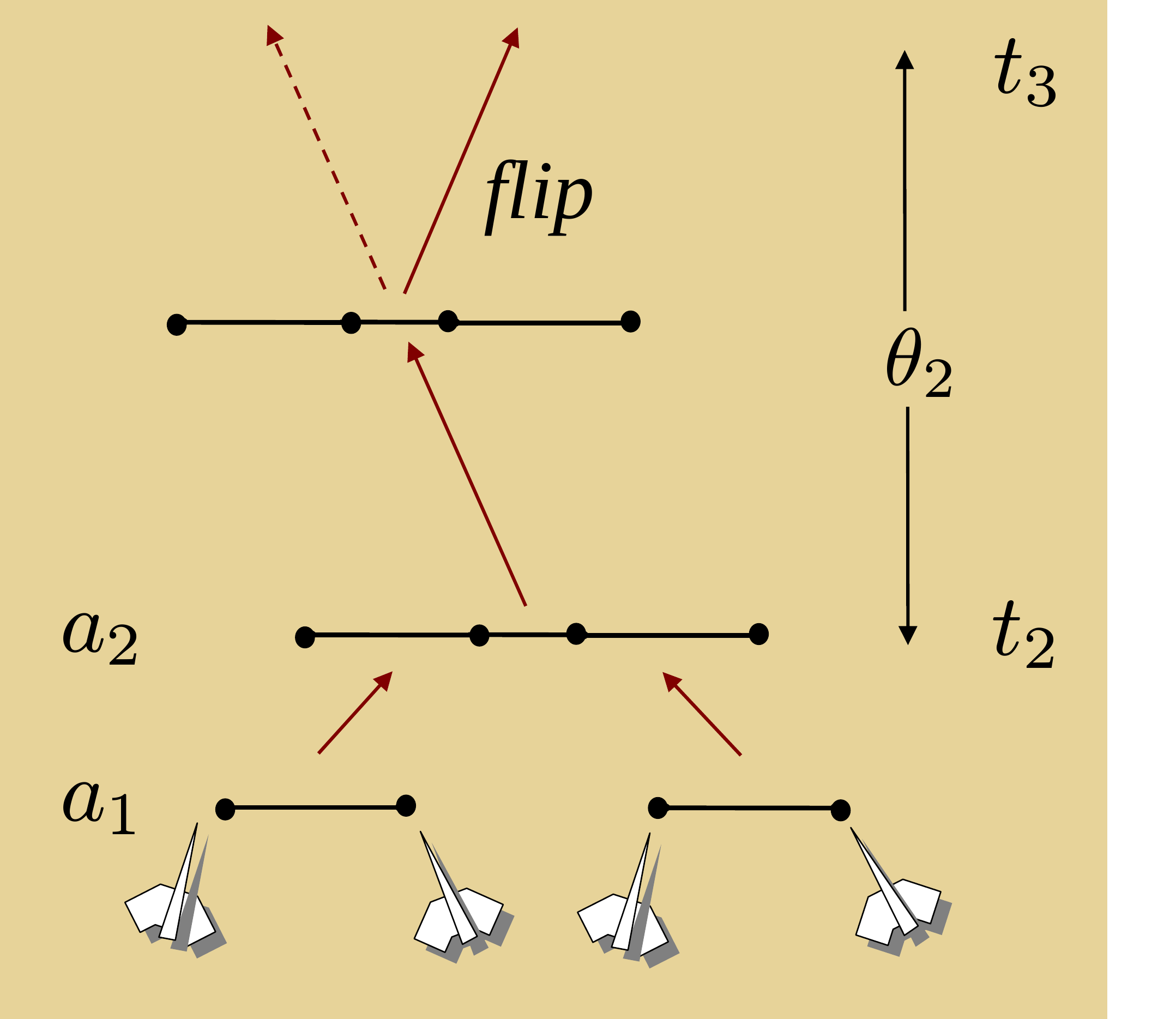}
\end{center}
\begin{quote}
\vspace{0cm}
\caption{\small 
The 4-bird flock at $a_2$ ``flips" at time $t_2+n^f$.
}
\label{fig-8bird}
\end{quote}
\end{figure}

\paragraph{Flipping Velocity at Height 2.}

Since $a_2$ is a flip node, the velocity vector
reverses sign after a lag of $n^f$ steps. 
Instead of redoing all the
calculations, we can apply a simple symmetry principle:
by linearity, the positions of the flock with
and without the flip average out to what it was
at time $t_2+n^f$ (Figure~\ref{fig-8bird}).
In other words, for $t_2+n^f<t\leq t_3$,
$$
\begin{pmatrix}
x_1(t_2+n^f)\\
\vdots\\
x_4(t_2+n^f)
\end{pmatrix}
= 
\frac{1}{2}
\left\{
\begin{pmatrix}
x_1(t)\\
\vdots\\
x_4(t)
\end{pmatrix}_{\!\!\!\text{\em\small flip}}
+ 
\begin{pmatrix}
x_1(t)\\
\vdots\\
x_4(t)
\end{pmatrix}_{\!\!\!\text{\em\small no-flip}}
\right\}.
$$
By~(\ref{x1-x4-noflip}), 
the position formula for the flock can be readily updated:
\begin{multline}\label{x1-x4-plus-flip}
\begin{pmatrix}
x_1(t)\\
\vdots\\
x_4(t)
\end{pmatrix}
=
\begin{pmatrix}
x_1(t_2)\\
\vdots\\
x_4(t_2)
\end{pmatrix}
+
{\mathbf m}_{a_{2}} (2n^f+t_2-t)
\begin{pmatrix}
1 \\
1 \\
1\\
1
\end{pmatrix}
+ \hbox{$\frac{1}{8}$} n^{-c}
\begin{pmatrix}
\,\,\,\, 11 \\
\,\,\,\, 5 \\
-5\\
-11
\end{pmatrix}
\\
+ n^{-c}
\Bigl(\frac{2}{3}\Bigr)^{n^f+1}
\Bigl(
2- \Bigl(\frac{2}{3}\Bigr)^{t-t_2-n^f} \,
\Bigr)
\begin{pmatrix}
-2 \\
-1 \\
\,\,\,\, 1\\
\,\,\,\, 2
\end{pmatrix}
\\
+ \hbox{$\frac{1}{24}$}
n^{-c}
(-3)^{-n^f} 
\Bigl(
2- (-3)^{t_2+n^f-t}
\Bigr)
\begin{pmatrix}
-1 \\
\,\,\,\, 1 \\
-1\\
\,\,\,\, 1
\end{pmatrix}.
\end{multline}
This proves that the lengths of the two end edges 
differ by what they were at $t_2$ by only $O(n^{-c})$.
Indeed, by~(\ref{x12(t2)}, \ref{x34(t2)}),
this implies that, for any $t\leq t_3$,
\begin{equation}\label{x2x4-allt-t3}
\begin{cases}
x_2(t)-x_1(t)=
\frac{2}{3}\pm O(n^{-c}) ; \\
x_4(t)-x_3(t)=
\frac{2}{3}\pm O(n^{-c}).
\end{cases}
\end{equation}
The middle edge has length 
\begin{multline*}
x_3(t)-x_2(t)=
x_3(t_2)-x_2(t_2)
- \hbox{$\frac{5}{4}$} n^{-c} 
\\
+ 
2 n^{-c} \Bigl(\frac{2}{3}\Bigr)^{n^f+1}
\Bigl(
2- \Bigl(\frac{2}{3}\Bigr)^{t-t_2-n^f} \,
\Bigr)
- \hbox{$\frac{1}{12}$}
n^{-c}
(-3)^{-n^f} 
\Bigl(
2- (-3)^{t_2+n^f-t}
\Bigr),
\end{multline*}
which, in view of~(\ref{x23(t2)}),
shows that, for $t_2+n^f<t\leq t_3$,
\begin{equation}\label{x32-t2-3}
1- O(n^{-c})
\leq 
x_3(t)-x_2(t)
\leq 1- \hbox{$\frac{3}{2}$} n^{-c} .
\end{equation}
This proves that the middle edge is safe
and the integrity of the entire 4-bird flock
is preserved.
Was it necessary to delay the flip by $n^f$? 
The particular choice of lag, $n^f$, will be justified
later by examining the bigger flocks, but we can see
right away that delaying the flip is mandatory.
Indeed, if we replace $n^f$ by $0$ in the expression above,
then, for $t=t_2+2$, we get
$$
x_3(t)-x_2(t) =
x_3(t_2)-x_2(t_2) 
+ \hbox{$\frac{2}{3}$} n^{-c} = 1+ \hbox{$\frac{1}{3}$} n^{-c},$$
which causes the flock to break apart.
The flock $({\mathcal B}_{5}, \ldots, {\mathcal B}_{8})$
follows the same 
trajectory as the 4-bird flock above,
shifted along the $X$-axis by 4
but with no velocity flip.
So, by~(\ref{x1(t)equals}, \ref{x1-x4-noflip}), we find that,
for $t_2+n^f<t\leq t_3$,
\begin{equation*}
\begin{cases}
x_5(t)
=
x_5(t_2)
+ 
{\mathbf m}_{a_{2}} (t-t_2)
+ \hbox{$\frac{11}{8}$} n^{-c}
-2 n^{-c}
(\frac{2}{3})^{t-t_2+1}
- \hbox{$\frac{1}{24}$} n^{-c}
(-3)^{t_2-t} ; \\
x_5(t_2)= x_1(t_2)+4= 
\frac{1}{2}n^{-c}\, (t_2+\hbox{$\frac{3}{4}$} 
+ \hbox{$\frac{1}{4}$} (-3)^{1-t_2})+4.
\end{cases}
\end{equation*}
At the same time, by~(\ref{x1-x4-plus-flip}),
\begin{multline*}
x_4(t)
=
x_4(t_2)
+
{\mathbf m}_{a_{2}} (2n^f+t_2-t)
- \hbox{$\frac{11}{8}$} n^{-c}
\\
+ 2 n^{-c}
\Bigl(\frac{2}{3}\Bigr)^{n^f+1}
\Bigl(
2- \Bigl(\frac{2}{3}\Bigr)^{t-t_2-n^f} \,
\Bigr)
+ \hbox{$\frac{1}{24}$}
n^{-c}
(-3)^{-n^f} 
\Bigl(
2- (-3)^{t_2+n^f-t}
\Bigr), 
\end{multline*}
where, by~(\ref{x3x4(t)brace}),
$$
x_4(t_2)
=
\hbox{$\frac{8}{3}$}
- \hbox{$\frac{1}{2}$}n^{-c}\,
(t_2-\hbox{$\frac{3}{4}$} - \hbox{$\frac{1}{4}$} (-3)^{1-t_2}).
$$
By~(\ref{theta1}), this shows that,
for $t_2+n^f<t\leq t_3$,
\begin{equation*}
\begin{split}
x_5(t)- x_4(t) 
&=
\hbox{$\frac{4}{3}$} + t_2 n^{-c}
+
2 {\mathbf m}_{a_{2}} (t-t_2-n^f)
+ \hbox{$\frac{11}{4}$} n^{-c} \\
& \hspace{5cm}
- 4 n^{-c}
\Bigl(\frac{2}{3}\Bigr)^{n^f+1}
-
\hbox{$\frac{1}{12}$}
n^{-c}
(-3)^{-n^f} \\
&= \hbox{$\frac{5}{3}$} 
+ (\hbox{$\frac{37}{12}$} \pm o(1)) n^{-c}
+ 2{\mathbf m}_{a_{2}} (t-t_2-n^f).
\end{split}
\end{equation*}
Recall from~(\ref{sigma2}) that 
${\mathbf m}_{a_{2}}$ is negative.
This allows the distance
$x_5(t)- x_4(t)$ to fall below 1. 
This happens at $t_3=t_2+\theta_2$, where 
$\theta_2= n^f+ \Theta(|{\mathbf m}_{a_{2}}^{-1}|)$.
Note that 
$|{\mathbf m}_{a_{2}}|$ is sufficiently
small for the newly formed edge $(4,5)$ to be safe at time $t_3$.
We can see that from~(\ref{sigma2}), which also shows that
\begin{equation}\label{theta2}
\theta_2 \geq  \Omega(|{\mathbf m}_{a_{2}}^{-1}|)
\geq e^{\Omega( {\mathbf m}_{a_{1}}^{-1} )}.
\end{equation}
We conclude this opening analysis with 
an estimation of the stationary velocity ${\mathbf m}_{a_3}$.
The flipping rule causes the velocity 
of the flock $({\mathcal B}_{1}, \ldots, {\mathcal B}_{4})$
to be reversed at time $t_2+n^f$. 
(It's a flip of type ``left,'' so named because it involves
a left child.)
Following the flip, the velocity of the 8-bird flock at $a_3$
is, at its creation,
\begin{equation*}
v^{a_3} =
\begin{pmatrix}
-P_{2}^{\theta_{2}}\, v^{a_{2}}
\\
\,\,\,\,   P_{2}^{\theta_{2}}\, v^{a_{2}}
\end{pmatrix}.
\end{equation*}
By~(\ref{Deltav2}, \ref{P2s-vec}),
$$
P_2^{\theta_2} \, v^{a_2}
= 
\hbox{$\frac{1}{12}$}
n^{-c}\left\{
-2 
\begin{pmatrix}
1 \\
1 \\
1 \\
1
\end{pmatrix}
(-3)^{1-\theta_1}
+
4
\Bigl(\frac{2}{3}\Bigr)^{\theta_2}
\begin{pmatrix}
\,\,\,\, 2 \\
\,\,\,\, 1 \\
-1 \\
-2
\end{pmatrix}
-2
(-3)^{-\theta_2}
\begin{pmatrix}
\,\,\,\, 1 \\
-1 \\
\,\,\,\, 1 \\
-1
\end{pmatrix}
\right\}.
$$
By~(\ref{StatVelSigmaj=}, \ref{theta2}),
therefore, 
\begin{equation*}
\begin{split}
{\mathbf m}_{a_3}
&=
\hbox{$ \frac{1}{7} $}
(\hbox{$\frac{1}{2}$},
1, 1, 1, 1, 1, 1, \hbox{$\frac{1}{2}$})
\begin{pmatrix}
- P_2^{\theta_2}\, v^{a_2}
\\
\,\,\,\, P_2^{\theta_2}\, v^{a_2}
\end{pmatrix} 
=
\hbox{$ \frac{1}{7} $}
(\hbox{$\frac{1}{2}$},
0, 0, -\hbox{$\frac{1}{2}$})
P_2^{\theta_2} \, v^{a_2}
\\
&=
\hbox{$\frac{1}{42}$}
n^{-c}
\Bigl(4\Bigl(\frac{2}{3}\Bigr)^{\theta_2}
- (-3)^{-\theta_2}\Bigr)
\leq  e^{- e^{\Omega( {\mathbf m}_{a_{1}}^{-1} )}}.
\end{split}
\end{equation*}
Since ${\mathbf m}_{a_3}>0$, the next flip
must be of type ``right,'' which happens to agree with 
the flipping rule.
Observe from~(\ref{sigma1}, \ref{sigma2}) 
how, as $j$ increases from $1$ to $3$,
the stationary velocity ${\mathbf m}_{a_{j}}$ decays
from polynomial to exponential
to doubly exponential.
To generalize this to further heights is not
difficult. What's tricky is to show that,
despite all the symmetries in the system, 
the stationary velocities never vanish.
For example, if we formed new flocks by
attaching to a smaller one its mirror image,
this would bring the drifting motion, and
hence the flocking, to an end.
We summarize our findings 
below~(\ref{sigma1}, \ref{theta1}, \ref{sigma2}, \ref{theta2}):

\begin{equation}\label{SummarySmallCases}
\begin{cases}
\, |{\mathbf m}_{a_{1}}|=  \hbox{$\frac{1}{2}$} n^{-c}
\hspace{1.7cm} \text{\&} \hspace{1.1cm} 
\theta_1 = \hbox{$\frac{1}{3}$} (n^{c} +1);
\\
\, |{\mathbf m}_{a_{2}}| 
\leq  e^{-\Omega( {\mathbf m}_{a_{1}}^{-1} )}
\hspace{1cm} \text{\&} \hspace{1.1cm} 
\theta_2 \geq \Omega(|{\mathbf m}_{a_{2}}^{-1}|)
\geq e^{ n^{c-1} }; 
\\
\, |{\mathbf m}_{a_3}|
\leq e^{- e^{ n^{c-1} }}.
\end{cases}
\end{equation}

\subsection{Velocity Analysis}

Our first task is to diagonalize the transition matrix $P_j$
given in~(\ref{P_j=1/3Matrix}). The Laplacian acting on a path
is akin to its acting on a folded cycle. Since the Fourier transform
over a finite cyclic group diagonalizes the one-dimensional Laplacian, we can
interpret the spectral shift as a linear operator acting 
on the Fourier coefficients. We explain why below.

\paragraph{The Folded Cycle.}

The Fourier transform over the additive group ${\mathbb Z}_m$ provides
the eigenvectors $y_1,\ldots, y_m$ of the linear map ${\mathcal M}$ defined 
by the circulant matrix

\begin{equation*}
\frac{1}{3}
\underset{m}{\underbrace{      
\begin{pmatrix}
1 & 1 & 0 & 0 &\dots & 1 \\
1 & 1 & 1 & 0 &\dots & 0 \\
0 & 1 & 1 & 1 &\dots & 0 \\
\vdots & \ddots &   &   &\ddots& \vdots \\
0 & \dots & 0 & 1 & 1   & 1 \\
1 & \dots & 0 & 0 & 1   & 1 
\end{pmatrix}
}} ;
\end{equation*}
namely,
$$
y_k= \Bigl( 1, e^{2\pi i(k-1)/m}, 
               \ldots ,  e^{2\pi i(k-1)(m-1)/m} \Bigr).
$$
The associated eigenvalue $\lambda_k$ is equal to 
$$\frac{1}{3}\Bigl(1+2\cos \frac{2\pi(k-1)}{m}\Bigr).$$
We shall see shortly why using the notation $\lambda_k$, 
reserved for the eigenvalue of $P_j$, is legitimate.
To see the relation with $P_j$, set $m=2n-2$ and $n=2^j$,
and note that the eigenvector coordinates $(y_k)_j$ and $(y_k)_{m+2-j}$
are conjugates. This implies that $\Re\, y_k$ is a real eigenvector
of ${\mathcal M}$ that lies in the $n$-dimensional linear subspace 
$${\mathcal F}= \bigcap_{j=2}^{m} \Bigl\{\, x_j-x_{m+2-j}=0 \, \Bigr\}.$$
Furthermore, it is immediate that $P_j$ is equivalent
to the restriction of ${\mathcal M}$ to ${\mathcal F}$;
in other words, folding the cycle in the middle by identifying
opposite sides creates the desired averaging
weights (in particular, $2/3$ at the end nodes)
and transform the Fourier vectors into right eigenvectors for $P_j$
(hence the valid choice of the notation $\lambda_k$).
It follows that, for $1<k\leq 2^j$,
\begin{equation*}
u_k   = 
        \Bigl(1,\, \cos\frac{\pi (k-1)}{n-1}, 
               \ldots , \, \cos\frac{\pi (k-1)(n-2)}{n-1}, \, 
                               (-1)^{k-1}\Bigr)^T
\end{equation*}
is the unique right eigenvector (up to scaling) of $P_j$ for
$\lambda_k$.  We note that, unlike for the $m$-cycle, the transition for the 
$n$-path has only simple eigenvalues.

\vspace{1cm}
\begin{figure}[htb]\label{fig-foldedcycle}
\begin{center}
\hspace{0.0cm}
\includegraphics[width=8.5cm]{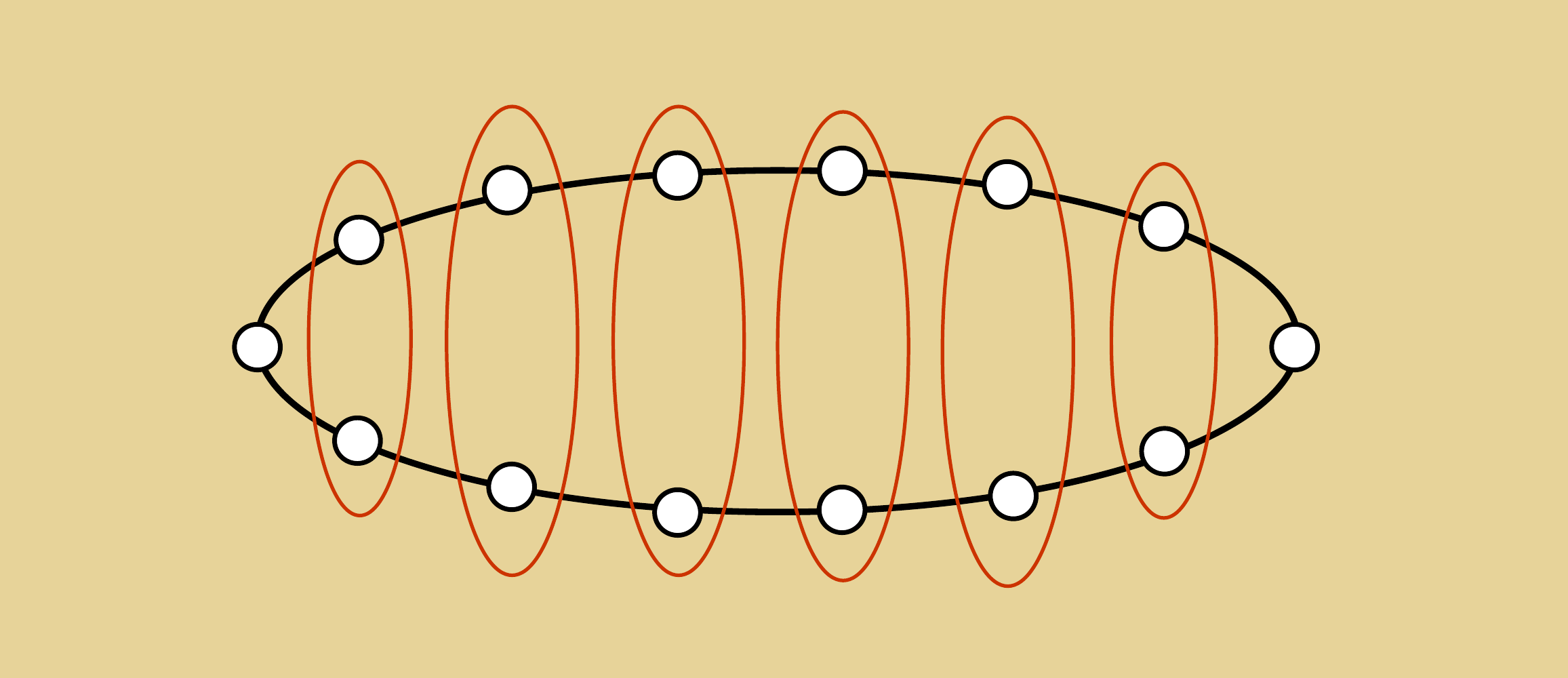}
\end{center}
\begin{quote}
\vspace{0.0cm}
\caption{\small 
The folded $m$-cycle. Identifying opposite nodes allows us to use
the harmonic analysis of the cyclic group to the path-shaped flock.
}
\end{quote}
\end{figure}

\noindent
Consider the evolution of a flock at 
node $a_j$ for $j\geq 1$. Let
\begin{equation}\label{pi-diagC}
\pi = 
\frac{1}{2^j-1}
(\, \overset{2^j}{\overbrace{
\hbox{$\frac{1}{2}$},1,\ldots, 1, \hbox{$\frac{1}{2}$}}}\,)^T
\ \ \ \ \ \text{and} \ \ \ \ \ 
\text{diag}\, C_j= 
\hbox{$\frac{1}{3}$}
(\, \overset{2^j}{\overbrace{
2,1,\ldots, 1, 2}}\,).
\end{equation}
For $s\geq 1$, we diagonalize the matrix
$P_j^s= {\mathbf 1} \pi^T + Q_j^s$, with,\footnote{We avoid
decorating $\pi$ and ${\mathbf 1}$ with subscripts
when their dimensionality is obvious from the context.
We use $v_k$ instead of the notation $u_k$ from~\S\ref{AlgGeom}.
One should be careful not to confuse these eigenvectors with the
velocities.}
by~(\ref{P_N}),
\begin{equation}\label{PjQjMuj}
Q_j^s= \sum_{k=2}^{2^j} \lambda_k^s
        C_j^{1/2} v_k v_k^T  C_j^{-1/2},
\end{equation}
where the right eigenvector $C_j^{1/2} v_k$ is proportional
to $u_k$ with the normalization condition, $\|v_k\|_2=1$.
By elementary trigonometry, it follows that, for any $1<k\leq 2^j$,
\begin{equation*}
\begin{cases}
\, \lambda_k = \,\frac{1}{3}+\frac{2}{3}\cos\frac{\pi (k-1)}{2^j-1} ,\\
\,  v_k   = \, \delta_k
        \Bigl(\frac{1}{\sqrt{2}},\, \cos\frac{\pi (k-1)}{2^j-1}, 
               \ldots , \, \cos\frac{\pi (k-1)(2^j-2)}{2^j-1}, \, 
                        \frac{(-1)^{k-1}}{\sqrt{2}}\Bigr)^T,
\end{cases}
\end{equation*}
where $\delta_k= \sqrt{2}\, (2^j-1)^{-1/2}$ for $1<k<2^j$ and 
$\delta_{2^j}= ({2^j-1})^{-1/2}$.
Recall that $\theta_j= t_{j+1}-t_j$ is the lifetime
of the flock $F_{a_j}$.
By the triangle inequality and
the submultiplicativity of the Frobenius norm,
for any $z$,
\begin{equation*}
\begin{split}
\|Q_j^s z\|_2
&\leq |\lambda_2|^s \sum_{k>1} 
             \| C_j^{1/2} v_k v_k^T  C_j^{-1/2} z\|_2
\leq |\lambda_2|^s \sum_{k>1} 
             \| C_j^{1/2}\|_F \|  C_j^{-1/2}\|_F \| z\|_2 \\
&\leq 2^{j+1} |\hbox{$\frac{1}{3}$} 
       +\hbox{$\frac{2}{3}\cos\frac{\pi }{2^j-1}$}|^s
\| z\|_2.
\end{split}
\end{equation*}
A Taylor series approximation shows that, for $j, s\geq 1$ and any $z$,
\begin{equation}\label{|Q_jz|}
\|Q_j^s z\|_2
\leq e^{j+1 -\Omega(s 4^{-j})} \|z\|_2.
\end{equation}

\vspace{1cm}
\begin{figure}[htb]
\begin{center}
\includegraphics[width=8cm]{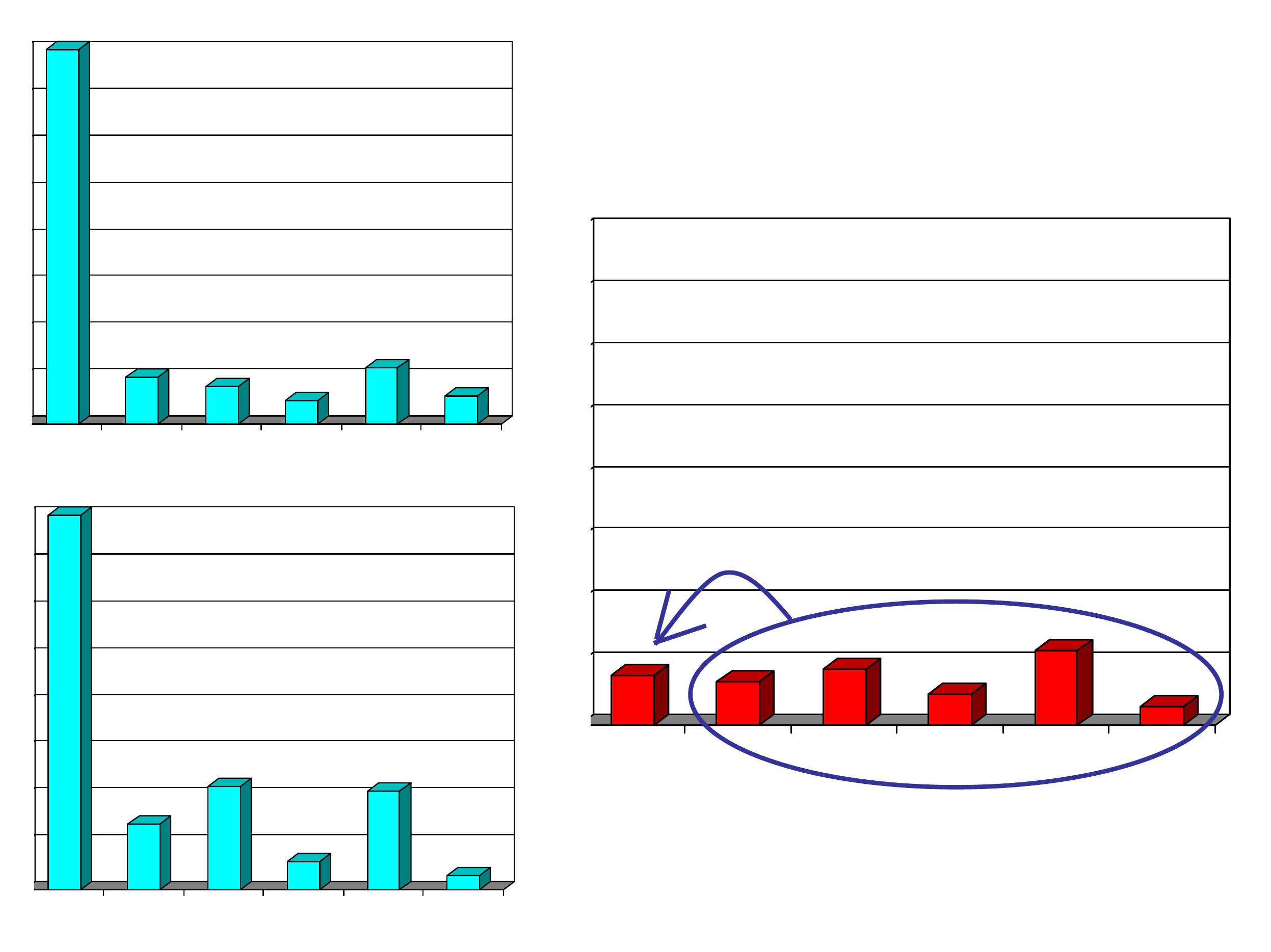}
\end{center}
\vspace{-.2cm}
\caption{\small 
The spectrum of two colliding flocks: to produce a tower-of-twos, the first
Fourier coefficients must cancel each other and be replaced by a linear
combination of the higher ones. This spectral shift must ensure that
the new first Fourier coefficient is nonzero. This will automatically
produce an exponentially decaying energy spectrum.
}
\label{fig-spectralshiftfourier}
\end{figure}

\paragraph{Spectral Shift as Energy Transfer.}

\begin{figure}[htb]
\begin{center}
\includegraphics[width=8cm]{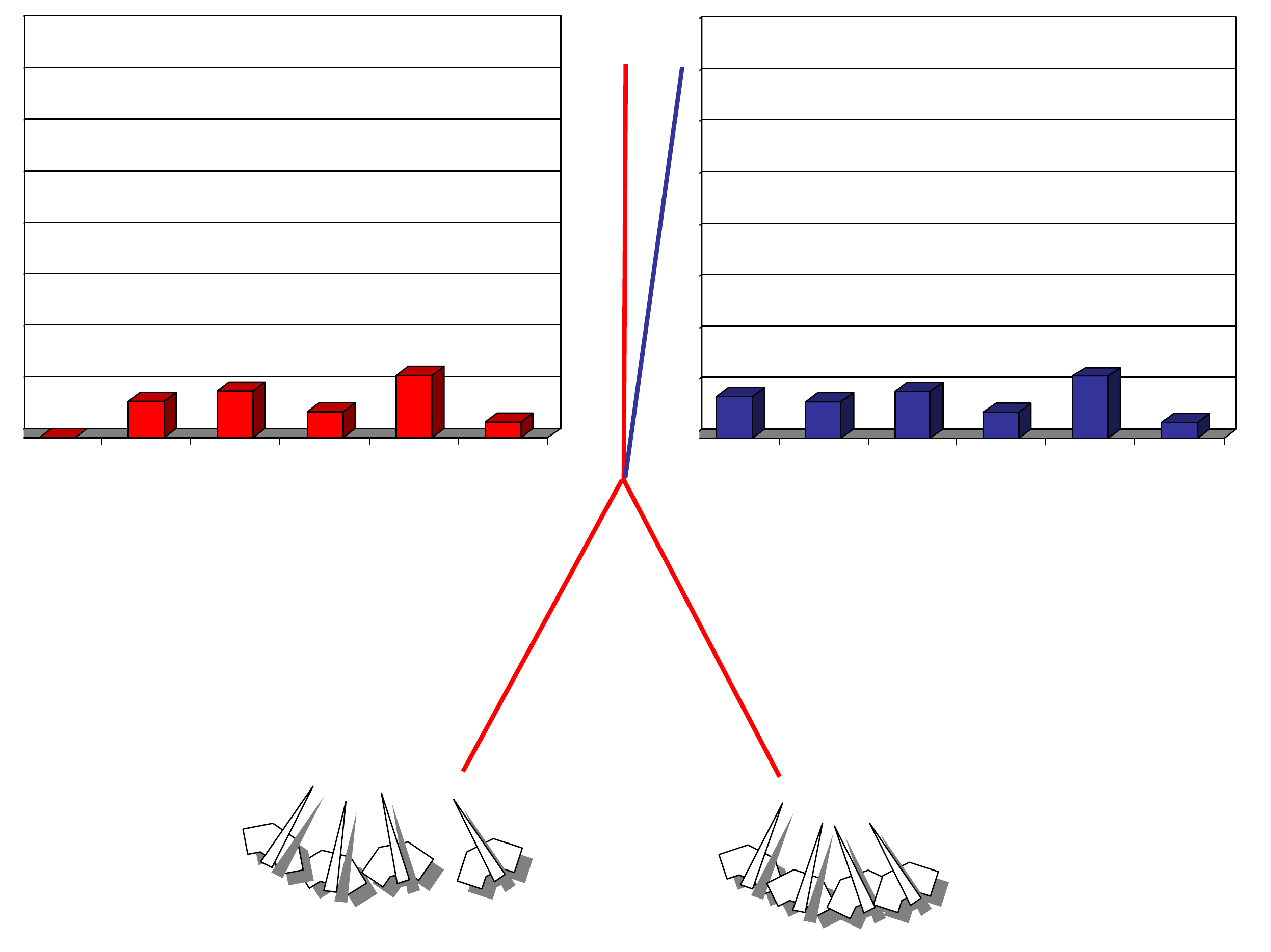}
\end{center}
\caption{\small 
Too much symmetry makes the first Fourier coefficient vanish (left box)
and produces a vertical flying direction. Too little symmetry produces 
an excessive stationary velocity and a polynomially small nonzero
angle with the $Y$ direction. The right amount of symmetry
produces an energy transfer from the decaying
higher Fourier coefficients to the first one, thus creating
an exponentially small angle (right box).
}
\label{fig-compare-fourier}
\end{figure}

After $s$ steps following its creation, the flock $F_j$
moves with velocity
$$v^{a_j}(s) =  P_{j}^{s}\, v^{a_{j}}
= {\mathbf m}_{a_{j}} {\mathbf 1} + 
\sum_{k=2}^{2^j} \alpha_k(s) C_j^{1/2} v_k ,
$$
where $\alpha_k(s)=  \lambda_k^s  v_k^T  C_j^{-1/2}v^{a_{j}}$.
For $k>1$, the Fourier coefficients $\alpha_k(s)$ decay exponentially fast with $s$
while the first one, the stationary velocity, remains constant.
What happens when another flock $G_j$ ``collides" with $F_j$?
A tower-of-twos growth requires two events: one is that,
within the algebraic expressions defining the new Fourier coefficients,
the stationary velocities should cancel out;
the other is that the new first Fourier coefficients should not be zero.
For example, consider a flock $G_j$ that is
mirror image to $F_j$ and heads straight toward it.
The two stationary velocities would cancel out, but the new
one would also be zero. Restoring the dimension $Y$
would produce the spectrum on the left
in Figure~\ref{fig-compare-fourier}
and, consequently, a vertical flying direction: this would dash
any hope of achieving a tower-of-twos.

The trick is to ensure that the energy contained in the higher
Fourier coefficients averages out in a way that produces a new
stationary velocity that is {\em nonzero}: in two dimensions, this will
create a direction close to vertical but not exactly so (right spectrum
in the figure).
The spectral shift can be viewed as a transfer of energy from
the $k$-th Fourier coefficients (for all $k>1$) to the first one.
The issue is not how to produce exponentially fast decay
but how to transfer strictly positive energy.
Too much symmetry wipes out all the energy in the first
Fourier coefficient, while too little symmetry produces a new
stationary velocity that is a nonzero average of the previous ones.
The first case prevents future collisions; the second one produces
a new flying direction that deviates from vertical by only
a polynomially small angle.

\vspace{1cm}

\paragraph{The Spectral Shift in Action.}

Since we are only concerned with velocities in this section, and not with positions,
we may assume without loss of generality that all flipping
is of the right type: in other words, we stipulate that the
flock of any right child of height at least 2 should get 
its velocity reversed after the prescribed lag time. To restore
the true flipping rule will then only be a matter of 
changing signs appropriately.
With this simplifying assumption, the aggregating formula of~(\ref{Deltav2})
becomes, for $j>1$,
\begin{equation}\label{Deltavj}
v^{a_j} =
\begin{pmatrix}
\,\,\,\, P_{j-1}^{\theta_{j-1}}\,
v^{a_{j-1}}
\\
- P_{j-1}^{\theta_{j-1}} \,
v^{a_{j-1}}
\end{pmatrix} 
= 
\begin{pmatrix}
\,\,\,\,  {\mathbf m}_{a_{j-1}} {\mathbf 1} + Q_{j-1}^{\theta_{j-1}}\,
       v^{a_{j-1}} \\
- 
{\mathbf m}_{a_{j-1}} {\mathbf 1} - Q_{j-1}^{\theta_{j-1}}\,
       v^{a_{j-1}}
\end{pmatrix}.
\end{equation}
The averaging operator $P_j$ cannot increase the 
maximum absolute value of the 
velocity coordinates; therefore,
by~(\ref{sigma1}),
$$\|v^{a_{j}}\|_2
\leq 2^{j/2} \|v^{a_{j}}\|_\infty
\leq 2^{j/2} \|v^{a_{1}}\|_\infty 
= 2^{j/2} (2n^{-c}).$$
In other words, for any $j\geq 1$,
\begin{equation}\label{Deltavj-l2}
\|v^{a_{j}}\|_2\leq 2^{j/2+1} n^{-c}.
\end{equation}

\begin{lemma}\label{|aj|-UB}
For any $j>1$, 
the stationary velocity of the flock
at node $a_j$ satisfies
$$
|{\mathbf m}_{a_j}|\leq e^{-\Omega(\theta_{j-1} 4^{-j} )}.
$$
\end{lemma}
\proof
The stationary distribution for 
a $2^{j-1}$-bird flock, being a left eigenvector,
is normal to the right eigenvectors;
hence to $Q_{j-1}^{\theta_{j-1}}\, v^{a_{j-1}}$.
By~(\ref{StatVelSigmaj=}, \ref{Deltavj}),
\begin{equation*}
\begin{split}
{\mathbf m}_{a^j}
&= 
\frac{1}{2^j-1}\,(\,\overset{2^j}{\overbrace{
      \hbox{$\frac{1}{2}$},1,\ldots,1,\hbox{$\frac{1}{2}$}}}\,)
v^{a_j} 
=
\frac{1}{2^j-1}\,(\,\overset{2^j}{\overbrace{
      \hbox{$\frac{1}{2}$},1,\ldots,1,\hbox{$\frac{1}{2}$}}}\,)
\begin{pmatrix}
\,\,\,  Q_{j-1}^{\theta_{j-1}}\, v^{a_{j-1}} \\
-       Q_{j-1}^{\theta_{j-1}}\, v^{a_{j-1}} 
\end{pmatrix}  \\
&= 
\frac{1}{2^j-1}\,(\,\overset{2^{j-1}}{\overbrace{
   \hbox{$\frac{1}{2}$}, 
   1,\ldots,1,\hbox{$\frac{1}{2}$}}},
                    \overset{2^{j-1}}{\overbrace{
      \hbox{$\frac{1}{2}$},1,\ldots,1, \hbox{$\frac{1}{2}$}, 
}}\,)
\begin{pmatrix}
\,\,\,  Q_{j-1}^{\theta_{j-1}}\, v^{a_{j-1}} \\
-       Q_{j-1}^{\theta_{j-1}}\, v^{a_{j-1}} 
\end{pmatrix}  \\
&
\hspace{4cm} +
\frac{1}{2^j-1}\,(\,\overset{2^{j-1}}{\overbrace{
      0,\ldots,0,\hbox{$\frac{1}{2}$}}},
                    \overset{2^{j-1}}{\overbrace{
      \hbox{$\frac{1}{2}$},0,\ldots,0}}\,)
\begin{pmatrix}
\,\,\,  Q_{j-1}^{\theta_{j-1}}\, v^{a_{j-1}} \\
-       Q_{j-1}^{\theta_{j-1}}\, v^{a_{j-1}} 
\end{pmatrix} \\
&= \frac{1}{2^{j+1}-2}\,
    \Bigl(\,(Q_{j-1}^{\theta_{j-1}}\, v^{a_{j-1}})_{2^{j-1}}
      -     (Q_{j-1}^{\theta_{j-1}}\, v^{a_{j-1}})_{1} \,\Bigr).
\end{split}
\end{equation*}
By~(\ref{Deltavj-l2}), therefore,
$\|v^{a_{j-1}}\|_2
\leq 2^{(j+1)/2} n^{-c}$ and, by~(\ref{|Q_jz|}), 
\begin{equation*}
|{\mathbf m}_{a^j}|
\leq 
\frac{1}{2^{j}-1}\,
\Bigl\|Q_{j-1}^{\theta_{j-1}}v^{a_{j-1}} \Bigr\|_\infty  
\leq \frac{1}{2^{j}-1}\,
\Bigl\|Q_{j-1}^{\theta_{j-1}}v^{a_{j-1}} \Bigr\|_2
\leq n^{2-c} e^{-\Omega(\theta_{j-1} 4^{-j})}.
\end{equation*}

\hfill $\Box$
\proofend

From the spectral decomposition
$$P_j^s \, v^{a_{j}} = 
      {\mathbf m}_{a_{j}} {\mathbf 1}
+ \sum_{k>1} \lambda_k^s
        C_j^{1/2} v_k v_k^T  C_j^{-1/2} v^{a_{j}},$$
we see that the stationary velocity 
${\mathbf m}_{a_{j}}$ is the first Fourier coefficient,
ie, the spectral coordinate associated with the dominant eigenvalue $1$.
The cancellations of the two copies of ${\mathbf m}_{a_{j-1}}$
in the computation of that coefficient 
has the effect of making ${\mathbf m}_{a_{j}}$ 
a linear combination of powers of higher eigenvalues.
That part of the spectrum being exponentially decaying,
the corresponding {\em spectral shift} implies a similar
exponential decay in the new first Fourier coefficient.
This is the key to the tower-of-twos growth.
Indeed, as we show next, the next inter-flock collision 
cannot occur before a number of steps inversely proportional
to that first Fourier coefficient.

\begin{lemma}\label{time-to-join-j}
For any $j\geq 1$, 
$\theta_{j} =  n^f + \Theta(\, |{\mathbf m}_{a_{j}}^{-1}|)$.
\end{lemma}
\proof
By~(\ref{SummarySmallCases}), we can assume that $j>1$.
For $t_j<t\leq t_{j+1}$, the velocity of the flock $F_{a_j}$
is of the form
$$ \pm P_j^{t-t_j} v^{a_j}
=  \pm ( {\mathbf m}_{a_{j}} {\mathbf 1} 
         + Q_{j}^{t-t_j}\, v^{a_j}),
$$
where the sign changes after a flip.
By~(\ref{|Q_jz|}, \ref{Deltavj-l2}),
\begin{equation*}
\| Q_{j}^{t-t_j}\, v^{a_j}\|_2
\leq e^{j+1 -\Omega((t-t_j) 4^{-j})} \|v^{a_j}\|_2
\leq  n^{3-c} e^{ -\Omega((t-t_j)n^{-2}) }.
\end{equation*}
Summing over all $t$, our choice of $c$ gives us the
conservative upper bound,
\begin{equation*}
\sum_{t>t_j} \| Q_{j}^{t-t_j}\, v^{a_j}\|_2
\leq \frac{1}{n} \, .
\end{equation*}
No bird belongs to more than $\log n$ different flocks,
so its entire motion is specified by the stationary
velocities of its flocks plus or minus 
an additive ``vibration" error of $o(1)$ on 
the bird's total displacement.

Until one of them flips, 
the flock $F_{a_j}$ and the one at its 
sibling node $a_j'$ are identical copies
that have moved in lockstep. The distance
between their leftmost birds at time $t_j+n^f$ is what
it was at time $0$, ie, $2^j$. We postpone the integrity analysis
for later and simply assume that the flocks
are, indeed, single-paths.
This implies that the diameter of
$F_{a_j}$ is at most $2^j-1$. By~(\ref{x2x4-allt-t3}),
its leftmost edge is of length $\frac{2}{3}\pm o(1)$ between
time $0$ and $t_3$. Since the first two birds always
share the same flock, the vibration bound above indicates
that they always remain within distance $\frac{2}{3}+ o(1)$ 
of each other. The same bound also shows that, 
at time $t_j+n^f$, both flocks have diameter at most
$2^j-\frac{4}{3}+ o(1)$. By our previous observation, they must
be at distance at least $\frac{4}{3}- o(1)$.
After flipping at time $t_j+n^f$, the two flocks head toward
each other\footnote{We must assume that the left flock flies to the right, so 
as to put it on a collision course with the other one, after flipping. 
Our argument is symmetric, however,
and would work just the same if directions and flip types were reversed.}
at a relative speed of $2 |{\mathbf m}_{a_{j}}|$,
plus or minus an error speed that contributes a displacement
of $o(1)$. This implies that the time between flipping and merging is
$|(6\pm o(1)){\mathbf m}_{a_{j}}|^{-1}$.
\hfill $\Box$
\proofend

\vspace{0.5cm}
\begin{figure}[htb]
\begin{center}
\includegraphics[width=8cm]{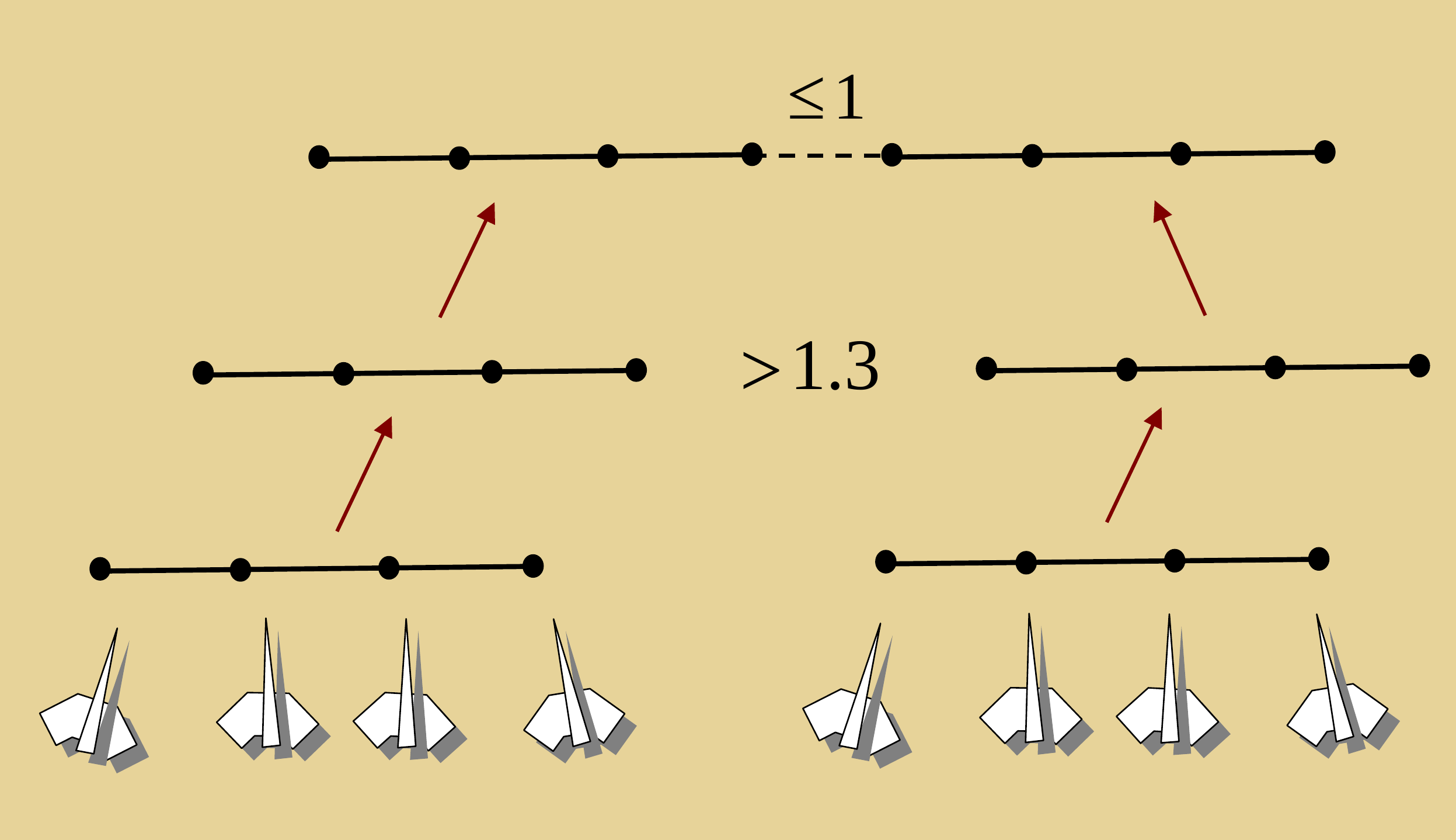}
\end{center}
\caption{\small 
Two flocks merge after a period inversely
proportional to their stationary velocities.
For convenience, we temporarily assume that all
flips are of right-type and that flocks
fly to the right after they are created: these conditions
will not always hold.
}
\label{fig-meetingtime}
\end{figure}

\bigskip
\noindent
For $j>1$, we find from
Lemmas~\ref{|aj|-UB} and~\ref{time-to-join-j} that
$$\theta_j \geq \Omega( e^{\Omega(\theta_{j-1} 4^{-j})}).$$
Since, by~(\ref{theta1}), $\theta_1> n^4$, it follows
immediately by induction that, for any $j\geq 1$,
\begin{equation}\label{thetaj-0}
\theta_j > n^4 \, \theta_{j-1}, 
\end{equation}
where, for convenience, we define $\theta_0=1$.
This allows us to rewrite our previous lower bound
in the slightly simpler fashion,
\begin{equation}\label{thetaj-exp}
\theta_j \geq e^{\Omega(\theta_{j-1} 4^{-j})},
\end{equation}
for any $j>1$. Note that 
the tower-of-twos lower bound on the flocking time
follows immediately from~(\ref{thetaj-exp}).
Indeed, let $\hat \theta_j = \sqrt{\theta_j}$.
By~(\ref{SummarySmallCases}), $\hat \theta_1> 2$ and, for $j>1$,
$\hat \theta_j \geq 2^{\hat \theta_{j-1}}$; therefore,
when $j$ reaches $\log n -1$,
$$\theta_j\geq \hat \theta_j > 2\uparrow\uparrow \log \frac{n}{2}\, ,$$
which establishes the main lower bound of this paper. 
\hfill $\Box$
\proofend

For future use, we state a weak bound on stationary velocities.
By Lemmas~\ref{|aj|-UB} and~\ref{time-to-join-j}, for $j>1$,
$$
|{\mathbf m}_{a_j}|\leq  e^{-\Omega(\theta_{j-1} 4^{-j} )}
\leq  e^{-\Omega( |{\mathbf m}_{a_{j-1}}^{-1}| 4^{-j} )}.$$
By~(\ref{Deltavj-l2}),
$|{\mathbf m}_{a_j}|= |\pi^T v^{a_{j}}|\leq 
\|v^{a_{j}}\|_\infty\leq 
\|v^{a_{j}}\|_2 < n^{1-c}$.
It then follows from~(\ref{sigma1}) that
\begin{equation}\label{sigma_vjn^6}
|{\mathbf m}_{a_{j}}|< 
\begin{cases}
\, n^{-c}; & \text{if $j=1$}; \\
\, n^{-c} \, |{\mathbf m}_{a_{j-1}}|   & \text{if $j>1$} . 
\end{cases}
\end{equation}

It remains for us to prove that stationary velocities never vanish
and that the flocks keep their structural integrity
during their lifetimes. Note that the former would not
be true if pairs of colliding flocks were mirror images of each other.
The proof must demonstrate that the symmetries needed for
the spectral shift do not cause more cancellations than needed.
But, first, let us see why the flips conform
to the noisy model. Both the number of perturbations
and their timing fall well within the admissible bounds.
The only nontrivial condition to check is that the 
change in velocity at flip time $t= t_j+n^f$ ($j>1$)
is $\frac{\log t}{t} e^{O(n^3)}$. 
The $\ell_2$ norm of the change is
$$
\delta= 2\|P_j^{t-t_j} v^{a_j}\|_2\leq 2\sqrt{n}\, \| v^{a_j}\|_2.
$$
We prove below~(\ref{Deltav_j+delta}) that
$$
\|v^{a_j}\|_2 \leq 2\sqrt{n}\, |{\mathbf m}_{a_{j-1}}|.
$$
Since $f=3$, by~(\ref{thetaj-0}), 
$$t=n^f+\theta_1+\cdots + \theta_{j-1}\leq 2 (\theta_{j-1} - n^f)\, ;$$
therefore, by Lemma~\ref{time-to-join-j},
$$\delta \leq 4 n|{\mathbf m}_{a_{j-1}}|
=
\frac{O(n)}{\theta_{j-1}-n^f} 
= O\Bigl(\frac{n}{t} \Bigr) 
\leq \frac{\log t}{t} e^{O(n^3)},$$
which establishes the conformity to the noisy model.

\medskip

To conclude the kinematic analysis, we must
prove that no stationary velocity ${\mathbf m}_a$
ever vanishes.
This is not entirely obvious in view of all the 
symmetries in the system: 
this would happen, for example,
if one flock were the mirror image of its sibling.

\paragraph{Nonvanishing Velocities.}

We need to take a closer look at the
dynamics of the system to show that flocks
never grind to a halt. In doing so, we will uncover
an iterated process of period 4 that allows
us to give a full description of the velocity 
vector at any time. 
Again, we assume that all flipping
is of type ``right,"
which affects only the flocks at right children
of height at least 2.

\begin{theorem}\label{ThmNotVanish}
For any $j\geq 1$, the stationary velocity
${\mathbf m}_{a_j}$ never vanishes.
Its direction is such that sibling flocks
head toward each other to form bigger flocks.
\end{theorem}
\proof
For $j\geq 1$,
define the $2^j$-by-$2^{j-1}$ matrix
$$
F_j= P_{j}^{\theta_j}
\left(
           \begin{pmatrix}
\,\,\,\, 1 \\
     -1    \end{pmatrix} 
\otimes I_{2^{j-1}} \right).
$$
We form $F_j$ by 
subtracting the right half of $P_{j}^{\theta_j}$ from its left half:
$$(F_j)_{k,l}= (P_{j}^{\theta_j})_{k,l}- 
                 (P_{j}^{\theta_j})_{k,l+2^{j-1}}.
$$
For example, if $j=3$ and $\theta_j=1$,
\begin{equation*}
F_j= 
\frac{1}{3}\begin{pmatrix}
\,\,\, 1 &\,\,\, 2 &\,\,\,   0 & \,\,\,  0 \\
\,\,\, 1 &\,\,\, 1 & \,\,\,  1 &  \,\,\, 0 \\
\,\,\, 0 & \,\,\, 1 & \,\,\,  1 &  \,\,\, 1 \\
-1       & \,\,\,  0 & \,\,\,  1 &  \,\,\, 1 \\
-1       & -1 & \,\,\,  0 &  \,\,\, 1 \\
-1       & -1 & -1 &  \,\,\, 0 \\
\,\,\, 0 & -1 & -1 & -1 \\
\,\,\, 0 & \,\,\,  0 & -2 & -1
\end{pmatrix}.
\end{equation*}
By~(\ref{Deltavj}), for $j>1$,
\begin{equation*}
v^{a_j} =
\begin{pmatrix}
\,\,\,\, P_{j-1}^{\theta_{j-1}} \,
v^{a_{j-1}}
\\
- P_{j-1}^{\theta_{j-1}} \,
v^{a_{j-1}}
\end{pmatrix}
\end{equation*}
and, at the end of its existence, the 
flock at $a_j$ has velocity (with right flips only):
\begin{equation}\label{PDelta_j}
P_{j}^{\theta_j}\, v^{a_j} 
= 
P_{j}^{\theta_j}
           \begin{pmatrix}
\,\,\,\,
       P_{{j-1}}^{\theta_{j-1}}\, v^{a_{j-1}} \\
     - P_{{j-1}}^{\theta_{j-1}}\, v^{a_{j-1}}
           \end{pmatrix} 
= F_j P_{{j-1}}^{\theta_{j-1}} \, v^{a_{j-1}} \\
= 
\Bigl( \prod_{i=j}^2  F_i \Bigr) 
P_{{1}}^{\theta_{1}} \, v^{a_{1}}.
\end{equation}
Note that indices run {\em down}, as the products
are not commutative.
By~(\ref{StatVelSigmaj=}), for $j>1$,
\begin{equation}\label{newsigma_v_j}
\begin{split}
{\mathbf m}_{a_j} 
&= 
\frac{1}{2^j-1}\,(\,\overset{2^j}{\overbrace{
\hbox{$\frac{1}{2}$},1,\ldots,1,\hbox{$\frac{1}{2}$}}}\,)
           \begin{pmatrix}
\,\,\,\,
       P_{j-1}^{\theta_{j-1}} \, v^{a_{j-1}} \\
     - P_{j-1}^{\theta_{j-1}} \, v^{a_{j-1}}
           \end{pmatrix} \\
&=
\frac{1}{2(1-2^{j})}\,(\,\overset{2^j}{\overbrace{
1,0,\ldots,0,1}}\,)
           \begin{pmatrix}
\,\,\,\,
       P_{{j-1}}^{\theta_{j-1}}\, v^{a_{j-1}} \\
     - P_{{j-1}}^{\theta_{j-1}}\, v^{a_{j-1}}
           \end{pmatrix}  \\
&= 
\frac{1}{2(1-2^{j})}\,(\,\overset{2^j}{\overbrace{
1,0,\ldots,0,1}}\,)
           \begin{pmatrix}
\,\,\,\,
\Bigl( \prod_{i=j-1}^2  F_i \Bigr) 
  P_{1}^{\theta_{1}}\, v^{a_{1}} \\
& \\
- 
\Bigl( \prod_{i=j-1}^2  F_i \Bigr) 
  P_{1}^{\theta_{1}}\, v^{a_{1}} 
           \end{pmatrix}  \\
&= 
\frac{1}{2(1-2^{j})} \, z_{j-1,1}^T\,
\Bigl( \prod_{i=j-1}^2  F_i \Bigr) 
P_{1}^{\theta_{1}}\, v^{a_{1}}, 
\end{split}
\end{equation}
where  $\prod_i=1$ if $j=2$ and 
$$z_{j,k}= (\, \overset{2^{j}}{\overbrace{
1,0,\ldots,0,(-1)^{k}}}\,)^T.$$
We now look more closely at the structure of $F_{j}$, going
back to the spectral decomposition of $P_{j}^{\theta_j}$.
By~(\ref{PjQjMuj}), for $j\geq 1$,
\begin{equation}\label{Spectral-Pmj^thet}
\begin{cases}
P_{j}^{\theta_j} = {\mathbf 1}_{\!2^j} \pi^T 
+ Q_{j}^{\theta_j}, \\
Q_{j}^{\theta_j}= 
\sum_{k=2}^{2^j}
\mu_{j,k} u_{j,k} (u_{j,k}- \hbox{$\frac{1}{2}$} z_{j,k-1})^T,
\end{cases}
\end{equation}
where, for notational convenience, we 
subscript ${\mathbf 1}$ to indicate its dimension;
for any $j\geq 1$ and $1< k\leq 2^j$,
\begin{equation}\label{mu-u(jk)}
\begin{cases}
\, \mu_{j,k}
= \, \frac{\varepsilon_{j,k}}{2^j-1}
\Bigl( \frac{1}{3}+\frac{2}{3}\cos\frac{\pi (k-1)}{2^j-1} 
      \Bigr)^{\theta_j} 
   \hbox{, with } \varepsilon_{j,k}= 2 \hbox{ if } k<2^j
\hbox{ and } \varepsilon_{j,2^j}= 1 ;
\\
\, u_{j,k}= \,
  \Bigl(1,\cos\hbox{$\frac{\pi (k-1)}{2^j-1}$}, 
   \ldots , \, \cos\hbox{$\frac{\pi (k-1)(2^j-2)}{2^j-1}$}, \, 
                        (-1)^{k-1}\Bigr)^T \in {\mathbb R}^{2^j}. 
\end{cases}
\end{equation}
Our algebraic approach requires bounds on eigenvalue gaps and
on the Frobenius norm of $Q_{j}^{\theta_j}$.
Note that $|\mu_{j,k}|<1$ for all $j\geq 1$ and $k\geq 2$.
We need much tighter bounds.
Recall that $n$ is assumed large enough and 
define $\mu_{0,2}=1$ for notational convenience.

\begin{lemma}\label{factEigenvalues-Q^theta-approx}
For any $j\geq 1$, both $|\mu_{j,2}/\mu_{j-1,2}^n|$ and 
$\|Q_{j}^{\theta_j}\|_F$ are less than~$e^{-n^{1.5}}$;
for $j>1$ and $k>2$, so is the ratio $|\mu_{j,k}/\mu_{j,2}|$.
\end{lemma}
\proof
We leave the bound on
$\|Q_{j}^{\theta_j}\|_F$ for last.
If $j=1$, then $\mu_{j,2}= (-3)^{-\theta_1}$
and, by~(\ref{thetaj-0}), 
$|\mu_{j,2}|< e^{-n^{4}}$. 
Since $\mu_{0,2}=1$, this proves the first upper bound
for $j=1$. Suppose now that $j>1$.
For $2\leq k\leq 2^j$,
$|1+2\cos\frac{\pi (k-1)}{2^j-1}|\leq  
 |1+2\cos\frac{\pi}{2^j-1}|$.
In view of the fact that
$j\leq \log n$ and, 
by~(\ref{thetaj-0}), 
$\theta_j >n^4$, for all $k\geq 2$,
\begin{equation}\label{mujk<e^n17}
|\mu_{j,k}|\leq |\mu_{j,2}| \leq
O(2^{-j})
    e^{-\Omega(\theta_j 4^{-j})}
< e^{-n^{1.7}} \, .
\end{equation}
By~(\ref{thetaj-0}), 
$$
|\mu_{j,2}| \leq
    e^{-\Omega(\theta_{j} 4^{-j})}
\leq
    e^{-\Omega(n^4\theta_{j-1} 4^{-j})}
\leq   e^{-\Omega(n^2)} e^{-\Omega(n^2\theta_{j-1})}
<   e^{-n^{1.5}}|\mu_{j-1,2}|^n.
$$
The last inequality follows from
the fact that $2^{1-j} 3^{-\theta_{j-1}}
\leq |\mu_{j-1,2}|< 1$.
To bound the ratio $|\mu_{j,k}/\mu_{j,2}|$
for $j>1$ and $k>2$, we 
begin with the case $j=2$ and verify directly
that $e^{-n^{3}}$ is a valid upper bound. Indeed,
\begin{equation*}
\mu_{2,k}
=
\begin{cases}
(\frac{2}{3})^{\theta_2 +1}
& \text{if $k=2$}; \\
0
& \text{if $k=3$}; \\
(-1)^{\theta_2}
(\frac{1}{3})^{\theta_2 +1}
& \text{if $k=4$}.
\end{cases}
\end{equation*}
Assume now that $j,k>2$. Then 
$-1\leq 1+2\cos\hbox{$\frac{\pi (k-1)}{2^j-1}$}\leq 
1+2\cos\hbox{$\frac{2\pi}{2^j-1}$}$.
Since $1+2\cos\frac{2\pi}{2^j-1}>1$,
$|1+2\cos\frac{\pi (k-1)}{2^j-1}|\leq  
 |1+2\cos\frac{2\pi}{2^j-1}|$; therefore,
$$
\Bigl|\frac{\mu_{j,k}}{\mu_{j,2}}\Bigr|
\leq
\left( \frac{1+2\cos \frac{2\pi}{2^j-1}}
            {1+2\cos \frac{\pi}{2^j-1}}
\right)^{\theta_j}
= 
( 2\cos \hbox{$\frac{\pi}{2^j-1}$} -1 )^{\theta_j}
= e^{- \Omega(\theta_j 4^{-j})} < e^{-n^{1.5}} \, .
$$
For all $j\geq 1$, by~(\ref{mujk<e^n17}) and
the submultiplicativity of the Frobenius norm,
\begin{equation*}
\begin{split}
\|Q_j^{\theta_j}\|_F 
&\leq 
\sum_{k=2}^{2^j} |\mu_{j,k}| \times
\| u_{j,k}\|_2 
\,(\|u_{j,k}\|_2
+ \hbox{$\frac{1}{2}$} \|z_{j,k-1}\|_2) \\
&\leq 2^{O(j)} |\mu_{j,2}| \leq 2^{O(j)}  e^{-n^{1.7}} < e^{-n^{1.5}} \, .
\end{split}
\end{equation*}
\hfill $\Box$
\proofend

For $j>1$, we express $F_{j}$, 
the ``folded'' half of $P_j^{\theta_j}$,
by subtracting the lower half of 
$u_{j,k}- \hbox{$\frac{1}{2}$} z_{j,k-1}$
from its upper half, forming
\begin{equation}\label{w_(j-1)k}
w_{j-1,k}=  (
\xi_1,\ldots, \xi_{2^{j-1}} )^T
 - \hbox{$\frac{1}{2}$} z_{j-1,k},
\end{equation}
where 
$$\xi_l= \cos \hbox{$\frac{\pi(k-1)(l-1)}{2^j-1}$}  
      -  \cos \hbox{$\frac{\pi(k-1)(2^{j-1}+l-1)}{2^j-1}$}.$$
It follows from~(\ref{pi-diagC}, \ref{Spectral-Pmj^thet})
that, for $j>1$,
\begin{equation}\label{F_jDefn}
F_j= 
\hbox{$\frac{1}{2(1-2^j)}$}{\mathbf 1}_{\!2^j} z_{j-1,1}^T
+ \sum_{k=2}^{2^j}
\mu_{j,k} u_{j,k} w_{j-1,k}^T.
\end{equation}
To tackle the formidable product 
$\prod_i F_i$ in~(\ref{PDelta_j}), we
begin with an approximation 
$\prod_i G_i$, where 
\begin{equation}\label{G_jDefn}
G_j= 
\hbox{$\frac{1}{2(1-2^j)}$}{\mathbf 1}_{\!2^j} z_{j-1,1}^T
+ \mu_{j,2} u_{j,2} w_{j-1,2}^T.
\end{equation}
Setting $k=2$, we find that
$$
u_{j,2}= 
     \Bigl(1,\cos\hbox{$\frac{\pi}{2^j-1}$},
            \ldots , \, \cos\hbox{$\frac{\pi(2^j-2)}{2^j-1}$},
             \, -1\Bigr)^T.
$$
For $0\leq l <2^{j-1}$,
$$\cos\hbox{$\frac{\pi l}{2^j-1}$}+ 
      \cos\hbox{$\frac{\pi (2^j-l-1)}{2^j-1}$}=  0 .$$
This extends to the case $j=1$, so that, 
for any $j\geq 1$,
\begin{equation}\label{uj2=}
\begin{cases}
\, 
u_{j,2} = \,
({\bar u}_1,\ldots,{\bar u}_{2^{j-1}},-
          {\bar u}_{2^{j-1}},\ldots,-{\bar u}_1)^T ; \\
\,
{\bar u}_l= \, \cos\frac{\pi (l-1)}{2^j-1} \, .
\end{cases}
\end{equation}
For $k=2$, we simplify $\xi_l$ into
$$\xi_l= \cos \hbox{$\frac{\pi(l-1)}{2^j-1}$}  
     + \sin \hbox{$\frac{\pi(l-\frac{1}{2})}{2^j-1}$},$$
for $1\leq l\leq 2^{j-1}$,
which shows that $\xi_l= \xi_{2^{j-1}+1-l}$; therefore,
for $j>1$,
\begin{equation}\label{wj2=}
\begin{cases}
\, 
w_{j-1,2} = \,
({\bar w}_1,\ldots,{\bar w}_{2^{j-2}},
          {\bar w}_{2^{j-2}},\ldots, {\bar w}_1)^T ; \\
\, 
{\bar w}_1= \,\frac{1}{2}
     + \sin \frac{\pi/2}{2^j-1}   ;  \\
\,
{\bar w}_l= \, \cos \frac{\pi(l-1)}{2^j-1}  
     + \sin \frac{\pi(l-\frac{1}{2})}{2^j-1}
\ \ \ \ \ (1<l\leq 2^{j-2}).
\end{cases}
\end{equation}
By~(\ref{G_jDefn}), for $j>2$,
\begin{equation}\label{G-j-1}
\prod_{i=j-1}^2  G_i
= \prod_{i=j-1}^2 
\Bigl\{
\hbox{$\frac{1}{2(1-2^i)}$} {\mathbf 1}_{\!2^i} z_{i-1,1}^T
+ \mu_{i,2} u_{i,2} w_{i-1,2}^T \Bigr\}. 
\end{equation}
Expanding this product is greatly simplified by
observing that, by~(\ref{uj2=}, \ref{wj2=}),
for any $j\geq 1$,
\begin{equation}\label{zw-innerprod}
\begin{cases}
\, z_{j,1}^T {\mathbf 1}_{\!2^{j}}
= w_{j,2}^T  u_{j,2} =0 ; \\
\,z_{j,1}^T u_{j,2} =2 ; \\
\, w_{j,2}^T {\mathbf 1}_{\!2^{j}} \,\, \defeq\,\,
                                        \gamma_j, \ \ \ 
\text{where } \ \ 2^{j-1}-1 < \gamma_j < 2^{j+1}-1.
\end{cases}
\end{equation}
To prove the bounds on $\gamma_j$,
we rely on~(\ref{wj2=}),
\begin{equation*}
\gamma_j  =
-1+2
\sum_{l=1}^{2^{j-1}}
\Bigl( \cos \hbox{$\frac{\pi(l-1)}{2^{j+1}-1}$}  
     + \sin \hbox{$\frac{\pi(l-\frac{1}{2})}{2^{j+1}-1}$} 
                   \Bigr),
\end{equation*}
and the fact 
that $\hbox{$\frac{\pi(l-1)}{2^{j+1}-1}$}\leq \frac{\pi}{3}$
and $\hbox{$\frac{\pi(l-1/2)}{2^{j+1}-1}$}< \frac{\pi}{2}$,
from which the two inequalities in~(\ref{zw-innerprod}) follow readily.
By~(\ref{G-j-1}), for $j>2$,
\begin{multline}\label{zGPDelta^v_0}
z_{j-1,1}^T\,
\Bigl( \prod_{i=j-1}^2  G_i \Bigr) 
P_{1}^{\theta_{1}} \, v^{a_{1}}  \\
=
z_{j-1,1}^T\,
\Bigl( 
\prod_{i=j-1}^2 
\Bigl\{
\hbox{$\frac{1}{2(1-2^i)}$} {\mathbf 1}_{\!2^i} z_{i-1,1}^T
+ \mu_{i,2} u_{i,2} w_{i-1,2}^T \Bigr\} 
\Bigr)
P_{1}^{\theta_{1}}\, v^{a_{1}}. 
\end{multline}
If we drop all sub/superscripts and expand the scalar
expression above, 
we find a sum of $2^{j-2}$ words 
$z a_{j-1} \cdots a_2 P_{1}^{\theta_{1}}\, v^{a_{1}}$, where
each $a_i$ is of the form $\mu uw$ or $1z$ (suitably scaled).
By~(\ref{zw-innerprod}), however,
the only nonzero word
is of the form $A= z(\mu uw)(1z)(\mu uw)(1z)\cdots P_{1}^{\theta_{1}}\, v^{a_{1}}$. 
This necessitates
distinguishing between even and odd values of $j$.

\begin{figure}[htb]\label{fig-ladderodd}
\vspace{0.5cm}
\begin{center}
\includegraphics[width=8.5cm]{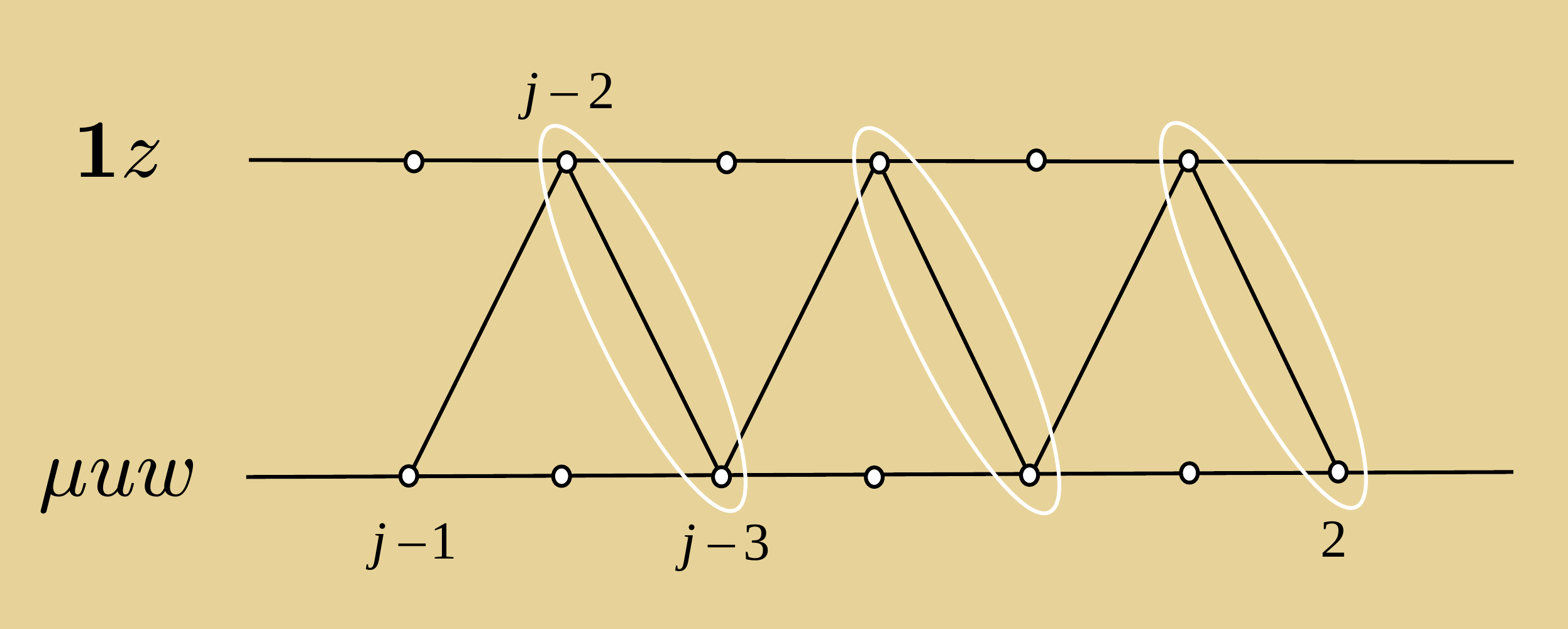}
\end{center}
\begin{quote}
\vspace{0cm}
\caption{\small 
If $j$ is odd, the word $A$ is of the form
$z(\mu uw)(1z)(\mu uw)(1z)\cdots (\mu uw) P_{1}^{\theta_{1}}\, v^{a_{1}}$. 
}
\end{quote}
\end{figure}

\bigskip\bigskip
\noindent
{\bf Case I.} ({\em odd $j>2$}): \ \
It follows from~(\ref{zw-innerprod}) that 
\begin{equation*}
\begin{split}
z_{j-1,1}^T\,
&\Bigl( \prod_{i=j-1}^2  G_i \Bigr) 
 \\
&=z_{j-1,1}^T
\mu_{j-1,2} u_{j-1,2} w_{j-2,2}^T
\prod_{\text{odd}\, i=j-2}^3 
\Bigl\{
\hbox{$\frac{1}{2(1-2^i)}$} {\mathbf 1}_{\!2^i} z_{i-1,1}^T
\mu_{i-1,2} u_{i-1,2} w_{i-2,2}^T \Bigr\} \\
&=
2 \mu_{j-1,2} w_{j-2,2}^T
\prod_{\text{odd}\, i=j-2}^3 
\Bigl\{
\hbox{$\frac{1}{1-2^i}$} {\mathbf 1}_{\!2^i} 
\mu_{i-1,2} w_{i-2,2}^T \Bigr\} 
= \alpha_j^{\text{\em odd}}\, w_{1,2}^T \, , 
\end{split}
\end{equation*}
where
\begin{equation}\label{alpha_jODD=Defn}
\alpha_j^{\text{\em odd}}= 
2 (-1)^{(j+1)/2}
\mu_{2,2} 
\prod_{\text{odd}\, i=j-2}^{3} 
\frac{\gamma_{i}\mu_{i+1,2}}{2^i-1}
\, .
\end{equation}
One must verify separately that this also holds for
the case $j=3$, where $\prod_i=1$.
Recall that, by~(\ref{sigma1}, \ref{wj2=}),
$w_{1,2}= (1,1)^T$ and 
$\|v^{a_{1}}\|_2= \sqrt{5}\, n^{-c}$.
By Lemma~\ref{factEigenvalues-Q^theta-approx} and 
the submultiplicativity of the Frobenius norm,
$$| w_{1,2}^T \,Q_{1}^{\theta_1} \, v^{a_{1}}|
\leq 
\|w_{1,2}\|_2 \|Q_{1}^{\theta_1}\|_F \|v^{a_{1}}\|_2
<  e^{-n^{1.5}}.$$
By~(\ref{Spectral-Pmj^thet}),
it follows that
\begin{equation}\label{zGiPD-Q1}
w_{1,2}^T P_{1}^{\theta_{1}}\, v^{a_{1}}
=  w_{1,2}^T 
      ( {\mathbf 1}_{\!2} \pi^T 
        + Q_{1}^{\theta_1} ) v^{a_{1}}
=  v^{a_{1}}_1 + v^{a_{1}}_2
              \pm O( e^{-n^{1.5}} )
\end{equation}
and 
\begin{equation}\label{zGPDElta-odd}
\begin{split}
A&= z_{j-1,1}^T\,
\Bigl( \prod_{i=j-1}^2  G_i \Bigr) 
P_{1}^{\theta_{1}}\, v^{a_{1}}
=  \alpha_j^{\text{\em odd}} \,
        w_{1,2}^T P_{1}^{\theta_{1}}\, v^{a_{1}} \\
&=  \alpha_j^{\text{\em odd}} 
        ( v^{a_{1}}_1 + v^{a_{1}}_2 )
              \pm O( \alpha_j^{\text{\em odd}} e^{-n^{1.5}} \,).
\end{split}
\end{equation}

\begin{figure}[htb]\label{fig-laddereven}
\vspace{0.5cm}
\begin{center}
\includegraphics[width=8.5cm]{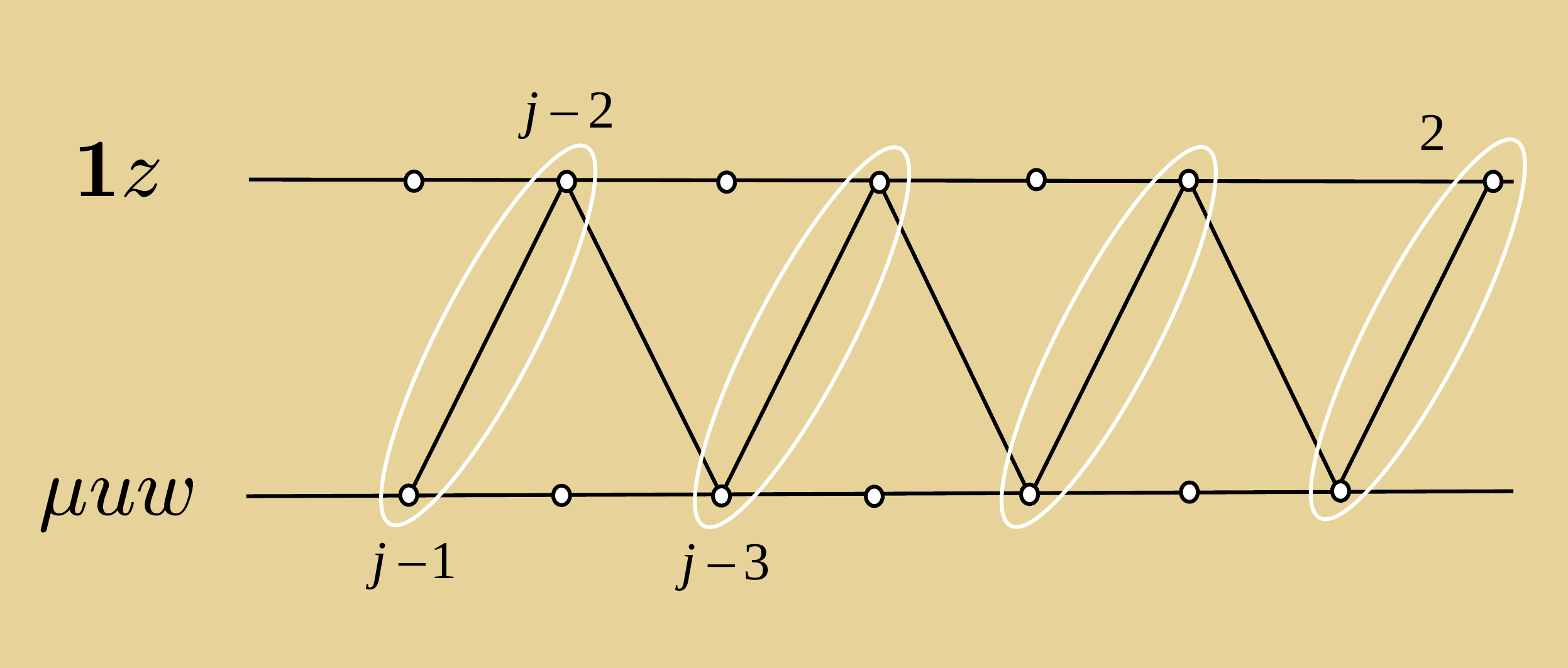}
\end{center}
\begin{quote}
\vspace{0cm}
\caption{\small 
If $j$ is even, the word $A$ is of the form
$z(\mu uw)(1z)(\mu uw)\cdots (1z) P_{1}^{\theta_{1}}\, v^{a_{1}}$. 
}
\end{quote}
\end{figure}

\bigskip\bigskip
\noindent
{\bf Case II.} ({\em even $j>2$}): 

\noindent
\begin{equation*}
\begin{split}
z_{j-1,1}^T\,
\Bigl( \prod_{i=j-1}^2  G_i \Bigr) 
&=
z_{j-1,1}^T
\prod_{\text{odd}\, i=j-1}^3 
\Bigl\{
\mu_{i,2} u_{i,2} w_{i-1,2}^T
\, (\hbox{$\frac{1}{2(1-2^{i-1})}$}) {\mathbf 1}_{\!2^{i-1}} 
z_{i-2,1}^T
 \Bigr\} \\
&=
z_{j-1,1}^T
\prod_{\text{odd}\, i=j-1}^3 
\Bigl\{
(\hbox{$\frac{1}{2(1-2^{i-1})}$}) 
\mu_{i,2} u_{i,2} \gamma_{i-1}
z_{i-2,1}^T
 \Bigr\}  
= \beta_j z_{1,1}^T,
\end{split}
\end{equation*}
where
$$
\beta_j= 
(-1)^{j/2+1}
\prod_{\text{odd}\, i=j-1}^{3} 
\frac{\gamma_{i-1}\mu_{i,2}}{2^{i-1}-1}
\, .
$$
It follows that 
$$
A= 
z_{j-1,1}^T\,
\Bigl( \prod_{i=j-1}^2  G_i \Bigr) 
P_{1}^{\theta_{1}} v^{a_{1}}
=  \beta_j z_{1,1}^T P_{1}^{\theta_{1}}\, v^{a_{1}}.
$$
By~(\ref{Spectral-Pmj^thet}),
\begin{equation}\label{z11Q1D-D}
z_{1,1}^T P_{1}^{\theta_1}\, v^{a_{1}} 
= 
z_{1,1}^T 
      ( {\mathbf 1}_{\!2} \pi^T 
        + Q_{1}^{\theta_1} ) v^{a_{1}}
=  z_{1,1}^T Q_{1}^{\theta_1}\, v^{a_{1}} 
= \mu_{1,2} ( v^{a_{1}}_1 - v^{a_{1}}_2 );
\end{equation}
therefore,
\begin{equation}\label{z11Q1D-alpha-even}
A= \alpha_j^{\text{\em even}}\,
     ( v^{a_{1}}_1 - v^{a_{1}}_2 ),
\end{equation}
where
\begin{equation}\label{alpha_jEVEN=Defn}
\alpha_j^{\text{\em even}}= 
(-1)^{j/2+1}
\mu_{1,2} 
\prod_{\text{odd}\, i=j-1}^{3} 
\frac{\gamma_{i-1}\mu_{i,2}}{2^{i-1}-1}
\, .
\end{equation}

This concludes the case analysis.
Next, we still assume that $j>2$ but 
we remove all restriction on parity.
Recall that $G_i$ is only an approximation of $F_i$ and,
instead of~(\ref{zGPDelta^v_0}), we must contend with 
\begin{multline}\label{zFPdeltaVSzGPdelta}
z_{j-1,1}^T\,
\Bigl( \prod_{i=j-1}^2  F_i \Bigr) 
P_{1}^{\theta_{1}}\, v^{a_{1}} 
\\ =
z_{j-1,1}^T\,
\Bigl( 
\prod_{i=j-1}^2 
\Bigl\{
\hbox{$\frac{1}{2(1-2^i)}$} {\mathbf 1}_{\!2^i} z_{i-1,1}^T
+ \sum_{k=2}^{2^i}
\mu_{i,k} u_{i,k} w_{i-1,k}^T \Bigr\} 
\Bigr)
P_{1}^{\theta_{1}}\, v^{a_{1}}. 
\end{multline}
If, again, we look at the expansion of the product
as a sum of words 
$$B= z a_{j-1} \cdots a_2 P_{1}^{\theta_{1}}\, v^{a_{1}},$$
then we see that each $B$-word is the form
$$z(\mu uw)\{1z,\mu uw\}\{1z,\mu uw\}\{1z,\mu uw\}\cdots 
     P_{1}^{\theta_{1}}\, v^{a_{1}},$$
where $\mu, u, w$ are now indexed by $k$.
Recall that previously the only word was of
the form $A= z(\mu uw)(1z)(\mu uw)(1z)\cdots P_{1}^{\theta_{1}}\, v^{a_{1}}$.
There is no need to go over the entire analysis again.
By showing that $|B|$ is always much smaller
than $|A|$, we prove

\begin{lemma}\label{zFPD}
For any $2<j\leq \log n$, 
\begin{equation*}
z_{j-1,1}^T\,
\Bigl( \prod_{i=j-1}^2  F_i \Bigr) 
P_{1}^{\theta_{1}}\, v^{a_{1}} 
=
\begin{cases}
(1+\eps_n)
(v^{a_{1}}_1 + v^{a_{1}}_2)
\alpha_j^{\text{odd}} 
& \ \text{if $j$ is odd}; \\
(1+\eps_n')
(v^{a_{1}}_1- v^{a_{1}}_2) 
\alpha_j^{\text{even}} 
& \ \text{else},
\end{cases}
\end{equation*}
where $\eps_n,\eps_n'$ are reals of
absolute value $O(e^{-n})$.
\end{lemma}
\proof
Note that, by~(\ref{zw-innerprod}),
$\gamma_i > 2^{i-1}-1$
for any $i\geq 1$. Also, by~(\ref{sigma1}),
$v^{a_{1}}_1+ v^{a_{1}}_2= n^{-c}$
and $v^{a_{1}}_2- v^{a_{1}}_1= 3n^{-c}$.
It follows from~(\ref{alpha_jODD=Defn},
\ref{zGPDElta-odd},
\ref{z11Q1D-alpha-even},
\ref{alpha_jEVEN=Defn}) that, 
for any $2< j\leq \log n$,
\begin{multline}\label{zGimumumu}
|A| 
= \Bigl| z_{j-1,1}^T\,
\Bigl( \prod_{i=j-1}^2  G_i \Bigr) 
P_{1}^{\theta_{1}}\, v^{a_{1}}  \Bigr| \\
\geq 
\Bigl(\frac{1}{n}\Bigr)^{c+1}
\begin{cases}
\, |\mu_{2,2}\, \mu_{4,2} \cdots\mu_{j-1,2}|
& \text{if $j$ is odd}; \\
\, |\mu_{1,2} \,\mu_{3,2} \cdots\mu_{j-1,2}|
& \text{else}.
\end{cases}
\end{multline}
We take absolute values on the right-hand side
for notational consistency: all the factors,
defined in~(\ref{mu-u(jk)}), are strictly positive, except for 
$\mu_{1,2}= (-3)^{-\theta_1}$ which, by~(\ref{theta1}),
is equal to $-3^{-\theta_1}<0$, ie, for $i>1$,
\begin{equation}\label{mui-2-sign}
\mu_{1,2}<0< \mu_{i,2}\, .
\end{equation}
Let's extend our notation by defining, for $i>1$,
\begin{equation*}
\begin{cases}
\, \mu_{i,1} = \frac{1}{2}(1-2^i)^{-1}\, ; \\
\, u_{i,1} = {\mathbf 1}_{\!2^i} \, ; \\
\, w_{i-1,1} = z_{i-1,1}\, .
\end{cases}
\end{equation*}
Then, any $B$-word is specified by an
index vector $(k_{j-1},\ldots,k_2)$:
$$
B_{k_{j-1},\ldots, k_2}
= w_{j-1,1}^T\,
\Bigl( 
\prod_{i=j-1}^2 
\mu_{i,k_i} u_{i,k_i} w_{i-1,k_i}^T 
\Bigr)
P_{1}^{\theta_{1}}\, v^{a_{1}}. 
$$
Observe that the $A$-word we considered earlier
is a particular $B$-word, ie,
$$
A= B_{\, \underset{j-2}{\underbrace{
          \text{\small 2,1,2,1,$\ldots$}}   }} \, .
$$
Since we wish to show that all the other $B$-words
are considerably smaller, we may 
ignore the settings of $k_i$ that
make a $B$-word vanish. All the conditions
on the index vector are summarized here:
\begin{equation}\label{conditions-k}
\begin{cases}
\, 1\leq k_i\leq 2^i \, ;\\
\, k_{j-1}\not = 1 \, ;\\
\, k_{i}k_{i-1}\not =1\ \ \ (2<i<j)\, .
\end{cases}
\end{equation}
By~(\ref{mu-u(jk)}, \ref{w_(j-1)k}),
for all $i>1$ and $k\geq 1$,
$\|u_{i,k}\|_2 \leq 2^{i/2}$
and for $i,k\geq 1$,
$\|w_{i,k}\|_2\leq 2^{i/2+2}$;
so, by Cauchy-Schwarz, for $i>2$ and $k,l\geq 1$,
$$|w_{i-1,k}^T u_{i-1,l}|\leq 2^{i+1}.
$$
Since $2< j\leq \log n$,
$$
\Bigl| w_{j-1,1}^T u_{j-1,k_{j-1}}
\prod_{i=j-2}^2 
w_{i,k_{i+1}}^T u_{i,k_{i}}
\Bigr|
\leq 2^{\frac{1}{2} (j+1)(j+2)}< n^{2\log n};
$$
therefore, 
\begin{equation}\label{Bk2-jUB}
|B_{k_{j-1},\ldots, k_2}|
\leq  n^{2\log n}
\Bigl( 
\prod_{i=j-1}^2 
\mu_{i,k_i} 
\Bigr)
|w_{1,k_2}^T 
P_{1}^{\theta_{1}}\, v^{a_{1}}|.
\end{equation}
We prove that all $B$-words are much smaller than $A$
in absolute value. 

\begin{lemma}\label{|B|vs|A|}
$\!\!\! .\,\,$
All $B$-words distinct from $A$ satisfy:
\begin{equation*}
|B_{k_{j-1},\ldots, k_2}|< e^{-n^{1.2}} |A|.
\end{equation*}
\end{lemma}
\proof
Since $P_1$ is stochastic, by~(\ref{sigma1}),
$$|w_{1,k_2}^T P_{1}^{\theta_{1}}\, v^{a_{1}}|
= O(\|v^{a_{1}}\|_\infty)
=O(n^{-c})<1, 
$$
and the upper bound~(\ref{Bk2-jUB}) becomes
\begin{equation}\label{B-vec-prodik}
|B_{k_{j-1},\ldots, k_2}|
\leq  n^{2\log n}
\prod_{i=j-1}^2 
\mu_{i,k_i}. 
\end{equation}

\begin{figure}[htb]\label{fig-BandA}
\vspace{0cm}
\begin{center}
\includegraphics[width=7cm]{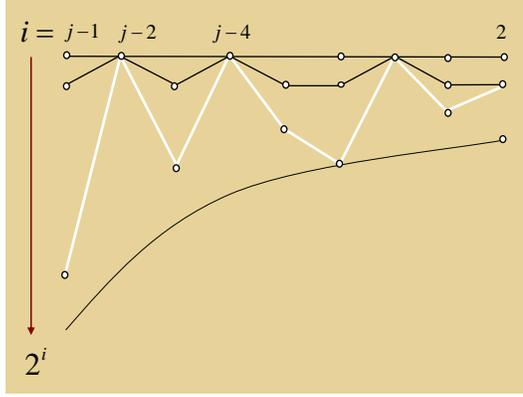}
\end{center}
\begin{quote}
\vspace{0cm}
\caption{\small 
The top horizontal line represents $k_i=1$.
The white dots below the line correspond to $k_i=2$.
The $B$-word in white is brought into canonical form (black jagged line) by 
setting all the indices $k_i>2$ to $2$. This cannot cause the
magnitude of $B$ to drop. We may also assume that the
end result is not the $A$-word, as this would cause
an exponential growth in line with the lemma.
}
\end{quote}
\end{figure}

To maximize the right-hand side of~(\ref{B-vec-prodik}),
we may replace any instance of 
$k_i> 2$ by $k_i= 2$ 
(Lemma~\ref{factEigenvalues-Q^theta-approx}).
This does not contradict 
conditions~(\ref{conditions-k}) since no index is set to 1.
Note the importance for this step of having removed all vectorial
presence from~(\ref{B-vec-prodik}).
We assume that the new $B$-word is not $A$, so its index vector is
not of the form $(2,1,2,1,\ldots)$; therefore, 
if we end up with this very pattern, and hence
with $A$, obviously
at least one index replacement must have taken place.
By Lemma~\ref{factEigenvalues-Q^theta-approx},
any such replacement causes an increase by a factor
of at least $e^{n^{1.5}}$ and Lemma~\ref{|B|vs|A|} follows. 
So, we may assume now that $k_i\in \{1,2\}$ and 
$$(k_{j-1}, k_{j-2}, \ldots, k_2)\not = (2,1,2,1,\ldots).$$
Scan the string $(k_{j-1}, \ldots, k_2)$ against
$(2,1,2,1,\ldots)$ from left to right and 
let $k_a$ be the first character that differs.
By~(\ref{conditions-k}), $k_{j-1}=2$, so
$2\leq a\leq j-2$; hence $j>3$.
Since we cannot have consecutive ones, $k_a=2$
and $j-a$ is even.
By~(\ref{zGimumumu}) and 
Lemma~\ref{factEigenvalues-Q^theta-approx},

\begin{equation*}
\begin{split}
\frac{|B_{k_{j-1},\ldots, k_2}|}{|A|}
&\leq  (n^{c+1} n^{2\log n})
\frac{|\mu_{j-1,2}\, \mu_{j-2,1} \, \mu_{j-3,2}
\cdots 
      \mu_{a+1,2} \, \mu_{a,2} \, \mu_{a-1,k_{a-1}}
           \cdots \mu_{2,k_2}|}
     {|\mu_{j-1,2}\, \mu_{j-3,2} \cdots \mu_{a+1,2}
      \, \mu_{a-1,2} \, \mu_{a-3,2} \cdots |} \\
&\leq n^{3\log n} \, 
\frac{|\mu_{j-2,1}\, \mu_{j-4,1} \cdots 
      \mu_{a+2,1} \, \mu_{a,2} \, \mu_{a-1,k_{a-1}}
           \cdots \mu_{2,k_2}|}
     {|\mu_{a-1,2} \, \mu_{a-3,2} \cdots |} \, .
\end{split}
\end{equation*}
The first numerator mirrors the index vector of the $B$-word accurately.
For the denominator, however, we use the lower bound of~(\ref{zGimumumu}).
The reason we can afford such a loose estimate is the presence
of the factor $\mu_{a,2}$, which plays the 
central role in the calculation by drowning out all the other
differences. Here are the details. All $\mu$'s are less than 1
and, by Lemma~\ref{factEigenvalues-Q^theta-approx},
$|\mu_{a-1,2}| \leq |\mu_{a-l,2}|$;
therefore,
$$
\frac{|B_{k_{j-1},\ldots, k_2}|}{|A|}
\leq 
n^{3\log n} \, 
\frac{|\mu_{a,2}|}{|\mu_{a-1,2}^{\log n}|}
< 
n^{3\log n} \, 
\frac{|\mu_{a,2}|}{|\mu_{a-1,2}^{n}|}
<  n^{3\log n}  e^{-n^{1.5}}\, .
$$
\medskip
\noindent
which proves Lemma~\ref{|B|vs|A|}.
\hfill $\Box$
\proofend

\begin{figure}[htb]\label{fig-countA-B}
\vspace{0cm}
\begin{center}
\includegraphics[width=8cm]{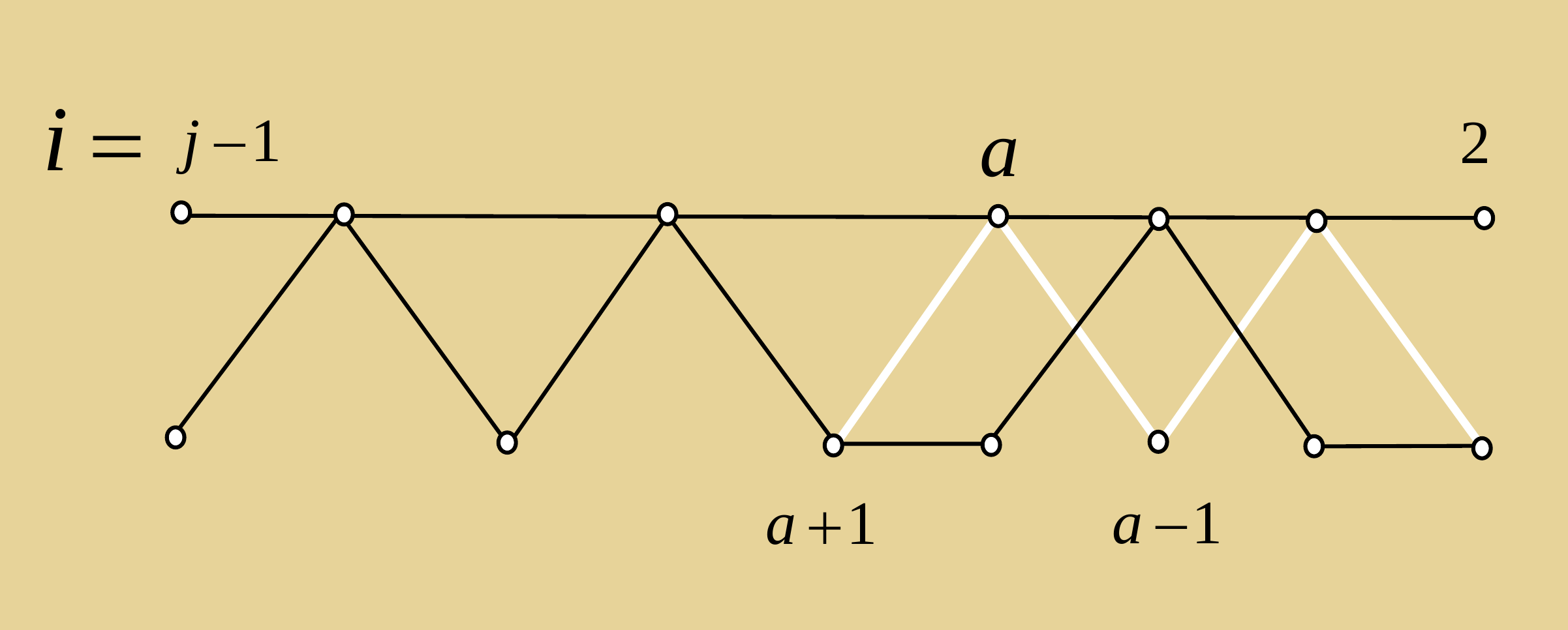}
\end{center}
\begin{quote}
\vspace{0cm}
\caption{\small 
We trace the index vectors of the
$A$ and $B$-words from left to right until they diverge
($i=a$). In this case, $j$ is odd and the index vector of
the $B$-word is $(2,1,2,1,2,2,1,2,2)$.
}
\end{quote}
\end{figure}

There are fewer than $n^{\log n}$ $B$-words; so,
by Lemma~\ref{|B|vs|A|},
their total contribution amounts to 
at most a fraction 
$n^{\log n} e^{-n^{1.2}}$ of $|A|$.
In other words, by~(\ref{zFPdeltaVSzGPdelta}), for $j>2$,
$$
z_{j-1,1}^T\,
\Bigl( \prod_{i=j-1}^2  F_i \Bigr) 
P_{1}^{\theta_{1}}\, v^{a_{1}} 
=
(1\pm O(e^{-n})) z_{j-1,1}^T\,
\Bigl( \prod_{i=j-1}^2  G_i \Bigr) 
P_{1}^{\theta_{1}}\, v^{a_{1}} \, , 
$$
and the proof of Lemma~\ref{zFPD} follows 
from~(\ref{sigma1}, \ref{zGPDElta-odd}, \ref{z11Q1D-alpha-even}).
\hfill $\Box$
\proofend

Recall from~(\ref{newsigma_v_j}) that, for $j>1$,
$$
{\mathbf m}_{a_j} 
= 
\frac{1}{2(1-2^{j})} \, z_{j-1,1}^T\,
\Bigl( \prod_{i=j-1}^2  F_i \Bigr) 
P_{1}^{\theta_{1}}\, v^{a_{1}} .
$$
We know from~(\ref{zw-innerprod},
\ref{alpha_jODD=Defn}, \ref{alpha_jEVEN=Defn},
\ref{mui-2-sign}) that neither $\alpha_j^{\text{\em even}}$
nor $\alpha_j^{\text{\em odd}}$ is null. By Lemma~\ref{zFPD}, 
it then follows that 
the stationary velocity ${\mathbf m}_{a_j}$ never vanishes for $j>2$.
By~(\ref{sigma1}, \ref{sigma2}), this is also the case for $j=1,2$.
To be nonnull is not enough, however: sibling flocks must
also head toward each other. This is what the
flipping rule ensures. We next show how.

\paragraph{Drifting Direction.}

\begin{figure}[htb]
\vspace{0cm}
\begin{center}
\includegraphics[width=7cm]{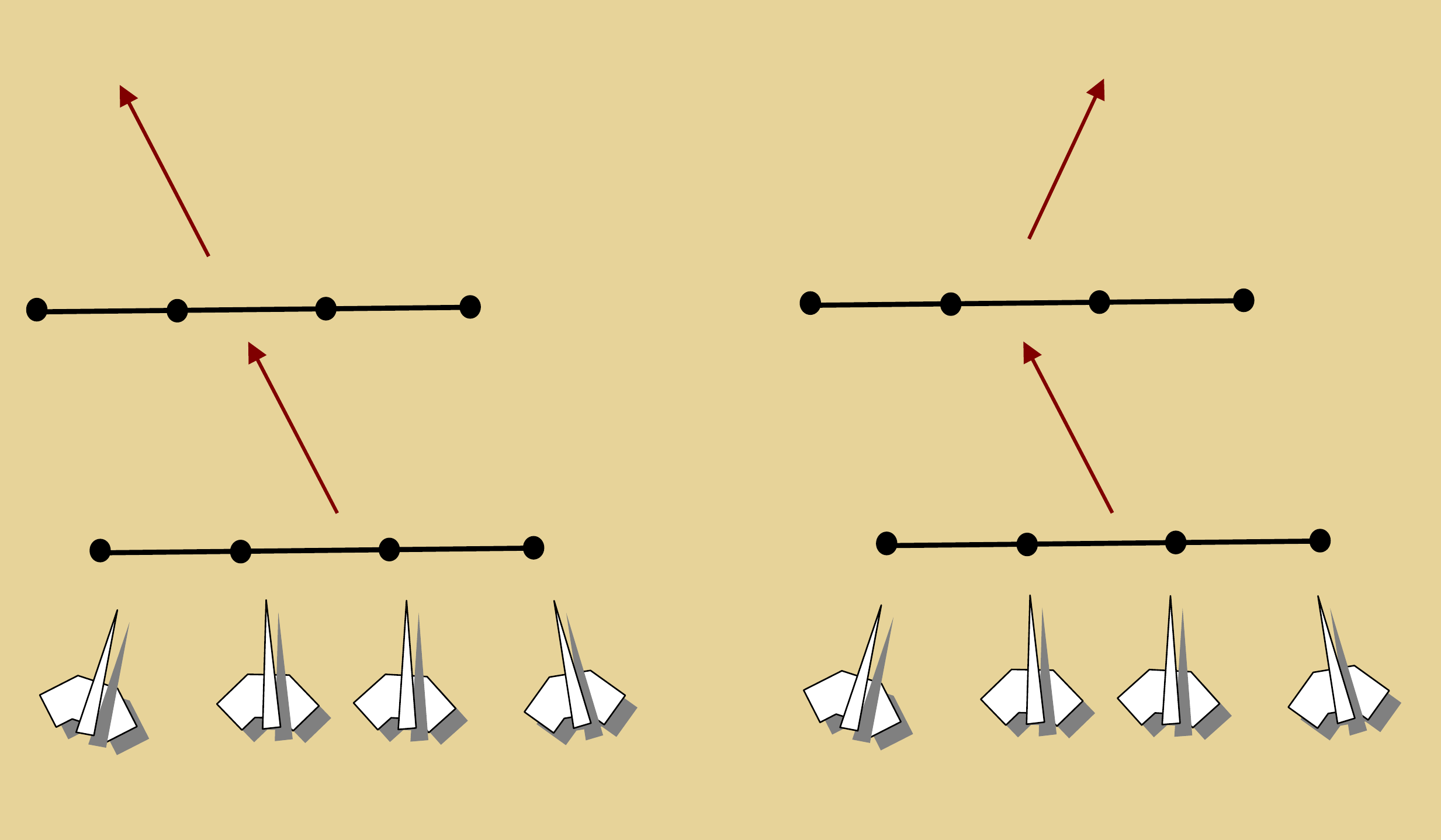}
\end{center}
\begin{quote}
\vspace{0cm}
\caption{\small 
A right-type flip would make the two 4-bird flocks
drift away from each other.}
\label{fig-23drift}
\end{quote}
\end{figure}

By~(\ref{sigma1}, \ref{sigma2}), 
${\mathbf m}_{a_{2}}<0< {\mathbf m}_{a_{1}}$.
By Lemma~\ref{zFPD}, for $j>2$,
\begin{equation}\label{maj-oddeven}
{\mathbf m}_{a_j}
= 
\frac{1}{2(1-2^{j})} \, 
\begin{cases}
(1+\eps_n)
(v^{a_{1}}_1 + v^{a_{1}}_2)
\alpha_j^{\text{odd}} 
& \ \text{if $j$ is odd}; \\
(1+\eps_n')
(v^{a_{1}}_1- v^{a_{1}}_2) 
\alpha_j^{\text{even}} 
& \ \text{else}.
\end{cases}
\end{equation}
We observed in~(\ref{mui-2-sign}) that
$\mu_{j,2}$ is positive for all $j\geq 1$, with the exception of $\mu_{1,2}<0$.
By~(\ref{alpha_jODD=Defn}),
the sign of $\alpha_j^{\text{\em odd}}$
is that of $(-1)^{(j+1)/2}$.
On the other hand, 
by~(\ref{alpha_jEVEN=Defn}),
the sign of $\alpha_j^{\text{\em even}}$ is
that of $(-1)^{j/2}$.
By~(\ref{sigma1}), this proves that, for $j>0$, the sign of
${\mathbf m}_{a_j}$ is positive if and only if
$j=0,1 \pmod{4}$.
Remember that this is what happens when all the flips
are confined to the right children 
of height $j\geq 2$, what we called right-type flips. 
The actual rule is more complex.
It applies to flocks at left children
of nodes of odd height at least 3
and to flocks at right children
of nodes of even height at least 4.
We verify that, after the appropriate flip, 
if any, every 
${\mathbf m}_{a_j}$ is positive, ie, all the flocks
along the left spine of the fusion tree~$\mathcal T$
drift to the right, as they should.
But, before we show this, let's convince ourselves that
right-type flips alone would not do: indeed, note that
${\mathbf m}_{a_2}<0$, so a right-type flip for the
right child of $a_3$ would send the two flocks flying
away from each other (Figure~\ref{fig-23drift}).

Here is a quick proof of the soundness of the true flipping rule.
Suppose we follow the right-type rule.
How do we then modify the velocities to end up
with the same sign assignment produced
by the true flipping rule? The answer is simple:
reverse the sign of the velocities 
of the flocks at both children
of nodes of odd height at least 3.
For $j\geq 2$, the velocity of the flock
at $a_j$ will be effectively reversed
a number of times equal to $\lfloor (j-1)/2 \rfloor$.
The velocity is effectively changed only when 
that number is odd, ie, when $j=0,3 \pmod{4}$.
Recall that ${\mathbf m}_{a_j}>0$ if $j=0,1 \pmod{4}$
and $j>0$.
That implies that ${\mathbf m}_{a_j}$ is now 
positive exactly when $j=1,3 \pmod{4}$, ie, $j$ is odd.
When $j$ is even, however, the node $a_j$, being
a left child of an odd-height node,
undergoes a flip, which therefore reverses
its stationary velocity and makes it positive.
So, in all cases, ${\mathbf m}_{a_j}$ is either positive
or made positive after the lag time for a flip: the corresponding flock
is then headed on a collision course with its sibling. 
Note that, as we observed in the footnote of the proof of
Lemma~\ref{time-to-join-j}, our previous analysis leading
to the tower-of-twos growth still holds despite the
restoration of the true flipping rule.
This concludes the proof of Theorem~\ref{ThmNotVanish}.
\hfill $\Box$
\proofend

It remains for us to establish the structural integrity of the flocks
throughout their lifetime. But, before we do so, it is useful
to revisit the spectral shift and its parity structure.

\paragraph{The Hidden Periodicity of the Spectral Shift.}

The formula for the stationary velocity in~(\ref{maj-oddeven})
reveals a built-in periodicity that
illustrates a fundamental aspect
of the spectral shift.
Looking at~(\ref{zGimumumu}), one may wonder
why the second largest eigenvalues all appear with the same
index parity: odd when $j$ is even and vice versa.
Think of the velocity of a flock
as being well approximated by 
$\sigma {\mathbf 1}+ \gamma {\mathbf u}$, where
$\sigma$ is the speed of its drift
and $\gamma {\mathbf u}$ is its {\em vibration} vector
pointing in the direction of the 
second right eigenvector scaled by 
a Fourier coefficient $\gamma$
decaying exponentially fast with time.
Take the time to be right before merging with 
the flock's sibling. Then the velocity of the new flock is of the form
\begin{equation*}
           \begin{pmatrix}
\,\,\,\,
    \sigma {\mathbf 1}+  \gamma {\mathbf u} \\
-   \sigma {\mathbf 1}- \gamma {\mathbf u} 
           \end{pmatrix} .
\end{equation*}
We approximate the 
transition matrix $P_j^{\theta_j}$ as 
${\mathbf 1} \pi^T + \mu_{j,2}{\mathbf R}$, where
${\mathbf R}$ is a fixed matrix of rank 1.
After $\theta_j$ steps, the velocity becomes roughly 
(ignoring time-independent factors):
\begin{equation*}
({\mathbf 1} \pi^T + \mu_{j,2}{\mathbf R})
           \begin{pmatrix}
\,\,\,\,
    \sigma {\mathbf 1} + \gamma {\mathbf u} \\
-   \sigma {\mathbf 1} - \gamma {\mathbf u} 
           \end{pmatrix} 
\approx  \gamma {\mathbf 1}  + \sigma \mu_{j,2}{\mathbf w},
\end{equation*}
where ${\mathbf w}$ is a unit vector. 
We ignore the lower-order term $\mu_{j,2} \gamma$.
It thus appears that the pair $(\sigma, \gamma)$ becomes
$(\gamma, \sigma \mu_{j,2})$ for the bigger flock.
Note the alternation between $(\sigma, \gamma)$ 
and $(\gamma, \sigma)$.
In particular, 
the switch of $\gamma$ from the right to the left position
in the pair
captures the spectral shift underlying the flocking process,
while the contrary motion of $\sigma$ indicates a re-injection
of the first Fourier coefficient into the spectral mix.
In general, we have the relation
$(\sigma_{j+1}, \gamma_{j+1}) =
(\gamma_j, \sigma_j \mu_{j,2})$;
hence,
$$(\sigma_{j+2}, \gamma_{j+2}) =
( \sigma_j \mu_{j,2}, \gamma_j \mu_{j+1,2}).$$
This shows that 
$\sigma_{j+2} =  (\mu_{j-2,2}\, \mu_{j,2})\sigma_{j-2}$,
which explains the parity-based 
grouping of~(\ref{alpha_jODD=Defn}, \ref{alpha_jEVEN=Defn}).
Of course, the hard part is to show that none of
these terms vanish. Note, in particular, that the vector
$$
{\mathbf 1} \pi^T 
           \begin{pmatrix}
\,\,\,\,
    \gamma {\mathbf u} \\
-   \gamma {\mathbf u} 
           \end{pmatrix} 
$$
comes frighteningly close to vanishing.
A little bit of symmetry
in the wrong place is enough to derail the spectral shift.
A uniform stationary distribution, for example,
would destroy the entire scheme;
so would a vector ${\mathbf u}$ with the same
first and last coordinates. 

\subsection{Integrity Analysis}

We saw in Section~\ref{TakeOff}
that the flocks of size $2$ and $4$
remain single paths during their lifetimes.
The following result establishes the integrity
of all the flocks. Though not stated explicitly,
the result also asserts that the 
birds ${\mathcal B}_1,\ldots, {\mathcal B}_{n}$
always appear in that order from left to right.

\begin{theorem}\label{DistB-LB-UB}
Any two adjacent birds
within the same flock lie at a distance
between $0.58$ and~$1$.
This holds over the entire lifetime of the flock,
whether it flips or not.
\end{theorem}
\proof
As is sometimes the case, it is simpler to prove 
a more complicated bound, from which the theorem follows.
For notational convenience, put
${\mathbf m}_{a_0}= \frac{1}{4}n^{-5}$ and 
define $h(i)$ as the height of the nearest common ancestor
of the two leaves associated with 
${\mathcal B}_i$ and ${\mathcal B}_{i+1}$; 
eg, $h(1)=1$ and $h(2)=2$.
We prove by induction on $j$ that, for any $1\leq j< \log n$, 
$t_j\leq t \leq t_{j+1}$, and $1\leq i< 2^j$,
\begin{multline}\label{BoundDist-rephrased}
1 - \hbox{$\frac{5}{3}$} 
(n^5+jn^{4}) |{\mathbf m}_{a_{h(i)-1}}| \leq
\text{\sc dist}_{t}({\mathcal B}_i, {\mathcal B}_{i+1})
\\ 
\leq 
\begin{cases}
1 & \text{if $i=2^{j-1}$ and $t=t_j$};\ \ \ \  \\
1 - \hbox{$\frac{1}{4}$} (1 - \hbox{$\frac{j}{n}$}) 
|{\mathbf m}_{a_{h(i)-1}}|
& \text{else}.
\end{cases}
\end{multline}
Recall that $a_0$, $a_1$, etc, constitute the left spine
of the fusion tree~$\mathcal T$.
By~(\ref{sigma_vjn^6}),
the upper and lower bounds above
fall between $0.58$ and $1$, so satisfying them
implies the integrity of the flocks along
the spine: indeed, the upper bound ensures the
existence of the desired edges, 
while the lower bound greater than $\frac{1}{2}$
rules out edges between nonconsecutive birds.
To extend this to all the flocks, and hence
prove the theorem, we establish~(\ref{BoundDist-rephrased})
for {\em nondeterministic} flipping, ie, assuming that 
any node may or may not flip regardless of what the 
true flipping rule dictates. The issue here is that
the left spine does not represent {\em all} flocks:
reversing velocities changes 
the positions of birds irreversibly, so technically
we should prove~(\ref{BoundDist-rephrased}) not just
along the left spine but along {\em any} path of~$\mathcal T$.
We can do this all at once by considering
both cases, flip and no-flip, at each node $a_j$.

We proceed by induction on $j$.
Before we get on with the proof, we should explain why
the upper bound of~(\ref{BoundDist-rephrased}) 
distinguishes between two cases.
In general, once two consecutive birds are joined in a flock,
they stay forever at a distance strictly less than 1.
There is only one exception to this rule: at the time $t$ when
they join, the only assurance we can give is that
their distance does not exceed 1; it could actually
be equal to 1, hence the difficulty of a nontrivial
upper bound when $t=t_j$ and $i=2^{j-1}$.
The case $j=1$ is special because
two-bird flocks never flip but are provided
with two different kinds of initial velocities;
therefore, we must check both $({\mathcal B}_1, {\mathcal B}_2)$
and $({\mathcal B}_3, {\mathcal B}_4)$.
We verify~(\ref{BoundDist-rephrased}) directly
from~(\ref{x12(t2)}, \ref{x34(t2)}). Indeed, 
for $0\leq t\leq t_2$,
\begin{equation*}
\hbox{$\frac{2}{3}$}- n^{-c}
\leq x_2(t)- x_1(t) \leq 
x_4(t)- x_3(t) \leq 
\hbox{$\frac{2}{3}$}+ n^{-c}.
\end{equation*}
Assume now that $j\geq 2$.
By applying successively~(\ref{|Q_jz|}, \ref{Deltavj-l2}), 
Lemma~\ref{time-to-join-j}, and~(\ref{sigma_vjn^6}),
we find that 
\begin{equation*}
\begin{split}
\|Q_{j-1}^{\theta_{j-1}}\, v^{a_{j-1}}\|_2 
&\leq e^{j -\Omega(\theta_{j-1} 4^{1-j})} 
\|v^{a_{j-1}}\|_2 
\leq e^{-\Omega(n^{-2}/|{\mathbf m}_{a_{j-1}}|)} \\
&\leq  e^{-2n - \Omega(n^{-2}/|{\mathbf m}_{a_{j-1}}|)}
< |{\mathbf m}_{a_{j-1}}| e^{-2n}.
\end{split}
\end{equation*}
By~(\ref{Deltavj}),
$$
v^{a_j} 
=
\pm        \begin{pmatrix}
\,\,\,\,
       P_{j-1}^{\theta_{j-1}}\, v^{a_{j-1}} \\
     - P_{j-1}^{\theta_{j-1}}\, v^{a_{j-1}}
           \end{pmatrix}
=
|{\mathbf m}_{a_{j-1}}|
           \begin{pmatrix}
\,\,\,\, 1 \\
     -1    \end{pmatrix} 
\otimes {\mathbf 1}_{\!2^{j-1}}
\pm         \begin{pmatrix}
\,\,\,\,
       Q_{j-1}^{\theta_{j-1}}\, v^{a_{j-1}} \\
     - Q_{j-1}^{\theta_{j-1}}\, v^{a_{j-1}}
           \end{pmatrix}.
$$
The $\pm$ leaves open the possibility of
a flip of either type, right or left, before
the $2^{j-1}$-bird flocks join at time $t_j$.
As we saw earlier, the choice of type ensures that 
the flock with the lower-indexed birds  
drifts to the right while its sibling, with the higher-indexed birds,
flies to the left; hence the certainty
that, after flipping, the ``fixed'' part of the velocity
vector $v^{a_j}$ is of the form
$|{\mathbf m}_{a_{j-1}}| (1,-1)^T \otimes {\mathbf 1}_{\!2^{j-1}}$.
(In fact, to achieve just this is the sole purpose
of flipping.)
It follows that 
\begin{equation}\label{Deltav_j+delta}
v^{a_j} 
=
|{\mathbf m}_{a_{j-1}}|
           \begin{pmatrix}
\,\,\,\, 1 \\
     -1    \end{pmatrix} 
\otimes {\mathbf 1}_{\!2^{j-1}}
+ \zeta,
\ \ \ \ \ 
\text{with}\ \ \|\zeta\|_2 < 
|{\mathbf m}_{a_{j-1}}| e^{-n} \,.
\end{equation}
For $1\leq i< 2^j$, define
$$\chi_i = 
(\,\underset{2^j}{\underbrace{
        \overset{i}{\overbrace{
0,\ldots,0,-1}},1,0,\ldots, 0}}\,)^T.$$
By~(\ref{PjQjMuj}), for $s\geq 1$,
$$\chi_i^T  P_{j}^{s}\, v^{a_j}
= 
{\mathbf m}_{a_{j}} \chi_i^T 
{\mathbf 1}_{\!2^{j}}
+ \chi_i^T  Q_{j}^{s}\, v^{a_j}
= \chi_i^T  Q_{j}^{s}\, v^{a_j};$$
hence, for $t_j<t\leq t_{j+1}$,
\begin{equation}\label{DistDiffPj}
\text{\sc dist}_{t}({\mathcal B}_i, {\mathcal B}_{i+1})
= \text{\sc dist}_{t_j}({\mathcal B}_i, {\mathcal B}_{i+1})
+ \sum_{s=1}^{t-t_j} (-1)^{f(s)} \chi_i^T  Q_{j}^{s}\, v^{a_j},
\end{equation}
where $f(s)=1$ if there is a flip and $s>n^f$, and $f(s)=0$ otherwise.
Note that there is no risk in using
$\text{\sc dist}_{t}({\mathcal B}_i, {\mathcal B}_{i+1})$,
instead of the signed version, $x_{i+1}(t)-x_i(t)$,
that birds might cross unnoticed: indeed, 
the bound in~(\ref{Deltavj-l2})
applies to all the velocities, so that 
distances cannot change by more than $O(n^{1-c})$ in
one step. This implies that a change of sign
for $x_{i+1}(t)-x_i(t)$ would be preceded by 
the drop of 
$\text{\sc dist}_{t}({\mathcal B}_i, {\mathcal B}_{i+1})$
below $\frac{1}{2}$ and a violation of~(\ref{BoundDist-rephrased}).
By Cauchy-Schwarz and~(\ref{|Q_jz|}, \ref{Deltav_j+delta}),
$$
|\chi_i^T  Q_{j}^{s} \, \zeta|
\leq 
\sqrt{2}\, \|Q_{j}^{s} \, \zeta\|_2 
\leq 
\sqrt{2} \, e^{j+1-\Omega(s4^{-j})} \|\zeta\|_2
\leq n^2 e^{-n-\Omega(s/n^2)} |{\mathbf m}_{a_{j-1}}|;$$
and, since $n$ is assumed large enough,
for $s\geq 1$,
\begin{equation}\label{chiQzetaUB}
|\chi_i^T  Q_{j}^{s} \zeta|< 
e^{-\frac{1}{2}n -sn^{-3}} |{\mathbf m}_{a_{j-1}}|.
\end{equation}
Likewise,
\begin{equation*}
\begin{split}
|\chi_i^T  Q_{j}^{s}\, v^{a_j}| 
&\leq 
\sqrt{2}\,\|Q_{j}^{s}\, v^{a_j}\|_2  
\leq  n^{1.45} e^{-\Omega(s/n^2)} \|v^{a_j}\|_2  \\
&\leq n^{1.45} e^{-\Omega(s/n^2)} (|{\mathbf m}_{a_{j-1}}|\sqrt{n}
+ \|\zeta\|_2).
\end{split}
\end{equation*}
For $s\geq 1$ and $1\leq i< 2^j$, by~(\ref{Deltav_j+delta}),
\begin{equation}\label{chiQDeltaUB}
|\chi_i^T  Q_{j}^{s}\, v^{a_j}| 
\leq n^{2} |{\mathbf m}_{a_{j-1}}| e^{-\Omega(s/n^2)}.
\end{equation}
Recall that $j\geq 2$. 
To prove~(\ref{BoundDist-rephrased}),
we distinguish between two cases:
whether the birds ${\mathcal B}_i, {\mathcal B}_{i+1}$
are joined at node $a_j$ or earlier.

\begin{figure}[htb]
\vspace{2cm}
\begin{center}
\includegraphics[width=7cm]{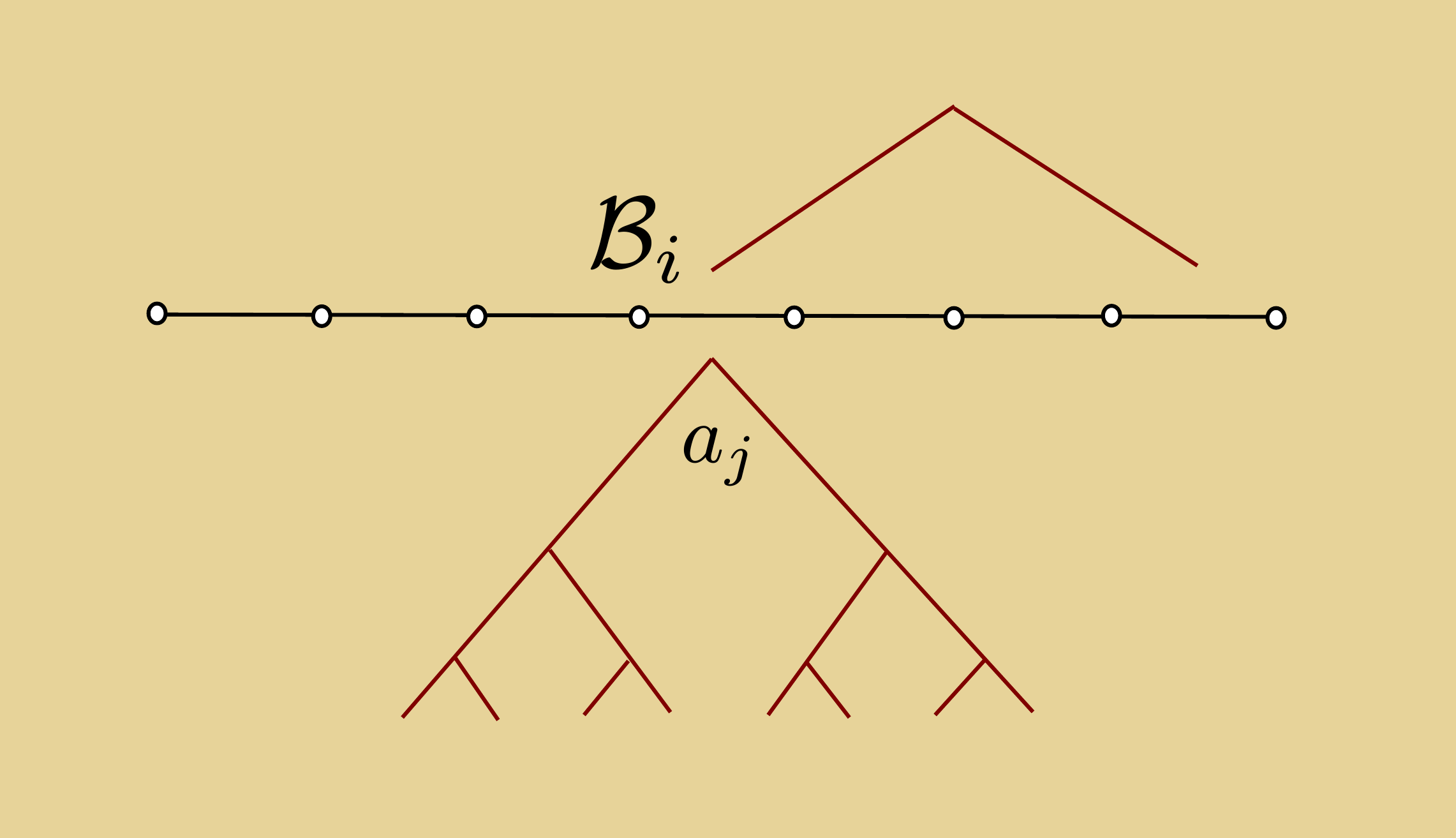}
\end{center}
\begin{quote}
\vspace{0cm}
\caption{\small 
The birds ${\mathcal B}_i$ and ${\mathcal B}_{i+1}$
are joined together at time $t_j$.
}
\label{fig-bijoin1}
\end{quote}
\end{figure}

\bigskip\bigskip
\noindent
{\bf Case I.} ($i=2^{j-1}$): \ \ The edge $(i,i+1)$ is created at node $a_j$ and 
$h(i)=j$, where $2\leq j< \log n$ (Figure~\ref{fig-bijoin1}).
We begin with the case $t=t_j$.
By construction, the upper bound in~(\ref{BoundDist-rephrased})
is equal to 1. To establish the lower bound, 
we observe that at time $t_{j}-1$
the two middle birds were more than one unit
of distance apart. By the expression of
the velocity given in~(\ref{Deltav_j+delta}),
which expresses the displacement
prior to $t_j$, neither bird moved by more than
$(1+ e^{-n})|{\mathbf m}_{a_{j-1}}|$
in that one step; therefore,
\begin{equation}\label{LB-dist-i-midedge}
\text{\sc dist}_{t_j}({\mathcal B}_i, {\mathcal B}_{i+1})
> 1- 3|{\mathbf m}_{a_{j-1}}|,
\end{equation}
which exceeds the lower bound of~(\ref{BoundDist-rephrased}), ie,
$1 - \hbox{$\frac{5}{3}$} 
(n^5+jn^{4}) |{\mathbf m}_{a_{h(i)-1}}|$.
Assume now that $t_j<t\leq t_{j+1}$.
Observe that 
$$ 
{\mathbf 1}_{\!2^{j}} \,
(\, \overset{2^j}{\overbrace{
\hbox{$\frac{1}{2}$},1,\ldots,1,\hbox{$\frac{1}{2}$}}}\,)
\Bigl\{
           \begin{pmatrix}
\,\,\,\, 1 \\
     -1    \end{pmatrix} 
\otimes {\mathbf 1}_{\!2^{j-1}}
\Bigr\}
= 0.
$$

\noindent
The $i$-th row of $P_{j}$
is the same as the $(2^j+1-i)$-th row read backwards.
This type of symmetry is closed under multiplication, 
so it is also true of $P_{j}^{s}$.
By~(\ref{PjQjMuj}), for any $s\geq 0$, it then follows that
$$
Q_{j}^{s} 
\Bigl\{
           \begin{pmatrix}
\,\,\,\, 1 \\
     -1    \end{pmatrix} 
\otimes {\mathbf 1}_{\!2^{j-1}}
\Bigr\}
= 
P_{j}^{s} 
\Bigl\{
           \begin{pmatrix}
\,\,\,\, 1 \\
     -1    \end{pmatrix} 
\otimes {\mathbf 1}_{\!2^{j-1}}
\Bigr\}
= 
\Bigl( b_1^{(s)},\ldots, b_{2^{j-1}}^{(s)},-
          b_{2^{j-1}}^{(s)},\ldots,-b_1^{(s)}\Bigr)^T.
$$
The following recurrence relation holds: 
$b_i^{(0)} = 1$ if $1\leq i\leq 2^{j-1}$,
and $b_i^{(0)} = -1$ else.
For $s\geq 0$, we get the identities below for $l\leq 2^{j-1}$,
plus an antisymmetric set for $l>2^{j-1}$:
\begin{equation*}
b_l^{(s+1)} = 
\frac{1}{3}
\begin{cases}
\ \ \  b_1^{(s)} + 2 b_2^{(s)}& \ \text{if $l=1$}; \\
\ \ \ b_{l-1}^{(s)} + b_l^{(s)} + b_{l+1}^{(s)}
& \ \text{if $1<l<2^{j-1}$}; \\
\ \ \ b_{2^{j-1}-1}^{(s)} & \ \text{if $l=2^{j-1}$}; \\
\ -b_{2^{j-1}-1}^{(s)} & \ \text{if $l=2^{j-1}+1$}; \\
\  - b_{2^j+2-l}^{(s)} - b_{2^j+1-l}^{(s)} - b_{2^j-l}^{(s)}
&\ \text{if $2^{j-1}+1<l<2^j$}; \\
\  -b_1^{(s)} - 2 b_2^{(s)}& \ \text{if $l=2^j$}\, .
\end{cases}
\end{equation*}
We find by induction that
\begin{equation*}
b_1^{(s)} \geq \cdots \geq b_{2^{j-1}}^{(s)} \geq 3^{-s}; 
\end{equation*}
therefore,
\begin{equation}\label{negativeChiQ-3^s}
\chi_{2^{j-1}}^T Q_{j}^{s} 
\Bigl\{    \begin{pmatrix}
\,\,\,\, 1 \\
     -1    \end{pmatrix} 
\otimes {\mathbf 1}_{\!2^{j-1}}
\Bigr\} 
= -2b_{2^{j-1}}^{(s)} < -3^{-s} \, .
\end{equation}
Since the two middle birds in the flock $F_{a_j}$
get attached in the flocking network
at time $t_j$, 
$\text{\sc dist}_{t_j}({\mathcal B}_i, {\mathcal B}_{i+1})\leq 1$.
Assume that $F_{a_j}$ does not undergo a flip.
Then, 
by~(\ref{Deltav_j+delta}, \ref{DistDiffPj}, \ref{chiQzetaUB}),
for $t_j< t\leq t_{j+1}$,
\begin{equation*}
\begin{split}
\text{\sc dist}_{t}({\mathcal B}_i, {\mathcal B}_{i+1})
&\leq 1 + 
\sum_{s=1}^{t-t_j} \chi_i^T  Q_{j}^{s}\, v^{a_j} \\
&\leq 1 + |{\mathbf m}_{a_{j-1}}|
\sum_{s=1}^{t-t_j}
\chi_{2^{j-1}}^T Q_{j}^{s} 
\Bigl\{    \begin{pmatrix}
\,\,\,\, 1 \\
     -1    \end{pmatrix} 
\otimes {\mathbf 1}_{\!2^{j-1}}
\Bigr\}
+ \sum_{s=1}^{t-t_j}
\chi_{2^{j-1}}^T Q_{j}^{s} \, \zeta \\
&\leq 
1- \hbox{$\frac{1}{3}$}|{\mathbf m}_{a_{j-1}}|
+ \sum_{s\geq 1}
|\chi_{2^{j-1}}^T Q_{j}^{s} \, \zeta|   \\
&\leq 
1- \hbox{$\frac{1}{3}$}|{\mathbf m}_{a_{j-1}}|
+ |{\mathbf m}_{a_{j-1}}|
\sum_{s\geq 1} e^{-\frac{1}{2}n -sn^{-3}} \\
&<
1- \hbox{$\frac{1}{3}$}(1-o(1))|{\mathbf m}_{a_{j-1}}|
= 1- \hbox{$\frac{1}{3}$}(1-o(1))|{\mathbf m}_{a_{h(i)-1}}|,
\end{split}
\end{equation*}
which proves the upper bound in~(\ref{BoundDist-rephrased})
for $i=2^{j-1}$.
The negative geometric series we obtain 
from~(\ref{negativeChiQ-3^s}) reflects the
``momentum'' (minus the vibrations) of the two flocks colliding and penetrating
into each other's zone of influence
before being stabilized.

Suppose now that the flock $F_{a_j}$ undergoes
a flip at time $t_j+n^f$. The previous analysis
holds for $t_j < t\leq t_j+n^f$; so assume that 
$t_j+n^f< t\leq t_{j+1}$. 
By~(\ref{chiQDeltaUB}) and $h(i)=j$,
$$
\sum_{s=1}^{t-t_j-n^f} |\chi_i^T  Q_{j}^{s+n^f}\, v^{a_j}| 
\leq 
\sum_{s=1}^{t-t_j-n^f} 
 n^{2} |{\mathbf m}_{a_{h(i)-1}}| e^{-\Omega(sn^{-2}+n^{f-2})} 
= o( |{\mathbf m}_{a_{h(i)-1}}|).
$$
By~(\ref{DistDiffPj}), therefore,
\begin{equation*}
\begin{split}
\text{\sc dist}_{t}({\mathcal B}_i, {\mathcal B}_{i+1})
&= \text{\sc dist}_{t_j}({\mathcal B}_i, {\mathcal B}_{i+1})
+ \sum_{s=1}^{n^f} \chi_i^T  Q_{j}^{s}\, v^{a_j}
- \sum_{s=n^f+1}^{t-t_j} \chi_i^T  Q_{j}^{s}\, v^{a_j} \\
&\leq \text{\sc dist}_{t_j+n^f}({\mathcal B}_i, {\mathcal B}_{i+1})
+ \sum_{s=1}^{t-t_j-n^f} |\chi_i^T  Q_{j}^{s+n^f}\, v^{a_j}| \\
&< 1- \hbox{$\frac{1}{3}$}(1-o(1))|{\mathbf m}_{a_{h(i)-1}}|
+ o( |{\mathbf m}_{a_{h(i)-1}}| )
< 1- \hbox{$\frac{1}{4}$}|{\mathbf m}_{a_{h(i)-1}}|.
\end{split}
\end{equation*}
This establishes
the upper bound in~(\ref{BoundDist-rephrased})
for $i= 2^{j-1}$, whether there is a flip or not.
We prove the lower bound as follows.
By~(\ref{DistDiffPj}, \ref{chiQDeltaUB}, \ref{LB-dist-i-midedge}),
for $t_j< t\leq t_{j+1}$,
\begin{equation*}
\begin{split}
\text{\sc dist}_{t}({\mathcal B}_i, {\mathcal B}_{i+1})
&\geq 
1- 3|{\mathbf m}_{a_{j-1}}|
- \sum_{s=1}^{t-t_j} |\chi_i^T  Q_{j}^{s}\, v^{a_j}|  \\
&\geq 
1- 3|{\mathbf m}_{a_{j-1}}|
- n^{2} |{\mathbf m}_{a_{j-1}}| 
\sum_{s\geq 1} e^{-\Omega(s/n^2)}  \\
&\geq 
1- n^5 |{\mathbf m}_{a_{j-1}}| = 1- n^5 |{\mathbf m}_{a_{h(i)-1}}|.
\end{split}
\end{equation*}
Note that this derivation still holds if 
the flock ``flips,'' ie, 
reverses the sign of $Q_{j}^{s}\, v^{a_j}$.
This establishes~(\ref{BoundDist-rephrased})
for $i=2^{j-1}$.

\begin{figure}[htb]
\vspace{2cm}
\begin{center}
\includegraphics[width=7cm]{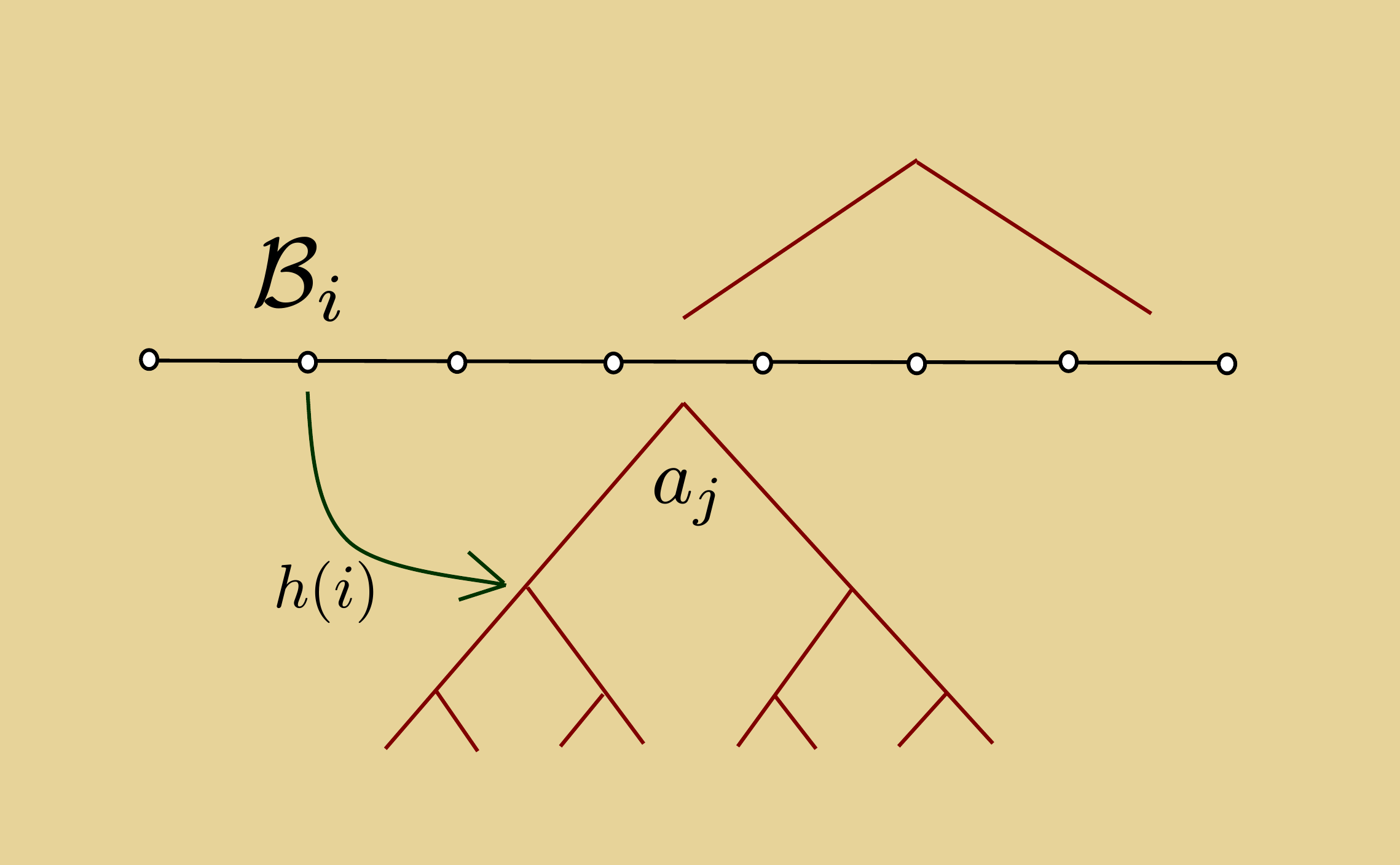}
\end{center}
\begin{quote}
\vspace{0cm}
\caption{\small 
The birds ${\mathcal B}_i$ and ${\mathcal B}_{i+1}$
are joined earlier than $t_j$. 
}
\label{fig-bijoin2}
\end{quote}
\end{figure}

\bigskip\bigskip
\noindent
{\bf Case II.} ($i<2^{j-1}$): \ \ This implies that $h(i)<j$
(Figure~\ref{fig-bijoin2}). Recall that $j\geq 2$.
We omit the case $i>2^{j-1}$, which is treated similarly.
The case $t=t_j$ follows by induction\footnote{If the reader 
is wondering why our induction invariant is defined over 
the interval $[t_j, t_{j+1}]$ and not
$(t_j, t_{j+1}]$, the benefit is a shorter presentation.}
for $j'=j-1$ and $t=t_{j'+1}$.
Note that $t\neq t_{j-1}$, so the inductive
use of~(\ref{BoundDist-rephrased})
does not provide 1 as an upper bound;
furthermore it provides even stronger bounds, as $j'<j$.
We assume now that $t_j< t\leq t_{j+1}$.
By~(\ref{DistDiffPj}, \ref{chiQDeltaUB}),
\begin{equation*}
\begin{split}
|\text{\sc dist}_{t}({\mathcal B}_i, {\mathcal B}_{i+1})
- \text{\sc dist}_{t_j}({\mathcal B}_i, {\mathcal B}_{i+1})|
&\leq 
\sum_{s\geq 1} |\chi_i^T  Q_{j}^{s}\, v^{a_j}| \\
&\leq n^{2} |{\mathbf m}_{a_{j-1}}| 
\sum_{s\geq 1} e^{-\Omega(s/n^2)}
\leq O( n^{4} |{\mathbf m}_{a_{j-1}}|). 
\end{split}
\end{equation*}

\noindent
We apply~(\ref{BoundDist-rephrased}) 
inductively once more for $j'=j-1$ and $t=t_{j'+1}$:
$$
1 - \hbox{$\frac{5}{3}$} 
(n^5+(j-1)n^{4}) |{\mathbf m}_{a_{h(i)-1}}| \leq
\text{\sc dist}_{t_j}({\mathcal B}_i, {\mathcal B}_{i+1})
\leq 
1 - \hbox{$\frac{1}{4}$} (1 - \hbox{$\frac{1}{n}$} 
(j-1)) |{\mathbf m}_{a_{h(i)-1}}|;
$$
hence, for $t_j < t\leq t_{j+1}$,
\begin{multline*}
1 - \hbox{$\frac{5}{3}$} 
(n^5+(j-1)n^{4}) |{\mathbf m}_{a_{h(i)-1}}| 
- O( n^{4} |{\mathbf m}_{a_{j-1}}|)
\leq
\text{\sc dist}_{t}({\mathcal B}_i, {\mathcal B}_{i+1})
\leq 
\\
1 - \hbox{$\frac{1}{4}$} (1 - \hbox{$\frac{1}{n}$} 
(j-1)) |{\mathbf m}_{a_{h(i)-1}}|
+ O( n^{4} |{\mathbf m}_{a_{j-1}}|).
\end{multline*}
Because $j>h(i)$, by~(\ref{sigma_vjn^6}), 
$|{\mathbf m}_{a_{j-1}}|<  n^{-c} |{\mathbf m}_{a_{h(i)-1}}|$,
for $h(i)>1$. In the case $h(i)=1$,
$$|{\mathbf m}_{a_{j-1}}|\leq |{\mathbf m}_{a_{1}}|< n^{-c} \leq 
4 n^{-6}|{\mathbf m}_{a_0}|= n^{-11},$$
for $c\geq 11$. This shows that, in all cases, 
$|{\mathbf m}_{a_{j-1}}|<  4 n^{-6} |{\mathbf m}_{a_{h(i)-1}}|$;
hence~(\ref{BoundDist-rephrased}).
Since sums involving velocities are immediately taken with
absolute values, the same derivation can be repeated
verbatim in the case of a flip.
\hfill $\Box$
\proofend

\section{Concluding Remarks}

We have established the first general convergence
bound for a standard neighbor-based flocking model.
We believe that it can be generalized to 
many of the metric and topological variants of
the Vicsek model. We have shown that
the spectral shift underpinning the slow convergence
is resistant to noise decaying with time. 
Without temporal decay, injecting a fixed amount of entropy
into the system at each step is likely to produce widely
different behaviors. Whether the techniques introduced 
in this work, in particular the geometric approach,
can shed light on phase transitions reported
experimentally in~\cite{chateGGR, vicsekCBCS95} is a fascinating
open question.

\subsection*{Acknowledgments}
I wish to thank 
Iain Couzin,
Joel Friedman, 
Phil Holmes,
Ali Jadbabaie,
and Naomi Leonard 
for helpful discussions.

\end{document}